\documentclass[botnum]{unmeethesis}
\pdfoutput=1

\title{Quantum Algorithms, Architecture, and Error Correction}
\author{Ciar\'{a}n Ryan-Anderson}
\date{December, 2018} 

\degreesubject{Ph.D., Physics}
\degree{Doctor of Philosophy \\ Physics}
\documenttype{Dissertation}
\previousdegrees{B.S., Physics, Missouri University of Science and Technology, 2009 \\
	M.S., Physics, Missouri University of Science and Technology, 2011}

\newcommand*{\figs}{figs}%
\newcommand*{\SurgeryFigs}{\figs/color_surgery}%
\newcommand*{\SimFigs}{\figs/stab_sim}%

\usepackage{caption}
\usepackage{subcaption}
\usepackage{graphicx}
\usepackage[bottom]{footmisc}

\usepackage{float}
\usepackage{pdfpages}
\usepackage{physics}
\usepackage{qcircuit}

\usepackage{amsmath}                          
\usepackage{amsfonts}                         
\usepackage{amssymb}                       
\usepackage{amscd}                            
\usepackage{amsthm}                          
\usepackage{mathrsfs}

\usepackage{dsfont}

\usepackage{enumerate}

\newcommand{\Tau}{\mathcal{T}}
\newcommand{\Cbold}{\mathds{C}}

\newcommand{\Rbold}{\mathds{R}}
\newcommand{\Hb}{{\cal H}}
\newcommand\CNOT{\ensuremath{\textit{CNOT\/}}}

\newcommand\SWAP{\ensuremath{\textit{SWAP\/}}}

\newcommand{\bigO}{{\cal O}}

\newcommand{\etal}{\textit{et~al.}}

\newcommand{\ie}{\textit{i.e.}}

\newcommand{\eg}{\textit{e.g.}}

\newcommand{\tens}[1]{%
  \mathbin{\mathop{\otimes}\limits_{#1}}%
}

\newtheorem{axiom}{Axiom}
\newtheorem{theorem}{Theorem}[chapter]
\newtheorem{lemma}{Lemma}[chapter]
\newtheorem{corollary}{Corollary}[theorem]

\usepackage[final]{listings}
\renewcommand{\lstlistingname}{Code Block}
\renewcommand{\lstlistlistingname}{List of \lstlistingname s}

\usepackage{xcolor}

\definecolor{DarkGray}{gray}{0.15}
\definecolor{LightGray}{gray}{0.80}
\definecolor{MediumGray}{gray}{0.60}
\definecolor{dkgreen}{rgb}{0,0.6,0}
\definecolor{gray}{rgb}{0.5, 0.5, 0.5}
\definecolor{mauve}{rgb}{0.58,0,0.82}
\definecolor{limegreen}{HTML}{8CC739}
\definecolor{skyblue}{HTML}{06A2CB}
\definecolor{Myorange}{HTML}{CC7832}
\definecolor{Mypurple}{HTML}{B10088}
\definecolor{emphcolor}{HTML}{E84A5F}
\definecolor{codehl}{HTML}{ECF0F3}

\usepackage{soul}

\lstdefinestyle{psedostyle}{
	language=Python,
	xleftmargin=5.0ex,
	moredelim = [s][\textit]{[}{]},
	basicstyle={\small\ttfamily},
	captionpos=b,
	columns=flexible,
 	numbers=left,
	stepnumber=1,
	numberstyle=\tiny\color{gray},
	escapechar=\%,
	breaklines=true,
	frame=single,
	tabsize=2,
	postbreak=\mbox{\textcolor{red}{$\hookrightarrow$}\space},
	captionpos=b,
	showspaces=false,
	showtabs=false,
}

\lstdefinestyle{pystyle}{
	language=Python,
	backgroundcolor=\color{DarkGray},
	framexleftmargin=5pt,
	basicstyle={\small\ttfamily\color{LightGray}},
	commentstyle=\itshape\color{MediumGray},
	otherkeywords={self,^},
	morekeywords = [5]{^},
	keywordstyle=\ttfamily\color{limegreen},
	keywords=[2]{True, False},
	keywords=[3]{pc, PECOS},
	keywords=[4]{circuit_converters, circuit_runners, circuits, decoders, error_gens, misc, outputs, qeccs, state_sims, tools},
	keywordstyle={[2]\ttfamily\color{yellow!80!orange}},
	keywordstyle={[3]\ttfamily\color{skyblue}},
	keywordstyle={[4]\ttfamily\color{Mypurple}},
	keywordstyle={[5]\bfseries\normalfont\color{LightGray}},
	emph={MyClass,__init__},
	emphstyle=\ttfamily\color{emphcolor},
	stringstyle=\color{Myorange},
	columns=flexible,
 	numbers=left,
	stepnumber=1,
	numberstyle=\tiny\color{gray},
	breaklines=true,
	tabsize=2,
	postbreak=\mbox{\textcolor{red}{$\hookrightarrow$}\space},
	captionpos=b,
	showspaces=false,
	showtabs=false,
}

\lstdefinestyle{pystyle2}{
	language=Python,
	backgroundcolor=\color{DarkGray},
	xleftmargin=5.0ex,
	moredelim = [s][\textit]{[}{]},
	basicstyle={\small\ttfamily\color{LightGray}},
	commentstyle=\itshape\color{MediumGray},
	otherkeywords={self,^},
	morekeywords = [5]{^},
	keywordstyle=\ttfamily\color{limegreen},
	keywords=[2]{True, False},
	keywords=[3]{pc, pecos},
	keywords=[4]{circuit_converters, circuit_runners, circuits, decoders, error_gens, misc, outputs, qeccs, state_sims, tools},
	keywordstyle={[2]\ttfamily\color{yellow!80!orange}},
	keywordstyle={[3]\ttfamily\color{skyblue}},
	keywordstyle={[4]\ttfamily\color{Mypurple}},
	keywordstyle={[5]\bfseries\normalfont\color{LightGray}},
	emph={MyClass,__init__},
	emphstyle=\ttfamily\color{emphcolor},
	stringstyle=\color{Myorange},
	captionpos=b,
	columns=flexible,
 	numbers=left,
	stepnumber=1,
	numberstyle=\tiny\color{gray},
	escapechar=\%,
	breaklines=true,
	frame=single,
	tabsize=2,
	captionpos=b,
	showspaces=false,
	showtabs=false,
}

\usepackage{tcolorbox}
\tcbuselibrary{listings,breakable}

\newtcblisting{mylisting}{
      arc=0pt,
      top=0mm,
      bottom=0mm,
      left=0mm,
      right=0mm,
      colback=white,
      breakable,
      boxrule=1pt,
      listing only,
      listing options={xleftmargin=1.5em,framexleftmargin=1em,framextopmargin=0.5ex}
}

\DeclareRobustCommand{\textcode}[1]{{\sethlcolor{codehl}\hl{\texttt{#1}}}}
\DeclareRobustCommand{\code}[1]{{\sethlcolor{codehl}\hl{\texttt{#1}}}}
\DeclareRobustCommand{\pack}[1]{\texttt{#1}\xspace}
\newcommand{\PECOS}{\pack{PECOS}}

\usepackage{hyperref}
\hypersetup{
    colorlinks=true,
    citecolor=blue,
    filecolor=blue,
    linkcolor=blue,
	urlcolor=blue,
}

\newcounter{tmp}

\usepackage{cleveref}

\usepackage{epigraph}
\usepackage{etoolbox}
\usepackage[backend=bibtex8, style=alphabetic, hyperref]{biblatex}

\usepackage{xspace}

\newcommand{\ptuple}{\textcode{tuple}\xspace}
\newcommand{\plist}{\textcode{list}\xspace}
\newcommand{\pdict}{\textcode{dict}\xspace}

\newcommand{\pappend}{\textcode{append}\xspace}
\newcommand{\pupdate}{\textcode{update}\xspace}

\newcommand{\QuantumCircuit}{\textcode{Quantum\-Circuit}\xspace}

\usepackage{cellspace}
\renewcommand{\arraystretch}{1.5}

\usepackage{array}
\newcolumntype{C}[1]{>{\centering\let\newline\\\arraybackslash\hspace{0pt}}m{#1}}

\begin{document}

\makeatletter

\makeatother

\frontmatter

\includepdf[pages={1}]{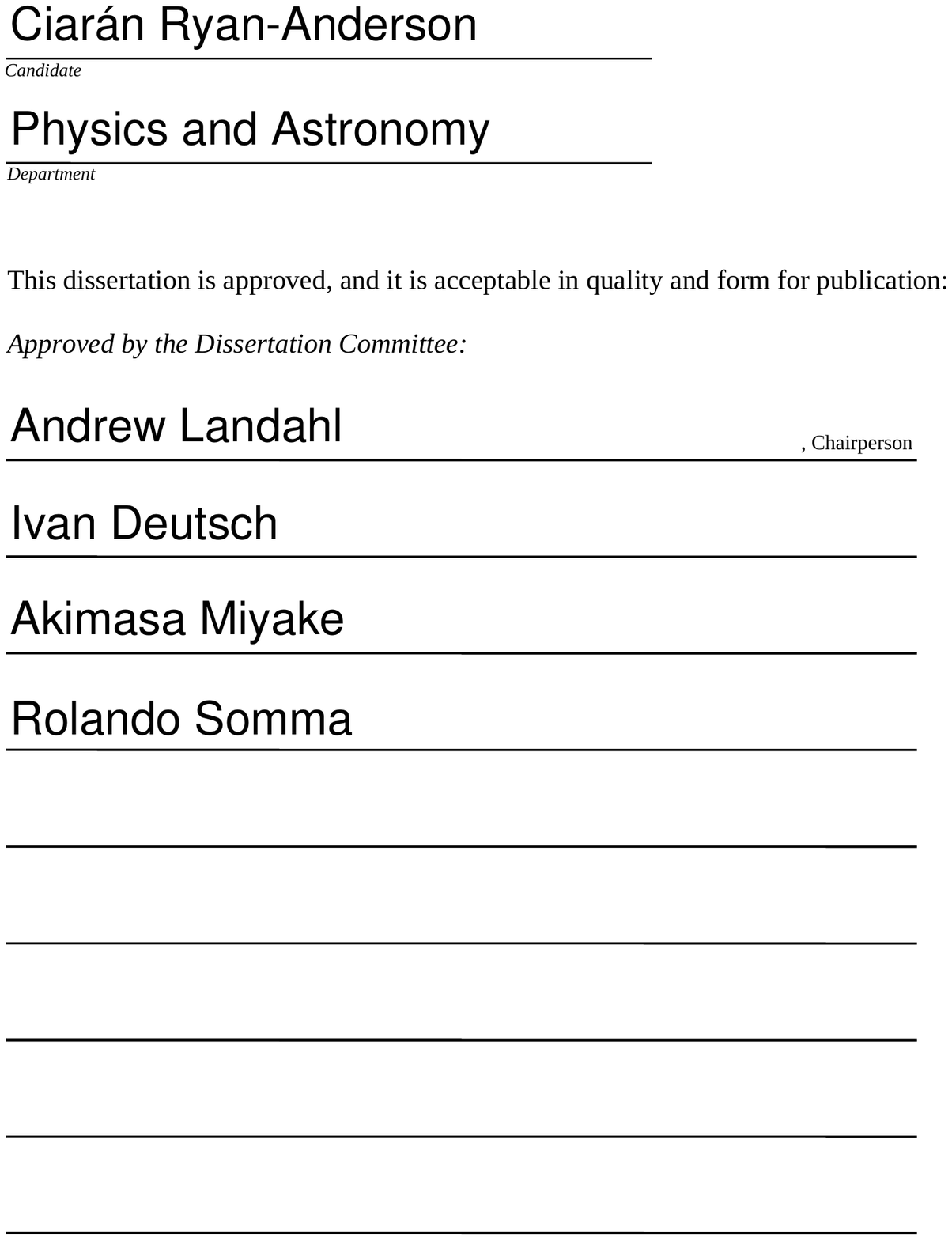}


\maketitle

\setcounter{page}{3}

\begin{dedication}
To my friends, family, and all who made this possible.
\end{dedication}

\begin{acknowledgments}
   \vspace{0.4in}
First and foremost, I would like to thank my advisor Andrew Landahl for his guidance and insight. He is an outstanding example that creativity, imagination, and passion are crucial for a successful scientist.

I am grateful to professors Carl Caves, Ivan Deutsch, Poul Jessen, Akimasa Miyake, and Elohim Becerra for providing the environment of CQuIC for students, like myself, to learn and grow as scientists. I would also like to thank my friends and fellow graduate students Lewis Chiang, Matt Curry, and Jaimie Stephens for many amusing discussions and adventures.

My time at Sandia National Laboratories has been an invaluable opportunity to learn from and work alongside many talented and diverse scientists. I would like to thank Ojas Parekh, Jonathan Moussa, Kenny Rudinger, Setso Metodi, Toby Jacobson, Anand Ganti, Uzoma Onunkwo, and many others at Sandia for this opportunity.

Finally, I would like to thank my parents Mairead Ryan-Anderson and Kevin Anderson, my brother Diarmait Ryan-Anderson, and my friend Gena Robertson for their continual encouragement and support.

The work in this dissertation was supported in part by the Laboratory Directed Research and Development program at Sandia National Laboratories. Sandia National Laboratories is a multimission laboratory managed and operated by National Technology \& Engineering Solutions of Sandia, LLC, a wholly owned subsidiary of Honeywell International Inc., for the U.S. Department of Energy’s National Nuclear Security Administration under contract \text{DE-NA0003525}.

This dissertation describes objective technical results and analysis. Any subjective views or opinions that might be expressed in this dissertation do not necessarily represent the views of the U.S. Department of Energy or the United States Government.
\end{acknowledgments}

\maketitleabstract

\begin{abstract}
Quantum algorithms have the potential to provide exponential speedups over some of the best known classical algorithms. These speedups may enable quantum devices to solve currently intractable problems such as those in the fields of optimization, material science, chemistry, and biology. Thus, the realization of large-scale, reliable quantum-computers will likely have a significant impact on the world. For this reason, the focus of this dissertation is on the development of quantum-computing applications and robust, scalable quantum-architectures. I begin by presenting an overview of the language of quantum computation. I then, in joint work with Ojas Parekh, analyze the performance of the quantum approximate optimization algorithm (QAOA) on a graph problem called Max Cut. Next, I present a new stabilizer simulation algorithm that gives improved runtime performance for topological stabilizer codes. After that, in joint work with Andrew Landahl, I present a new set of procedures for performing logical operations called ``color-code lattice-surgery.'' Finally, I describe a software package I developed for studying, developing, and evaluating quantum error-correcting codes under realistic noise.
\clearpage
\end{abstract}

\tableofcontents

\listoffigures
\listoftables
\lstlistoflistings

\mainmatter

\addcontentsline{toc}{chapter}{\lstlistlistingname}

\chapter{Introduction}
\chaptermark{TESTESTESTSET}

\setlength\epigraphwidth{5cm}
\epigraph{\textit{``DON'T PANIC''}}{--- \textup{Douglas Adams \cite{adams1995hitchhikers}}}

Quantum computation (QC) is the application of quantum mechanics for the purpose of computing. Arguably, the origins of QC began in the early 1980s. In 1980 (article received in June 1979), Paul Benioff described how to simulate Turing machines with quantum Hamiltonians \cite{Benioff1980}. At the ``First Conference of Physics and Computation'' (1981), Richard Feynman gave the keynote speech titled ``Simulating Physics with Computers'' \cite{Feynman:1982a}. In this speech, Feynman argued that to efficiently simulate quantum mechanical phenomena one needs computers that employ the rules of quantum mechanics. In 1985, David Deutsch gave the first definition of a universal quantum computer \cite{Deutsch97}. Later, David Deutsch and Richard Jozsa developed the first quantum algorithm with a speedup over known classical algorithms in 1992 \cite{Deutsch553}. Then two years later, Peter Shor published his famous quantum algorithm, which gave an exponential speedup for taking discrete logarithms and performing factorization \cite{Shor:1994b,Shor:1997a}. Despite the apparent power of quantum computing, some argued that quantum computers could not be realized in practice due to the fragility of qubits \cite{Landauer:1996a,DreamNight}. To overcome these concerns, Peter Shor, Robert Calderbank, and Andrew Steane developed the first quantum error-correcting codes in 1995 \cite{Shor:1995a,CS95,Steane96,Steane96b}. In the following couple of years, fault-tolerant quantum computing was developed \cite{Shor:1996a,Aharonov:1997a,Kitaev:1997b,Steane:1997a,Aharonov:1997a,Knill:1998a,Preskill:1998a,Preskill:1998c,GotQC,Aharonov:1999a}, which not only promises that errors that arise during quantum computations maybe corrected but that guarantees, so long as errors are sufficiently innocuous, that arbitrarily many operations can be reliably performed during quantum computations.

In the following decades, the potential for quantum computers to solve significant problems spurred academics and industry to develop quantum computers and their applications \cite{PreskillNISQ}. This potential has resulted in the development of many new quantum algorithms\cite{Jordan:2011AlgoSite} and quantum error correction protocols \cite{LB13_lidar_brun_2013}. Multiple substrates for quantum computing are also being investigated, including ion traps \cite{Blatt2008,SPMIZCS09,HTMMB18}, superconductors \cite{MSS00,Wendin16,HTMMB18}, semiconductors \cite{LD997,ZDMSHKRCE12,GZ18,HTMMB18}, neutral atoms \cite{BCJD98,Jaksch:1999a,MLXSPCIS15}, nitrogen vacancies in diamonds \cite{DCJTMJZHL07,Yao:2012b,childress_hanson_2013}, and optics \cite{KMNRDM05,Barz2014}. Even numerous quantum programming languages and software have been created \cite{LaRose18,Fingerhuth18,CQC18}. 

The motivation for the work that appears in this dissertation is to contribute to the realization of large-scale, robust quantum computers and to further develop the applications of these devices. The following will be the structure of this work: In Chapter~\ref{ch.qi_intro}, I introduce the notation that I will use in this dissertation and discuss the language of quantum computation. In Chapter~\ref{chp:qaoa}, I discuss joint work with Ojas Parekh in which we show how the quantum adiabatic optimization algorithm (QAOA) \cite{Farhi:2014a} can be described as a discretization of adiabatic quantum computing. I then analyze the performance of QAOA on the classical problem Max Cut and derive a closed-form expression for the performance of QAOA on this problem. From this closed-form expression, we argue that QAOA beats the best known classical approximation algorithm for Max Cut on triangle-free $k$-regular graphs. In Chapter~\ref{ch.stab_sim}, I introduce stabilizer codes and present a new classical algorithm for simulating these codes. I provide an analysis of the typical runtime complexity of the new algorithm and find that the runtimes are improved for simulations of topological stabilizer codes compared to previous algorithms. I then show that in implementation, the practical runtime of these algorithms support the complexity analysis. In Chapter~\ref{ch.lattice_surgery}, in joint work with Andrew Landahl, I introduce a new set of logical operations called \textit{color-code lattice-surgery}. I also optimize surface-code lattice-surgery, which was initially proposed by Horsman \textit{et al.} \cite{Horsman:2012a}. In Chapter~\ref{ch.pecos}, I present a Python package I developed that provides a standardized framework for studying and evaluating quantum error correction. The package allows users to quickly represent QECCs in Python and provides a library of flexible classes for developing tools to evaluate QECCs and develop intuition. Finally, in Chapter~\ref{chp:conclusion}, I provide a summary and outlook for this dissertation.

The following is the list of the technical results that appear in the dissertation:

\begin{itemize}
\item Sevag Gharibian, Ojas Parekh, and Ciar\'{a}n Ryan-Anderson. \textit{Approximate, Constraint Satisfaction in the Quantum Setting}. Paper in preparation. Part of which is described in Chapter~\ref{chp:qaoa}. Contribution: This chapter focuses on the work contributed by me, in joint work with Ojas Parekh. My technical contributions include developing the proofs in this chapter with Ojas Parekh. I also wrote the introductory material and made final edits.

\item Ciar\'{a}n Ryan-Anderson. \textit{Improved Stabilizer-Simulation for Topological Stabilizer Codes}. Paper in preparation. Described in Chapter~\ref{ch.stab_sim}. Contribution: I developed and wrote all of this work.

\vfill
\pagebreak[4]
\item Andrew Landahl and Ciar\'{a}n Ryan-Anderson. \textit{Quantum Computing by Color-code Lattice-Surgery}, arXiv:1407.5103 \cite{colorsurgery}. Described in Chapter~\ref{ch.lattice_surgery}. Contribution: I predominately developed the lattice-surgery procedures described in this work, while Andrew Landahl largely contributed the performance analysis and writing.

\item Ciar\'{a}n Ryan-Anderson. \textit{Performance Estimator of Codes On Surfaces}. Paper in preparation. Described in Chapter~\ref{ch.pecos}. Contribution: I developed and wrote all of this work.
\end{itemize}

\chapter{\label{ch.qi_intro}A Brief Introduction to Quantum Computation}

\setlength\epigraphwidth{9cm}
\epigraph{\textit{``It's turtles all the way down!''}}{--- \textup{Often attributed to Bertrand Russell \cite{hawking1998brief}}}

This chapter presents the notation used in this dissertation by giving a brief introduction to the language of quantum computation (QC)---quantum mechanics. As the primary intent of this chapter is to present notation, the style of this chapter will be that of giving a high-level survey rather than discussing the subject in a rigorous, axiomatic fashion. Therefore, this chapter may serve as a quick overview of quantum computation for those outside the field.

For curious readers who might be interested in arguments about why nature's choice of the axioms of quantum mechanics might make sense, see Aaronson's book \cite{aaronson_2013}. For short introductions to QC for non-physicists see \cite{RP:1998}, \cite{Nannicini17}, and \cite{Landsberg18}. More comprehensive presentations of QC can be found in works such as \cite{MerminBook,RieffelBook,QCQI,Wilde2013,Preskill:1998b}.

\pagebreak[4]

The organization of this chapter is as follows: In Section~\ref{sec:qintro:qubit}, I introduce the representation of noiseless classical and quantum states, \textit{i.e.}, bits and qubits. Then, I discuss the evolution and measurement of qubits. In Section~\ref{sec:qintro:qcirc}, I present how quantum states and operations are represented pictorially. In Section~\ref{sec:qintro:noise}, I discuss a formalism for describing noisy quantum-systems. Finally, on Section~\ref{sec:intro:compcompare} I give a brief comparison between classical and quantum computation.

\section{\label{sec:qintro:qubit}Ideal Classical and Quantum Systems}

\subsection{Classical States}

Before discussing quantum systems, I will start by presenting how the states of digital computers\footnote{By ``digital computer,'' unless otherwise indicated, I will mean a binary, transistor-based computing device with finite memory.}, which are so ubiquitous in modern life, are represented. The fundamental unit of information in classical information theory is the binary digit, or \textit{bit}. This unit takes the value of 0 or 1; therefore, a single bit is represented as a one-dimensional vector of the set $\{0, 1\}$. Alternatively, a bit is represented as an element of $\{\text{TRUE}, \text{FALSE}\}$ or $\{+1, -1\}$. 

Physically, bits can be represented by two-level classical systems such as the on-off states of switches, the up-down spins of magnetic domains, or the not-charged (1) or charged (0) states of floating gate transistors in NAND solid-state drives (SSDs). 

In the parlance of theoretical computer science, digital computers are finite-state machines (although given the large size of a typical digital computer's memory, they are practically equivalent to Turing machines, whose memory is formally infinite). The state of these devices is represented as a \textit{binary vector}, or \textit{binary string}, $z \in \{0,1\}^n$. Here $\{0,1\}^n$ is the set formed by taking the $n$-fold Cartesian product of $\{0,1\}$. Often, a sequence of bits is presented as a concatenation  

\begin{equation}
z = z_1 z_2 \cdots z_n,
\end{equation} 

\noindent where $z_i \in \{0, 1\}$. For example, $z = 101010 = (1,0,1,0,1,0) \in \{0,1\}^6$.
 
\subsection{Quantum States\label{intro.sec.qstate}} 
 
The quantum analog of the bit is the quantum bit, or \textit{qubit}. A pure qubit state is represented by a two-dimensional vector in $\Cbold^2$. As I discuss later, a norm is defined for the complex vector spaces of which quantum states are elements. Thus, such a space is a Hilbert space (commonly denoted as $\Hb$). Qubits can represent any physical two-level quantum-system, which include the spin of an electron, the ground and excited energy-levels of an atom, or the polarization of a photon.

Qubits may be generalized to $d$-level quantum states, known as \textit{qudits}, or even to non-discrete, infinite-dimensional states; however, this work focuses on qubit systems.

A standard choice for the basis states of a two-level quantum system is described by the notation $\ket{0}$ and $\ket{1}$, known as the \textit{computational basis}. Here the ket, $\ket{\cdot}$, of the Dirac notation \cite{dirac_1939} is used to represent vectors. The notation was developed to facilitate expressing the linear algebra of Hilbert space. Note that the Dirac notation can be used to represent binary vectors as well. For example, a bit can be represented as an element in $\{\ket{0}, \ket{1}\}$, and a bit string $z$ is equivalent to $\ket{z}$. However, kets tend to be reserved for indicating that a vector is a member of a Hilbert space rather than just an element of $\{0,1\}^n$.

\pagebreak[4]
Besides the ket of the Dirac notation, computational-basis states can also be represented as column vectors, where

\begin{equation}
\renewcommand{\arraystretch}{0.75}
\ket{0} = \mqty[1 \\ 0] \text{ and } \ket{1} = \begin{bmatrix}0 \\ 1\end{bmatrix}.
\end{equation}

Written in terms of these basis states, a general pure state $\ket{\psi}$ can then be described as a vector

\begin{equation}
\label{eq.ch2.psi}
\renewcommand{\arraystretch}{0.75}
\ket{\psi} = \alpha \ket{0} + \beta \ket{1} = \begin{bmatrix}\alpha\\ \beta\end{bmatrix},
\end{equation}

\noindent where the coefficients $\alpha, \beta \in \Cbold$ and $\abs{\alpha}^2 + \abs{\beta}^2 = 1$. The last condition is known as \textit{normalization}. As we will discuss later, normalization is a useful condition that allows the complex coefficients of the basis states to be related to the probability of measurement outcomes. Note that these coefficients are often referred to as \textit{amplitudes}.

The computational basis is not the only possible basis. I discuss two more common bases when introducing the Pauli matrices in Section~\ref{sec:intro:singleuni}.  

A collection of $n$ qubits can be represented as unit a vector in the vector space $\Cbold^{2^{n}}	\cong \left(\Cbold^2\right)^{\otimes n}$---the $n$-fold tensor product of the vector space $\Cbold^2$ ($\Cbold \times \Cbold$). A basis for such a vector space is the $n$-fold tensor product of single-qubit computational-basis states. So, for example, a basis for a two-qubit state would be

\begin{equation}
\label{eq.2qubitbasis}
\ket{0}_1\otimes\ket{0}_2, \; \ket{0}_1\otimes\ket{1}_2, \; \ket{1}_1\otimes\ket{0}_2, \text{ and } \ket{1}_1\otimes\ket{1}_2,
\end{equation}

\noindent where the subscripts identify qubits. It is common to drop the tensor product and allow the order of the vectors to identify qubits. Thus, the set of basis vectors in Eq.~\ref{eq.2qubitbasis} are often presented as  

\begin{equation}
\label{eq.2qubitbasisB}
\ket{00}, \; \ket{01}, \; \ket{10}, \text{ and } \ket{11}.
\end{equation}

\noindent In terms of the $n$-qubit computational basis, a general $n$-qubit state is described as a vector

\begin{equation}
\label{eq:genstate}
\renewcommand{\arraystretch}{0.75}
\ket{\psi} = \sum_{z \in \{0,1\}^n} \alpha_z \ket{z} = \mqty[\alpha_{00\cdots 00} \\ \alpha_{00\cdots 01} \\ \vdots \\ \alpha_{11\cdots 11}],
\end{equation}

\noindent where for all the amplitudes $\alpha_z \in \Cbold$ and the normalization condition is 

\begin{equation}
\label{eq:normsumsimp}
\sum_{z \in \{0,1\}^n} \abs{\alpha_z}^2 = 1.
\end{equation}

\noindent Note, unless otherwise stated, the vectors and matrices presented in this work are written in terms of the computational basis using standard positional notation, where bit strings labeling the computational-basis states start at zero, the value of each successive string increases by one, and the significance of the bits increases from right-to-left. Also note, that Eq.~\ref{eq:genstate} could just have easily been written in terms of any basis that spans the space.

The dual of a qubit vector is represented in Dirac notation as a \textit{bra}, $\bra{\cdot}$.  The bra of a state $\ket{\psi}$ is the conjugate transpose of $\ket{\psi}$. For example, the dual of the state $\ket{\psi}$ as defined in Eq.~\ref{eq:genstate} is the bra or row vector

\begin{equation}
\renewcommand{\arraystretch}{0.75}
\bra{\psi} \equiv \ket{\psi}^{\dagger} = \left(\ket{\psi}^{*}\right)^{T} = \sum_{z \in \{0,1\}^n} \alpha_z^{*} \bra{z} = \mqty[\alpha^{*}_{00\cdots 00} & \alpha^{*}_{00\cdots 01} & \cdots & \alpha^{*}_{11\cdots 11}],
\end{equation}

\noindent where $*$ is the complex conjugate, $T$ is the transpose operator, and $\dagger$ is the conjugate transpose.

In Dirac notation, an inner product between states $\ket{\psi}$ and $\ket{\phi}$ is notated as $\ip{\psi}{\phi}$, where, given two vectors $\ket{\psi} = \sum_j \psi_j \ket{j}$ and $\ket{\phi} = \sum_j \phi_j \ket{j}$ expressed in the same basis, the inner product is defined as

\begin{equation}
\label{eq:outer}
\renewcommand{\arraystretch}{0.75}
\ip{\psi}{\phi} \equiv \bra{\psi} \ket{\phi} = \mqty[\psi_1^{*} & \psi_2^{*} &  \cdots &  \psi_n^{*}] \mqty[\phi_1 \\ \phi_2 \\ \vdots \\  \phi_n] =  \sum_j \psi_j^{*}\phi_j.
\end{equation}

As we will see, the inner product is useful in defining things such as normalization and probability of measurement outcomes. Note that since $\bra{\psi}$ is a bra and $\ket{\phi}$ is a ket, the inner product $\ip{\lambda}{\psi}$ is also known as a \textit{bra-ket}. This is an example of physicist notation-humor. 

Like the inner product, an outer product, also known as a dyad, is simply defined as

\begin{equation}
\renewcommand{\arraystretch}{0.75}
\dyad{\psi}{\phi} \equiv \mqty[\psi_1 \\ \vdots \\  \psi_n] \mqty[\phi_1^{*} & \cdots &  \phi_n^{*}] = 
\mqty[\psi_1\phi_1^{*} & \psi_1\phi_1^{*} & \cdots & \psi_1\phi_n^{*} \\
      \psi_2\phi_1^{*} & \psi_2\phi_2^{*} & \cdots & \psi_2\phi_n^{*} \\
      \vdots           & \vdots           & \ddots & \vdots           \\
      \psi_n\phi_1^{*} & \psi_n\phi_2^{*} & \cdots & \psi_n\phi_n^{*}],
\end{equation}

\noindent where the vectors $\ket{\psi}$ and $\ket{\phi}$ are defined as they were in Eq.~\ref{eq:outer}.

Given a set of basis vectors $\{\ket{j}\}$ that span a Hilbert space $\Hb = \expval{\{ \ket{j} \}}_{\Cbold}$, where $\expval{\cdot}_{\Cbold}$ is the span over the field $\Cbold$, it is clear that the identity matrix for $\Hb$ is

\begin{equation}
I_{\Hb} = \sum_j \dyad{j}.
\end{equation}

\noindent This equation is known as the \textit{resolution of the identity}. We can see that by linearity $I_{\Hb}$ sends any vector in $\Hb$ to itself. Often, the subscript indicating the Hilbert space of an identity is dropped and is understood by context. Also, it is common to equate $I$ with the scalar $1$.

Using the inner product, the Euclidean norm, also known as the $2$-norm, of a vector $\ket{\psi}$ is written as

\begin{equation}
\norm{\ket{\psi}}_2 \equiv \sqrt{\ip{\psi}{\psi}}.
\end{equation}

\noindent Note, other norms can be defined and are distinguished by subscripts  such as $2$ for the Euclidean norm. In this work, if a norm for a vector is not identified by a subscript, it can be assumed that the Euclidean norm is being used.

Likewise, norms of operators can be defined. If a norm of any operator $A$ on some Hilbert space $\Hb$ is not specified in this work, then $\norm{A}$ is taken to be the \textit{operator norm} defined as

\begin{equation}
\norm{A}_{\text{op}} = \text{sup}\{\norm{A\ket{\psi}} \; | \; \forall \ket{\psi} \in \Hb, \norm{\ket{\psi}} = 1 \}.
\end{equation}  

\pagebreak[4]
The Euclidean norm is used to define a \textit{unit vector}. Such a vector $\ket{\psi}$ has the property

\begin{equation}
\label{eq:intro:ketnorm}
\norm{\ket{\psi}} = 1.
\end{equation} 

\noindent Note that Eq.~\ref{eq:intro:ketnorm} is equivalent to Eq.~\ref{eq:normsumsimp}.

Two vectors $\ket{\psi}$ and $\ket{\phi}$ are said to be \textit{orthogonal} if and only if 

\begin{equation}
\ip{\psi}{\phi} = 0.
\end{equation}

A set of distinct vectors are said to be \textit{orthonormal} if and only if for any pair of vectors $\ket{j}$ and $\ket{k}$ in the set

\begin{equation}
\ip{j}{k}=\delta_{ij}.
\end{equation}

\subsection{\label{sec:qintro:dyn}Dynamics}

Now that I have presented a representation of pure states, I now discuss how states evolve through time. That is, how the amplitudes of states change. The evolution of states is described by operators that act on states. As states can be represented as complex vectors, linear operators can be represented as complex matrices. See \cite{AL98,aaronson05,aaronson_2013} for arguments why quantum mechanics is linear.

\pagebreak[4]
\subsubsection{Schr{\"o}dinger's equation}

The evolution of a closed quantum system is described by Schr{\"o}dinger's equation

\begin{equation}
\label{eq.schr}
H(t)\;\ket{\psi} = i \hbar \dv{\ket{\psi}}{t},
\end{equation}

\noindent where $H$ is an operator known as a Hamiltonian. A \textit{Hamiltonian} is a Hermitian operator that describes the energy of the system. A \textit{Hermitian operator} $A$ is a matrix such that $A = A^{\dagger}$. A Hamiltonian is defined as 

\begin{equation}
\label{eq:intro:Hamdef}
H(t) \equiv \sum_j E_j(t) \; \dyad{j},
\end{equation}

\noindent where $E_j(t) \in \Rbold$ and the vectors $\{ \ket{j} \}$ form an orthonormal basis, which are known as energy eigenvectors. Note, in this dissertation I will assume that basis vectors are time-independent; however, one may also consider basis vectors that are time-dependent. The (possibly time-dependent) eigenvalues $E_j(t)$ are referred to as energy eigenvalues. The expectation of the Hamiltonian $H(t)$ of a system is the classical energy of the system (see Section~\ref{sec:qintro:meas} for how to calculate an expectation value).

Note that Eq.~\ref{eq:intro:Hamdef} is a natural definition for a Hamiltonian operator since it assigns real numbers $E_j(t)$, representing amounts of energy, to eigenvectors. Given this definition, the Hamiltonian $H$ must be Hermitian since

\begin{equation}
H(t)^{\dagger} = \sum_j E_j(t)^{*} \dyad{j}^{\dagger} = \sum_j E_j(t) \dyad{j} = H(t).
\end{equation} 

\pagebreak[4]
Note, by the spectral decomposition theorem (see Chapter 2 of \cite{QCQI}), all Hermitian operators can be described by Eq.~\ref{eq:intro:Hamdef}. Thus, all Hermitian matrices can be identified as Hamiltonians.

\subsubsection{Unitary Evolution\label{sec:intro:unitaries}}

While dynamics are physically implemented through Hamiltonians, it is often useful to think about dynamics in terms of the evolution operators that Hamiltonians induce. Such evolution operators that take a state $\ket{\psi(t_0)}$ to state $\ket{\psi(t)}$ are define as

\begin{equation}
\ket{\psi(t)} = U(t,t_0)\ket{\psi(t_0)},
\end{equation}

\noindent where $U(t, t_0)$ is a unitary operator. A \textit{unitary operator} $U$ is represented as a matrix such that $U^{-1} = U^{\dagger}$. Thus, $UU^{\dagger} = U^{\dagger}U = I$. Note, here I have suppressed the time dependence for the unitaries. I will often drop the dependency when it is unnecessary or cumbersome. 

The class of unitary operators is the class of square matrices that leaves the norm of kets unchanged since

\begin{equation}
\norm{U\ket{\psi}} = \sqrt{\ev{U^{\dagger}U}{\psi}} = \sqrt{\ip{\psi}{\psi}} = \norm{\ket{\psi}}. 
\end{equation}

\noindent This implies that the association of probabilities to the square of amplitudes of basis vectors still holds after the evolution of a state by a unitary operator.

\pagebreak[4]
The differential formula that relates a Hamiltonian $H$ to the unitary operator $U$ is given as

\begin{equation}
\label{eq:hamunit}
H(t)\; U(t,t_0) = i \hbar \dv{U(t,t_0)}{t}
\end{equation}

\noindent with the boundary condition that $U(t_0,t_0)=I$.

The solution to Eq.~\ref{eq:hamunit} depends on the time-dependence and the commutativity of the Hamiltonian with itself. The general solution to Eq~\ref{eq:hamunit} is

\begin{align}
\label{eq:hamtimdep1}
U(t, t_0) & = \Tau \; \text{exp}({- \frac{i}{\hbar} \int_{t_0}^{t} d\tau \; H(\tau) }) \\
& = 1 + \sum_{n=1}^{\infty} \left(\frac{-i}{\hbar}\right)^n \int_{t_0}^{t} d\tau_n \int_{t_0}^{\tau_n} d\tau_{n-1}\cdots \int_{t_0}^{\tau_2} d\tau_{1} H(\tau_i)H(\tau_2)\cdots H(\tau_n),
\end{align}

\noindent where $\Tau$ is the time-ordering operator, which is defined as

\begin{equation}
\label{eq:intro:timeorder}
\Tau A(t_2)B(t_1) = 
 \begin{cases}
 A(t_2)B(t_1) & \text{if } t_2 > t_1\\
 B(t_1)A(t_2) & \text{otherwise}. 
 \end{cases}
\end{equation}

Two operators $A$ and $B$ are said to commute if and only if the commutator

\begin{equation}
[A, B] \equiv AB - BA
\end{equation}

\noindent is equal to zero.

\pagebreak[4]
If for all times $t_1$ and $t_2$, $[H(t_1),H(t_2)] = 0$, then effectively the time-ordering operator (Eq~\ref{eq:intro:timeorder}) acts like identity. Thus in such a case, Eq.~\ref{eq:hamtimdep1} simplifies to

\begin{equation}
\label{eq:hamtimdep2}
U(t, t_0) = \text{exp}\{- \frac{i}{\hbar} \int_{t_0}^{t}d\tau \; H(\tau) \}.
\end{equation}

It is clear that if the Hamiltonian is time-independent, then Eq.~\ref{eq:hamtimdep2} further simplifies to

\begin{equation}
\label{eq:intro:hamindep}
U(t, t_0) = \text{exp}\{- \frac{i}{\hbar} H \; [t-t_0] \}.
\end{equation}

\noindent Note, the interval between two times is often notated as $\Delta t = t - t_0$.  

One can easily prove that $U(t,t_0)$, as defined by equation \ref{eq:hamtimdep1}, \ref{eq:hamtimdep2}, or \ref{eq:intro:hamindep}, must be unitary if $H$ is Hermitian (which it is by definition). 

Note, so far, I have explicitly included Planck's constant $\hbar \approx 1.0546 \text{ x } 10^{-34}$ Joule seconds; however, in QC it is common to choose units such that $\hbar = 1$. In this work, I often follow this convention. 

In the following sections, I discuss unitaries that are commonly used in the study of quantum information.

\subsubsection{Single-qubit Unitaries\label{sec:intro:singleuni}}

A particularly useful set of single-qubit unitaries are the Pauli matrices, which are

\begin{equation}
\renewcommand{\arraystretch}{0.75}
\sigma_1 = X \equiv \mqty[0&1\\1&0], \; \sigma_2 = Y \equiv \mqty[0&-i\\i&0], \text{ and } \sigma_3 = Z \equiv \mqty[1&0\\0&-1]. 
\end{equation} 

Although not technically a member of the Pauli matrices, the single-qubit identity

\begin{equation}
\renewcommand{\arraystretch}{0.75}
\sigma_0 = I \equiv \mqty[1&0\\0&1] 
\end{equation}

\noindent is often informally considered a Pauli matrix. In this work, I also include $I$ as a member of the Pauli matrices.

As these matrices are ubiquitous in quantum information text, it is useful to highlight some of their properties. The Pauli matrices are Hermitian matrices that square to identity. Excluding identity, the Pauli matrices anticommute with each other. That is,

\begin{equation}
\sigma_i \sigma_j = - \sigma_j \sigma_i.
\end{equation}

\noindent In terms of the anticommutator

\begin{equation}
\{\sigma_i, \sigma_j\} \equiv \sigma_i \sigma_j + \sigma_j \sigma_i,
\end{equation}

\noindent this can be alternatively stated that for all $i$ and $j$, $\{\sigma_i, \sigma_j\} $ is equal to zero when $i \neq j$ and $i,j \neq 0$.

The Pauli matrices may serve as observables in measurements. The eigenvectors associated with the non-identity Pauli measurements are commonly used basis-vector sets. 

\pagebreak[4]
The Pauli $Z$ operator is defined as

\begin{equation}
Z \equiv \dyad{0} - \dyad{1}.
\end{equation}

\noindent Thus, the associated basis vectors of $Z$ are the eigenvectors $\ket{0}$ and $\ket{1}$ with eigenvalues $+1$ and $-1$, respectively. Therefore, the computational basis is also known as the \textit{$Z$-basis}. As the unitary $Z$ adds a $-1$ phase to a $\ket{1}$ state and leave a $\ket{0}$ alone, $Z$ is also known as the \textit{phase-flip operator}.

The Pauli $X$ operator is

\begin{equation}
X \equiv \dyad{+} - \dyad{-},
\end{equation}

\noindent where the eigenvectors

\begin{equation}
\ket{+} \equiv \frac{1}{\sqrt{2}}\left(\ket{0}+\ket{1}\right) \text{ and } \ket{-} \equiv \frac{1}{\sqrt{2}}\left(\ket{0}-\ket{1}\right)
\end{equation}

\noindent with eigenvalues $+1$ and $-1$, respectively. This basis is therefore known as the \textit{$X$-basis}. The operator $X$ can also be written as 

\begin{equation}
X = \dyad{0}{1} + \dyad{1}{0}
\end{equation}

\noindent Therefore, $X$ is known as the \textit{bit-flip operator}.

The Pauli $Y$ operator is given as

\begin{equation}
Y \equiv \dyad{+i} - \dyad{-i},
\end{equation}

\noindent where the eigenvectors 

\begin{equation}
\ket{+i} \equiv \frac{1}{\sqrt{2}}\left(\ket{0}+i\ket{1}\right) \text{ and } \ket{-i} \equiv \frac{1}{\sqrt{2}}\left(\ket{0}-i\ket{1}\right) 
\end{equation}

\noindent with eigenvalues $+1$ and $-1$, respectively. Since $Y = i XZ$, the operator $Y$ performs both a bit-flip and phase-flip.

Another common operator seen in quantum information is the \textit{Hadamard gate}, which is defined as

\begin{equation}
\renewcommand{\arraystretch}{0.75}
H \equiv \dyad{0}{+} + \dyad{1}{-} = \dyad{+}{0} + \dyad{-}{1} = \frac{1}{\sqrt{2}} \mqty[1 & 1 \\ 1 & -1].
\end{equation}

\noindent Thus, the Hadamard exchanges the basis states $\ket{0}$ and $\ket{+}$ as well as the basis states $\ket{1}$ and $\ket{-}$.

Two other common unitaries used in quantum information are roots of $Z$. The first is known as the \textit{phase gate}, $S$, which is defined as

\begin{equation}
\renewcommand{\arraystretch}{0.75}
S \equiv \sqrt{Z} = \mqty[1 & 0 \\ 0 & i] = \dyad{0} + i \dyad{1}.
\end{equation}

\noindent The other commonly used root of $Z$ is the $T$ operator, 

\begin{equation}
\renewcommand{\arraystretch}{0.75}
T \equiv \sqrt[4]{Z} = \mqty[1 & 0 \\ 0 & e^{i\pi / 4}] = \dyad{0} + e^{i\pi / 4} \dyad{1}.
\end{equation}

\pagebreak[4]
\subsubsection{Two-qubit Unitaries}

One of the simplest two-qubit unitaries is the SWAP operator. The SWAP operator simply exchanges the state of two qubits. That is, in the computational basis SWAP maps 

\begin{equation}
\ket{00} \rightarrow \ket{00}, \ket{01} \rightarrow \ket{10}, \ket{10} \rightarrow \ket{01}, \text{ and } \ket{11} \rightarrow \ket{11}.  
\end{equation}

The SWAP gate is equivalent to the matrix

\begin{equation}
\renewcommand{\arraystretch}{0.75}
\text{SWAP} \equiv \mqty[1 & 0 & 0 & 0 \\ 0 & 0 & 1 & 0 \\ 0 & 1 & 0 & 0 \\ 0 & 0 & 0 & 1],
\end{equation}

\noindent with respect to the computational basis.

Other common two-qubit operations are controlled unitaries. For each of these operations, one qubit is called the ``target'' qubit and the other, the ``control'' qubit. These controlled unitaries have the property that if the control qubit is in the $\ket{0}$ state, then identity is applied to the target qubit; however, if the control qubit is in the $\ket{1}$ state, a single-qubit unitary is applied. This can be expressed as

\begin{equation}
\text{CU} \equiv \dyad{0} \otimes I + \dyad{1} \otimes U,
\end{equation}

\noindent where the tensor product has been suppressed.   

One of the most commonly discussed two-qubit operations is the CNOT or controlled-$X$ (CX). This unitary is equivalent to

\begin{equation}
\renewcommand{\arraystretch}{0.75}
\text{CNOT} \equiv \dyad{0} \otimes I + \dyad{1} \otimes X = \mqty[1 & 0 & 0 & 0 \\ 0 & 1 & 0 & 0 \\ 0 & 0 & 0 & 1 \\ 0 & 0 &
 1 & 0],
\end{equation}

\noindent in the computational basis.

A related controlled unitary is the controlled-$Z$ (CZ). This unitary is described as  

\begin{equation}
\renewcommand{\arraystretch}{0.75}
\text{CZ} \equiv \dyad{0} \otimes I + \dyad{1} \otimes Z = \mqty[1 & 0 & 0 & 0 \\ 0 & 1 & 0 & 0 \\ 0 & 0 & 1 & 0 \\ 0 & 0 & 0 & -1],
\end{equation}

\noindent in the computational basis. It is easy to prove that 

\begin{equation}
\label{eq:czsym}
\renewcommand{\arraystretch}{0.75}
CZ = \dyad{0} \otimes I + \dyad{1} \otimes Z = I \otimes \dyad{0} + Z \otimes \dyad{1}.
\end{equation}

\noindent That is, either qubit can be considered the control qubit.

\subsubsection{Three-qubit Unitaries}

In quantum information, perhaps the most common three-qubit unitary discussed is the Toffoli gate. This gate is similar to the CNOT except that the Toffoli gate has two controls instead of one; thus, the Toffoli gate is also known as the CCNOT gate. The Toffoli gate can therefore be represented as

\begin{equation}
\renewcommand{\arraystretch}{0.75}
\text{Toffoli} \equiv (II - \dyad{11}) \otimes I + \dyad{11} \otimes X = \mqty[
1 & 0 & 0 & 0 & 0 & 0 & 0 & 0 \\
0 & 1 & 0 & 0 & 0 & 0 & 0 & 0 \\
0 & 0 & 1 & 0 & 0 & 0 & 0 & 0 \\
0 & 0 & 0 & 1 & 0 & 0 & 0 & 0 \\
0 & 0 & 0 & 0 & 1 & 0 & 0 & 0 \\
0 & 0 & 0 & 0 & 0 & 1 & 0 & 0 \\
0 & 0 & 0 & 0 & 0 & 0 & 0 & 1 \\
0 & 0 & 0 & 0 & 0 & 0 & 1 & 0 ],
\end{equation}

\noindent in the computational basis.

The CZ analog of the Toffoli gate is the CCZ gate, that is, the doubly-controlled $Z$ gate. The CCZ gate can be written as

\begin{equation}
\renewcommand{\arraystretch}{0.75}
\text{CCZ} \equiv (II - \dyad{11}) \otimes I + \dyad{11} \otimes Z = \mqty[
1 & 0 & 0 & 0 & 0 & 0 & 0 & 0 \\
0 & 1 & 0 & 0 & 0 & 0 & 0 & 0 \\
0 & 0 & 1 & 0 & 0 & 0 & 0 & 0 \\
0 & 0 & 0 & 1 & 0 & 0 & 0 & 0 \\
0 & 0 & 0 & 0 & 1 & 0 & 0 & 0 \\
0 & 0 & 0 & 0 & 0 & 1 & 0 & 0 \\
0 & 0 & 0 & 0 & 0 & 0 & 1 & 0 \\
0 & 0 & 0 & 0 & 0 & 0 & 0 & -1 ],
\end{equation}

\subsection{\label{sec:qintro:meas}Projective Measurements}

Normalization allows the complex coefficients of a ket to be associated with probabilities. For example, if the state $\ket{\psi} = \alpha \ket{0} + \beta \ket{1}$ is projectively measured in the computational basis, then after the measurement the qubit is found in either the state $\ket{0}$ with probability $\abs{\alpha}^2$ or the state $\ket{1}$ with probability $\abs{\alpha}^2$. Likewise, if the state $\ket{\psi} = \sum_{z = \{0,1\}^n} \alpha_z \ket{z}$ is measured in the computational basis, then the probability of finding the state as $\ket{z}$ after measurement is $\abs{\alpha_z}^2$. The identification of the square of amplitudes to probabilities is known as the \textit{Born rule} \cite{Born1926}.

In general, a \textit{projective measurement}, also known as a \textit{von Neumann measurement}, is associated with an \textit{observable} $M$, which is written as

\begin{equation}
M = \sum_{\lambda} \lambda P_{\lambda},
\end{equation}

\noindent where $P_{\lambda}$ is a projector onto an eigenspace of $M$ with eigenvalues $\lambda$. The projector $P_{\lambda}$ is a sum

\begin{equation}
\label{eq:intro:proj}
P_{\lambda} = \sum_{j} \dyad{j},
\end{equation}

\noindent where the vectors $\{ \ket{j} \}$ are a set of basis vectors that span the $\lambda$-eigenspace.

Note that a projector $P_{\lambda}$ has the property

\begin{equation}
P_{\lambda}^2 = P_{\lambda}
\end{equation}

\noindent since

\begin{equation}
P_{\lambda}^2 = \sum_{j} \dyad{j}\sum_{k} \dyad{k} = \sum_{j} \sum_{k} \ket{j}\bra{j}\ket{k}\bra{k} = \sum_{j} \sum_{k} \delta_{j,k}\dyad{j}{k} =\sum_{j} \dyad{j} = P_{\lambda},
\end{equation}

\noindent where the sums run over the same basis.

\pagebreak[4]
Also, since the projectors are a sum of basis vectors that span the $\lambda$-eigenspaces, the projectors have the property that

\begin{equation}
\label{eq:intro:projsum}
\sum_{\lambda} P_{\lambda} = I.
\end{equation}  

The probability that the state $\ket{\psi}$ is projected into the $\lambda$-eigenspace is 

\begin{equation}
\label{eq:into:projectiveprobdef}
p(\lambda) = \ev{P_{\lambda}}{\psi}.
\end{equation}

Note that by identifying probabilities with  Eq.~\ref{eq:into:projectiveprobdef} and assuming the state $\ket{\psi}$ is normalized, $0 \leq p(\lambda) \leq 1$, as we would want for a quantity representing probabilities. 

We see Eq.~\ref{eq:intro:proj} and Eq.~\ref{eq:into:projectiveprobdef} guarantees that $p(\lambda) \geq 0$ since

\begin{equation}
p(\lambda) = \ev{P_{\lambda}}{\psi} = \sum_{j} \ip{\psi}{j}\ip{j}{\psi} = \sum_{j} \abs{\ip{j}{\psi}}^2 \geq 0.
\end{equation}

Further, normalization of the pure states and Eq.~\ref{eq:intro:projsum} mean that $p(\lambda) \leq 1$ since

\begin{equation}
\label{eq:intro:normdensity}
\sum_{\lambda} p(\lambda) = \sum_{\lambda} \ev{P_{\lambda}}{\psi} =  \ev{\sum_{\lambda} P_{\lambda}}{\psi} = \ip{\psi} = 1.
\end{equation}

Upon measuring, a measurement device outputs a result corresponding to the eigenvalue of the eigenspace that the state was projected to, and the state $\ket{\psi}$ becomes

\begin{equation}
\ket{\psi '} = \frac{P_{\lambda}\ket{\psi}}{\sqrt{p(\lambda)}}.
\end{equation}

Note that if two states $\ket{\psi}$ and $\ket{\phi}$ differ by a \textit{global phase}, \textit{i.e.}, $\ket{\psi}=e^{i\theta}\ket{\phi}$ where $\theta \in \Rbold$, then the measurement statistics of the two states are the same. For this reason, global phases are said to be physically meaningless.

The average eigenvalue found when measuring an observable $M$ is expected to be

\begin{equation}
\mathds{E}[M] \equiv \ev{M}{\psi} = \sum_{\lambda} \lambda \ev{P_{\lambda}}{\psi} = \sum_{\lambda} \lambda \; p(\lambda),
\end{equation}

\noindent where $\ket{\psi}$ is the state being measured. This value is known as an \textit{expectation value}.

So far, I have only discussed projective measurements. These type of measurements take pure states to pure states. However, a more general type of quantum measurement is known as the \textit{Positive Operator-Valued Measure} (POVM), which I discuss later in Section~\ref{sec:intromeas}.

\section{\label{sec:qintro:qcirc}Quantum Circuits}

A \textit{quantum circuit} is a space-time diagram depicting the sequence of quantum gates. In this text, I use the term ``gate'' to refer to any quantum operation including unitaries, measurements, and state preparations. An example of a quantum circuit is shown in the diagram:

\begin{figure}[ht]
{\small \begin{center}
    \begin{tabular}{c} 
			\Qcircuit @C=1em @R=2em {
\lstick{\ket{\psi}} & \qw        & \qw       & \ctrl{1} & \gate{H}     & \measureD{Z}  & \control \cw\\
\lstick{\ket{0}}    & \gate{H}   & \ctrl{1}     & \targ    & \measureD{Z}  & \control \cw & \cwx\\
\lstick{\ket{0}}    &  \qw  & \targ & \qw      & \qw & \gate{X} \cwx & \gate{Z} \cwx & \rstick{\ket{\psi}} \qw
}
    \end{tabular}
\end{center}}
\caption{An example of a teleportation quantum circuit depicting input states, unitaries, and measurements.}
\label{fig:qcirc1}
\end{figure}
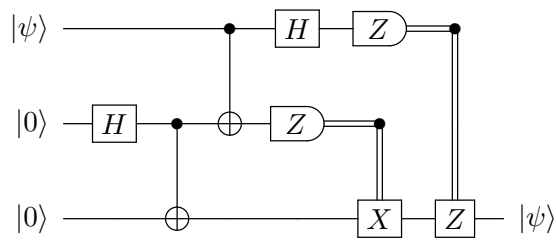 

\vfill
\pagebreak[4]
In these circuits, time moves from left-to-right and each wire corresponds to the world-line of a qubit. Input states generally appear on the far left, while output states appear on the far right.    

A measurement is usually depicted as a stylistic drawing of a meter or as a cap-shape as seen in Fig.~\ref{fig:qcirc1}. A label may be included in the drawing to indicate what basis the measurement is being made in. If no basis is indicated, then it often assumed that the measurement is in the $Z$ basis. Double lines drawn exiting the right-side of a measurement are used to represent classical information determined from measuring. On such lines, a filled in circle is used to indicate classical control of unitaries as seen in Fig~\ref{fig:qcirc1}.    

\begin{figure}[ht]
\begin{minipage}{14cm}
    \centering
    \begin{subfigure}[b]{1.6cm}
        \centering
			\Qcircuit @C=1em @R=2em {
& \ctrl{1} & \qw \\
& \gate{U} & \qw
}
        \caption{CU} \label{fig:aa}
    \end{subfigure}
    \hfill
    \begin{subfigure}[b]{2cm}
        \centering
			\Qcircuit @C=1em @R=2.5em {
& \ctrl{1} & \qw \\
& \targ & \qw
}
        \caption{CNOT} \label{fig:aaa}
    \end{subfigure}
    \hfill
    \begin{subfigure}[b]{1.6cm}
        \centering
			\Qcircuit @C=1em @R=3.1em {
& \ctrl{1} & \qw \\
& \ctrl{-1} & \qw
}
        \caption{CZ} \label{fig:a}
    \end{subfigure}
    \hfill
    \begin{subfigure}[b]{2cm}
        \centering
			\Qcircuit @C=1em @R=3.5em {
& \qswap & \qw \\
& \qswap \qwx & \qw
}
        \caption{SWAP} \label{fig:c}
    \end{subfigure}
    \hfill
    \begin{subfigure}[b]{2cm}
        \centering
			\Qcircuit @C=1em @R=2em {
& \ctrl{1} & \qw \\
& \ctrl{1} & \qw \\
& \targ & \qw
}
        \caption{Toffoli} \label{fig:b}
    \end{subfigure}
    \hfill
    \begin{subfigure}[b]{2cm}
        \centering
			\Qcircuit @C=1em @R=2.3em {
& \ctrl{1} & \qw \\
& \ctrl{1} & \qw \\
& \ctrl{-1} & \qw
}
        \caption{CCZ} \label{fig:ccz}
    \end{subfigure}
\end{minipage}
    
\caption{\label{fig:twocircs} Quantum-circuit diagrams for commonly used multi-qubit unitaries. In these figures the filled-in circles represent controls, and the hollow circles represent targets for which Pauli $X$ may be applied to the qubit. Note that the $CZ$ symbol shown in (c) is depicted with filled-in circles on both qubits. This is to indicate that either qubit can be treated as the control in the manner described by Eq.~\ref{eq:czsym}.}
\end{figure}
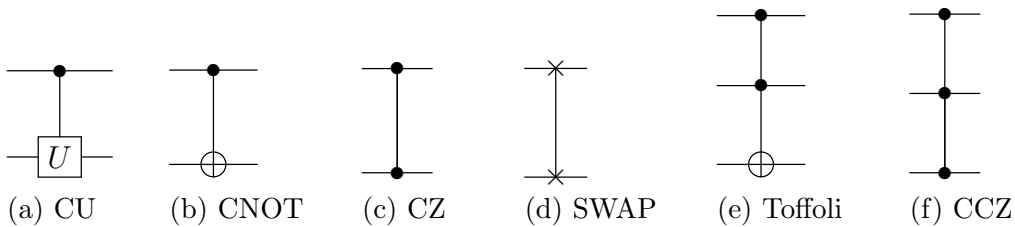

Unitaries are represented as boxes with symbols indicating the unitary that is being applied. Examples of the Pauli $X$, Pauli $Z$, and Hadamard gates are depicted in quantum circuit seen in Fig.~\ref{fig:qcirc1}. The lines representing the qubits being affected by a unitary will enter from the left of the unitary and exit from the right. Some multi-qubit gates are indicated by unique symbols rather than boxes. Examples such symbols are seen in Fig.~\ref{fig:twocircs}.

\section{\label{sec:qintro:noise}Noisy Quantum Systems}

So far, I have discussed the quantum mechanical formalism for pure states, which are ideal states that we assume we have perfect knowledge of. I now briefly describe the density-matrix formalism, which allows one to describe quantum systems where there is some lack of knowledge of the state of the system as well as its evolutions and measurements. Ideally, we would avoid using such a formalism, but in practice our knowledge and control of any system is never perfect. 

\subsection{Density Matrix}

We begin by considering an ensemble of pure states 

\begin{equation}
\label{eq:density}
\mathcal{E} \equiv \{p(i), \ket{\psi_i}\},
\end{equation}

\noindent where $i$ indexes the pure states $\ket{\psi_i}$. For such an ensemble, it is not necessary that states $\ket{\psi_i}$ are orthogonal to each other; however, since the $p_i$ represent probabilities, there is the requirement that

\begin{equation}
\label{eq:psum}
\sum_i p(i) = 1.
\end{equation}

A \textit{density matrix}, also known as a \textit{density operator}, is used to describe the state of such an ensemble of pure states. A density operator is defined as

\begin{equation}
\label{eq:intro:generaldensity}
\rho \equiv \sum_i p(i) \dyad{\psi_i}.
\end{equation}

The probabilities $p(i)$ of the density matrix also have the requirement of Eq.~\ref{eq:psum} and the states $\ket{\psi_i}$ are normalized.

The density matrix is an alternative representation of a state. Any pure state $\ket{\psi}$ can be represented as a density operator

\begin{equation}
\rho = \dyad{\psi}.
\end{equation}

For the density matrix representation, the trace is often useful when defining such things as normalization, measurements, and expectation values. Using the Dirac notation, the trace is written as

\begin{equation}
\tr(A) = \sum_i \ev{A}{i} = \sum_i A_{ii},
\end{equation} 

\noindent where the vectors $\ket{i}$ form a complete, orthonormal basis appropriate for the space that the operator $A$ belongs to.

\pagebreak[4]
A useful property of the trace to note is

\begin{equation}
\tr(AB) = \sum_i \sum_j A_{ij} B_{ji} = \sum_j \sum_i B_{ji} A_{ij} = \tr(BA).
\end{equation} 

It is straightforward to show that the trace is invariant for general cyclic permutations, \textit{e.g.},

\begin{equation}
\tr(ABC) = \tr(BCA) = \tr(CAB).
\end{equation}

Using the trace, given that a density operator is an ensemble of normalized pure-states and that the probabilities of these states sum to one, it is natural to express the normalization of density operator as

\begin{align}
\tr(\rho) & = \sum_i \ev{\rho}{i} = \sum_i \sum_j p(j) \bra{i}\ket{\psi_j}\bra{\psi_j}\ket{i} = \sum_j p(j) \bra{\psi_j}\left(\sum_i \dyad{i}\right) \ket{\psi_j} \nonumber \\
& = \sum_j p(j)\ev{I}{\psi_j}= \sum_j p(j) = 1.
\end{align}

Later, I discuss how to use the trace to express measurements and expectation values of density matrices.

Other frequently cited properties of the density matrix, which one can show by the representation's definition, is that

\begin{equation}
\rho^{\dagger} = \sum_i p(i)^{*} \dyad{\psi_i}^{\dagger} = \sum_i p(i) \dyad{\psi_i} = \rho
\end{equation} 

\noindent and

\begin{equation}
\forall \ket{\phi} \in \Hb, \ev{\rho}{\phi} \geq 0
\end{equation}

\noindent since for any state $\ket{\phi}$ we find

\begin{equation}
\ev{\rho}{\phi} = \sum_i p(i) \ip{\phi}{\psi_i}\ip{\psi_i}{\phi} = \sum_i p(i) \abs{\ip{\psi_i}{\phi}}^2 \geq 0.
\end{equation}

Now that density matrices have been defined, one might be curious about whether one can easily determine if a density matrix represents a pure state or not. It would be even better if there was a method for evaluating the degree of purity.

We can start to answer these questions by first considering the ensembles representing pure states and maximally-mixed states.

Any pure state is represented as a normalized state $\ket{\psi}$, and thus the ensemble

\begin{equation}
\mathcal{E}_{\text{pure}} = \{p=1, \ket{\psi}\}
\end{equation}

\noindent and the density matrix

\begin{equation}
\rho_{\text{pure}} = \dyad{\psi}.
\end{equation}

A maximally-mixed state for a Hilbert space $\Hb$ spanned by $d$ basis states $\{\ket{j}\}$ is represented by the ensemble 

\begin{equation}
\mathcal{E}_{\text{max-mix}} = \{p(j)=\frac{1}{d}, \ket{j}\}
\end{equation} 

\noindent and the density matrix

\begin{equation}
\rho_{\text{max-mix}} = \frac{1}{d}\sum_j \dyad{j} = \frac{1}{d}I.
\end{equation}

\noindent That is, the maximally-mixed state has an equal probability of being in any of the basis states $\{\ket{j}\}$.

The purity of a density matrix $\rho$ is defined as

\begin{equation}
P(\rho) = \tr(\rho^2).
\end{equation}

If we evaluate the purity of $\rho_{\text{pure}}$ and $\rho_{\text{max-mix}}$ we find

\begin{equation}
P(\rho_{\text{pure}}) = 1
\end{equation}

\noindent and

\begin{equation}
P(\rho_{\text{max-mix}}) = \tr(\frac{I}{d^2}) = \frac{1}{d}.
\end{equation}

For a general density matrix $\rho$ (Eq.~\ref{eq:intro:generaldensity}) we have

\begin{align}
P(\rho) &= \sum_k \bra{k} \left[ \sum_i p(i) \dyad{\psi_i} \sum_j p(j) \dyad{\psi_j} \right] \ket{k} \nonumber \\
&= \sum_{i,j} p(i) p(j) \abs{\ip{\psi_i}{\psi_j}} = \sum_i p(i)^2.
\end{align}

\noindent Thus, we see that $\frac{1}{d} \leq P(\rho) \leq 1$.

\subsection{Dynamics}

Evolution of an ensemble of pure states as described by  Eq.~\ref{eq:density} evolves by a unitary $U$ according to 

\begin{equation}
\mathcal{E} = \{p(i), \ket{\psi_i}\} \rightarrow \mathcal{E}' = \{p(i), U\ket{\psi_i}\}.
\end{equation}

\noindent Thus, the evolution of a density operator from $\rho$ to $\rho '$ by a unitary $U$ is described by

\begin{equation}
\rho ' = U\rho U^{\dagger}.
\end{equation}

Note that unitaries also preserve the normalization of a density matrix since

\begin{equation}
\tr(U\rho U^{\dagger}) = \tr(\rho U^{\dagger} U) = \tr(\rho).
\end{equation}

\subsection{Measurements\label{sec:intromeas}}

The previously described projective measurements are ideal measurements that take pure states to pure states. In this section, I now discuss a more general measurement called a Positive Operator-Valued Measure (POVM). A POVM may take pure states to mixed states.

A POVM is described by a set of measurement operators $\{M_i\}$. These operators must be such that

\begin{equation}
\label{eq:intro:measopconstraint}
\sum_{i} M_i^{\dagger} M_i = I.
\end{equation}

Given such a set, the probability of measuring a density operator $\rho$ and obtaining the result $i$ after the measurement is

\begin{equation}
\label{eq:intro:densityprob}
p(i) = \tr(M_i\rho M_i^{\dagger}).
\end{equation}

\noindent Note that just as the probability $p(\lambda)$ as defined for projective measurements (Eq.~\ref{eq:into:projectiveprobdef}) could be shown to be restricted to lie in $[0,1]$, it is simple to show that the same holds for $p(i)$ of Eq.~\ref{eq:intro:densityprob} given the normalization constrain of the density matrix Eq.\ref{eq:intro:normdensity} and measurement operator constraint Eq.~\ref{eq:intro:measopconstraint}.

Immediately after the measurement, the density operator becomes

\begin{equation}
\rho ' = \frac{M_i \; \rho \; M_i^{\dagger}}{p(i)}.
\end{equation}  

The expectation of measuring an operator $A$ of a density matrix $\rho$ is defined as

\begin{align}
\mathds{E}[A] & \equiv \tr(\rho A) = \sum_i \ev{\rho A}{i} = \sum_i \sum_j p(j) \bra{i}\ket{\psi_j}\bra{\psi_j}A\ket{i} \nonumber \\
& = \sum_j p(j) \bra{\psi_j}A \left(\sum_i \dyad{i}\right) \ket{\psi_j} = \sum_j p(j)\ev{A}{\psi_j}.
\end{align}

\pagebreak[4]
\subsection{Quantum Channels}

More general evolutions of density matrices are described by a \textit{quantum channel}

\begin{equation}
\rho ' = \mathcal{E}(\rho) = \sum_i E_i \rho E_i^{\dagger},
\end{equation}

\noindent where the set of operators $\{E_i\}$, called \textit{Kraus operators}, are such that

\begin{equation}
\label{Eq:Eiii}
\sum_i E_i^{\dagger} E_i = I.
\end{equation} 

\noindent Such a transformation turns out to be the most general map that is consistent with quantum mechanics that takes a valid density matrix to another valid density matrix (see Section 4.6 of \cite{Wilde2013}). This transformation can be interpreted replacing the state $\rho$ with 

\begin{equation}
\frac{E_i \rho E_i ^{\dagger}}{\tr(E_i \rho E_i ^{\dagger})}
\end{equation}

\noindent with probability

\begin{equation}
\label{Eq:piii}
p(i) = \tr(E_i \rho E_i ^{\dagger}).
\end{equation}

\pagebreak[4]
\subsubsection{Error Channels\label{sec:intro:errors}}

Quantum channels are commonly used to describe quantum noise. In this section, I will outline some commonly considered quantum error-channels.

Digital computers sometimes experience bit flip errors ($0 \leftrightarrow 1$). The analogous quantum error-channel is also known as the \textit{bit-flip channel}. On a single qubit, this channel acts as

\begin{equation}
\mathcal{\rho} = (1-p) \rho + p X\rho X,
\end{equation} 

\noindent since $X\ket{0} = \ket{1}$ and $X\ket{1} = \ket{0}$. We see that this channel applies $X$ on the qubit with probability $p$ and otherwise leaves the qubit unchanged.

Unlike digital computers, qubits can also experience phase flips, which occur when $Z$ is applied as an error with probability $p$, as well as both phase and bit flips, which occurs when the error is $Y$ instead.

A symmetric application of these three errors ($X$, $Y$, and $Z$) is the often studied channel known as the \text{symmetric depolarizing-channel}, which acts on single qubits as

\begin{equation}
\mathcal{\rho} = (1-p) \rho + \frac{p}{3} (X\rho X + Y\rho Y + Z\rho Z).
\end{equation} 

\noindent This channel can be interpreted as a process that leaves the system untouched with probability $1-p$ and applies either $X$, $Y$, or $Z$ with an equal probability of $p/3$. Note that on two qubits, errors are chosen equally from the set $\{I, X, Y, Z\}^{\otimes 2} \; \setminus \; I\otimes I$.

Finally, the amplitude-damping channel corresponds to qubits spontaneously decaying to the ground-state $\ket{0}$ with probability $p$. The Kraus operators for this channel is

\begin{equation}
E_0 = \sqrt{p} \dyad{0}{1}  \text{ and } E_1 = \dyad{0} + \sqrt{1-p}\dyad{1}.
\end{equation}

Later, in Chapter~\ref{ch.stab_sim}, I discuss a class of quantum error-correcting codes known as stabilizer codes. For those who are curious about the broader field of quantum error correction, see the textbook \cite{LB13_lidar_brun_2013} as well as Preskill's lecture notes \cite{Preskill:1998b}. For a book on how classical errors affect the architecture of modern CMOS processors, see \cite{Mukherjee:2008}.

\section{A Brief Comparison between Classical and Quantum Computation\label{sec:intro:compcompare}}

I started this chapter by first considering the states of digital computers. I now end this chapter by making some comparisons between classical and quantum computation.

In classical (binary) computation, it is known that any Boolean function on $n$ bits can be written as a circuit composed of only NAND gates ($00 \rightarrow 1$, $01 \rightarrow 1$, $10 \rightarrow 1$, and $11 \rightarrow 0$) and fanout operations \cite{Sheffer1913} (assuming bit preparation and readout). That is, $\{\text{NAND}, \text{fanout}\}$ is a universal set of gates for finite-state machines (digital computers) since the composition of gates from this set will take any state $z\in \{0,1\}^n$ to any other state $z'\in \{0,1\}^n$. Note that a NAND gate can be formed from two transistors (see, for example, Chapter 2 of Feynman's lecture notes on computation \cite{Feynman:CompLectures}).

\pagebreak[4]
Given that $\{\text{NAND}, \text{fanout}\}$ is universal for classical computation, one might wonder whether there are simple sets of gates that can be composed to form any quantum circuit. Fortunately, universal sets of quantum gates that allow any unitary in $U(2^n)$ to be efficiently approximated do exist. The ``standard set'' of universal gates is $\{H, S, \text{CNOT}, T\}$ \cite{QCQI} (assuming state preparations and measurements in the computational basis). Other universal gate sets (making the same assumptions) include:

\begin{enumerate}
\item The three-qubit Deutsch gate \cite{Deutsch:1989a}
\item All two-qubit gates \cite{DiVincenzo:1995a}
\item CNOT and all single-qubit $U(2)$ \cite{Barenco:1995a}
\item $H$, $T$, and CNOT \cite{Boykin:1999a}
\item Controlled single-qubit rotation \cite{RG02}
\item Toffoli and $H$ \cite{Shi:2003a,Aharonov03}
\end{enumerate} 

\noindent For proofs and discussions on universal quantum-gate sets, see Chapter 4 of \cite{QCQI} and Chapter 6 of \cite{Preskill:1998b}.

The fanout gate is often implicitly assumed and is rather natural for digital circuits. It is achieved by splitting a wire to produce multiple bits from one. In QC, the analogous operation is copying a general quantum state. The \textit{no-cloning theorem} \cite{Wootters:1982a,Dieks:1982a,Jozsa02} states that no unitary will allow such copying perfectly. That is, for any general state $\ket{\psi}$ and a fiducial state $\ket{\phi}$ (\textit{e.g.}, $\ket{0}$), there is no unitary that takes $\ket{\psi}\otimes\ket{\phi} \rightarrow \ket{\psi}\otimes\ket{\psi}$. If cloning were possible, it can be shown that it would allow for superluminal communication \cite{Dieks:1982a}. For discussions about the no-cloning theorem see Box 12.1 of \cite{QCQI}, Chapter 3 of \cite{Wilde2013}, and Chapter 4 of \cite{Preskill:1998b}.

Another important difference between classical and quantum circuits, which we have already implicitly discussed, is how measurements affect the state of a computer. In digital computers one can measure any bit as many times as one likes and not affect the state of the computer. However, as we have discussed in Section~\ref{sec:qintro:meas}, even for ideal pure states and ideal projective measurements a general measurement will collapse a state onto a subspace. Thus, we must be careful with what measurements we make during the execution of quantum algorithm as measurements \textit{can} result in one loosing information that was encoded in the quantum state. Together, measurement collapse combined with no-cloning means that, in the middle of a quantum computation, one can't just arbitrarily read-off and see what the state is.

However, if done wisely, it is sometimes advantageous to make measurements while running a quantum algorithm. For example, some quantum error-correction protocols enact judicially chosen measurements to digitize error and determine recovery operations (\textit{e.g.}, see Chapter~\ref{ch.stab_sim}). Further, there is a quantum computational model entirely based on measurements known as \textit{measurement-based quantum computation} (MQC or MBQC) \cite{Raussendorf:2001a,Jozsa05meas,Briegel09a,}.

Although measurements collapse quantum states, one may wonder whether it is possible to extract more information out of $n$ qubits than $n$ bits. Holevo's theorem \cite{Holevo1973} proves that this is impossible. One may obtain at most 1 bit of information for every 1 qubit measured. Thus, the notion often heard in popular science reporting that quantum computation is really a form of massively-parallel computation is simplistic, at best. 

The unique state-space and evolutions allowed by quantum mechanics result in quantum computers being able to manipulate probability distributions of states in ways seemingly unavailable to classical computers. Even without access to large-scale reliable quantum computes, researchers have developed numerous quantum algorithms that outperform (sometimes exponentially so) certain best known classical algorithms. In Chapter~\ref{chp:qaoa}, I will discuss a quantum algorithm that can serve as framework for finding approximate solutions to classical combinational satisfaction problems. For a comprehensive collection of quantum algorithms, see the site maintained by Jordan \cite{Jordan:2011AlgoSite}. For introductions to quantum algorithms see \cite{Shor00,MM:NC:2011,Montanaro15,QAlgoBegin}.

Finally, unlike modern digital computers, which are predominately based on complementary metal-oxide-semiconductor (CMOS) technology, there is currently no dominent substrate for quantum computation. Technologies that are currently being explored include ion traps \cite{Blatt2008,SPMIZCS09,HTMMB18}, superconductors \cite{MSS00,Wendin16,HTMMB18}, semiconductors \cite{LD997,ZDMSHKRCE12,GZ18,HTMMB18}, neutral atoms \cite{BCJD98,Jaksch:1999a,MLXSPCIS15}, nitrogen vacancies in diamonds \cite{DCJTMJZHL07,Yao:2012b,childress_hanson_2013}, and optics \cite{KMNRDM05,Barz2014}. Each of these quantum-computing substrates has its own advantages and disadvantages (see the works cited in the previous sentence).

The field of quantum computation is still rapidly evolving. In the following chapters, I present my contributions to the study of quantum algorithms and the development of large-scale, reliable quantum-computing architectures.

\chapter{An Analysis of the Quantum Adiabatic Optimization Algorithm\label{chp:qaoa}}

\setlength\epigraphwidth{14cm}
\epigraph{\textit{``truth...is much too complicated to allow anything but approximations''}}{--- \textup{John von Neumann (more or less) \cite{vonNeumann:1947}}}

Currently, large-scale, fault-tolerant, universally-programmable quantum-computers do not exist. It is, therefore, useful to consider what algorithms one potentially wants to run on noisy intermediate-scale quantum (NISQ) devices \cite{PreskillNISQ} when designing architectures. These algorithms inform the architect about the type of interactions and connectivity that are needed and how difficult it may be to map the desired applications to the constraints of the hardware. Further analysis of the target algorithms and hardware implementation may also reveal the type and degree of noise the computation must be protected against to obtain satisfactory outputs.

The focus of this chapter is on evaluating a quantum algorithm that is of potential interest. More specifically, this chapter presents joint work with Ojas Parekh on analyzing the performance of the Quantum Approximate Optimization Algorithm (QAOA) applied to the Max Cut problem. QAOA was designed to find approximate solutions to classical constraint satisfaction problems (CSPs). QAOA is an algorithm of interest for NISQ hardware since QAOA provides a natural tradeoff between quality of solution and circuit depth. Here we argue that QAOA outperforms the currently known best classical approximation algorithm for the Max Cut problem on certain restricted graphs. 

Several sections of this chapter were originally written by Ojas Parekh; however, I have made the final edit, included additional background information, and rewrote some proofs to be more applicable to other classical problems before specializing the proofs to the problem of Max Cut. A paper that includes the work presented in this chapter, as well as additional results, is currently being prepared. This paper consists of the work by Sevag Gharibian, Ojas Parekh, and myself; however, in this chapter, I have restricted the discussion to work I contributed to, along with Ojas Parekh. My technical contributions include developing the proofs in this chapter with Ojas Parekh.

\section{Introduction}

QAOA was originally defined by Farhi, Goldstone, and Gutmann in 2014 \cite{Farhi:2014a} and is the first known quantum approximation algorithm for classical constraint satisfaction problems (CSPs). Specifically, QAOA is a quantum-algorithmic framework for finding approximate solutions to discrete optimization problems \cite{Farhi:2014a,Farhi:2014b}. QAOA may be viewed as a discretized simulation of adiabatic quantum computation (AQC) and, as with AQC, is a universal model of quantum computation. Farhi \textit{et al.} demonstrated worst-case bounds on the performance of QAOA for Max Cut in 3-regular graphs \cite{Farhi:2014a} and Max-3-XOR \cite{Farhi:2014b} (\textit{i.e.}, each clause is the XOR of 3 Boolean literals). QAOA has since been generalized by others \cite{JRW17,HWGRVB17,MW18}.

In this work, we explore the performance of QAOA and present a closed-form expression for the expectation of QAOA on Max Cut for any graph instance. Using this result, we show that QAOA for Max Cut on $k$-regular triangular-free graphs outperforms the currently best known classical approximation algorithms.

This chapter is organized as follows: In Section~\ref{sec.qaoa.back}, we present background on the QAOA algorithm and the Max Cut problem. In Section~\ref{sec.qaoa.prev}, we discuss the previous results of QAOA on the Max Cut problem. In Section~\ref{sec:qaoa:sampopt}, we make a brief comment on the difference between sampling and optimization. In Section~\ref{sec.qaoa.res}, we discuss the specific results of this work as well as give an overview of the techniques used to derive the results. In Section~\ref{sec.qaoa.analysis}, we give the proof of our results. Finally, in Section~\ref{sec.qaoa.concl} we conclude.

\section{Background\label{sec.qaoa.back}}

In this section, we will introduce CSPs and discuss how one may arrive at QAOA from an approximate simulation of AQC. We focus on the version of QAOA first described by Farhi \textit{et al.} in \cite{Farhi:2014a}.

\subsection{Constraint Satisfaction Problems\label{qaoa.sec.csp}}

Following the definition of a CSPs as defined in other works on QAOA (\textit{e.g.}, see \cite{Farhi:2014a} and \cite{FH16}), we now define the constraint satisfaction problem confined to Boolean function constraints.\footnote{\noindent One can define CSPs more generally, \textit{e.g.}, see \cite{LZ16}.} A CSP is specified by $n$, the size of bit strings to be considered, and a set $\{C_{\alpha}(z)\}$ of $m$ clauses (also known as constraints). Each clause is of the form
 
 \begin{equation}
 C_{\alpha}(z) = 
 \begin{cases}
 1 & \text{if } z \text{ satisfies the clause}\\
 0 & \text{otherwise}, 
 \end{cases}
 \end{equation}

\noindent where for QAOA we let $z \in \{+1, -1\}^n$ (the eigenvalues of Pauli $Z$). Typically, clauses only evaluate a few bits; therefore, we will restrict all $C_{\alpha}(z)$ to evaluate the satisfiability of at most $k$ bits, where $k$ is some fixed integer. In other words, each clause is $k$-local. 

A CSP is solved when a bit string $z$ is found that satisfies as many clauses as possible. To accomplish this goal, we can define an objective function

\begin{equation}
\label{eq:qaoa:objective}
C(z) = \sum_{\alpha} C_{\alpha} (z)
\end{equation}

\noindent and maximize $C(z)$ for all $z$. Since $C(z)$ merely counts the number of satisfied clauses, a $z$ that maximizes $C(z)$ is a solution to the CSP.

The Max Cut problem is an example of a CSP and can be defined as follows: Given a set of vertices $V$ and a set of edges $E$ between the vertices in $V$, Max Cut on a simple graph $G=(V,E)$ is the problem of finding a bi-partition of $V$ that maximizes the number of edges that run between the two partitions. Note that the edge between vertices $i$ and $j$ will be indicated by $ij$. Since we will only consider undirected graphs, $ij$ is equivalent to the set $\{i, j\}$. If we label one partition $0$ and the other $1$, then one choice of objective function for Max Cut is 

\begin{equation}
C(z) = \sum_{ij \in E} C_{ij}(z), 
\end{equation}  

\noindent where 

\begin{equation}
\label{eq:qaoa:classicalclause}
 C_{ij}(z) = \frac{1}{2}(1 - z_iz_j).
\end{equation}

\noindent Here the value of bit $z_u$ indicates which partition vertex $u$ belongs to. On inspection, we see that for an edge $ij$ the clause $C_{ij}(z)$ assigns the value $1$ if $z_i \neq z_j$ and $0$ if $z_i = z_j$. Thus, $C(z)$ is maximized when the number of edges whose vertices are not in the same partition is maximized.

\subsection{Adiabatic Quantum Computation}

QAOA can be described as a discrete approximation of AQC. Arguably, understanding the connection between AQC and QAOA provides useful insight into QAOA. Therefore, in this section we will briefly review AQC.

AQC is an application of the adiabatic approximation for the purposes of quantum computation. Loosely speaking, the adiabatic approximation says that if a state is an instantaneous eigenstate of a time-dependent Hamiltonian, then the state will remain in the instantaneous eigenstate of the Hamiltonian as long as the evolution of the Hamiltonian is slow enough and there is always an energy gap between the instanteous eigenstate and the rest of eigenstates \cite{Born:1928a}. There are numerous proofs of varying rigor and refinements of the adiabatic approximation. Many of these proofs as well as a review of AQC are given in \cite{AL16} by Albash and Lidar.

AQC is defined by two time-independent Hamiltonians $H_D$ (the ``driver Hamiltonian'') and $H_P$ (the ``problem Hamiltonian'').  These Hamiltonians are chosen to be $k$-local and, therefore, can be written as

\begin{equation}
\label{eq:qaoa:localham}
H = \sum_{\substack{K \subseteq [n]\\|K|\leq k}} H_K,
\end{equation}  

\noindent where $[n]\equiv \{1, 2, \cdots, n\}$ and each $H_K$ only acts on the qubits indexed by the set $K$. 

The Hamiltonian $H_D$ is the initial Hamiltonian of the computation and is chosen so that the ground state of $H_D$ is easy to prepare. The Hamiltonian $H_P$ is the final Hamiltonian and is chosen so that the ground state of $H_P$ encodes the solution to a problem of interest.

Inspired by the adiabatic approximation, AQC is performed by initializing in the (not necessarily unique) ground state of $H_D$ and adiabatically interpolating to $H_P$. The state then remains in the ground state throughout the evolution with high probability. After the evolution, the final state is measured in the computational basis to determine the encoded solution as defined by $H_P$. 

The interpolation is given by time-dependent Hamiltonian

\begin{equation}
\label{eq.hinterp}
H(t) = (1-s(t)) H_{D} + s(t) H_{P},
\end{equation}

\noindent where $s(t)$ is a smooth function of time $t$ such that $s(t=0)=0$ and $s(t=T)=1$. Here $T$ is the time interval during which the interpolation between the driver and problem Hamiltonian occurs. The function $s(t)$ is often referred to as the \textit{schedule} of the computation.

Note, assuming an AQC interpolation $H(t)$ given in Eq.~\ref{eq.hinterp} between two time-independent Hamiltonians $H_D$ and $H_P$, a necessary condition for a finite gap to exist during a non-trivial computation is that $H_D$ and $H_P$ must not commute \cite{qcstackWoot}. As described by the simultaneous diagonalization theorem (see Theorem 2.2 of \cite{QCQI}), if $[H_D, H_P]=0$ and both $H_D$ and $H_P$ act non-trivially over an entire joint Hilbert space, then $H_D$ and $H_P$ share the same set of energy eigenvectors. Therefore, assuming the two Hamiltonians are not the same, $H_D$ and $H_P$ assign different energy eigenvalues to at least some of the energy eigenvectors. Assuming the initial ground-state is not the solution to the problem, which would make the computation effectively trivial, the final ground state is not the same as the initial ground state. Since the initial ground-state is an eigenstate of $H_D$ and the final eigenstate is an eigenstate of $H_P$, then for the computation to be non-trivial, the energy eigenvalues of the two states must cross during the computation. Thus, there is no gap if $[H_D, H_P]=0$ and the computation is non-trivial. 

Different formulations of the adiabatic theorem place different constraints on the schedule $s(t)$ such as the derivatives of the function. To meet the assumptions of the adiabatic approximation, $T$ is often chosen to be $T = \mathcal{O}(1/g^2_{\text{min}})$,  where $g_{\text{min}}$ is the minimum energy gap between the ground state and the first excited-state, during the schedule for times on this order, the Euclidean norm between the actual final ground-state and the ideal ground-state of $H_P$ (the ``error'') can be made arbitrarily small \cite{AL16}. Selecting a version of the adiabatic theorem in order to specify the constraints on $s(t)$ and $T$ that will ensure a target error is unnecessary to understand the connection between AQC and QAOA. For a discussion on AQC's robustness to experimental error, see \cite{Childs:2001a}.

AQC has been proven to be a quantum computation model that is universal \cite{ADKLLR04}. A version of AQC that is restricted to solving optimization problems is known as the quantum adiabatic algorithm (QAA or, alternatively, QADI---for \textbf{Q}uantum \textbf{ADI}abatic algorithm) \cite{Farhi:2000a,Farhi:2001a}. We will now consider QAA.

\vfill
\pagebreak[4]
Given a CSP, it is relatively simple to construct a problem Hamiltonian. The a clause $C_{\alpha}(z)$ as defined by Eq.~\ref{eq:qaoa:objective} can be described by a matrix

\begin{equation}
\label{eq.qaoa.calpha1}
C_{\alpha} \equiv \sum_{z\in \{0,1\}^n} C_{\alpha}(z) \dyad{z},  
\end{equation}

\noindent in the computational basis.

Restricting each clause $C_{\alpha}(z)$ to act on at most $k$ bits and recalling that $\dyad{0} = \frac{1}{2}\left(I+Z\right)$ and $\dyad{1} = \frac{1}{2}\left(I-Z\right)$, we can re-express Eq.~\ref{eq.qaoa.calpha1} as

\begin{equation}
\label{eq.qaoa.calpha2}
C_{\alpha} = \sum_{\substack{K\subseteq [n]\\|K|\leq k}} W_{\alpha, K} Z^K,  
\end{equation}

\noindent where $W_{\alpha, K} \in \mathbb{R}$ and we define 

\begin{equation}
M^K \equiv \tens{j\in K}M_j.
\end{equation}

\noindent Here $M_j$ is a $2 \text{x} 2$ matrix acting on qubit $j$ and it is assumed that $M^K$ acts as $I$ on any qubit not indexed by $K$. Also, we define $M^{\varnothing} \equiv I$.

Each clause $C_{\alpha}(z)$ can be considered a Boolean function $f:\{0,1\}^n \rightarrow \{0,1\}$. The conversion of Boolean functions to operators of the form defined in Eq.~\ref{eq.qaoa.calpha2} is described in \cite{Biamonte:2008a} and \cite{Hadf2018}.

\pagebreak[4]
From  Eq.~\ref{eq:qaoa:objective}, Eq.~\ref{eq.qaoa.calpha2}, and Eq.~\ref{eq.qaoa.calpha1}, it follows that the matrix or operator form of the objective function in the computational basis is

\begin{equation}
\label{eq:qaoa:C}
C \equiv \sum_{\alpha} C_{\alpha} = \sum_{z\in \{0,1\}^n} C(z) \dyad{z} = \sum_{\substack{K\subseteq [n]\\|K|\leq k}} W_{K} Z^K,
\end{equation}

\noindent where $W_{K}=\sum_{\alpha} W_{\alpha, K}$ and $C(z)$ is given in Eq.~\ref{eq:qaoa:objective}.

Since

\begin{equation}
C^{\dagger} = \sum_{z\in \{0,1\}^n} C(z)^* \dyad{z}^{\dagger} = \sum_{z\in \{0,1\}^n} C(z) \dyad{z} = C,
\end{equation}

\noindent $C$ is Hermitian and could be identified as a Hamiltonian. However, solutions to the CSP defined by $C$ corresponds to eigenvectors with the maximum eigenvalue. Since AQC is normally defined to track the ground state, for QAA we take 

\begin{equation}
\label{eq:qaoa:HP}
H_P = - C = - \sum_{\substack{K\subseteq [n]\\|K|\leq k}} W_{K} Z^K.
\end{equation} 

Also, for QAA problems, $H_D$ is taken to be 

\begin{equation}
\label{eq:qaoa:HD}
H_D = -B,
\end{equation}

\noindent where

\begin{equation}
\label{eq:qaoa:B}
B \equiv \sum_{j \in [n]} X_j.
\end{equation}

\pagebreak[4]
\noindent The ground state of $H_D$, in which we initialize the state of our computation, is 

\begin{equation}
\label{eq:qaoa:sstate}
\ket{s} \equiv \ket{+}^{\otimes n} = \frac{1}{2^{n/2}}\sum_{z\in \{0,1\}^n} \ket{z}. 
\end{equation}

The minimum energy gap between the ground state and first excited state is unknown for general AQC problems; however, since the $H(s)$ as defined by Eq.~\ref{eq:qaoa:HP} and Eq.~\ref{eq:qaoa:HD} is stoquastic (a Hamiltonian with off-diagonal elements only in $\mathbb{R}_{\leq 0}$), QAA always has a non-vanishing gap, which is assured by the Perron-Frobenius theorem \cite{Perron1907,Frobenius1912}.

If we consider Max Cut on a simple graph $G=(V,E)$ as defined in Section~\ref{qaoa.sec.csp}, then the matrix form of a clause $C_{ij}(z)$ as described by Eq.~\ref{eq:qaoa:classicalclause} is given as

\begin{equation}
 C_{ij} = \dyad{01}_{ij}+\dyad{10}_{ij} = \frac{1}{2}(I - Z_iZ_j).
\end{equation}

Thus, $C$ for Max Cut is

\begin{equation}
\label{eq.Cmaxcut}
C = \sum_{ij\in E} \dyad{01}_{ij}+\dyad{10}_{ij} = \sum_{ij\in E} \frac{1}{2}(I - Z_iZ_j).
\end{equation}

\subsection{Discretizing Adiabatic Quantum Computation\label{sec:qaoa:discaqc}}

The continuous evolution of Hamiltonians can be simulated by sequences of discrete, $k$-local unitaries for fixed $k$ \cite{Feynman:1982a,Lloyd:1996a}. One algorithm used to find such simulations is commonly known as ``Trotterization.'' It is through the Trotterization of AQC that we present the connection between QAOA and AQC.

\pagebreak[4]
We start by considering Trotterization since the AQC Hamiltonian is time-dependent and generally the Hamiltonian will not commute with itself at different points in time, the unitary induced by AQC is Eq.~\ref{eq:hamtimdep1}, that is $U(T, 0) = \Tau \text{exp}\{-i \int_0^T H(t) \;  dt\}$. Besides being potentially intractable to calculate, this AQC unitary may be difficult to experimentally implement. We can begin to simplify the unitary by first noting that we can break it up as follows:

\begin{align}
\label{eq:qaoa:allunitaries}
U(T,0) & = U(T, T - \Delta t) U(T - \Delta t, T - 2 \Delta t) \cdots U(\Delta t, 0) \nonumber \\
& = \prod^p_{j=1} U(j \Delta t, (j-1) \Delta t),
\end{align}

\noindent where $p \Delta t = T$.

Then, if we choose $\Delta t$ to be small enough so that $H(t)$ is approximately constant over the time interval $[(j-1) \Delta t, j \Delta t]$ for all $1 \leq j \leq p$, we can use the time-independent unitary solution $U(t, t_0) = \text{exp}\{-i H \left(t-t_0\right)\}$ (Eq.~\ref{eq:intro:hamindep}) to approximate the unitaries in Eq.~\ref{eq:qaoa:allunitaries} as 

\begin{equation}
\label{eq:qaoa:approxsingleunitary}
U(j \Delta t, (j-1) \Delta t) \approx e^{-i \; H(j \Delta t) \; \Delta t}.
\end{equation}

\noindent Applying Eq.~\ref{eq:qaoa:approxsingleunitary} to  Eq.~\ref{eq:qaoa:allunitaries}, we now write

\begin{equation}
\label{eq:Htimedep}
U(T,0) \approx \prod^p_{j=1}  e^{-i \; H(j \Delta t) \; \Delta t}. 
\end{equation}

\noindent Note, for a version of Trotterization that accounts for time-dependence, see \cite{PQSV11}.

Given Eq.~\ref{eq:Htimedep}, We can further simplify the unitary sequence by recalling that the AQC Hamiltonian (Eq.~\ref{eq:qaoa:localham}) has the form $H=\sum_{\substack{K\subseteq [n]\\|K|\leq k}} H_K$  and that the Lie-Trotter product formula \cite{Trotter59} is

\begin{equation}
\label{eq:Trotter}
e^{i \left(A+B\right) x} = \lim_{p\rightarrow \infty} \left(e^{iA x/p}e^{iB x/p}\right)^p,
\end{equation}

\noindent where $x \in \mathbb{C}$ and the matrices $A$ and $B$ are real or complex.

Suzuki derived recursive formulas for calculating approximations of Eq.~\ref{eq:Trotter} \cite{Suzuki91,HS05}, known as Lie-Trotter-Suzuki decompositions. It is common to use the lower order decompositions

\begin{equation}
\label{eq:qaoa:simpletrotter}
e^{i\left(A+B\right) x} = e^{i A x}e^{i B x} + \mathcal{O}(x^2)
\end{equation}

\noindent or

\begin{equation}
e^{i \left(A+B\right)x} = e^{i B x/2}e^{i A x}e^{i B x/2} + \mathcal{O}(x^3)
\end{equation}

\noindent for simulating Hamiltonians.

Combining Eq.~\ref{eq:Htimedep} and the lower-order approximation of Eq.~\ref{eq:qaoa:simpletrotter}, we arrive at

\begin{equation}
\label{eq:qaoa.HamApprox}
U(T,0) \approx \prod_{j=1}^p \prod_{\substack{K\subseteq [n] \\ |K|\leq k}} e^{-i \; H_K \left(j\Delta t \right) \; \Delta t}.
\end{equation}

\pagebreak[4]
\noindent We now have an approximation of the AQC unitary consisting of unitaries that each act on at most $k$ qubits. If necessary, we might further simplify an instance of Eq.~\ref{eq:qaoa.HamApprox} by decomposing the unitaries $e^{-i \; H_K \left(j\Delta t \right) \; \Delta t}$ into sequences of simpler gates.

Note, for an analysis of the error of Trotterizations described by Eq.~\ref{eq:qaoa.HamApprox} and higher-order decompositions, see \cite{WBHS08}. For a discussion of the stability of the Lie-Trotter-Suzuki decomposition due to imprecisions in applying unitaries, see \cite{DS14,KM15}. 

If we now turn to approximating the QAA Hamiltonian defined by the $H_P$ of Eq.~\ref{eq:qaoa:HP} and the $H_D$ of Eq.~\ref{eq:qaoa:HD} with the approximate unitary given by Eq.~\ref{eq:qaoa.HamApprox}, we get

\begin{equation}
\label{eq:qaoa:qaaform}
U(T,0) \approx \prod_{j=1}^p  \text{exp}\{i \; [1-s(j\Delta t)] \; \Delta t \; B \} \; \text{exp}\{i \; s(j\Delta t) \; \Delta t \; C \}.
\end{equation}

\subsection{Quantum Approximate Optimization Algorithm}

The number of steps $p$ needed to adequately approximate a QAA computation with Eq.~\ref{eq:qaoa:qaaform} may be large in general. However, rather than finding an exact solution to a CSP, one may be satisfied with finding a high-quality solution that obtains a large fraction of the maximum number of satisfiable clauses. One may then be curious to what degree one can reduce the quality of a unitary approximation of QAA and still produce adequate results. This question naturally leads to the formulation of QAOA.

\pagebreak[4]
Given a CSP that is specified by an objective operator $C$ (defined by Eq.~\ref{eq:qaoa:C}), QAOA generates states of the form 

\begin{equation}
\ket{\boldsymbol{\gamma},\boldsymbol{\beta}} \equiv U_{\text{QAOA}_p}(\boldsymbol{\gamma},\boldsymbol{\beta}) \ket{s},
\end{equation}

\noindent where $\boldsymbol{\gamma}\equiv (\gamma_1, \gamma_2, \cdots, \gamma_p)$ and $\boldsymbol{\beta}\equiv (\beta_1, \beta_2, \cdots, \beta_p)$ are vectors of angles, the state $\ket{s}$ is the equal superposition of the computational basis-states as defined in Eq.~\ref{eq:qaoa:sstate}, and

\begin{equation}
\label{eq:qaoa:qaoaunip}
U_{\text{QAOA}_p}(\boldsymbol{\gamma},\boldsymbol{\beta}) \equiv \prod_{j=1}^p  e^{-i \; \beta_j B } e^{-i \; \gamma_j C }.
\end{equation} 

\noindent Here $B$ is the sum of Pauli $X$s as defined in Eq.~\ref{eq:qaoa:B}, and by $\text{QAOA}_p$ we mean the subclass of QAOA where a fixed $p$ specifies the number of $e^{-i \; \beta_j \; B } e^{-i \; \gamma_j \; C }$ products present in $U_{\text{QAOA}_p}$.

We see that if we identify $\gamma_j$ as $-s(j\Delta t) \; \Delta t$ and $\beta_j$ as $-[1-s(j\Delta t)] \; \Delta t$, then the $\text{QAOA}_p$ unitary Eq.~\ref{eq:qaoa:qaoaunip} is equivalent to the QAA unitary Eq.~\ref{eq:qaoa:qaaform} for the same value of $p$. While the QAA and QAOA unitaries are of similar form and both the angles and $p$ may be chosen so that they are equal, there are two differences.

First, since QAA is an AQC algorithm, QAA must turn the problem of finding a state that maximizes $C$ to finding a ground state that minimizes $-C$. However, QAOA approximately simulates a process that instead initializes in an eigenstate with the \textit{maximum} eigenvalue of the driver Hamiltonian and remains in the instantaneous eigenstate that has the \textit{maximum} eigenvalue of the interpolating Hamiltonian $H(t) = (1-s(t))H_D + s(t)H_P$, where for QAOA we let $H_D=B$ and $H_P=C$. While this is inconsistent with ground-state AQC, the adiabatic theorem merely requires an energy gap between the instantaneous state and the rest of the energy spectrum. 

The other distinction to note between the QAA and QAOA unitaries is that for QAOA, typically $p$ is chosen to be some small integer. Currently, $p=1$ or $p=2$ is frequently considered. 

To compensate for such an extremely crude approximation, Farhi \textit{et al.} \cite{Farhi:2014a} showed that a hybrid classical and quantum gradient-search algorithm can be used to efficiently choose the optimal values for $\boldsymbol{\gamma}$ and $\boldsymbol{\beta}$ given a fixed $p$, where optimal means obtaining the value  

\begin{equation}
M_p = \max_{\boldsymbol{\gamma},\boldsymbol{\beta}} \ev{C}{\boldsymbol{\gamma},\boldsymbol{\beta}}.
\end{equation}

Given a CSP, the goal of QAOA obtain a high approximation ratio 

\begin{equation}
\alpha_p \equiv \frac{M_p}{C_{\text{max}}},
\end{equation}

\noindent where   

\begin{equation}
C_{\text{max}} \equiv \max_{z \in \{0,1\}^n} C(z).
\end{equation}

Note, for a given class of problems, $C_{\text{max}}$ is not always know \textit{a priori}.

Once good angles are found, the state $\ket{\boldsymbol{\gamma},\boldsymbol{\beta}}$ is prepared and measured in the computational basis until a measurement output $z$ is found such that $C(z) \geq M_p$.

\pagebreak[4]
If desired, the approximation ratio can be increased by raising $p$ to obtain better solutions, since 

\begin{equation}
\lim_{p \rightarrow \infty} M_p = C_{\text{max}}.
\end{equation}

\noindent This is because as $p$ tends to infinity, then $U_{\text{QAOA}_p}$ can at the very least simulate QAA and, due to the Perron-Frobenius theorem, QAA will in principle be able to find the solution to the CSP (although $T$ might be \textit{very} large). Further,

\begin{equation}
M_{p} \geq M_{p-1}
\end{equation} 

\noindent since $U_{\text{QAOA}_p} = e^{-i \; \beta_p \; B } e^{-i \; \gamma_p \; C} U_{\text{QAOA}_{p-1}}$ implies that finding optimal angles for $p-1$ means that we can at least obtain $M_p = M_{p-1}$ since we can set $\gamma_p, \beta_p = 0$. Therefore, $M_p$ monotonically increases.

$\text{QAOA}$ then provides a natural trade-off between depth of the resulting quantum circuit and the quality of solution produced. While we may increase $p$ to produce better solutions, precise analysis of $M_p$ versus $p$ poses a challenge. However, Wang \textit{et al.} \cite{WHJR17} obtained an analytic expression for $\text{QAOA}_p$ for Max Cut on a simple cycle.

\section{Previous Work\label{sec.qaoa.prev}}

Previously, Farhi \textit{et al.}'s analysis of $\text{QAOA}_1$ for Max-3-XOR in 2014 showed that $\text{QAOA}_1$ is expected to satisfy at least $\left(\frac{1}{2}+\frac{\mathcal{O}(1)}{d^{3/4}}\right)m$  clauses on instances with $m$ clauses where each variable occurs in at most $d$ of them. This surprising quantum result improved upon a longstanding classical approximation algorithm by H{\aa}stad \cite{Hastad:2000} from 2000, guaranteeing only at least $\left(\frac{1}{2}+\frac{\mathcal{O}(1)}{d}\right)m$ clauses. In 2015 Barak \textit{et al.} \cite{BMDRRSTVWW} subsequently gave a classical randomized approximation algorithm for Max-$k$-XOR with an expected performance of at least $\left(\frac{1}{2}+\frac{\mathcal{O}(1)}{d^{1/2}}\right)m$  clauses. Farhi \textit{et al.} \cite{Farhi:2014b} soon thereafter provided an improved analysis showing that $\text{QAOA}_1$ matches this bound within a $\mathcal{O}(\ln(d))$ factor, \textit{i.e.}, $\left(\frac{1}{2}+\frac{\mathcal{O}(1)}{d^{1/2} \ln(d)}\right)m$. Substantial improvements in these results are unlikely since Trevisan showed in 2001 that there is a constant $c > 0$ such that a $\left(\frac{1}{2}+\frac{c}{d^{1/2}}\right)$-approximation is NP-hard \cite{Trevisan2001}.

For Max Cut in 3-regular graphs, $\text{QAOA}_1$ is a 0.6924-approximation \cite{Farhi:2014a}; however, Halperin, Livnat, and Zwick \cite{Halperin2002} give a classical 0.9326-approximation based upon an improved SDP relaxation. Yet it is still possible that $\text{QAOA}_1$ may outperform classical algorithms on specific instances of 3-regular Max Cut.

\section{A Comment on Sampling versus Optimization\label{sec:qaoa:sampopt}}

Comparing the performance of QAOA against classical algorithms requires a bit of care. Farhi and Harrow showed that $\text{QAOA}_1$, for $C$ corresponding to a classical 2-local CSP, can produce quantum states that are hard to sample from classically \cite{FH16}. However, QAOA is foremost an optimization algorithm, and as such the most relevant performance measure is $\ev{C}{\gamma,\beta}$, as we seek a state that maximizes $C$. Thus, although a classical algorithm may not be able to sample from $\ket{\gamma,\beta}$, it can still outperform QAOA as an optimization algorithm if its performance exceeds $\ev{C}{\gamma,\beta}$. 

To further illustrate the distinction between sampling and optimization in the context of QAOA, observe that for $C$ defined above corresponding to Max Cut, we may multiply any term $\frac{1}{2}(I-Z_iZ_j)$ by $(1+2\pi l/\gamma)$ for an integer $l$ without altering the resulting state $\ket{\gamma,\beta}$. This observation can be used to show that the states used by Farhi and Harrow to establish the classical hardness of sampling from QAOA correspond to a multitude of weighted optimization problems, including trivial ones.

\section{Results\label{sec.qaoa.res}}

We now summarize the results of our analysis found later in Section~\ref{sec.qaoa.analysis}. Note, in our analysis we consider QAOA's simplest form ($\text{QAOA}_1$):

\begin{align}
\label{qaoa.qn.def1}
\ket{\gamma,\beta} = e^{-i\beta B} e^{-i \gamma C}\ket{s}.
\end{align}

\begin{theorem}[Shown formally by Theorem~\ref{qaoa.thm.4}.]
\label{qaoa.thm.1}
For any 3-regular graph, there is a deterministic linear-time classical algorithm that delivers at least as large a cut as $\text{QAOA}_1$ is expected to.
\end{theorem} 

On the other hand, we show:

\begin{theorem}[Shown formally by Theorem~\ref{qaoa.thm.3}.]
\label{qaoa.thm.2}
The expected number of edges cut by $\text{QAOA}_1$ on a $k$-regular triangle-free graph with $m$ edges is

\begin{align}
\expval{C} = \displaystyle \left( \frac{1}{2}+\frac{1}{2\sqrt{k}} \left( 1-\frac{1}{k}\right) ^{\frac{k-1}{2}} \right) m \geq \left( \frac{1}{2}+\frac{0.3032}{\sqrt{k}}\right) m.
\end{align}
\end{theorem}

This improves upon classical randomized algorithms developed by Hirvonen, Rybicki, Schmid, and Suomela \cite{HRSS14} in 2014, with performance $(\frac{1}{2}+\frac{0.2812}{\sqrt{k}})m$, and by Shearer \cite{Shearer92} in 1992, with performance $(\frac{1}{2}+\frac{0.177}{\sqrt{k}})m$. 

\pagebreak[4]
Our results here are enabled by a new exact closed-form expression for the expectation of $\text{QAOA}_1$ on any Max Cut instance, $G = (V,E)$. The expectation for an edge $ij \in E$ is

\begin{align}
\label{eq:qaoa:closedform}
\expval{C_{ij}} & = \displaystyle \frac{1}{2} - \frac{1}{4} \sin^2(2\beta) \cos (\gamma)^{\sigma_i+\sigma_j-2(n_{ij}+1)}(1-\cos(2\gamma)^{n_{ij}}) \nonumber \\
\displaystyle & + \frac{1}{4}\sin (4\beta) \sin(\gamma)\left[ \cos(\gamma)^{\sigma_i-1}+\cos(\gamma)^{\sigma_j-1}\right]
\end{align}

\noindent (the proof of which is given in Lemma~\ref{qaoa.lemma.5}), where $\sigma_u$ is the degree of a vertex $u \in V$, and $n_{ij}$ are the numbers of common neighbors of $i$ and $j$, which is also the number of triangles in $G$ containing the edge $ij$. The parameters $\beta$ and  $\gamma$ are specified as input to $\text{QAOA}_1$, and one typically chooses them to maximize the expectation. We give precise values for $\beta$ and $\gamma$ that maximize the expectation of $\text{QAOA}_1$ on $k$-regular triangle-free instances of Max Cut. Our formula helps shed light on the features of a graph that influence the performance of $\text{QAOA}_1$. We note that the above result was obtained simultaneously and independently by Wang \text{et al.} \cite{WHJR17}, who applied it in a different context.

\section{Analysis of QAOA for Max Cut\label{sec.qaoa.analysis}}

For Max Cut on general graphs we obtain an exact closed-form expression for the expectation of
$\text{QAOA}_1$. Since the objective function of a classical CSP can be written as a sum of tensor products of Pauli $Z$s, we start our analysis by considering the following lemma. 

\pagebreak[4]
\begin{lemma}
\label{qaoa.lemma.general}
For $\ket{\gamma,\beta}$ as defined as in Eq.~\ref{qaoa.qn.def1} with $C$ as defined in Eq.~\ref{eq:qaoa:C} and $B=\sum_{j\in [n]} X_j$ as defined in Eq.~\ref{eq:qaoa:B},

\begin{align}
\label{eq.gen.exp}
& \displaystyle \ev{Z^K}{\gamma,\beta} = \nonumber \\
& \ev{Z^K \left[\sum_{L \subseteq K} (i)^{|L|} \cos(2\beta)^{|K|-|L|} \sin(2\beta)^{|L|} X^{L} \prod_{\substack{M \subseteq [n]\\ |M \cap L| \text{ is odd}}}\exp(-2 i\gamma W_M Z^M)\right]}{s}
\end{align}

\begin{proof}

First note

\begin{align}
\ev{Z^K}{\gamma,\beta} &= \ev{e^{i\gamma C}e^{i \beta B} Z^K e^{-i \beta B}e^{-i\gamma C}}{s} \nonumber \\
&= \ev{e^{i\gamma C} e^{i \beta B} \left[ \tens{u\in K} Z_u \right] e^{-i \beta B} e^{-i\gamma C}}{s} \nonumber \\
&= \ev{e^{i\gamma C} \left[ \tens{u\in K} e^{i \beta B} Z_u e^{-i \beta B} \right] e^{-i\gamma C}}{s} \nonumber \\
&= \ev{e^{i\gamma C} \left[ \tens{u\in K} Z_u e^{-2 i \beta X_u} \right] e^{-i\gamma C}}{s} \nonumber \\
&= \ev{e^{i\gamma C} \left[ \tens{u\in K} Z_u \right] \left[\tens{u\in K} e^{-2 i \beta X_u} \right] e^{-i\gamma C}}{s} \nonumber \\
&= \ev{Z^K e^{i\gamma C} \left[ \tens{u\in K} e^{-2 i \beta X_u} \right] e^{-i\gamma C}}{s},
\end{align}

\noindent since $e^{-i\beta X_u}$ and $Z_i$ commute if $u\neq i$ and anticommute otherwise. Also, both $Z^K$ and $e^{-i\gamma C}$ commute since both are diagonal matrices. Continuing, we have

\begin{align}
\renewcommand{\arraystretch}{1.5}
& \ev{Z^K e^{i\gamma C} \left[ \tens{u\in K} e^{-2 i \beta X_u} \right] e^{-i\gamma C}}{s} = \nonumber \\
& \ev{Z^K e^{i\gamma C} \left[ \tens{u\in K} \left(\cos(2\beta)-i X_u \sin(2\beta) \right) \right] e^{-i\gamma C}}{s} = \nonumber \\
& \ev{Z^K e^{i\gamma C} \left[ \sum_{L \subseteq K} \left((i)^{|L|} \cos(2\beta)^{|K|-|L|}\sin(2\beta)^{|L|} X^L \right) \right] e^{-i\gamma C}}{s} = \nonumber \\
& \ev{Z^K \left[ \sum_{L \subseteq K} (i)^{|L|} \cos(2\beta)^{|K|-|L|}\sin(2\beta)^{|L|} e^{i\gamma C}X^L e^{-i\gamma C} \right]}{s} = \nonumber \\
& \bra{s}Z^K \left[ \sum_{L \subseteq K} (i)^{|L|} \cos(2\beta)^{|K|-|L|}\sin(2\beta)^{|L|} \right. \nonumber \\
& \left. \left(\prod_{M \subseteq [n]} e^{i\gamma W_M Z^M}\right) X^L \left(\prod_{M \subseteq [n]} e^{-i\gamma W_M Z^M}\right) \right]\ket{s} = \nonumber \\
& \ev{Z^K \left[\sum_{L \subseteq K} (i)^{|L|} \cos(2\beta)^{|K|-|L|}\sin(2\beta)^{|L|} X^{L} \prod_{\substack{M \subseteq [n]\\ |M \cap L| \text{ is odd}}}\exp(-2 i\gamma W_M Z^M)\right]}{s},
\end{align}

\noindent where the identity $e^{i \theta A} = I \cos(\theta)+ i A \sin(\theta)$ (assuming $A^2 = I$ for matrix $A$) was used. Note that if $M = \varnothing$, then $|M \cap L| = 0$; thus, $M = \varnothing$ is not evaluated in the product. 
\end{proof}
\end{lemma}

\pagebreak[4]
Restricting the claim of Lemma~\ref{qaoa.lemma.general} to the case of the Max Cut problem results in the following lemma.

\begin{lemma}
\label{qaoa.lemma.5}

Let $\sigma_i$ be the degree of a vertex $i \in V$, and let $n_{ij}$ be the number of common neighbors of $i$ and $j$. For $\ket{\gamma,\beta}$ as defined as in Eq.~\ref{qaoa.qn.def1} with $C=\sum_{ij\in E} \frac{1}{2}(I-Z_iZ_j)$ and $B=\sum_{i\in V} X_i$,

\begin{align}
\displaystyle \ev{Z_iZ_j}{\gamma,\beta} & = \frac{1}{2}\sin^2(2\beta)\cos(\gamma)^{\sigma_i+\sigma_j-2(n_{ij}+1)}(1-cos(2\gamma)^{n_{ij}}) \nonumber \\
\displaystyle & - \frac{1}{2} \sin(4\beta)\sin(\gamma)\left(\cos(\gamma)^{\sigma_i-1}+\cos(\gamma)^{\sigma_j-1}\right),
\end{align}

\noindent when $ij \in E$.

\end{lemma}

\begin{proof}
Note that Eq.~\ref{eq.Cmaxcut} can be written as 

\begin{equation}
C = \frac{m}{2}I - \sum_{ij\in E} \frac{1}{2}Z_iZ_j.
\end{equation}

\noindent We then first consider evaluating $\ev{Z_iZ_j}{\gamma,\beta}$. Applying Eq.~\ref{eq.gen.exp} of Lemma~\ref{qaoa.lemma.general} to $\ev{Z_iZ_j}{\gamma,\beta}$ for Max Cut, we get

\begin{align}
\displaystyle \ev{Z_iZ_j}{\gamma,\beta} & = \bra{s}\cos^2(2\beta) Z_iZ_j  \prod_{\substack{uv \in E\\ |uv \cap \varnothing | \text{ is odd}}}\exp(i\gamma Z_u Z_v) \\
&- \sin(2\beta)\cos(2\beta) Z_iY_j \prod_{\substack{uv \in E\\ |uv \cap \{j\}| \text{ is odd}}}\exp(i\gamma Z_u Z_v) \\
&- \sin(2\beta)\cos(2\beta) Y_iZ_j \prod_{\substack{uv \in E\\ |uv \cap \{i\}| \text{ is odd}}}\exp(i\gamma Z_u Z_v) \\
&+ \sin^2(2\beta) Y_iY_j \prod_{\substack{uv \in E\\ |uv \cap \{i, j\}| \text{ is odd}}}\exp(i\gamma Z_u Z_v)\ket{s}, 
\end{align}

\noindent where we have used the identity $Y=iXZ$ to simplify the expression. 

To simplify further derivations, we defining the edge set

\begin{equation}
\Delta_S \equiv \{uv \in E | \; |\{u,v\}\cap S| \text{ is odd }\} \text{ where } S\subseteq V.
\end{equation}

\noindent For clarity, Fig.~\ref{fig:graphs} depicts examples of edges sets $\Delta_{\{i\}}$ and $\Delta_{\{i,j\}}$.

\begin{figure}[H]
	\begin{subfigure}[t]{4.5cm}
		\includegraphics[width=4.3cm]{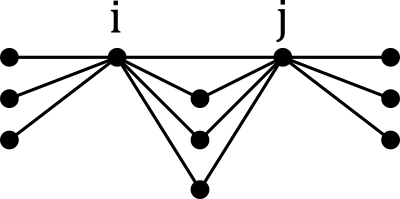}
		\caption{\label{fig:graph1} Example Graph}
	\end{subfigure}
	\hfill
	\begin{subfigure}[t]{4.5cm}
		\includegraphics[width=4.3cm]{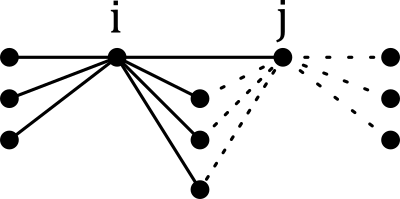}
		\caption{\label{fig:graph2} Edges in $\Delta_{\{i\}}$}
	\end{subfigure}
	\hfill
		\begin{subfigure}[t]{4.5cm}
		\includegraphics[width=4.3cm]{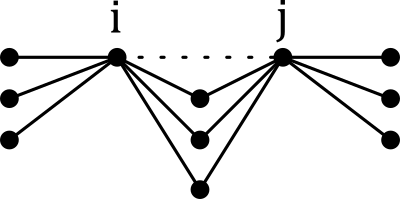}
		\caption{\label{fig:graph3} Edges in $\Delta_{\{i,j\}}$}
	\end{subfigure}
\caption{\label{fig:graphs} Given the graph (a), the solid edges in (b) are the set of edges in $\Delta_{\{i\}}$ and the solid edges in (c) are the set of edges in $\Delta_{\{i,j\}}$. That is, (a) and (b) are graphical depictions of the set of edges ($\Delta_S$) that are incident on the corresponding set of vertices ($S$) an odd number of times.}
\end{figure}

We then obtain

\begin{align}
\label{qaoa.eq.16}
\displaystyle \ev{Z_iZ_j}{\gamma,\beta} & = \bra{s}\cos^2(2\beta) Z_iZ_j \nonumber \\
&- \sin(2\beta)\cos(2\beta) Z_iY_j \prod_{uv \in \Delta_{\{j\}}}\exp(i\gamma Z_uZ_v) \nonumber \\
&- \sin(2\beta)\cos(2\beta) Y_iZ_j \prod_{uv \in \Delta_{\{i\}}}\exp(i\gamma Z_uZ_v) \nonumber \\
&+ \sin^2(2\beta) Y_iY_j \prod_{uv \in \Delta_{\{i,j\}}}\exp(i\gamma Z_uZ_v)\ket{s}. 
\end{align}

\noindent Note, the first term drops out since $\ev{Z_iZ_j}{s} = 0$ for $i \neq j$. 

\pagebreak[4]
We next derive closed-form expressions for $\ev{Y_iY_j\left[\prod_{uv \in \Delta_{\{i,j\}}}e^{(i\gamma Z_uZ_v)}\right]}{s}$ and $\ev{Z_iY_j\left[\prod_{uv \in \Delta_{\{j\}}}e^{(i\gamma Z_uZ_v)}\right]}{s}$; an expression $\ev{Y_iZ_j\left[\prod_{uv \in \Delta_{\{i\}}}e^{(i\gamma Z_uZ_v)}\right]}{s}$ will follow from the analysis of the latter.

We consider $\ev{Z_iY_j\left[\prod_{uv \in \Delta_{\{j\}}}e^{(i\gamma Z_uZ_v)}\right]}{s}$ first. Letting 

\begin{equation}
\alpha_F \equiv (i\sin(\gamma))^{|F|}\cos(\gamma)^{|\Delta_{\{j\}}|-|F|}
\end{equation} 

\noindent for $F \subseteq \Delta_{\{j\}}$, we have

\begin{align}
\label{qaoa.eq.18}
\displaystyle \ev{Z_iY_j\left[\prod_{uv\in \Delta_{\{j\}}}e^{i\gamma(Z_uZ_v)}\right]}{s} &= \ev{Z_iY_j\left[\prod_{uj\in \Delta_{\{j\}}}e^{i\gamma(Z_jZ_u)}\right]}{s} \nonumber \\
& \displaystyle = \ev{Z_iY_j\prod_{uj\in \Delta_{\{j\}}}(\cos(\gamma)I+i\sin(\gamma)Z_jZ_u)}{s} \nonumber \\
 & \displaystyle = \ev{Z_iY_j\sum_{F\subseteq \Delta_{\{j\}}}\alpha_F \prod_{uj\in F}Z_jZ_u}{s} \nonumber \\
& \displaystyle = \sum_{F\subseteq \Delta_{\{j\}}}\alpha_F \ev{Z_iY_j \prod_{uj\in F}Z_jZ_u}{s} \nonumber \\
& \displaystyle = \sum_{F\subseteq \Delta_{\{j\}}}\alpha_F \Tr \left( \left( \otimes_{i \in V} \frac{1}{2}(I+X_i)\right)Z_iY_j \prod_{uj\in F}Z_jZ_u \right).
\end{align}

\noindent Note, here we have used the fact that $\dyad{+}=(I+X)/2$ to replace $\dyad{s}$ with $\otimes_{i \in V} \frac{1}{2}(I+X_i)$. 

The operator within the trace is a sum of tensor product of Pauli operators on each qubit, hence the trace is proportional to the coefficient of the $I$ term. The only way to obtain such a term is when $F$ is chosen so that it contains an odd number of edges incident upon each of $i$ and $j$, and an even number of edges incident upon all other vertices. The only such $F \subseteq \Delta_{\{j\}}$ is $F=\{\{i,j\}\}$.

In other words, it is a necessary condition for non-vanishing terms that the product $\prod_{uj\in F}Z_jZ_u$ yields the value $Z_iZ_j$ to cancel the $Z_iZ_j$ introduced by $Z_iY_j$ in the expression in the last line of Eq.~\ref{qaoa.eq.18}. Given that sum on $F$ is restricted to $F\subseteq \Delta_{\{j\}}$ (\textit{e.g.}, see Fig.~\ref{fig:graph2}), the sufficient condition that $\prod_{uj\in F}Z_jZ_u$ yields $Z_iZ_j$ is only met for $F=\{\{i,j\}\}$. Terms where $F \neq \{\{i,j\}\}$ necessarily vanish.

Restricting the sum to $F=\{\{i,j\}\}$ leads to

\begin{align}
& \displaystyle \sum_{F\subseteq \Delta_{\{j\}}}\alpha_F \Tr \left( \left( \otimes_{i\in V} \frac{1}{2}(I+X_i)\right)Z_iY_j \prod_{uj\in F}Z_jZ_u \right) \nonumber \\
& = \displaystyle \alpha_{\{ij\}} \Tr \left( \left( \otimes_{i\in V} \frac{1}{2}(I+X_i)\right)Z_iY_j Z_jZ_i \right) \nonumber \\
& = \displaystyle \alpha_{\{ij\}} \Tr \left( \left( \otimes_{i \in V} \frac{1}{2}(I+X_i)\right)iX_j \right) \nonumber \\
& = i \alpha_{\{ij\}} \nonumber \\
& = -\sin(\gamma)\cos(\gamma)^{|\Delta_{\{j\}}|-1}.
\end{align}

The above, in conjunction with Eq. \ref{qaoa.eq.18}, yields

\begin{align}
\label{qaoa.eq.19}
\ev{Z_iY_j\left[\prod_{uv\in \Delta_{\{j\}}}e^{i\gamma(Z_uZ_v)}\right]}{s} = -\sin(\gamma)\cos(\gamma)^{|\Delta_{\{j\}}|-1}.
\end{align}

The following may be derived analogously.

\begin{align}
\label{qaoa.eq.20}
\ev{Y_iZ_j\left[\prod_{uv\in \Delta_{\{i\}}}e^{i\gamma(Z_uZ_v)}\right]}{s} = -\sin(\gamma)\cos(\gamma)^{|\Delta_{\{i\}}|-1}.
\end{align}

\vfill
\pagebreak[4]
It remains to analyze $\ev{Y_iY_j\left[\prod_{uv\in \Delta_{\{i,j\}}}e^{i\gamma(Z_uZ_v)}\right]}{s}$. Following the derivation of Eq. \ref{qaoa.eq.18}:

\begin{align}
\label{qaoa.eq.21}
& \ev{Y_iY_j\left[\prod_{uv\in \Delta_{\{i,j\}}}e^{i\gamma(Z_uZ_v)}\right]}{s} \nonumber \\
& \displaystyle = \ev{Y_iY_j\prod_{uv\in \Delta_{\{i,j\}}}(\cos(\gamma)I+i\sin(\gamma)Z_uZ_v)}{s} \nonumber \\
& \displaystyle = \sum_{F\subseteq\Delta_{\{i,j\}}}\alpha_F\ev{Y_iY_j\prod_{uv\in F}Z_uZ_v}{s} \nonumber \\
& \displaystyle = \sum_{F\subseteq\Delta_{\{i,j\}}}\alpha_F \Tr ((\otimes_{i\in V}\frac{1}{2}(I+X_i))Y_iY_j\prod_{uv\in F}Z_uZ_v).
\end{align}

\noindent As with our previous derivation, the only terms of the sum that do not vanish are those for which  $F \subseteq \Delta_{\{ij\}}$ (for reference see, \textit{e.g.}, Fig.~\ref{fig:graph3}) contains an odd number of edges incident upon each of $i$ and $j$, and an even number of edges incident upon all other vertices. Or, in other words, non-vanishing terms only arise when the product $\prod_{uv\in F}Z_uZ_v$ yields the value $Z_iZ_j$ to cancel the $Z_iZ_j$ introduced by $Y_iY_j$. Note that $ij \notin \Delta_{\{i, j\}}$; however, any path of length two from $i$ to $j$ results in a non-vanishing term in the sum. So, for example, the product $Z_iZ_a$ times $Z_aZ_j$ (\textit(i.e.,) $F = \{\{i,a\}, \{a, j\}\}$) would survive as it would yield $Z_iZ_j$. Moreover, a union of an odd number of such paths also yields a non-vanishing term. Let $W_{ij} \subseteq V$ be the set of common neighbors of both $i$ and $j$ (\textit{e.g.}, see Fig.~\ref{fig:graphwedge}). Then each path of length two between $i$ and $j$, $(ik; kj)$, is in one-to-one correspondence with a $k \in W_{ij}$. Finally, observe that $\alpha_F$ only depends on $|F|$; we will use $\alpha_{l}$ as shorthand for an $\alpha_F$ with $|F|=l$.

\begin{figure}[H]
\centering
		\includegraphics[width=4.3cm]{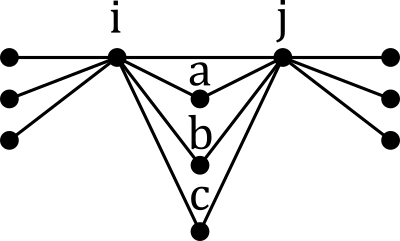}
\caption{\label{fig:graphwedge} An example of $W_{ij}$. Here $W_{ij} = \{a, b, c\}$. That is, $W_{ij}$ is the set of vertices that are common neighbors to $i$ and $j$.}
\end{figure}

We consequently have

\begin{align}
& \displaystyle \sum_{F\subseteq\Delta_{\{i,j\}}}\alpha_F \Tr \left(\left(\otimes_{i\in V}\frac{1}{2}\left(I+X_i\right)\right)Y_iY_j\prod_{uv\in F}Z_uZ_v\right) \nonumber \\
& = \displaystyle \sum_{\substack{U\subseteq W_{ij}: \\ |U| \text{ odd}}}\alpha_{2|U|} \Tr \left(\left(\otimes_{i\in V}\frac{1}{2}\left(I+X_i\right)\right)i X_iX_j\right) \nonumber \\
& \displaystyle = - \sum_{\substack{U\subseteq W_{ij}: \\ |U| \text{ odd}}} \alpha_{2|U|} = - \sum_{\substack{1\leq l \leq |W_{ij}|: \\ \text{l odd}}} \binom{|W_{ij}|}{l} \alpha_{2l} \nonumber \\
& \displaystyle = - \sum_{\substack{1\leq l \leq |W_{ij}|: \\ l \text{ odd}}} \binom{|W_{ij}|}{l}(i \sin(\gamma))^{2l} \cos(\gamma)^{|\Delta_{i,j}|-2l} \nonumber \\
& \displaystyle = \cos(\gamma)^{|\Delta_{\{i,j}|-2|W_{ij}|} \sum_{\substack{1\leq l \leq |W_{ij}|: \\ \text{l odd}}} \binom{|W_{ij}|}{l}\left(\sin^2(\gamma)\right)^{l} \left(\cos^2(\gamma)\right)^{|W_{ij}|-l} \nonumber \\
& \displaystyle = \cos(\gamma)^{|\Delta_{\{i,j}|-2|W_{ij}|} \frac{1-\left(\cos^2(\gamma)-\sin^2(\gamma)\right)^{|W_{ij}|}}{2} \nonumber \\
& \displaystyle= \frac{1}{2}\cos(\gamma)^{|\Delta_{\{i,j}|-2|W_{ij}|}\left(1-\cos(2\gamma)^{|W_{ij}|}\right),
\end{align}

\noindent where, as mentioned, we use $\alpha_{l}$ as shorthand for an $\alpha_F$ with $|F|=l$ and the penultimate equality follows from the identity 

\begin{equation}
\sum_{l \text{ odd}} \binom{n}{l} q^l p^{n-l} = ((p+q)^n - (p-q)^n)/2.
\end{equation}

\noindent By Eq.~\ref{qaoa.eq.21}, we have

\begin{align}
\label{qaoa.eq.22}
\ev{e^{-i\gamma C}Y_iY_je^{i\gamma C}}{s} = \frac{1}{2} \cos(\gamma)^{|\Delta_{\{i,j\}}|-2|W_{ij}|}\left(1-\cos(2\gamma)^{|W_{ij}|}\right).
\end{align}

Combining Eq.~\ref{qaoa.eq.16} with Equations \ref{qaoa.eq.19}, \ref{qaoa.eq.20}, and \ref{qaoa.eq.22} yields the claim of the lemma, since $ij\in E$ and $|\Delta_{\{i,j\}}|=\delta_i + \delta_j - 2$ (the number of edges shared between $i$ and $j$ not including the edge $ij$). Recall that $\sigma_u$ is the degree of vertex $u$ and $|W_{ij}| = n_{ij}$ is the number of common neighbors between $i$ and $j$.
\end{proof}

\begin{corollary}
\label{qaoa.corol.11}
The expected number of edges cut by $\text{QAOA}_1$ on a $k$-regular graph with $m$ edges is,

\begin{align}
& \expval{C} = \nonumber \\ 
& \left(\frac{1}{2}-\frac{1}{4}\sin^2(2\beta) \cos(\gamma)^{2k-2(n_{ij}+1)}\left(1-\cos(2\gamma)^{n_{ij}}\right)+\frac{1}{2}\sin(4\beta)\sin(\gamma)\cos(\gamma)^{k-1}\right)m.
\end{align}
\end{corollary}

\begin{proof}
This follows from Lemma~\ref{qaoa.lemma.5} since 

\begin{equation}
\ev{C}{\gamma,\beta} = \frac{m}{2}-\frac{1}{2}\sum_{ij\in E} \ev{Z_iZ_j}{\gamma,\beta}.
\end{equation}
\end{proof}

\begin{theorem}
\label{qaoa.thm.3}
We may set $\beta$ and $\gamma$ so that the expected number of edges cut by $\text{QAOA}_1$ on a $k$-regular triangle-free graph with $m$ edges is

\begin{align}
\expval{C} = \displaystyle \left(\frac{1}{2} + \frac{1}{2\sqrt{k}}\left(1-\frac{1}{k}\right)^{\frac{k-1}{2}}\right) m \geq \left( \frac{1}{2} + \frac{0.3032}{\sqrt{k}}\right) m.
\end{align} 
\end{theorem}
\begin{proof}
Since $n_{ij} = 0$ for a triangle-free graph, Corollary~\ref{qaoa.corol.11} yields

\begin{align}
\displaystyle \ev{C}{\gamma,\beta} = \left( \frac{1}{2} + \frac{1}{2}\sin(4\beta)\sin(\gamma)\cos(\gamma)^{k-1}\right) m.
\end{align}

\noindent We select $\beta = \pi/8$ to maximize the above. The quantity $\sin(\gamma)\cos(\gamma)^{k-1}$ is maximized when $\sin(\gamma)=\frac{1}{\sqrt{k}}$ and $\cos(\gamma)=\sqrt{\frac{k-1}{k}}$ so that,

\begin{align}
\displaystyle \ev{C}{\gamma,\beta} &= \left(\frac{1}{2}+\frac{1}{2\sqrt{k}}\left(1-\frac{1}{k}\right)^{\frac{k-1}{2}}\right) m \nonumber \\
&\geq \left(\frac{1}{2}+\frac{1}{2\sqrt{e}}\frac{1}{\sqrt{k}}\right) m \geq \left(\frac{1}{2}+\frac{0.3032}{\sqrt{k}}\right)m.
\end{align}

\noindent where the first inequality follows since for $k\geq 1$, $(1-1/k)^{k-1}\geq 1 / e$ (see, \textit{e.g.}, page 435 of Ref.~\cite{MR95}).
\end{proof}

\begin{lemma}
\label{qaoa.lemma.6}
For a 3-regular graph with $m$ edges, the expectation of $\text{QAOA}_1$ is at most,

\begin{align}
\displaystyle \left(\frac{1}{2}+\frac{1}{3\sqrt{3}}\right) m \leq 0.6925 m.
\end{align}
\end{lemma}

\begin{proof}
For a $k$-regular graph, we observe that the $\frac{1}{4} \sin^2(2\beta)\cos(\gamma)^{2k-2(n_{ij}+1)}(1-\cos(2\gamma)^{n_{ij}})$ term from Corollary~\ref{qaoa.corol.11} is non-negative, hence 

\begin{align}
\ev{C}{\gamma,\beta} \leq \left(\frac{1}{2}+\frac{1}{2}\sin(4\beta)\sin(\gamma)\cos(\gamma)^{k-1}\right)m.
\end{align} 

\noindent We obtain the result by using the values of $\beta$ and $\gamma$ from the proof of Theorem~\ref{qaoa.thm.3} for $k = 3$.
\end{proof}

\begin{theorem}
\label{qaoa.thm.4}
For any 3-regular graph, there is a linear-time deterministic classical algorithm that delivers at least as large a cut as $\text{QAOA}_1$ is expected to.
\end{theorem}

\begin{proof}
Locke \cite{Locke82} gives a deterministic algorithm that yields a cut with at least $\frac{7}{9}m$ edges in any 3-regular graph except $K_4$. By Lemma~\ref{qaoa.lemma.6}, Locke's algorithm outperforms $\text{QAOA}_1$ on 3-regular graphs, except possibly $K_4$. The most time-consuming step of Locke's algorithm for 3-regular graphs is finding a Brooks' coloring, which can be done in linear time \cite{Skulra02}.
\end{proof} 

\section{Conclusion\label{sec.qaoa.concl}}

In conclusion, we analytically evaluated the performance of $\text{QAOA}_1$ on the Max Cut problem. In so doing, we developed a closed-form expression for the expectation of $\text{QAOA}_1$ for Max Cut on any simple graph (see Lemma~\ref{qaoa.lemma.5}). Restricting $\text{QAOA}_1$ on Max Cut to $k$-regular triangle-free graphs with $m$ edges, we found the optimal values for the angles $\beta$ and $\gamma$. These angles allowed us to determine that for $k$-regular triangle-free graphs $\ev{C} = \left( \frac{1}{2}+\frac{1}{2\sqrt{k}} \left( 1-\frac{1}{k}\right) ^{\frac{k-1}{2}} \right) m \geq \left( \frac{1}{2}+\frac{0.3032}{\sqrt{k}}\right) m$. These results were obtained simultaneously and independently by Wang \textit{et al.} \cite{WHJR17}; however, our derivation differs from theirs and may provide further insight into future analyses of QAOA. We observe that our analysis of performance of $\text{QAOA}_1$ on Max Cut for $k$-regular triangle-free graphs improves upon the currently known best classical approximation algorithm for these graphs for $k > 3$, which was developed by Hirvonen \textit{et al.} \cite{HRSS14} in 2014. This classical randomized algorithm has a performance of $\overline{C(z)} = (\frac{1}{2}+\frac{0.2812}{\sqrt{k}})m$.

\chapter{Improved Stabilizer-Simulation for Topological Stabilizer Codes\label{ch.stab_sim}}

\setlength\epigraphwidth{14cm}
\epigraph{\textit{``Science, my lad, has been built upon many errors; but they are errors which it was good to fall into, for they led to the truth.''}}{--- \textup{Jules Verne \cite{AJTTCOE}}}

While some quantum algorithms may have a degree of resilience to experimental noise and find application in noisy intermediate-scale quantum (NISQ) devices, to realize the full potential of quantum computing we will need to employ fault-tolerant quantum-computing protocols \cite{PreskillNISQ}. To this end, in this chapter, I introduce stabilizer codes, a class of quantum error-correcting codes (QECCs). I also review stabilizer simulation algorithms and present a new algorithm with improved runtime for simulations of topological stabilizer codes. I developed this new algorithm while optimizing the software package \PECOS, which will be discussed in Chapter~\ref{ch.pecos}. 

This chapter was written in preparation to publish a paper on this work, to which I am the sole author of the paper. Thus, all the technical work is my own.

\section{Introduction}

Large-scale quantum computers have the potential to solve certain problems that are currently intractable. As intensive experimental efforts to build such devices have demonstrated \cite{IonColor,RPHBVSD14,SuperCon,GNRMB14,TCEABCCG16,Woot16,ExLif16,DLFLWM16,Monz1068}, qubits are significantly more fragile than their classical counterparts. Due to this fragility, it is likely that large-scale quantum computers will need to employ fault-tolerant quantum computing (FTQC) protocols to reach acceptable levels of reliability. While at the end of the design process, a quantum computer architect ideally would run FTQC protocols, directly on a target device to characterize performance, during the development of FTQC protocols classical simulation is a useful tool. Classical simulation can provide a controlled environment to allow a quantum architect to verify designs and to gain insight into vulnerabilities by subjecting protocols to specific error models.

Currently, the time and/or space complexity of classical simulations of general quantum systems grow exponentially in the size of the system being simulated. Better classical simulation complexities can be achieved for quantum circuits formed from restricted sets of quantum operations. For example, the Gottesman-Knill theorem \cite{GotKnill} states that for \textit{stabilizer circuits}, which are circuits containing only Clifford gates and Pauli-basis initializations and measurements, each operation can be simulated in polynomial time. In fact, the algorithm associated with the Gottesman-Knill theorem \cite{GotQC,Gottesman:1997a,GotKnill} has a worst-case time-complexity of only $\mathcal{O}(n^3)$ per operation, where $n$ is the number of qubits being simulated. Conveniently, many protocols, including numerous quantum error correcting codes (QECCs) \cite{Shor:1995a,Steane96,Bacon05,SurfCode,Bombin:2006b,TriangleSurf}, are implemented using stabilizer circuits. 

\vfill
\pagebreak[4]
More efficient stabilizer-circuit simulation algorithms were developed by Aaronson and Gottesman in \cite{CHP} and later by Anders and Briegel in \cite{GraphSim}. These algorithms were implemented as the programs \pack{CHP} and \pack{GraphSim}, respectively. The time complexities often reported for these works are $\mathcal{O}(n^2)$ and $\mathcal{O}(n \log{n})$, respectively. Here $\mathcal{O}(n \log{n})$ was argued as a typical complexity expected when simulating QECC protocols, entanglement purification, and other practical applications.

In this work, I will present an amortized runtime analysis of the Aaronson-Gottesman and Anders-Briegel algorithms. This analysis shows that for topological codes, a class of QECCs that have been a target for implementation in the near-term, both algorithms achieve an average runtime per operation that scales as $\mathcal{O}(n)$. I will also present results of numerical experiments that demonstrate that implementations of these algorithms have operations that run on average close to linear time. I will further introduce a new algorithm, which was inspired by the Aaronson-Gottesman algorithm, that has average operation-runtimes that scale as $\mathcal{O}(\sqrt{n})$ for topological-code protocols. I will also present an implementation of this new algorithm (called \pack{SparseSim}) and show a result demonstrating near $\mathcal{O}(\sqrt{n})$ average runtime per operation is achieved. As the new algorithm is designed to take advantage of sparsity in stabilizer/destabilizer representations, similar speed improvements for the simulation of other quantum low-density parity-check (LDPC) codes are expected.

\begin{figure}[H]
	\centering
	\includegraphics[width=0.98\textwidth]{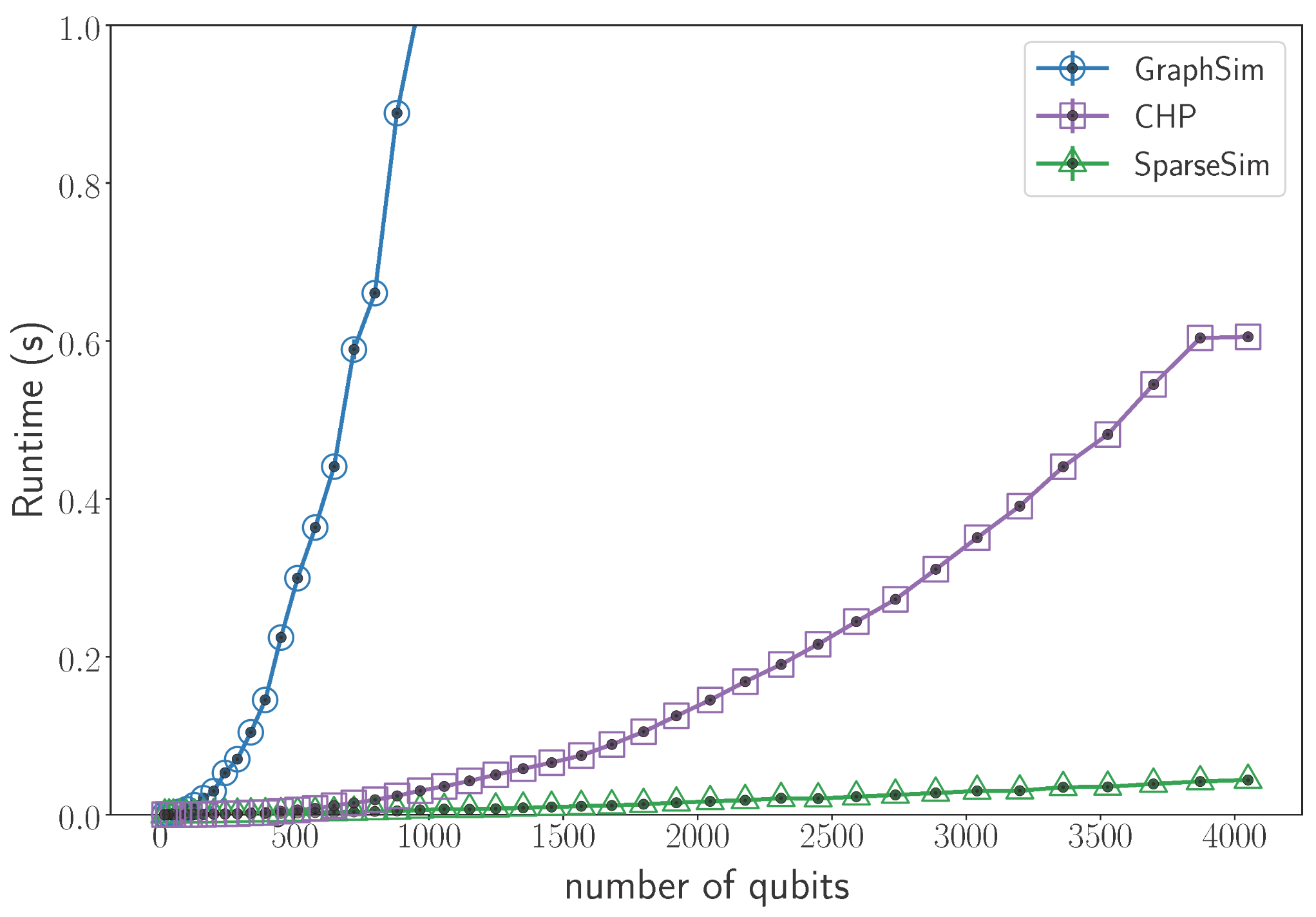}	
	\caption{Medial surface-code \cite{Bombin:2007d}: Average runtime to simulate a single round of syndrome extraction on a length-$n$ code. Error bars represent one standard deviation from Monte Carlo sampling; they are hardly perceptible on this plot. Here, \pack{CHP} is the implementation of the Aaronson-Gottensman algorithm, \pack{GraphSim} is the implementation of the Anders-Briegel, and \pack{SparseSim} is the implementation of the new algorithm.}
	\label{fig:TotalRuntimeOfMedialSurfaceCode}
\end{figure}

\begin{figure}[H]
	\centering
	\includegraphics[width=0.98\textwidth]{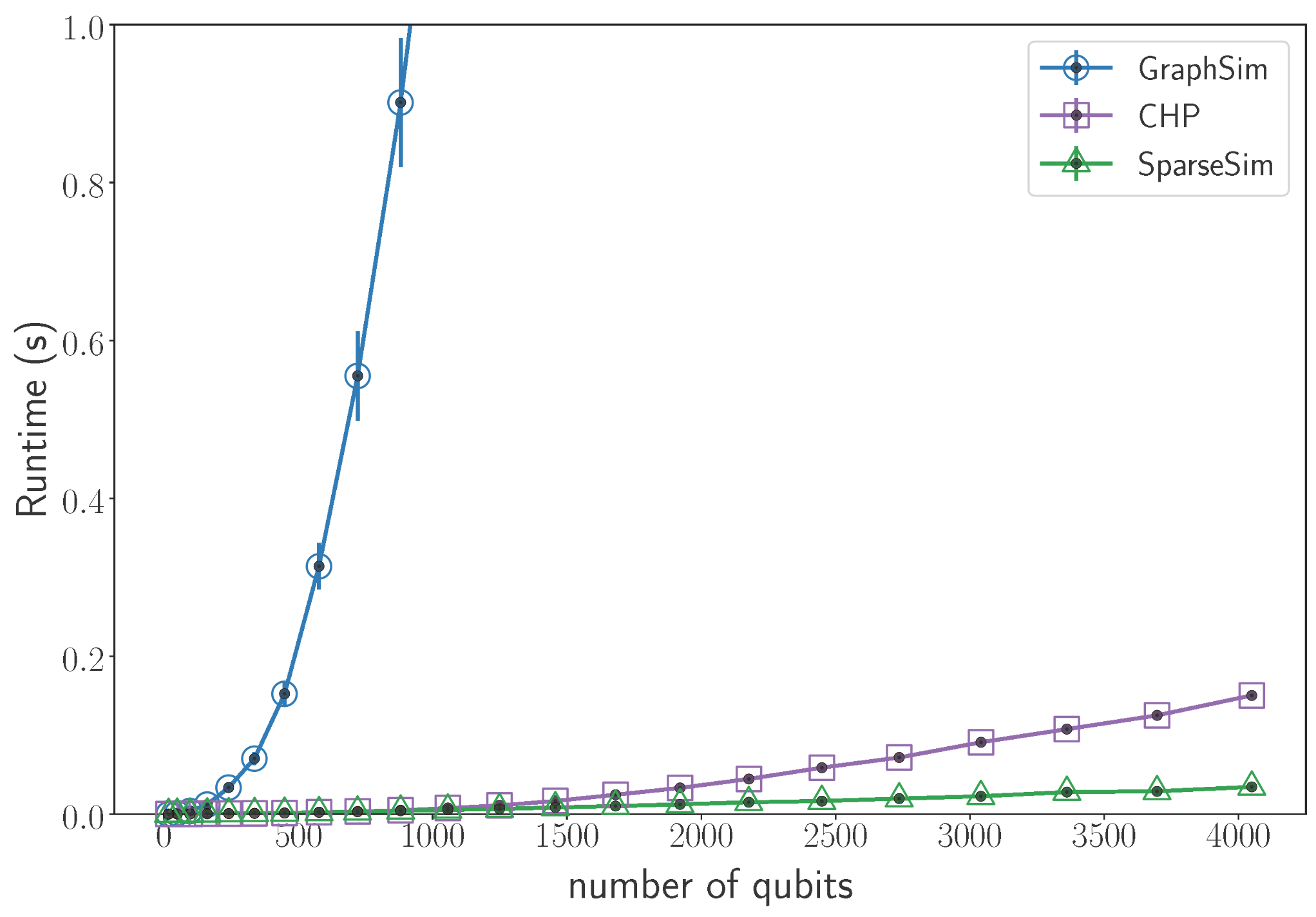}	
	\caption{\label{fig:runtimeColor}4.8.8 Color code \cite{Bombin:2006b}: Average runtime to simulate a single round of syndrome extraction. Error bars represent one standard deviation from Monte Carlo sampling. Here, \pack{CHP} is the implementation of the Aaronson-Gottensman algorithm, \pack{GraphSim} is the implementation of the Anders-Briegel, and \pack{SparseSim} is the implementation of the new algorithm.}
\end{figure}

As a preview, Figures \ref{fig:TotalRuntimeOfMedialSurfaceCode} and \ref{fig:runtimeColor} show the runtime for simulating one round of syndrome extraction for the 2D medial surface-code \cite{Bombin:2007d} and 2D color code \cite{Bombin:2006b} as a function of code length, respectively.

This chapter is organized as follows: I begin by reviewing the stabilizer formalism in Section~\ref{sec:stab-form}. I then discuss previous stabilizer simulation algorithms in Section~\ref{sec:prev-sims}. There, I present tighter time complexities for circuit operations than had previously been presented for these algorithms. In Section~\ref{sec:our-alg}, I present a new algorithm that takes advantage of the sparsity found in stabilizer/destabilizer representations. In this section, I present pseudo-code for this algorithm. In Section~\ref{sec:topo-complex}, I consider complexities expected when simulating syndrome extraction for topological stabilizer codes and quantum LDPC codes with similar sparsity. I present my implementation of the new algorithm in Section~\ref{numer-exp}. In this section, I also present the results of numerical experiments demonstrating that the predicted complexities for topological stabilizer codes are achieved. Finally, in Section~\ref{sec:concl}, I conclude.

\vfill

\section{\label{sec:stab-form}Stabilizer Formalism}

To facilitate discussions of stabilizer-circuit simulations, I begin by introducing the stabilizer formalism. 

\subsection{Stabilizer States, Groups, and Generators}

A unitary operator $S$ is said to be a stabilizer of a state $\ket{\psi}$ iff $S\ket{\psi} = \ket{\psi}$. That is, the state $\ket{\psi}$ is a $+1$ eigenstate of $S$. The set of all stabilizers that mutually stabilizes a state forms a group under matrix multiplication known as a \textit{stabilizer group}. It is straightforward to show that the group axioms hold for a complete set of stabilizers. Closure of the group is guaranteed since if $S_i$ and $S_j$ are both stabilizers of a state $\ket{\psi}$, then $S_i S_j$ is also a stabilizer of the state since $\ket{\psi} = S_i\ket{\psi} = S_i S_j\ket{\psi}$. The inverse $S^{-1}$ of a stabilizer $S$ is contained in the stabilizer group since $S^{-1}\ket{\psi}=S^{-1}S\ket{\psi} = \ket{\psi}$. The identity element is just the identity matrix, which stabilizes any state. Finally, associativity trivially holds as stabilizers can be represented as matrices.

Not every subgroup of $U(2^n)$ is a stabilizer group. The elements of a stabilizer group must commute, \textit{i.e.}, a stabilizer group is an Abelian group. This is required for stabilizers to mutually stabilize a state. Also, stabilizer groups may not contain $-I$ since no state can be stabilized by $-I$ given $-I\ket{\psi} = \ket{\psi} \Leftrightarrow \ket{\psi}=0$.

Rather than listing all the elements of a stabilizer group, a stabilizer group $\mathcal{S}$ can be more compactly represented by a subset of stabilizers $\mathscr{S} \subseteq \mathcal{S}$ that group multiply to generate $\mathcal{S}$. That $\mathscr{S}$ generates $\mathcal{S}$ is expressed as $\mathcal{S} = \langle \mathscr{S}_1, \cdots, \mathscr{S}_g  \rangle$, where $\mathscr{S} = \{\mathscr{S}_1, \cdots, \mathscr{S}_g\}$. While any generating subset of stabilizers can be used, when I refer to stabilizer generators I will assume that the generating set contains the minimal number of elements needed to generate the group. Note, that such a minimal generating set is not unique since elements of the generating set can be multiplied together to form other generating sets of the same size.

\pagebreak[4]
A stabilizer group $\mathcal{S}$ identifies a joint $+1$ eigenspace called a \textit{stabilizer subspace}. Such a stabilizer subspace $\mathcal{C}$ can be defined as

\begin{equation}
\mathcal{C} = \expval{\{ \ket{\psi} \in \mathscr{H} \; | \; \ket{\psi} = S_i \ket{\psi} \; \forall S_i \in \mathcal{S} \}}_{\mathbb{C}},
\end{equation}

\noindent where $\mathscr{H}$ is a Hilbert space and $\expval{\cdot}_{\mathbb{C}}$ indicates a span over the complex numbers. When $\mathcal{S}$ stabilizes a unique state rather than a span of states, this state is a \textit{stabilizer state}. 

The central idea of stabilizer simulations is to turn to the Heisenberg picture \cite{GotKnill, Messiah:1999a} and use stabilizer generators to represent states, since for certain restricted classes of stabilizer states we can use far fewer stabilizer generators than the $2^n$ amplitudes needed to represent general states. This is the case for stabilizer groups that are subsets of the Pauli group. As we will see, Pauli stabilizer groups need only $n$ generators to represent $n$-qubit stabilizer-states, and these states are exactly those that are prepared and evolved by stabilizer circuits. 

\subsubsection{Pauli Group and Stabilizers}

As the scope of this work is the simulation of stabilizer circuits, I will now restrict future discussions (unless otherwise stated) to only Pauli-stabilizer groups, and for brevity I will often drop the adjective Pauli when referring to Pauli stabilizers. 

The Pauli group on $n$ qubits, $\mathcal{P}_n$, can be written as

\begin{equation}
\mathcal{P}_n = i^k \{I, X, Y, Z\}^{\otimes n} \text{ and } k \in \mathbb{Z}_4,
\end{equation}

\noindent where $I$, $X$, $Y$, and $Z$ are the usual single-qubit Pauli matrices. A stabilizer group $\mathcal{S}$ is a Pauli-stabilizer group iff $\mathcal{S} \subseteq \mathcal{P}_n$. 

Note that a stabilizer group cannot have elements in the set

\begin{equation}
\mathcal{P}^{\pm i}_n = \pm i \{I, X, Y, Z\}^{\otimes n}
\end{equation}

\noindent since if $P\in \mathcal{P}^{\pm i}_n$, then $P^2 = -I$. The phase $\pm i$ is only needed to preserve group closure and make the Pauli group a group \cite{Cliff}.

Each generator $\mathscr{S}_i$ of a stabilizer generating set is an independent and commuting Pauli operator. Each independent Pauli operator $\mathscr{S}_i$ divides the Hilbert space orthogonally in half (with one half associated with $\expval{\mathscr{S}_i} = +1$ and the other, associated with $\expval{\mathscr{S}_i} = -1$). If there are $g$ generators, then the dimension of the corresponding stabilizer space is $2^{n-g}$. A stabilizer state is then uniquely identified when $g = n$.    

\subsection{\label{sec:stab-rules}Stabilizer Update Rules}

So far, we have discussed that stabilizer generators can be used to represent states. To simulate the dynamics of stabilizer circuits we must also know how to identify new generating sets to represent the state as it evolves. I will now review how unitaries, initializations, and measurements modify stabilizer generators. 

\subsubsection{Unitary Evolution}

If a unitary $U$ is applied to a state $\ket{\psi}$ and $S$ is a stabilizer of $\ket{\psi}$, then $U\ket{\psi} = US\ket{\psi} = (USU^{\dagger}) U\ket{\psi}$. Thus, if the stabilizer generators $\mathscr{S}$ stabilizes the state $\ket{\psi}$, then after $U$ is applied to $\ket{\psi}$ the stabilizer generators $U\mathscr{S} U^{\dagger}$ stabilize $U\ket{\psi}$. Since conjugation by unitaries is a homomorphism of the group---that is the product of any two stabilizers $S_i S_j$ is mapped by unitary conjugation by a unitary $U$ as $US_i S_j U^{\dagger} = (US_i U^{\dagger}) (US_j U^{\dagger})$---the algebraic structure of the stabilizer group is preserved. This means that a stabilizer group will remain a stabilizer group under unitary conjugation; however, in general, conjugation by a unitary will not keep the stabilizer group a subset of $\mathcal{P}_n$. We will discuss the class of unitaries that does so next.

As mentioned previously, stabilizer circuits are composed of Clifford gates as well as initializations and measurements in a Pauli basis. The Clifford gates are the subgroup of the unitaries map Pauli operators to Pauli operators through conjugation. More formally the Clifford group on $n$ qubits can be written as

\begin{equation}
\mathscr{C}_n = \{C \in U(2^n) \mid C\mathcal{P}_n C^{-1} = \mathcal{P}_n\}. 
\end{equation}

\noindent Therefore, the set of all Clifford conjugations can be thought of as the set of all permutations of the Pauli group that are allowed by unitary maps. For this reason, Clifford gates are exactly the set of all the unitaries that map Pauli stabilizer groups to any other Pauli stabilizer group.

A useful fact to note is that the Clifford group can be generated by the gates

\begin{align}
H &= \dyad{+}{0} + \dyad{-}{1} \\
S &= \dyad{0}{0} + i\dyad{1}{1} \\
CNOT & = \dyad{0}{0} \otimes I + \dyad{1}{1} \otimes X,
\end{align}

\noindent as discussed in \cite{Gottesman:1997a}, where $\ket{\pm} = \frac{1}{\sqrt{2}}(\ket{0}\pm\ket{1})$.

\subsubsection{Initialization\label{sec:stab-init}}

It is a common practice to initialize quantum circuits in the state $\ket{0}^{\otimes n}$. This practice is continued in stabilizer-simulation algorithms discussed in this paper. The state $\ket{0}$ is stabilized by $Z$. Thus, initializing in $\ket{0}^{\otimes n}$ corresponds to starting with the generating set $\mathcal{Z}_n = \{Z_1, \cdots, Z_n\}$, where $Z_i$ is Pauli $Z$ on qubit $i$ and identity everywhere else. Since the Clifford group will map any $n$-qubit stabilizer group to any other $n$-qubit stabilizer group, we have the following theorem.

\begin{theorem} 
Any $n$-qubit stabilizer state $\ket{\psi}$ can be obtained from $\ket{0}^{\otimes n}$ by applying $n$-qubit Clifford-gates.
\end{theorem}

It is now convenient to introduce another important group for the discussion of stabilizer simulations. This group is known as the \textit{destabilzer group}. I first introduce this group by noting that set $\mathcal{X}_n = \{X_1, \cdots, X_n\}$ and $\mathcal{Z}_n$ together generate the $n$-qubit Pauli group, up to phases. Additionally, $\mathcal{X}_n$ and $\mathcal{Z}_n$ have the structure that each pair of elements $X_i\in \mathcal{X}_n$ and $Z_i \in \mathcal{Z}_n$ will anticommute with each other and commute with all other elements in $\mathcal{X}_n$ or $\mathcal{Z}_n$. The destabilizer group is the group we would get if we evolved $\mathcal{X}_n$ (\textit{i.e.}, the state $\ket{+}^{\otimes n}$) alongside $\mathcal{Z}_n$. 

One can see that the destabilizer group is also a stabilizer group; however, it is defined in relation to another stabilizer group. Since Clifford conjugation is homomorphic, applying the same Clifford conjugations to both a stabilizer group and its destabilizer group will preserve the algebraic relations between them. These relationships can be used advantageously in stabilizer simulations, which was a key insight of  Aaronson and Gottesman in \cite{CHP}. 

\subsubsection{Measurements\label{subsub:meas}}

I will now discuss how measurements update the set of stabilizer generators $\mathscr{S}$ and the set of destabilizer generators $\mathscr{D}$.  

When a Pauli operator $P$ is measured, there are two cases to consider:

\begin{enumerate}[(i)]
\item The Pauli operator $P$ commutes with all of the stabilizer generators. In this case either $+P$ or $-P$ must be in $\mathscr{S}$. The reason for this is as follows: Since  $\mathscr{S}$ and $\mathscr{D}$ together generate $\mathcal{P}_n$ up to phases, $P$ (up to a sign) must be in $\mathscr{S}$, $\mathscr{D}$, or a product of an element from $\mathscr{S}$ and an element from $\mathscr{D}$. Since each element of $\mathscr{D}$ anticommutes with at least one element in $\mathscr{S}$, then if $P$ commutes with $\mathscr{S}$ it must be contained in $\mathscr{S}$. The measurement is therefore determined. The only task left is to find the sign of associated stabilizer and return it as the measurement outcome. The details of how this is accomplished depends on the specific stabilizer algorithm, which will be discussed later.

\item The operator $P$ does not commute with  at least one stabilizer generator. Following the argument in case ($i$), one finds that if $P$ does not commute with $\mathscr{S}$, then it is not in $\mathscr{S}$. Since $P$ is not yet a stabilizer, then the measurement outcome is undetermined. The measurement outcome $m$ is $\pm 1$ with equal probability. 

After the measurement is made, $m P$ stabilizes the state and can be chosen as a new stabilizer generator. The previous stabilizers that anticommute with this new stabilizer generator no longer stabilize the state and must be removed from the generating set. To accomplish this, a single stabilizer generator, $R$, that anticommutes with $mP$ is removed from the set of stabilizer generators. Any other remaining stabilizer generators that anticommute with $mP$ are then multiplied by $R$. This ensures that a new set of commuting generators has been chosen. 

To update the destabilizer generators, we must ensure that the relationship between the stabilizer and destabilizer generators is preserved. We start by replacing the destabilizer generator  corresponding to the new stabilizer generator $mP$ with the stabilizer $R$. Since $R$ was a stabilizer generators, the new destabilizer generator $R$ anticommutes only with the new stabilizer generator $mP$. However, the other dstabilizer generators do not necessarily commute with the new stabilizer generator $mP$. Therefore, any destabilizer generator that anticommutes with $mP$ (excluding the destabilizer generator $R$) is multiplied by $R$ to ensure that all the destabilizer generators commute with $mP$.

\end{enumerate}

\section{\label{sec:prev-sims}Previous Stabilizer Simulators and Their Complexities}

Now that some background about stabilizer states and their evolution has been discussed, I will now give an overview of the previous approaches to stabilizer simulations to place the algorithm I will introduce in context. While reviewing these algorithms, I will also discuss and give tighter bounds on their complexities.

\subsection{Gottesman\label{GotAlg}}

In \cite{GotQC,Gottesman:1997a,GotKnill} Gottesman discussed different representations of stabilizer generators as well as the stabilizer update rules.  I will refer to the algorithm that was implied by \cite{GotQC,Gottesman:1997a,GotKnill} as the Gottesman algorithm. 

\begin{figure}[H]
\begin{minipage}{13cm}
    \centering
    \begin{subfigure}[b]{6cm}
        \centering
        \includegraphics[width=0.85\linewidth]{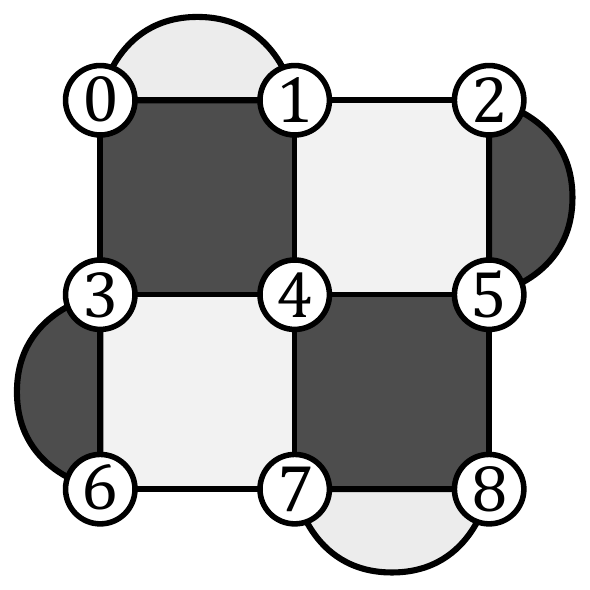} 
        \caption{Medial surface-code} \label{fig:surface}
    \end{subfigure}
    \hfill
    \begin{subfigure}[b]{6cm}
        \centering
\resizebox{0.98\linewidth}{!}{
\begin{tabular}[b]{c|cccccccccc}
\hline\hline
Name &  \multicolumn{10}{c}{Operator} \\ \hline
\it
$g_0$ & $+$ & $Z$ & $Z$ & $I$ & $I$ & $I$  & $I$ & $I$ & $I$ & $I$ \\
$g_1$ & $+$ & $I$ & $I$ & $I$  & $I$ & $I$ & $I$ & $I$ & $Z$ & $Z$ \\
$g_2$ & $+$ & $I$ & $I$ & $X$  & $I$ & $I$ & $X$ & $I$ & $I$ & $I$ \\
$g_3$ & $+$ & $I$ & $I$ & $I$  & $X$ & $I$ & $I$ & $X$ & $I$ & $I$ \\
$g_4$ & $+$ & $I$ & $Z$ & $Z$  & $I$ & $Z$ & $Z$ & $I$ & $I$ & $I$ \\
$g_5$ & $+$ & $I$ & $I$ & $I$  & $Z$ & $Z$ & $I$ & $Z$ & $Z$ & $I$ \\
$g_6$ & $+$ & $X$ & $X$ & $I$  & $X$ & $X$ & $I$ & $I$ & $I$ & $I$ \\
$g_7$ & $+$ & $I$ & $I$ & $I$  & $I$ & $X$ & $X$ & $I$ & $X$ & $X$ \\
$g_8$ & $+$ & $Z$ & $I$ & $I$  & $Z$ & $I$ & $I$ & $Z$ & $I$ & $I$ \\
\end{tabular}}
        \caption{Stabilizer generators} \label{fig:tableau}
    \end{subfigure}
\end{minipage}
    
\caption{\label{fig:surf_tab} (a) is a distance-three, medial surface-code where data qubits are indicated as white circles, stabilizers that are tensor products of Xs are represented by dark-grey polygons, and stabilizers that are tensor products of Zs are represented by white polygons. (b) is a complete set of stabilizer generators for a logical $\ket{0}$ stabilizer-state of the code depicted in (a). The last stabilizer generator is a logical $Z$ operator. Note that the tensor product symbol has been dropped.}
\end{figure}

The Gottesman algorithm does not include the use of destabilizers but instead only uses stabilizer generators to represent stabilizer states.  One choice of stabilizer generators to represent a logical $\ket{0}$ state of a distance-three medial (or rotated) surface-code \cite{Bombin:2007d} is shown in Fig.~\ref{fig:surf_tab} (see also Section~\ref{sec.pecos.qeccs}).

Following \cite{Gottesman:1997a}, the stabilizer generators of Fig.~\ref{fig:tableau} may be represented using a binary matrix as seen in Fig.~\ref{fig:stab-bin}; therefore, the data structure of the Gottesman algorithm takes $\mathcal{O}(n^2)$ space. 

Like Fig.~\ref{fig:tableau}, each row in this data structure represents a stabilizer generator. If $b$ is such a stabilizer binary matrix, then for the $i$th stabilizer generator and the $j$th qubit $b_{ij} = 0$ and $b_{i(j+n)} = 0$ indicates $I_j$, $b_{ij} = 1$ and $b_{i(j+n)} = 0$ indicates $X_j$, $b_{ij} = 1$ and $b_{i(j+n)} = 1$ indicates $Y_j$, and $b_{ij} = 0$ and $b_{i(j+n)} = 1$ indicates $Z_j$. The final column corresponds to the sign of the stabilizer generator. The element indicating a sign is 0 if the stabilizer has a $+1$ phase and 1 if the stabilizer has a $-1$ phase.

\begin{figure}[H]
\centering
\resizebox{8cm}{!}{
$\left(\begin{array}{ccccccccc|ccccccccc|c} 
0 & 0 & 0 & 0 & 0 & 0 & 0 & 0 & 0     & 1 & 1 & 0 & 0 & 0 & 0 & 0 & 0 & 0    & 0 \\ 
0 & 0 & 0 & 0 & 0 & 0 & 0 & 0 & 0     & 0 & 0 & 0 & 0 & 0 & 0 & 0 & 1 & 1    & 0 \\ 
0 & 0 & 1 & 0 & 0 & 1 & 0 & 0 & 0     & 0 & 0 & 0 & 0 & 0 & 0 & 0 & 0 & 0    & 0 \\ 
0 & 0 & 0 & 1 & 0 & 0 & 1 & 0 & 0     & 0 & 0 & 0 & 0 & 0 & 0 & 0 & 0 & 0    & 0 \\ 
0 & 0 & 0 & 0 & 0 & 0 & 0 & 0 & 0     & 0 & 1 & 1 & 0 & 1 & 1 & 0 & 0 & 0    & 0 \\ 
0 & 0 & 0 & 0 & 0 & 0 & 0 & 0 & 0     & 0 & 0 & 0 & 1 & 1 & 0 & 1 & 1 & 0    & 0 \\ 
1 & 1 & 0 & 1 & 1 & 0 & 0 & 0 & 0     & 0 & 0 & 0 & 0 & 0 & 0 & 0 & 0 & 0    & 0 \\ 
0 & 0 & 0 & 0 & 1 & 1 & 0 & 1 & 1     & 0 & 0 & 0 & 0 & 0 & 0 & 0 & 0 & 0    & 0 \\ 
0 & 0 & 0 & 0 & 0 & 0 & 0 & 0 & 0     & 1 & 0 & 0 & 1 & 0 & 0 & 1 & 0 & 0    & 0 \\ 
\end{array}\right)$
} \nonumber
\caption{\label{fig:stab-bin} A $n$ by $2n+1$ binary matrix representing the stabilizer generators in Fig.~\ref{fig:tableau}.}
\end{figure}

The circuit operations update the generators according to the stabilizer update rules as described in Section~\ref{sec:stab-rules}. Clifford gates are accomplished by running over all $n$ stabilizer generators and updating each generator as appropriate. Assuming the Gottesman algorithm contains only gates that act on at most some constant $c$ number of qubits, then each update of a generator takes a constant time. Therefore, a Clifford gate needs a total of $\mathcal{O}(n)$ time steps to complete for this algorithm.

To show that each update of a $n$-qubit generator by a Clifford gate that acts on $c$ qubits only takes constant time we note that since $\mathcal{X}_n$ and $\mathcal{Z}_n$ generate the $n$-qubit Pauli group, to know how any $c$-qubit Clifford-gate transforms a Pauli operator, it is sufficient describe how it transforms the $2c$ generators of $\mathcal{X}_c$ and $\mathcal{Z}_c$. For example, the $CNOT$ can be thought of as the map

\begin{align}
\label{eq:cnotmap}
& X \otimes I \rightarrow X \otimes X \nonumber \\
& I \otimes X \rightarrow I \otimes X \nonumber \\
& Z \otimes I \rightarrow Z \otimes I \nonumber \\
& I \otimes Z \rightarrow I \otimes Z,
\end{align}

\noindent where the first qubit is the control and the second is the target. So, for example, if $CNOT$ is applied to the Pauli operator $X\otimes Y$ the following transformation occurs:

\begin{align}
& CNOT(X\otimes Y) CNOT = \nonumber \\
& CNOT(X\otimes I)CNOT^{\dagger}\; CNOT(I\otimes Y)CNOT^{\dagger} = \nonumber \\
& i \; (X\otimes X) \; CNOT(I\otimes X)CNOT^{\dagger}\; CNOT(I\otimes Z)CNOT^{\dagger} = \nonumber \\
& i \; (X\otimes X) (I \otimes X) (Z\otimes Z) = \nonumber \\
& Y \otimes Z.
\end{align}

While discussing measurements, I will constrain the discussion to single-qubit measurements in the computational basis. This is a reasonable choice for a stabilizer simulator since many quantum algorithms require only single-qubit, $Z$-basis measurements. If multi-qubit measurements or measurements in a basis other than $Z$ are needed, they can be constructed from appropriate Clifford gates, the introduction of ancillas, and single-qubit $Z$ measurements.

Following the stabilizer update rules for measurements as described earlier in Section \ref{subsub:meas}, to perform a measurement of $Z$ on qubit $j$, whether any stabilizers anticommute with $Z_j$ is first checked by running through the rows of the binary matrix. If the $i$th element of a row is 1, then the stabilizer generator corresponding to that row anticommutes with $Z_j$. This process takes $\mathcal{O}(n)$ steps.  

If all the stabilizer generators commute, then the measurement is determined and Gaussian elimination is used to put the generators into a form to find the sign of the stabilizer equivalent to $Z_j$. Gaussian elimination takes $\mathcal{O}(n^3)$ time steps to complete, which causes deterministic measurements to take time $\mathcal{O}(n^3)$ as well.

If there are stabilizer generators that do not commute, the measurement is undetermined. An anticommuting stabilizer generator is then removed and replaced with $m Z_j$, where $m$ is a random measurement outcome. Each remaining stabilizer generator that anticommutes with $Z_j$ is then multiplied by the removed generator. Since each stabilizer generator is an $n$-qubit Pauli operator, each multiplication takes $\mathcal{O}(n)$ time. Therefore, the time-complexity for non-deterministic measurements on qubit $j$ is $\mathcal{O}(n|q^{s,x}_j|)$, where $|q^{s,x}_j|$ is the number of stabilizer generators that are 1 in the $j$th column of the binary matrix, \textit{i.e.}, anticommute with $Z_j$. Note, I use the symbol $|q^{s,x}_j|$ to be consistent with the notation used later in this paper when discussing the new algorithm in Section~\ref{sec:our-struct}.

\subsection{Aaronson-Gottesman\label{sec:aar-got}}

The algorithm introduced by Aaronson and Gottesman in \cite{CHP} simulates circuits using just the operations $CNOT$, $H$, and $S$ (which they label as $P$) as well as measurement in the $Z$-basis. Because of the set of Clifford gates used, the implementation of the algorithm introduced by Aaronson and Gottesman is called \texttt{CHP}. 

The Aaronson-Gottesman algorithm extended the previous algorithm by adding destabilizers generators to the simulation of stabilizer states. To represent these new generators, the data structure of the Aaronson-Gottesman algorithm prepended the binary matrix of the Gottesman algorithm with $n$ additional rows, where each row represented a destabilizer generator in the same manner as a stabilizer generator. The prepended destabilizers are arranged so that the destabilizer generator in the $i$th row is the anticommuting partner of the stabilizer generator in the $(i+n)$th row. An example of this data structure is seen in Fig.~\ref{fig:stabdestab-bin}. As the addition of destabilizer generators increased the binary matrix representation from a $n$ by $2n+1$ matrix to a $2n$ by $2n+1$, the space complexity of this data structure remained $\mathcal{O}(n^2)$.

\begin{figure}[H]
\centering
\resizebox{8cm}{!}{
$\left(\begin{array}{ccccccccc|ccccccccc|c}
0 & 1 & 0 & 0 & 0 & 1 & 0 & 0 & 0     & 0 & 0 & 0 & 0 & 0 & 0 & 0 & 0 & 0    & 0 \\
0 & 0 & 0 & 0 & 0 & 0 & 0 & 0 & 1     & 0 & 0 & 0 & 0 & 0 & 0 & 0 & 0 & 0    & 0 \\
0 & 0 & 0 & 0 & 0 & 0 & 0 & 0 & 0     & 0 & 0 & 1 & 0 & 0 & 0 & 0 & 0 & 0    & 0 \\
0 & 0 & 0 & 0 & 0 & 0 & 0 & 0 & 0     & 1 & 0 & 0 & 1 & 0 & 0 & 0 & 0 & 0    & 0 \\
0 & 0 & 0 & 0 & 0 & 1 & 0 & 0 & 0     & 0 & 0 & 0 & 0 & 0 & 0 & 0 & 0 & 0    & 0 \\
0 & 0 & 0 & 0 & 0 & 0 & 0 & 1 & 1     & 0 & 0 & 0 & 0 & 0 & 0 & 0 & 0 & 0    & 0 \\
0 & 0 & 0 & 0 & 0 & 0 & 0 & 0 & 0     & 1 & 0 & 0 & 0 & 0 & 0 & 0 & 0 & 0    & 0 \\
0 & 0 & 0 & 0 & 0 & 0 & 0 & 0 & 0     & 1 & 0 & 0 & 0 & 1 & 0 & 0 & 0 & 0    & 0 \\
0 & 0 & 0 & 0 & 0 & 0 & 1 & 1 & 1     & 0 & 0 & 0 & 0 & 0 & 0 & 0 & 0 & 0    & 0 \\
\hline  
0 & 0 & 0 & 0 & 0 & 0 & 0 & 0 & 0     & 1 & 1 & 0 & 0 & 0 & 0 & 0 & 0 & 0    & 0 \\ 
0 & 0 & 0 & 0 & 0 & 0 & 0 & 0 & 0     & 0 & 0 & 0 & 0 & 0 & 0 & 0 & 1 & 1    & 0 \\ 
0 & 0 & 1 & 0 & 0 & 1 & 0 & 0 & 0     & 0 & 0 & 0 & 0 & 0 & 0 & 0 & 0 & 0    & 0 \\ 
0 & 0 & 0 & 1 & 0 & 0 & 1 & 0 & 0     & 0 & 0 & 0 & 0 & 0 & 0 & 0 & 0 & 0    & 0 \\ 
0 & 0 & 0 & 0 & 0 & 0 & 0 & 0 & 0     & 0 & 1 & 1 & 0 & 1 & 1 & 0 & 0 & 0    & 0 \\ 
0 & 0 & 0 & 0 & 0 & 0 & 0 & 0 & 0     & 0 & 0 & 0 & 1 & 1 & 0 & 1 & 1 & 0    & 0 \\ 
1 & 1 & 0 & 1 & 1 & 0 & 0 & 0 & 0     & 0 & 0 & 0 & 0 & 0 & 0 & 0 & 0 & 0    & 0 \\ 
0 & 0 & 0 & 0 & 1 & 1 & 0 & 1 & 1     & 0 & 0 & 0 & 0 & 0 & 0 & 0 & 0 & 0    & 0 \\ 
0 & 0 & 0 & 0 & 0 & 0 & 0 & 0 & 0     & 1 & 0 & 0 & 1 & 0 & 0 & 1 & 0 & 0    & 0 \\
\end{array}\right)$
} \nonumber
\caption{\label{fig:stabdestab-bin} A $2n$ by $2n+1$ binary matrix representing the stabilizers in generators in Fig.~\ref{fig:tableau} as well as destabilizer generators.}
\end{figure}

The Clifford gates are performed in the same manner as the Gottesman algorithm as outlined previously in Section~\ref{GotAlg} except that the gates must update the destabilizer generators as well. As the total number of $n$-qubit generators has increased from $n$ to $2n$, the time complexity of performing a Clifford gate remains $\mathcal{O}(n)$.

For deterministic measurements, instead of using Gaussian elimination to find the outcome of measuring $Z$ on qubit $j$, the Aaronson-Gottesman algorithm made use of the commutation relation between stabilizer and destabilizer generators. Ignoring overall signs, $Z_j$ is in the stabilizer group and can be represented as a product of stabilizer generators. Since the destabilizer generators only anticommute with their anticommuting stabilizer partners, if the destabilizer generator of the $i$th row anticommutes with $Z_j$ (i.e the $j$th element of the $i$th row is 1), then the stabilizer generator of the $(i+n)$th row is one of the stabilizer generators that multiplies to give $Z_j$. 

To find the sign of $Z_j$ and determine the measurement output, the algorithm copies one of the stabilizer generators that multiply to give $Z_j$, and then multiplies this copied generator with all the other stabilizer generators that $Z_j$ factors into. Given $|q^{d,x}_j|$ destabilizers that anticommute with $Z_j$ and since each multiplication takes $\mathcal{O}(n)$, the time to determine the measurement outcome is $\mathcal{O}(n |q^{d,x}_j|)$. Once again, I choose the symbol $|q^{d,x}_j|$ to be consistent with the notation in Section~\ref{sec:our-struct}.

The Aaronson-Gottesman algorithm handles non-deterministic measurement case in an almost identical manner as outlined previously in Section~\ref{GotAlg}. The only difference is that the destabilizer generators must now be updated. As mentioned in Section~\ref{subsub:meas}, destabilizers that anticommute with $Z_j$ are multiplied by a stabilizer generator that is removed. The complexity for this measurement case therefore becomes $\mathcal{O}(n (|q^{s,x}_j|+|q^{d,x}_j|))$, where as before $|q^{s,x}_j|$ ($|q^{d,x}_j|$) is the number of stabilizer (destabilizer) generators that anticommute with $Z_j$. 

The operation with the worst-case time-complexity for this algorithm is the non-deterministic $Z$ measurement, which needs $\mathcal{O}(n (|q^{s,x}_j|+|q^{d,x}_j|))$ time steps to complete when applied to qubit $j$. As both $1\leq |q^{s,x}_j| \leq n$ and $1\leq |q^{d,x}_j| \leq n$, the worst-case time-complexity that is usually reported for the Aaronson-Gottesman algorithm is $\mathcal{O}(n^2)$. However, we will see that when the stabilizer/destabilizer representations of the states in certain quantum protocols, such as syndrome extraction of topological stabilizer codes, is analyzed, that one finds that for all $j$, $|q^{s,x}_j|+|q^{d,x}_j| = O(1)$. Thus, in many practical applications the operations of the Aaronson-Gottesman algorithm will run in $\mathcal{O}(n)$ time.

\subsection{\label{sec:anders-breigel}Anders-Briegel}

In \cite{GraphSim} Anders and Briegel presented an algorithm that uses the graph-state representation to represent stabilizer states. This representation was introduced in \cite{BR00} and is used in the study of resource states in measurement-based quantum computing (MBQC). In \cite{GraphSim} an implementation called \texttt{GraphSim} was also presented, which contained all 24 single-qubit Clifford operations, measurements in the $X$, $Y$, and $Z$ basis, and the two-qubit gates $CNOT$ and controlled-$Z$. 

\begin{figure}[H]
    \centering
    \begin{subfigure}[b]{4cm}
        \centering{
        \includegraphics[width=0.85\linewidth]{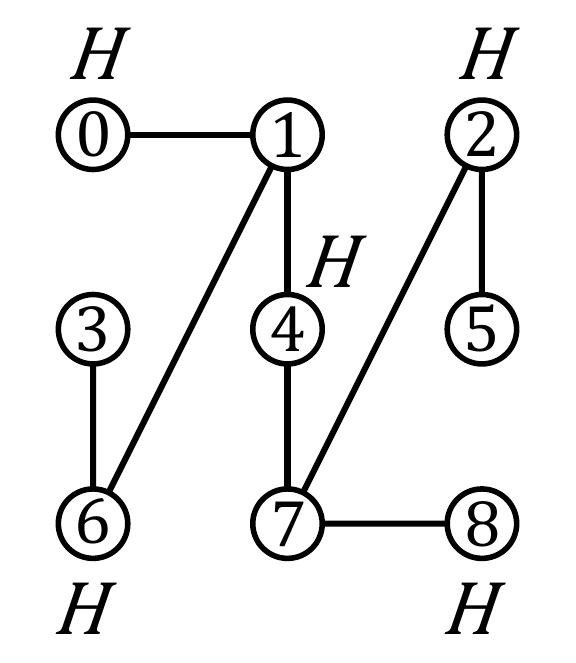} 
        \caption{Graph representing a graph state.} \label{fig:graph_surface}
				}
    \end{subfigure}
    \hfill
    \begin{subfigure}[b]{4cm}
        \centering
\resizebox{0.98\linewidth}{!}{
\begin{tabular}[b]{c|ccccccccc}
\hline\hline
Name &  \multicolumn{9}{c}{Operator} \\ \hline
\it
$g_0$ & $X$ & $Z$ & $I$  & $I$ & $I$ & $I$ & $I$ & $I$ & $I$ \\
$g_1$ & $Z$ & $X$ & $I$  & $I$ & $Z$ & $I$ & $Z$ & $I$ & $I$ \\
$g_2$ & $I$ & $I$ & $X$  & $I$ & $I$ & $Z$ & $I$ & $Z$ & $I$ \\
$g_3$ & $I$ & $I$ & $I$  & $X$ & $I$ & $I$ & $Z$ & $I$ & $I$ \\
$g_4$ & $I$ & $Z$ & $I$  & $I$ & $X$ & $I$ & $I$ & $Z$ & $I$ \\
$g_5$ & $I$ & $I$ & $Z$  & $I$ & $I$ & $X$ & $I$ & $I$ & $Z$ \\
$g_6$ & $I$ & $Z$ & $I$  & $Z$ & $I$ & $I$ & $X$ & $I$ & $I$ \\
$g_7$ & $I$ & $I$ & $Z$  & $I$ & $Z$ & $I$ & $I$ & $X$ & $Z$ \\
$g_8$ & $I$ & $I$ & $I$  & $I$ & $I$ & $I$ & $I$ & $Z$ & $X$ \\
\hline
 LC   & $H$ & $I$ & $H$  & $I$ & $H$ & $I$ & $H$ & $I$ & $H$ \\
\end{tabular}}
        \caption{Stabilizer generators with LC operations.} \label{fig:graph_tableau}
    \end{subfigure}
    \hfill
    \begin{subfigure}[b]{3.5cm}
		\hspace{.5cm}
        \centering
\resizebox{0.98\linewidth}{!}{
\begin{tabular}[b]{c|c|l}
\hline\hline
Vertex & LC & Adjacency list \\ \hline
\it
$0$ & $H$ & $1$ \\
$1$ & $I$ & $0, 4, 6$ \\
$2$ & $H$ & $5, 7$ \\
$3$ & $I$ & $6$ \\
$4$ & $H$ & $1, 7$ \\
$5$ & $I$ & $2$ \\
$6$ & $H$ & $1, 3$ \\
$7$ & $I$ & $2, 4, 8$ \\
$8$ & $H$ & $7$ \\
\end{tabular}}
        \caption{\label{fig:graph_adjacency} Graph-state data-structure.} 
    \end{subfigure}
    
\caption{(a) is a graph associated with a graph-state representation of the stabilizer state in Fig.~\ref{fig:tableau}. (b) is set of stabilizer generators that is equivalent to those in Fig.~\ref{fig:surf_tab} up to group multiplication and local Clifford operations. Note that the signs of the stabilizers are accounted for by LC operator. (c) is a sparse data-structure used to represent (a) and (b).}
\end{figure}

Any graph-state representation can be described by a stabilizer representation and vice versa \cite{Schl02,GKR07}. A graph-state representation is a choice of stabilizer generators such that, up to local Clifford (LC) operations, a stabilizer tableau used to represent stabilizer generators can be written with Pauli $X$s along the diagonal and Pauli $Z$s or $I$s everywhere else. For example, the stabilizer tableau Fig.~\ref{fig:tableau} in can be represented using the set of graph-state stabilizer generators in Fig.~\ref{fig:graph_tableau}. It should be noted that while a graph-state representation can be viewed as a restricted choice of stabilizer generators, the graph-state representation of a state is not unique. Starting from a graph-state representation, the application of an operation called \textit{local complementation} on the representation will result in another LC equivalent graph-state representation ~\cite{NDM04,HDERNB06}. 

A graph can be constructed from the stabilizer generators of the graph-state representation. In this graph each vertex corresponds to a qubit. Associated with each vertex is the stabilizer generator that acts like $X$ only on the vertex. The vertex is connected by an edge to each qubit that the associated stabilizer generator acts like $Z$ on. An example of such a graph is seen in Fig.~\ref{fig:graph_surface}.

Anders and Briegel used a data structure, such as the one seen in Fig.~\ref{fig:graph_adjacency}, to represent these graphs. In the data structure, for each vertex the LC applied to vertex as well as a list of nearest neighbors is stored. Therefore, the data structure has a space complexity of $\mathcal{O}(n\overline{\Delta})$, where $\overline{\Delta}$ is the average vertex degree of the graph. Given the close connection between the graph-state representation and the stabilizer representation, one can view the data structure used by the Anders-Briegel algorithm as a sparse representation of a stabilizer matrix that is compressed row-wise, where the choice of stabilizer generators must conform to the rules mentioned in the previous paragraph.

To apply single-qubit Clifford-gates, the Anders-Briegel algorithm only needs to update the LC associated with the qubit being acted on and can do so by using a lookup table; therefore, single-qubit Clifford-gates can be performed in $\Theta(1)$ steps.

\pagebreak[4]
To apply a two-qubit gate, up to five local complementation operations are performed. This operation requires at most $\mathcal{O}(\Delta^2)$ time steps to complete, where $\Delta$ is the maximum degree of the graph. Because of this, two-qubit gates take $\mathcal{O}(\Delta^2)$ steps to complete.

To measure the operator $Z_j$ the Anders-Briegel algorithm removes the edges from qubit $j$. This operation then requires at most $\mathcal{O}(\Delta)$ steps. Since the measurements of $X_j$ or $Y_j$ can be accomplish by measuring $Z_j$ and applying the appropriate single-qubit Clifford-gates before and after the measurement, a measurement in any single-qubit Pauli basis on qubit $j$ can be performed in $\mathcal{O}(\Delta)$ time.

As we can see, two-qubit gates require the most time to complete; therefore, the overall worst-case time-complexity is $\mathcal{O}(\Delta^2)$. It was argued in \cite{GraphSim} that in many applications $\Delta=\mathcal{O}(\log{n})$; however, as we will discuss later, for planar topological codes $\Delta=\mathcal{O}(\sqrt{n})$. Further, I will show that for a standard choice of stabilizer generators to represent logical basis-states of these codes that $\overline{\Delta}  = \mathcal{O}(\sqrt{n})$. Therefore, in practice, when simulating topological codes we find that the slowest operations of the Anders-Briegel algorithm will in time $\mathcal{O}(\Delta^2)=\mathcal{O}(n)$.

\section{\label{sec:our-alg}Algorithm}

We have discussed how the data structure of the Anders-Briegel algorithm can be seen as a row-wise-compressed representation of a stabilizer tableau, albeit with a choice of stabilizer generators that must fit the form required by the graph-state representation. This sparse representation was utilized by the Anders-Briegel algorithm to speed up the runtime of circuit operations. I have also noted that the Aaronson-Gottesman algorithm can in practice run faster than $\mathcal{O}(n^2)$ if the binary matrix columns representing the $X$ action of the stabilizers and destabilizers are sparse. A natural question then arises as to what extent can an algorithm take advantage of the sparsity of a stabilizer/destabilizer representation. In the following, I present an algorithm that is a result of pursuing this question.

\subsection{Data Structure\label{sec:our-struct}}

The data structure that I will now present can be seen as representing two copies of the binary-matrix data-structure used in the Aaronson-Gottesman algorithm. One copy compresses the matrix in a row-wise direction and the other, in a column-wise direction. To facilitate computation, I will further break up these representations of the matrix according to whether the generators belong to stabilizer or a destabilizer group as well as whether the generators act as Pauli $X$ or $Z$.

I will now discuss some basic notation used to describe the data structure employed by the new algorithm. Each qubit, stabilizer generator, and destabilizer generator is uniquely associated with an index in $K = \{0, \cdots, n-1 \}$. The notation $m^{t,p}$ will be used to represent a sequence of sets where $m \in \{g, q\}$, $t\in \{s,d\}$, $p \in \{x, z\}$. Each sequence $m^{t,p}$ contains $n$ sets. $m^{t,p}_i$ is the $i$th set of $m^{t,p}$. 

I will now discuss the part of the data structure that is analogous to compressing the binary matrix in the row-wise direction. The stabilizer generators will be represented using two sequences $g^{s, x}$ and $g^{s, z}$. The $i$th element of these sequences, \textit{i.e.} $g_i^{s,x}$ ($g_i^{s,z}$), is the set of qubit indices for which the $i$th stabilizer acts as Pauli $X$ ($Z$) on the qubits. Note, that if a qubit index is in $g^{s,x}_i \cap g^{s,z}_i$ this indicates the the $i$th stabilizer generator acts like $XZ$ on the qubit, not $Y=iXZ$. 

The overall phases of the stabilizer generators are stored in two sets, $c^{(re)}$ and $c^{(im)}$. If the $j$th stabilizer has an overall phase of $-1$ or $-i$, then $j\in c^{(re)}$, otherwise $j\notin c^{(re)}$. Likewise, if the $j$th stabilizer has an overall phase $i$ or $-i$, then $j\in c^{(im)}$, otherwise $j\notin c^{(im)}$. So for example, if $I\otimes X \otimes Z \otimes Y=i\left(I\otimes X\otimes I \otimes X\right) \left(I \otimes I \otimes Z \otimes Z\right)$ is the $0$th stabilizer generator, then $g^{s,x}_0 = \{1, 3\}$, $g^{s,z}_0 = \{2, 3\}$, $0\notin c^{(re)}$ and $0\in c^{(im)}$.

The destabilizer generators are represented in a similar fashion except the label $d$ replaces the label $s$. Also, the destabilizers are indexed so that they match their anticommuting stabilizer generator partner. Therefore, for all $i, j$, $(|g^{s,x}_i \cap g^{d,z}_j| + |g^{s,z}_i \cap g^{d,x}_j|) \text{ mod } 2 = \delta_{i,j}$. We do not track the signs of destabilizer generators since the destabilizers are only used for their commutation relation with the stabilizer generators.

I will now discuss how the binary matrix is represented a sparse column-wise manner. To do this I track indices of stabilizers that act on each qubit. That is, for the stabilizers I introduce two new sequences $q^{s, x}$ and $q^{s, z}$, each size $n$. The $i$th element of $q^{s, x}$ ($q^{s, z}$), that is $q^{s,x}_i$ ($q^{s,z}_i$), is a set that contains $j$ iff the $j$th stabilizer generator acts as Pauli $X$ ($Z$) on the $i$th qubit. Therefore, $i \in g^{s,x}_j \Longleftrightarrow j \in  q^{s,x}_i$  and $i \in g^{s,z}_j \Longleftrightarrow j \in  q^{s,z}_i$, where $i, j \in K$.

For the destabilizer generators we have a similar collection of sequences except with the labels $s$ replaced with $d$, to indicate that the sequences and sets describe destabilizer generators.

\subsection{Operations}

So far, we have discussed the data structure used in the new algorithm. In the following, I will describe how circuit operations act on this data structure. The new algorithm is similar to the Aaronson-Gottesman algorithm; however, I use the row-wise ($g^{t,p}$) and column-wise ($q^{t,p}$) data structures to avoid running over $n$ elements. While we gain this benefit, we have the disadvantage that we must be careful to properly update two representations of the generators ($g^{t,p}$ and $q^{t,p}$). 

I have implemented all 24 single-qubit Clifford-gates as well as $CNOT$, controlled-$Z$, and $SWAP$ gates. For brevity, I will present the pseudo-code for the Hadamard, phase, and $CNOT$ gates,  since these gates generate any $n$-qubit Clifford-gate. In the pseudo-code, I will use the symmetric difference $A\triangle B := (A\cup B) - (A \cap B)$ in the operation $A \leftarrow A \triangle \{i\}$, \textit{i.e.}, remove element $i$ from $A$ if $i\in A$, otherwise add $i$ to $A$. I will assume that the functions listed in the pseudo-code that follow have full access to and can directly modify the data structure mentioned previously in Section~\ref{sec:our-struct}.

\subsubsection{Hadamard}
As discussed in Section \ref{GotAlg}, to know how a gate modifies Pauli operators it is sufficient to describe how conjugation by the gate maps $\mathcal{X}_n$ and $\mathcal{Z}_n$ to other Pauli operators.  

Conjugation by the Hadamard $H$ gate maps 

\begin{align}
\label{eq:hadamardmap}
& X \rightarrow Z \nonumber \\                                                                  
& Z \rightarrow X. 
\end{align}

\pagebreak[4]
\noindent Therefore, $H$ acting on the $i$th qubit is accomplish using the pseudo-code in following:

\vspace{0.5cm}
\begin{minipage}{0.95\linewidth}
\begin{lstlisting}[label={code:hadamard}, caption={The Hadamard gate on qubit $i$.},style=psedostyle]
hadamard(%$i$%):
	for %$j \in q^{s,x}_i$%:
		if %$j \in q^{s,z}_i$%:
			%$c^{(re)} \leftarrow c^{(re)} \triangle \{j\}$%
			
	for %$j \in q^{s,x}_i$%:
		if %$j \notin q^{s,z}_i$%:
			%$g^{s,x}_j \leftarrow g^{s,x}_j - \{i\}$%
			%$g^{s,z}_j \leftarrow g^{s,z}_j \cup \{i\}$%
			
	for %$j \in q^{s,z}$%:
		if %$j \notin q^{s,x}$%:
			%$g^{s,z} \leftarrow g^{s,z} - \{i\}$%
			%$g^{s,x} \leftarrow g^{s,x} \cup \{i\}$%
				
	%$q^{s,x}_j \leftrightarrow q^{s,z}_j$%	
	
	[Repeat lines 5 to 13 for the destabilizer generators]
\end{lstlisting}
\end{minipage}

While the change of stabilizer phases in lines 1 to 3 in Code Block \ref{code:hadamard} is not directly indicated by Eq.~\ref{eq:hadamardmap}, it is needed as $H$ switches the order of $XZ$ on qubit $i$. 

\pagebreak[4]
\subsubsection{Phase gate}

Conjugation by the phase gate $S$ is the map

\begin{align}
\label{eq:phasemap}
& X \rightarrow Y \nonumber \\
& Z \rightarrow Z, 
\end{align}

\noindent which when applied to the $i$th qubit can be accomplished by:

\vspace{0.5cm}
\begin{minipage}{0.95\linewidth}
\begin{lstlisting}[label={code:phase}, caption={The $S$ gate applied to qubit $i$.},style=psedostyle]
phase(%$i$%):
	for %$j \in q^{s,x}_i$%:
		if %$j \in c^{(im)}$%:
			%$c^{(re)} \leftarrow c^{(re)} \triangle \{j\}$%
			
		%$c^{(im)} \leftarrow c^{(im)} \triangle \{j\}$%
		
	for %$j \in q^{s,x}_i$%:
		%$g^{s,z}_j \leftarrow g^{s,z}_j \triangle \{i\}$%
		%$q^{s,z}_i \leftarrow q^{s,z}_i \triangle \{j\}$%
	[Repeat lines 7 to 9 for the destabilizer generators]
\end{lstlisting}
\end{minipage}

\vfill
\pagebreak[4]
\subsubsection{CNOT}

The conjugation by the controlled-X gate $CNOT$ is the map given in Eq.~\ref{eq:cnotmap}. The pseudo-code to accomplish this is as follows:

\vspace{0.5cm}
\begin{minipage}{0.95\linewidth}
\begin{lstlisting}[label={code:cnot}, caption={The $CNOT$ gate with target qubit $t$ and control qubit $c$.},style=psedostyle]
cnot(%$t$%, %$c$%):
	for %$j \in q^{s,x}_t$%:
		%$g^{s,x}_j \leftarrow g^{s,x}_j \triangle \{c\}$%
		%$q^{s,x}_c \leftarrow q^{s,x}_c \triangle \{j\}$%
		
	for %$j \in q^{s,z}_c$%:
		%$g^{s,z}_j \leftarrow g^{s,z}_j \triangle \{t\}$%
		%$q^{s,z}_t \leftarrow q^{s,z}_t \triangle \{j\}$%
	[Repeat all for the destabilizer generators]
\end{lstlisting}
\end{minipage}

One might note that the $CNOT$ pseudo-code does not include a sign update. This is due to the choice of using $W=XZ$ instead of $Y=iXZ$ in representing stabilizer generators. Removing the need to update signs provides a slight runtime advantage when simulating quantum circuits that are composed of many $CNOT$ gates, which is common for QECC and many other protocols.  

\vfill
\pagebreak[4]
\subsection{Measurements}

As measurement in the $X$ or $Y$ basis can be accomplished by a measurement in the $Z$-basis with the proper Clifford rotations, I will restrict the presentation to the pseudo-code for a $Z$-basis measurement on qubit $i$:

\vspace{0.5cm}
\begin{minipage}{0.95\linewidth}
\begin{lstlisting}[label={code:meas}, caption={Measurement of the operator $Z_i$. The variable \texttt{random\_outcome} $\in \{-1,0,1\}$. If \texttt{random\_outcome} is equal to $-1$, then a non-deterministic measurement will return a random outcome, otherwise it will return \texttt{random\_outcome}. Deterministic measurements are unaffected.},style=psedostyle]
measure(%$i$%, random_outcome):
	if %$i \notin q^{s,x}_i$%:
		outcome %$\leftarrow$% meas_determined(%$i$%)
		[For function see Code Block %\ref{code:measdetermin}%.]
		
	else:
		outcome %$\leftarrow$% meas_undetermined(%$i$%, random_outcome)
		[For function see Code Block %\ref{code:measindetermin}%.]
		
	return outcome
\end{lstlisting}
\end{minipage}

\vfill
\pagebreak[4]
The following pseudo-code handles the deterministic measurement case:

\vspace{0.5cm}
\begin{minipage}{0.95\linewidth}
\begin{lstlisting}[label={code:measdetermin}, caption={Pseudo-code for a deterministic measurement of $Z_i$.},style=psedostyle]
meas_determined(%$i$%):
	num_minuses %$\leftarrow 0$%
	num_is %$\leftarrow 0$%
	cumulative_x %$\leftarrow \{\}$%
	
	for %$j \in q^{d,x}_i$%:	
		if %$j \in c^{(re)}$%:
			num_minus %$\leftarrow$% num_minus + 1
		
		if %$j \in c^{(im)}$%:
			num_is %$\leftarrow$% num_is + 1	
		
		[Sign update due to left-multiplying %$X$% by %$Z$%:]
		for %$k \in g^{s,z}_j$%:
			if %$k \in$% cumulative_x:
				num_minus %$\leftarrow$% num_minus + 1
				
		for %$k \in g^{s,x}_j$%:
			cumulative_x %$\leftarrow$% cumulative_x%$\triangle \{k\}$%
					
	if num_is mod 4 == 2:
		num_minus %$\leftarrow$% num_minus + 1	
	return num_minuses mod 2	
\end{lstlisting}
\end{minipage}

\vfill
\pagebreak[4]
The non-deterministic measurement case is handled by the pseudo-code:

\begin{minipage}{0.95\linewidth}
\begin{lstlisting}[label={code:measindetermin},style=psedostyle]
meas_undetermined(%$i$%, random_outcome):	
	[Stabilizer to remove:]
	%$r \leftarrow$% [Choose a %$j\in q^{s,x}_i$% such that %$|g^{s,x}_j| + |g^{d,x}_j|$% is minimized.]				
	%$a^{s,x} \leftarrow q^{s,x}_i - \{r\}$%
	%$a^{d,x} \leftarrow q^{d,x}_i - \{r\}$%
	%$x_r \leftarrow g^{s,x}_r$%
	%$z_r \leftarrow g^{s,z}_r$%	
	if random_outcome == -1:
		out %$\leftarrow$% [0 or 1 with equal probability.]
	else:
		out %$\leftarrow$% random_outcome		
	
	update_signs(%$c^{(im)}$%, %$c^{(re)}$%, %$a^{s,x}$%, %$x_r$%,  %$z_r$%, %$g^{s,x}$%, out)
	[For update_signs see Code Block %\ref{code:updatesigns}%.]			
	update_gens(%$a^{s,x}$%, %$x_r$%,  %$z_r$%, %$g^{s,x}$%, %$g^{s,z}$%, %$q^{s,x}$%, %$q^{s,z}$%)
	update_gens(%$a^{d,x}$%, %$x_r$%,  %$z_r$%, %$g^{d,x}$%, %$g^{d,z}$%, %$q^{d,x}$%, %$q^{d,z}$%)
	[For update_gens see Code Block %\ref{code:updategens}%.]	
	
	[Remove the %$r$%th destabilizer generator and replace it with the %$r$%th stabilizer generator:]
	for %$j \in g^{d,x}_r$%:
		%$q^{d,x}_j \leftarrow q^{d,x}_j-\{r\}$%
	for %$j \in x_r$%:
		%$q^{d,x}_j \leftarrow q^{d,x}_j\cup\{r\}$%
	for %$j \in g^{d,z}_r$%:
		%$q^{d,z}_j \leftarrow q^{d,z}_j-\{r\}$%
	for %$j \in z_r$%:
		%$q^{d,z}_j \leftarrow q^{d,z}_j\cup\{r\}$%
	
\end{lstlisting}
\end{minipage}

\vfill
\begin{minipage}{0.95\linewidth}
\begin{lstlisting}[nolol,firstnumber=28,style=psedostyle,caption={Pseudo-code a non-deterministic measurement of $Z_i$.}]
	%$g^{d,x}_r\leftarrow x_r$%
	%$g^{d,z}_r\leftarrow z_r$%	
	
	[Remove the %$r$%th stabilizer generator and replace it with %$Z$% on qubit %$i$%:]
	for %$j \in x_r$%:
		%$q^{s,x}_j\leftarrow q^{s,x}_j - \{r\}$%
	for %$j \in z_r$%:
		%$q^{s,z}_j\leftarrow q^{s,z}_j - \{r\}$%	
		
	%$g^{s,z}_r\leftarrow \{i\}$%
	%$g^{s,x}_r\leftarrow \{\}$%
	%$q^{s,z}_i\leftarrow \{r\}$%		
	return out	
\end{lstlisting}
\end{minipage}

\vfill
\pagebreak[4]
The following updates stabilizer signs during a measurement:

\begin{minipage}{0.95\linewidth}
\begin{lstlisting}[label={code:updatesigns}, caption={Pseudo-code to update signs during a measurement.},style=psedostyle]
update_signs(%$c^{(im)}$%, %$c^{(re)}$%, %$a^{s,x}$%, %$x_r$%,  %$z_r$%, %$g^{s,x}$%, out):
	[Update signs due to multiplying stabilizer generators by the removed stabilizer:]
	if %$r \in c^{(re)}$%:
		for %$j \in a^{s,x}$%:
			%$c^{(re)} \leftarrow c^{(re)} \triangle \{j\}$%
	if %$r \in c^{(im)}$%:
		for %$j \in a^{s,x}$%:
			if %$j \in c^{(im)}$%:
				%$c^{(re)} \leftarrow c^{(re)} \triangle \{j\}$%
			%$c^{(im)} \leftarrow c^{(im)} \triangle \{j\}$%
	[Sign due to left multiplying %$X$% by %$Z$%:]
	for %$j \in a^{s,x}$%:
		num_minuses %$\leftarrow 0$%
		for %$k \in z_r$%:
			if %$k \in g^{s,x}_j$%:
				num_minuses %$\leftarrow$% num_minuses + 1
		if num_minuses mod 2 == 1:
			%$c^{(re)} \leftarrow c^{(re)} \triangle \{j\}$%
	[Update signs due to measurement outcome:]
	if outcome != 0:
		%$c^{(re)} \leftarrow c^{(re)} \cup \{r\}$%
	else:
		%$c^{(re)} \leftarrow c^{(re)} - \{r\}$%
	if %$r \in c^{(im)}$%:
		%$c^{(im)} \leftarrow c^{(im)} - \{r\}$%
\end{lstlisting}
\end{minipage}

\pagebreak[4]
We have the function below to multiply generators that anticommuted with $Z_i$ with the removed generator:

\begin{minipage}{0.95\linewidth}
\begin{lstlisting}[label={code:updategens}, caption={Pseudo-code that multiply anticommuting generators with the removed stabilizer.},style=psedostyle]
update_gens(%$a$%, %$x_r$%,  %$z_r$%, %$g^{t,x}$%, %$g^{t,z}$%, %$q^{t,x}$%, %$q^{t,z}$%):
	for %$j \in a$%:
		for %$k \in x_r$%:
			%$q^{t,x}_k\leftarrow q^{t,x}_k  \triangle \{j\}$%
			%$g^{t,x}_j\leftarrow g^{t,x}_j  \triangle \{k\}$%
		for %$k \in z_r$%:
			%$q^{t,z}_k\leftarrow q^{t,z}_k  \triangle \{j\}$%
			%$g^{t,z}_j\leftarrow g^{t,z}_j  \triangle \{k\}$%
\end{lstlisting}
\end{minipage}

\subsection{Complexity}

The space complexity used by the data structure in the new algorithm is similar to the complexity seen the Anders-Briegel algorithm. That is, each sequence of sets $m^{t,p}$ uses $\mathcal{O}(n \; \overline{m^{t,p}})$ space, where as mentioned before $m \in \{g, q\}$, $t\in \{s,d\}$, $p \in \{x, z\}$, and $\overline{m^{t,p}}$ is the average size of the sets contained in $m^{t,p}$. The complexities of $c^{(im)}$ and $c^{(re)}$ are at worst $\mathcal{O}(n)$. Therefore, the overall space complexity of the data structure is $\mathcal{O}(n \sum \overline{m^{t,p}})$, where the summation runs over the sets $m$, $t$, and $p$.

Now, I will turn the discussion to the time complexity of the algorithm. It is reasonable to model operations such as determining if an element is or is not in a set, adding an element to a set, or removing an element from a set as being able to be done in worst case $O(1)$ time. For example, this is the case when a set is represented using a hash table with a perfect hash function, which can be identity for integers, and enough buckets to avoid collisions. Therefore, by inspection of the pseudo-code provided in this paper, we see that the time complexity for a single-qubit Clifford-gate and a $CNOT$ is

\begin{equation}\label{eq:comp-single}
O(|q^{s,x}|+|q^{s,z}|+|q^{d,x}|+|q^{d,z}|),
\end{equation}

\noindent for a deterministic measurement it is 

\begin{equation}\label{eq:comp-determ}
O(|q^{d,x}|\{{|g^{s,x}|}+{|g^{s,z}|}\}),
\end{equation} 

\noindent and for a non-deterministic measurement it is 

\begin{equation}\label{eq:comp-meas-indeterm} 
O(\{|q^{s,x}|+|q^{d,x}|\}\{{|g^{s,x}|}+{|g^{s,z}|}\} +{|g^{d,x}|}+{|g^{d,z}|}).
\end{equation}

\section{\label{sec:topo-complex}Amortized Analysis of Simulating Topological Stabilizer Codes}

\subsection{Overview}

The goal of this section is to establish practical runtimes for simulating QECCs. A worst-case analysis of an algorithm can often be an overly pessimistic bound and not reflect the practical performance of the algorithm. I will therefore perform an amortized analysis \cite{Tarjan85}. One performs such an analysis on a sequence of operations that are applied to data in a data structure. This analysis consists of determining the worst-case bound on the runtime of each operation in the sequence and averaging over these runtimes. Since an amortized analysis gives an upper bound on the average performance of each operation, an amortized runtime gives a tighter bound on the runtimes experienced in practice. 

The application that I will focus on for our amortized analysis of stabilizer simulations will be the syndrome extraction circuit of the surface-code \cite{SurfCode}, a code targeted for experimental implementation in the near term~\cite{SuperCon,GNRMB14,TCEABCCG16,Woot16}.

\begin{figure}[ht]
		\centering
			\includegraphics[width=10cm]{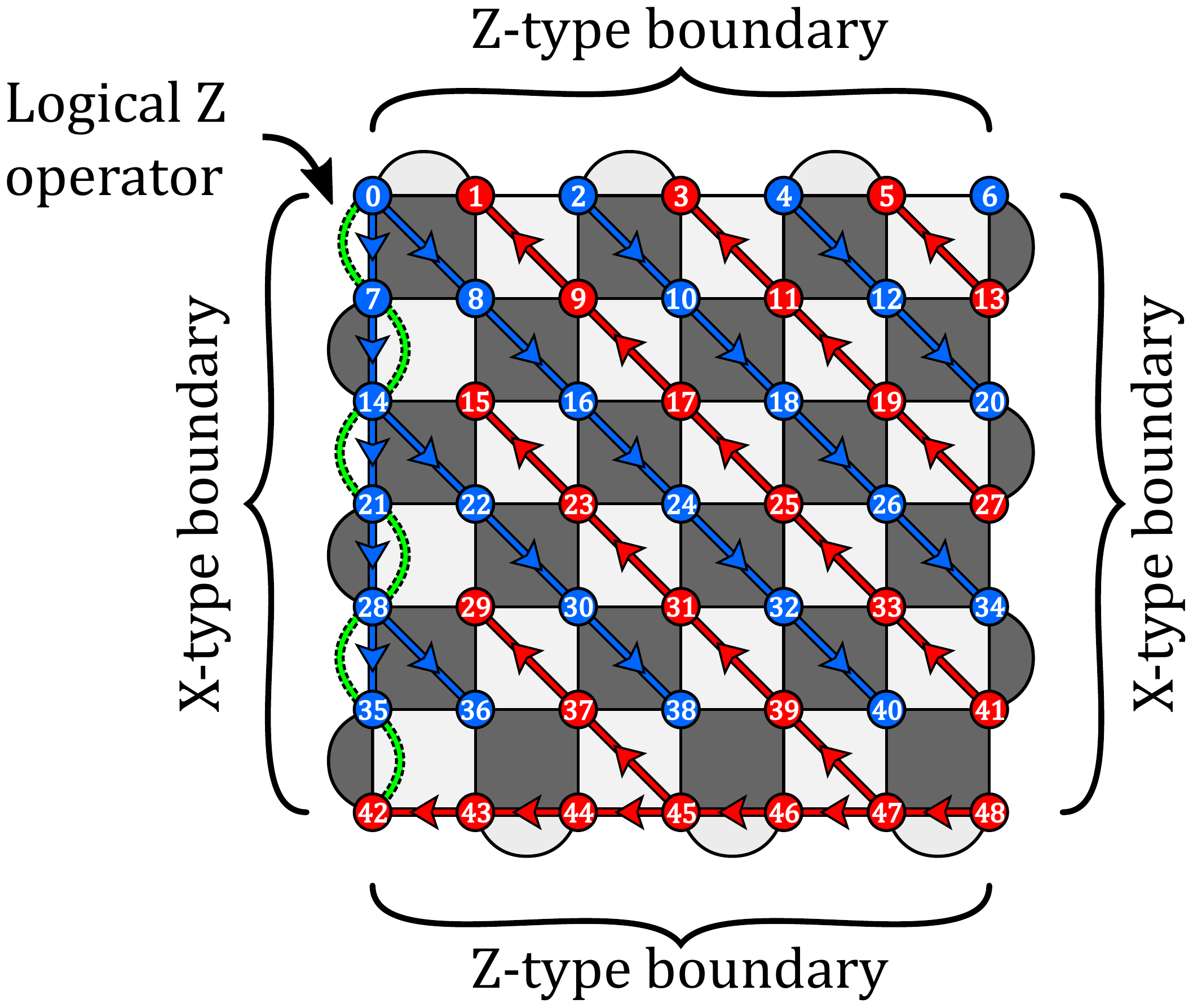}	
		\caption{\label{fig:MedialDestabs} A logical zero-state encoded in a distance-seven, medial surface-code patch. Circles represent data qubits. The curved, green string running along the left-edge of the patch is a logical $Z$ operator, which acts as Pauli $Z$ on the qubit the string touches. Since we are encoding a logical zero-state, this logical $Z$ operator is included as a stabilizer. Destabilizer generators of type $X$ ($Z$) run along the red (blue) paths. Each destabilizer string starts of a boundary of its type, continues along the appropriate path in the direction indicated by arrows, and ends on a stabilizer generator of the opposite type.}
	\end{figure}

I will focus our study on the simulating square patches of the ``medial'' (or ``rotated'') version surface-code which encodes a single logical zero-state, such as shown in Fig.~\ref{fig:MedialDestabs}. A similar analysis for other logical basis-states can be made, and the same amortized runtimes can be found. Although I focus on the surface code, these arguments extend to other codes with similar structures such as other topological stabilizer codes as well as many codes that belong to the broader class of quantum LDPC codes.

The surface-code syndrome-extraction circuit I will consider is composed of $H$, $CNOT$ and $Z$-basis measurement/initialization, where initialization in the $Z$ basis is performed by making a measurement in the $Z$ basis and applying $X$ if the outcome results in $\ket{1}$. Each of these operations is performed $\mathcal{O}(n)$ times during a round of syndrome extraction, where $n$ is the number of qubits in the code; therefore, an amortized runtime for an entire syndrome extraction circuit will be equivalent to $\mathcal{O}(n)$ times the runtime of the circuit operation with the worst amortized runtime. 

Because the runtime of each operation depends directly on the sparsity of the state representation, and because each type of operation is applied uniformly to all the data and/or ancilla qubits, the amortized runtime analysis of operations amounts to determining the average sparsity of data structures. To perform the amortized analysis, I will present what will be considered standard choices for stabilizer/destabilizer generators and then argue for the sparsity of the state representations given these choices.

Before analyzing the runtime of circuit operations, I will first present a few useful definitions that will be used in the following discussion. The weight of a Pauli operator is the number of qubits on which the operator acts non-trivially. If a generator's non-trivial action for each qubit is only as one Pauli type $P$, I refer to the generator as a $P$-type generator. Calderbank-Shor-Steane (CSS) codes \cite{CS95,Steane96b} are stabilizer codes that can be described by stabilizer generators that are $X$ and/or $Z$-type only. A string \cite{BombinTopo} is an operator that acts non-trivially on a path. A string may be on an open or closed path. If strings of type $P$ can begin or end on a code boundary without anticommuting with check on the boundary, that boundary is referred to as a $P$-type boundary. 

\pagebreak[4]
A stabilizer subspace of a stabilizer QECC is often called a \textit{codespace}. The codespace of a code can be used to encode logical qubits, which are logically equivalent to physical qubits. Logical qubits also have logical operators that act like the corresponding physical operators on the logical qubits. For example, logical $Z$ applied to logical $\ket{0}$ gives logical $\ket{0}$, while logical $Z$ applied to logical $\ket{1}$ gives logical $-\ket{1}$. Logical operators and states are often distinguished by adding a bar above the symbols. So logical $Z$ is often written as $\overline{Z}$, and logical $\ket{0}$ is often written as $\ket{\overline{0}}$. As there is an equivalence between logical and physical operators, logical operators have the same algebra associated with them. For example, $\overline{X}\;\overline{Z} = -\overline{Z}\;\overline{X}$. If a stabilizer state is a logical basis-state, then the corresponding logical operator that stabilizes the logical basis-state is included in the stabilizers of the state. For example, $\ket{\overline{0}}$ is stabilized by $\overline{Z}$, so $\overline{Z}$ is included as a stabilizer of $\ket{\overline{0}}$.

\begin{figure}[H]
	\begin{subfigure}[t]{7cm}
		\includegraphics[width=7cm]{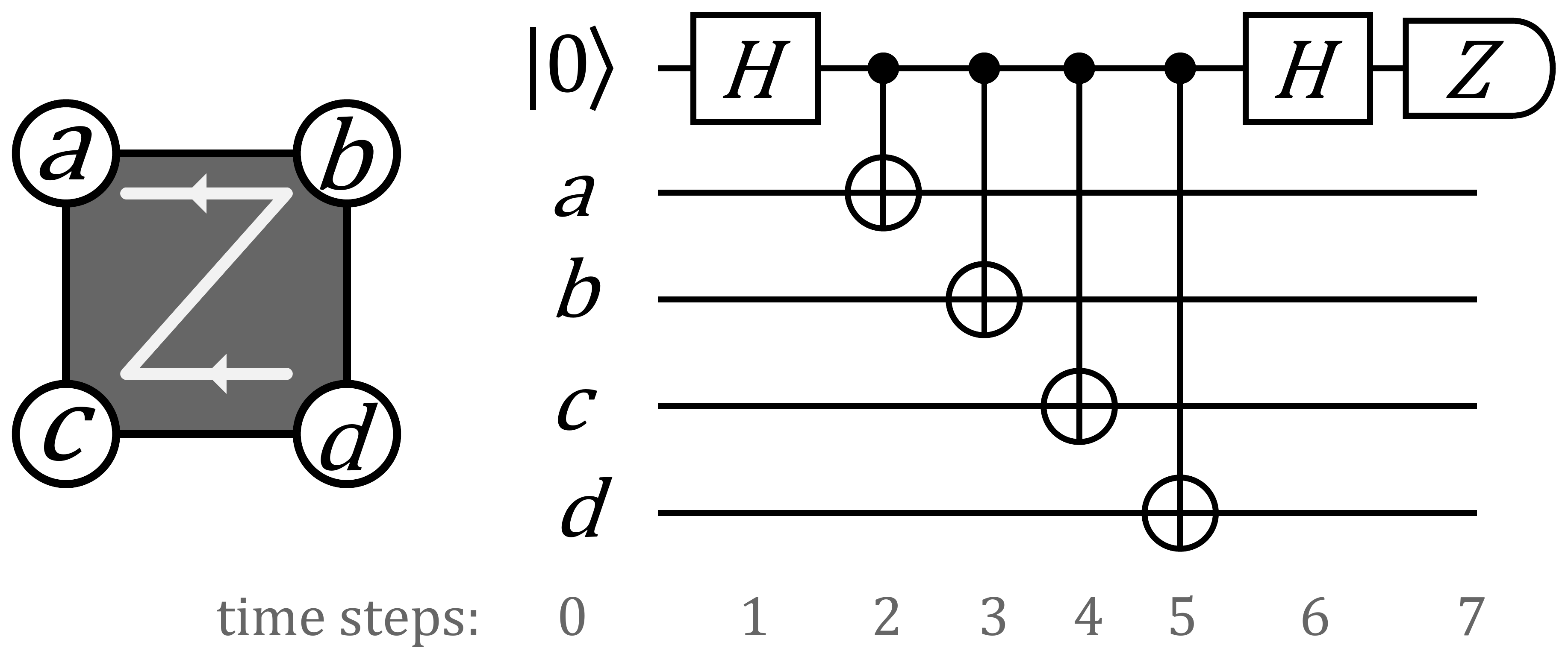}
		\caption{\label{fig:xcircuits} $X$-check}
	\end{subfigure}
	\begin{subfigure}[t]{7cm}
		\includegraphics[width=7cm]{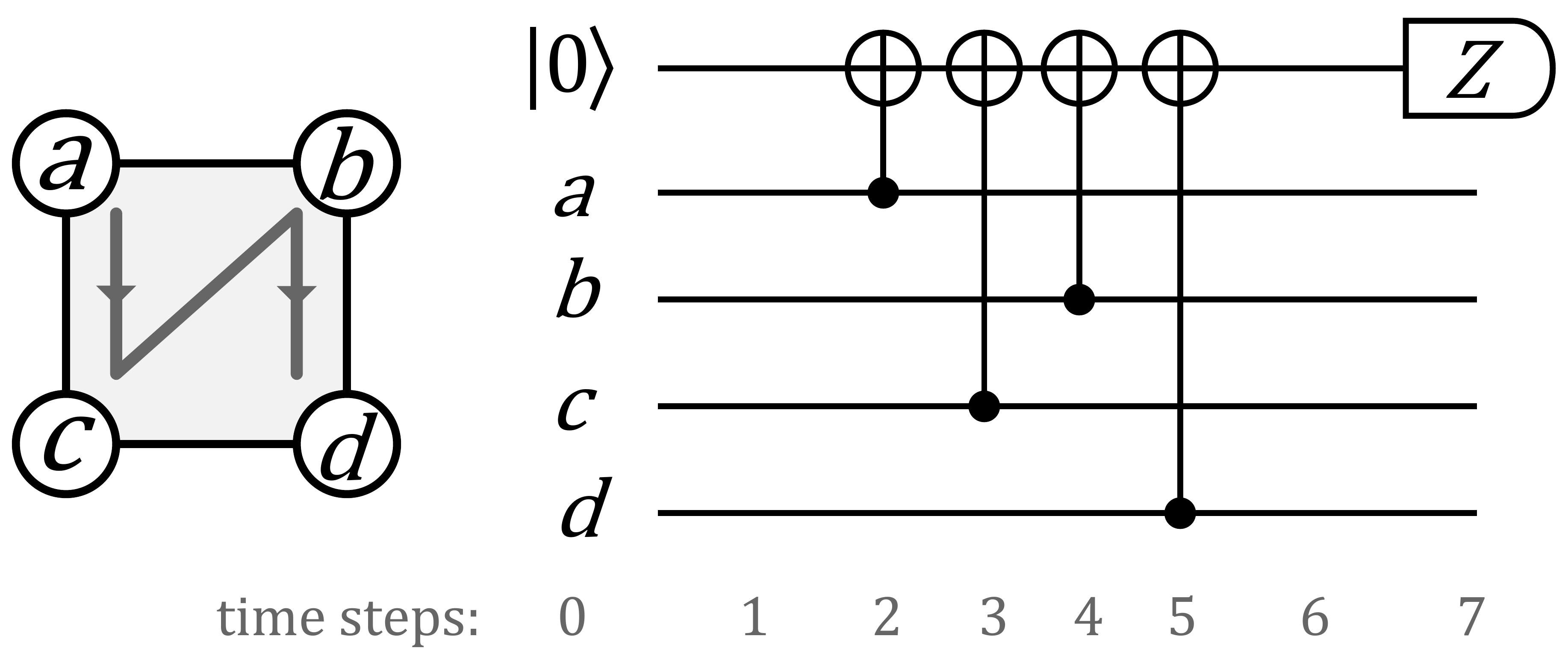}
		\caption{\label{fig:zcircuits} $Z$-check}
	\end{subfigure}
\caption{\label{fig:surf-checks} The circuits used to measure the stabilizer operators (a) $X_a\otimes X_b\otimes X_c\otimes X_d$ and (b) $Z_a\otimes Z_b\otimes Z_c\otimes Z_d$. At the center of each check is an ancilla, which is used to output the result of measuring the stabilizer operator.}
\end{figure}

I will now continue discussing the analysis of syndrome extraction for the surface code. A round of syndrome extraction consists of measuring stabilizer generators. These stabilizer measurements are known as \textit{checks}, and I will consider enacting these checks through the circuits shown in Fig.~\ref{fig:surf-checks}, where there is one ancilla per check. Note, for general QECCs, such check circuits as show in Fig.~\ref{fig:surf-checks} do not guarantee fault-tolerance; however, these check circuits do allow the surface-code to be fault-tolerant. In a surface-code patch with $n_{data}$ data qubits, to encode one logical qubit there are $n_{data} - 1$ independent checks. For the check circuits I am considering, there is one ancilla qubit per check; thus, there are $n_{ancilla} = n_{data} - 1$ ancilla qubits. The total number of physical qubits is therefore $n = n_{data} + n_{ancilla} = 2 n_{data} - 1$. Note, unlike previous discussions of stabilizer and destabilizer generators of QECCs in this paper, I will now include ancilla qubits in addition to data qubits.

\subsection{Analysis of the Gottesman, the Aaronson-Gottesman, and the New Algorithm}

I now define a standard choice of stabilizer and destabilizer generators to initially represent a logical zero-state before syndrome extraction. I then determine the upper bound on the average sparsities of this choice and argue how the sparsities of the generators change during a single or multiple rounds of syndrome extraction. Because the Gottesman, the Aaronson-Gottesman, and the new algorithm can use the same choice of stabilizer/destabilizer generators and because the data structures employed in these algorithms are similar, I will analyze the runtimes of these algorithms together. Following this analysis, I will define a standard choice of graph-state stabilizer generators and argue for the amortized runtime complexity of the Anders-Briegel algorithm.

To help with clarity and for brevity, in the following discussions I will refer to stabilizer/destabilizer generators that initially only had support on data (ancilla) qubits as data (ancilla) stabilizer and destabilizer generators.

First, I will discuss the ancilla stabilizer and destabilizer generators. Each ancilla $i$ is initialized with the stabilizer $Z_i$, \textit{i.e.}, each starts in the state $\ket{0}_i$. The corresponding destabilizer generator on qubit $i$ is $X_i$. Therefore, at initialization, the weight of the generators is one and the maximum number of generators acting on any qubit is also one. Thus, initially, for ancilla generators $|q^{s,x}|=|q^{s,z}|=|q^{d,x}|=|q^{d,z}|=1$ and $|g^{s,x}|=|g^{s,z}|=|g^{d,x}|=|g^{d,z}|=1$. I discuss later how these sparsities change as circuit operations are applied to them. 

I now discuss the initial data stabilizer-generators. Since each check measures an independent stabilizer, I will choose these measured stabilizers as a standard set of stabilizer generators. Because I am simulating a logical zero-state, logical $Z$ must be a stabilizer. I include logical $Z$ as a stabilizer generator and choose a representation of the operator that has a minimum weight. In particular, I will choose a representation that runs along the left edge of the patch like the logical $Z$ operator in Fig.~\ref{fig:surf-checks}. The minimum weight of the logical operator is called the \textit{distance} $d$ of the code and is equal the number of data qubits along an edge of a patch. Due to the relationship between the distance and geometry of the code, $d = \mathcal{O}(\sqrt{n})$.

From these arguments, we see that, initially, the stabilizer generators of the data qubits $|q^{s,x}|=|q^{s,z}|=|g^{s,x}|=\Theta(1)$. One sees that $|g^{s,z}|=\mathcal{O}(\sqrt{n})$ due to the stabilizer generator, logical $Z$; however, the average size of the sets in $g^{s,z}$, which I will notate as $\overline{g^{s,z}}$, is equal to $\mathcal{O}(\frac{c (n-1) + \sqrt{n}}{n}) = \mathcal{O}(1)$, where $c$ is some constant. We then have $\overline{q^{s,x}}=\overline{q^{s,z}}=\overline{g^{s,x}}=\overline{g^{s,z}}=\mathcal{O}(1)$ for the data stabilizer-generators.

We will now turn our attention to choosing a collection of suitable data destabilizer-generators. Each stabilizer generator must have one and only one anticommuting, destabilizer-generator partner. Like stabilizer generators, destabilizer generators must also commute with each other. To achieve these commutation relations for the destabilizer generators of the data qubits, I will choose a collection of destabilizers such as those seen in Fig.~\ref{fig:MedialDestabs}. Each of these destabilizer generators is either a string of type $X$ or $Z$ that starts at a boundary of its type and ends on the stabilizer-generator that it anticommutes with. In Fig.~\ref{fig:MedialDestabs} $X$ ($Z$)-type destabilizers lie along the red (blue) paths. For example, on the red path that begins on qubit 27, there are four $X$-type destabilizer strings that act on qubits $\{27\}$, $\{27, 19\}$, $\{27, 19, 11\}$, and $\{27, 19, 11, 3\}$. By inspection one can see that destabilizer generators such as those indicated by Fig.~\ref{fig:MedialDestabs} have weights that are at most $\mathcal{O}(d)=\mathcal{O}(\sqrt{n})$.  

I will now discuss the average number of data destabilizer-generators incident on a qubit. To do this I will first focus on the data $X$-type destabilizer-generators. Many of the destabilizer generators in Fig.~\ref{fig:MedialDestabs} lie along purely diagonal paths that originate on a boundary. There are at most $\mathcal{O}(d)$ destabilizer generators that lie along any one of these paths; therefore, at most $\mathcal{O}(d)$ destabilizer generators are incident on a data qubit in these diagonal paths. There is one path that only runs along the bottom boundary and $\mathcal{O}(\frac{d-1}{2})$ paths that branch off to run diagonal, \textit{e.g.}, from the points corresponding to qubits 47, 45, and 43 in Fig.~\ref{fig:MedialDestabs}. Each of these paths have at most $\mathcal{O}(d)$ destabilizer generators that lie along them. Therefore, at most $\mathcal{O}(\frac{d-1}{2}d) = \mathcal{O}(d^2)$ destabilizer generators are incident on any data qubit along the bottom boundary, and at most $\mathcal{O}(d)$ destabilizer generators are incident on any data qubit in the diagonal portions of the paths. Since along the bottom boundary there are $d$ data qubits and there are $\mathcal{O}(n)$ data qubits in the rest of the patch, we find for $X$-type data destabilizer-generators $\overline{g^{d,x}}= \mathcal{O}(\frac{d^3 + n d}{n})= \mathcal{O}(\frac{n+2 n ^{3/2}}{n})=\mathcal{O}(\sqrt{n})$. 

A similar argument can be made for the $Z$-type data destabilizer generators to give $\overline{g^{d,z}}=\mathcal{O}(\sqrt{n})$.

So far, I have argued for the average sparsities for an initial standard choice of stabilizer and destabilizer generators. I will now analyze how they evolve as syndrome-extraction circuit-operations are applied to these generators.

 By applying the circuits of Fig.~\ref{fig:surf-checks}, one finds that during odd applications (where the counting start with one) of syndrome extraction, right before the ancilla qubits are measured, the stabilizer generators of each ancilla qubit is multiplied by the data stabilizer-generator corresponds to that ancilla qubit. After an even application, the ancilla stabilizer generator will once again become the original weight one operator on the ancilla qubit. We therefore see that the average weight of the ancilla stabilizer generators remains constant. Also, the ancilla stabilizer generators are incident on at most a constant number of data and ancilla qubits. After applying a check circuit to an ancilla destabilizer-generator one finds that this generator remains a weight-one generator acting on the original ancilla. At most, the Pauli type of a destabilizer generator changes when Hadamard gates are applied. From this discussion I have shown that the ancilla generators do not change the upper bound on the average sparsities that we have discussed previously.

I now discuss how data stabilizer and destabilizer generators evolve as syndrome extraction operations are applied. A data generator is not modified by checks measuring the same Pauli type as the generator. Because a data generator is incident on any check of the opposite Pauli type, during repeated applications of syndrome extraction, one finds by applying the check circuits of Fig.~\ref{fig:surf-checks} that a data generator will at most be multiplied by Pauli operators acting on the ancillas of checks of the opposite Pauli type that data generators are incident on. Data stabilizer generators touch at most two checks of the opposite Pauli type; therefore, data generators only ever have a constant increase in weight and contribute to only a constant increase in the number of stabilizer generators incident on any ancilla qubit. 

Each data destabilizer string is of weight $\mathcal{O}(d)$ and is incident on $\mathcal{O}(d)$ checks of the opposite Pauli type. The weight of the data destabilizer-strings, therefore, only at most increases by a multiplicative constant amount and remains $\mathcal{O}(d)$. Since each data destabilizer-string may also be incident on the ancilla qubits associated with the checks that touch the destabilizer generator, as with the data qubits, we now find for the ancilla qubits that $\overline{q^{d,x}} = \overline{q^{d,z}} =\overline{g^{d,x}} = \overline{g^{d,z}} =\mathcal{O}(\sqrt{n})$.

With our standard choice of stabilizer and destabilizer generators, we have found that throughout multiple applications of syndrome-extraction operations 

\begin{equation}
\overline{q^{s,x}}=\overline{q^{s,z}}=\overline{g^{s,x}}=\overline{g^{s,z}}=\mathcal{O}(1)
\end{equation}

\noindent and 

\begin{equation}
\overline{q^{d,x}}=\overline{q^{d,z}}=\overline{g^{d,x}}=\overline{g^{d,z}}=\mathcal{O}(\sqrt{n}) 
\end{equation}

\noindent for both data and ancilla qubits.

Using Equations \ref{eq:comp-single} to \ref{eq:comp-meas-indeterm}, we find that if we represent a stabilizer state using these generators, then the practical runtime for each circuit operation for the new algorithm is as listed in Table~\ref{tb:TheoryComplex}. If we use the proposed generators for the Aaronson-Gottesman algorithm, the runtime complexity for non-deterministic measurements will be $\mathcal{O}(n^{3/2})$ while other operations will be at most $\mathcal{O}(n)$. The runtime of non-deterministic measurements can be improved by noting that the runtime of this operation depends on $|q^{s,x}|+|q^{d,x}|$ and that the $X$-type destabilizers can be multiplied together to form a collection of destabilizer generators that are single-qubit $X$ operators. To maintain the commutation relations between stabilizer and destabilizers generators, if we multiply a destabilizer $d_1$ with the destabilizer $d_2$ to get $d_1\rightarrow d_1 d_2$ while letting $d_2 \rightarrow d_2$, then the corresponding stabilizer generators $s_1$ and $s_2$ will be updated as $s_1\rightarrow s_1$ and $s_2 \rightarrow s_1 s_2$. Because of this, to get weight-one $X$-type destabilizer generators, the $Z$-type stabilizer generators must be multiplied together. Using similar arguments as above, it can be shown that such a choice of stabilizer and destabilizer generators will result in   

\begin{equation}
\overline{q^{s,x}}=\overline{q^{d,x}}=\overline{g^{s,x}}=\overline{g^{d,x}}=\mathcal{O}(1)
\end{equation}

\noindent and 

\begin{equation}
\overline{q^{s,z}}=\overline{q^{d,z}}=\overline{g^{s,z}}=\overline{g^{d,z}}=\mathcal{O}(\sqrt{n}). 
\end{equation}

This second choice of stabilizer and destabilizer generators gives more favorable results for the complexities for both the Aaronson-Gottesman and the new algorithm, and is what I report in the Table~\ref{tb:TheoryComplex}. If, as we do in numerical experiments below, one initializes the logical zero-state by first initializing in the computational state $\ket{0}^{\otimes n}$ and then measuring the checks to project the state to the stabilizer state of the surface code, then both the Aaronson-Gottesman and the new algorithm will choose the generators that are very similar to this second choice of generators. For the Gottesman algorithm, only the runtime of non-deterministic measurements are affected by the sparsity of data structure; however, deterministic measurement runtimes dominate and take $\mathcal{O}(n^3)$ steps.

\subsection{Analysis of the Anders-Briegel Algorithm}

I now discuss the runtime complexities for the Anders-Briegel algorithm. Since we are simulating a logical zero-state, the logical $Z$ must be included as a stabilizer. The minimum weight of the logical operator is $d=\mathcal{O}(\sqrt{n})$ and cannot be reduced further. As mentioned previously in Section~\ref{sec:anders-breigel}, a graph-state representation can be viewed as a particular choice of stabilizer generators. Since different choices of stabilizer generators can not reduce the minimum weight of the logical operator, the maximum degree of the graph represented in Anders-Briegel algorithm must be $\Delta=\mathcal{O}(d)=\mathcal{O}(\sqrt{n})$. 

For CSS states, such as a surface-code state, a typical way to convert the stabilizer generators into graph-state generators is to use a protocol discussed in \cite{CL04} that makes use of Gaussian elimination. Our standard choice for the graph-state representation will be the graph-state representation resulting from applying this procedure to our standard choice of stabilizer generators.

\begin{figure}[ht]
		\centering
			\includegraphics[width=0.8\textwidth]{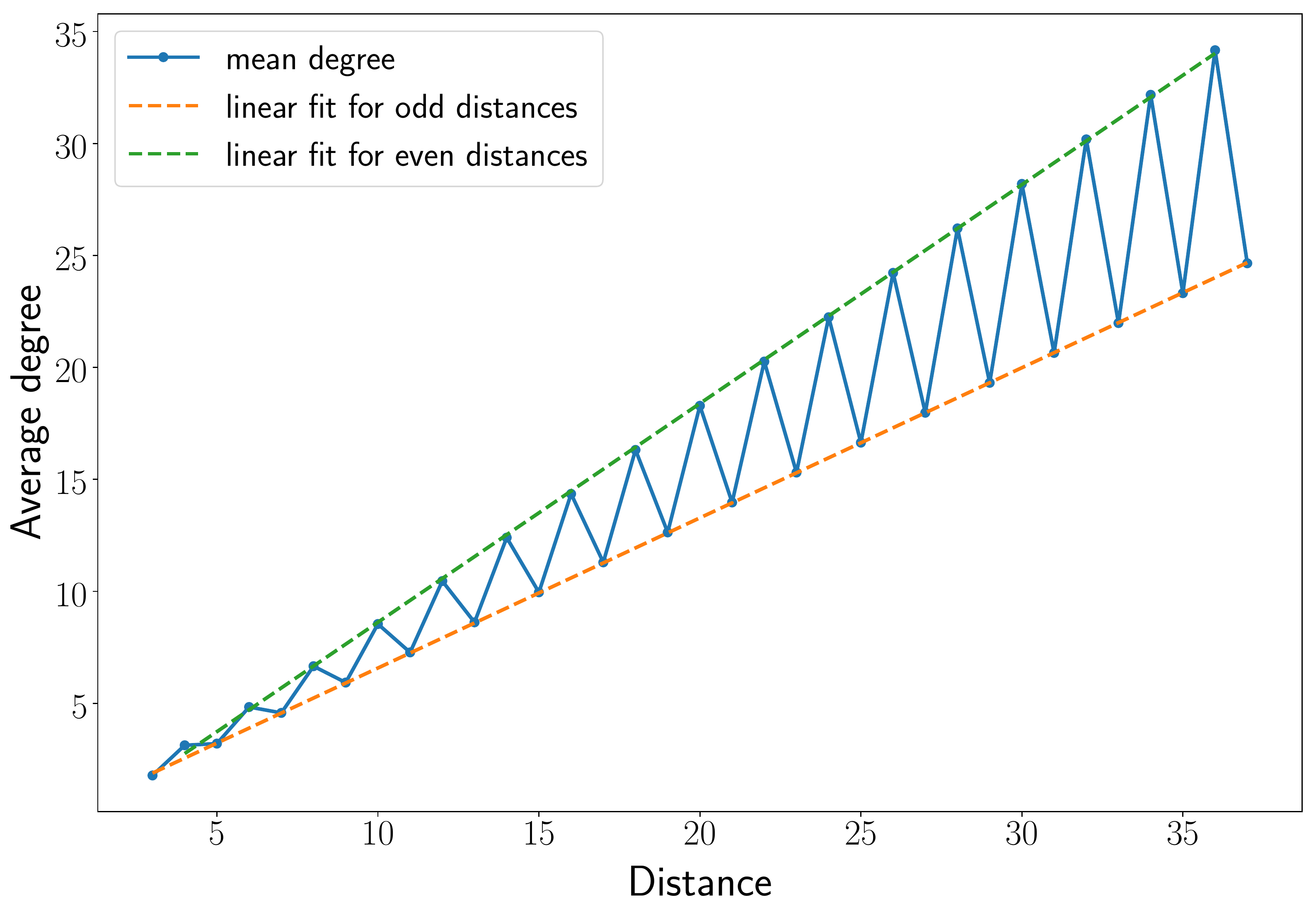}	
		\caption{The average degree of the graph-state stabilizer-generators on data qubits for a standard choice of graph state representing a logical $\ket{0}$ planar medial surface-code. The average degree corresponding to odd and even distance codes grows linearly with the distance of the code. The zig-zag nature of the date warrants further study. \label{fig:stand-graph}}
\end{figure}

By using Gaussian elimination to create the standard choice of graph-state representation, we find for the initial state the that the graph-state stabilizer generators on the data qubits have $\overline{\Delta}=\mathcal{O}(d)=\mathcal{O}(\sqrt{n})$. This result is shown in Fig.~\ref{fig:stand-graph}. 

As with the other algorithms, the stabilizers on the ancilla qubits are initialized as Pauli $Z$ on the qubits. During syndrome extraction, each ancilla qubit is entangled by a CNOT gate with the data qubits of the check associated with the ancilla qubit. Each application of a CNOT between the ancilla and data qubit adds an edge between these qubits in the associated graph of the state. After a measurement of an ancilla qubit, the stabilizer of the ancilla qubit is once again a weight-one operator on the ancilla qubit. Because of this, the degree on the ancilla qubits is at most four, or $\Theta (1)$, during syndrome extraction. Since each data qubit is involved in a constant number of checks, the degree of the data qubits remains $\mathcal{O}(\sqrt{n})$. 

We therefore find that our standard choice of graph-state representation results in the complexities listed in Table~\ref{tb:TheoryComplex}.

\subsection{Conclusion}

The results of my analysis are seen in Table~\ref{tb:TheoryComplex}. From this table we see that the new algorithm has an overall square-root improvement in the amortized worst-case runtime for circuit operations.

\begin{table}[ht]
    \centering
\resizebox{14cm}{!}{
\begin{tabular}[b]{|l|c|c|c|c|}
\hline
Algorithm & SU(2) & $CNOT$ & $Z$ measurement & Maximum runtime \\ \hline
Gottesman & $\mathcal{O}(n)$ &  $\mathcal{O}(n)$ & $\mathcal{O}(n^3)$ & $\mathcal{O}(n^3)$ \\ \hline
Aaronson-Gottesman & $\mathcal{O}(n)$ & $\mathcal{O}(n)$ & $\mathcal{O}(n)$ & $\mathcal{O}(n)$ \\ \hline
Anders-Briegel & $\Theta (1)$ & $\mathcal{O}(n)$ &  $\Theta(1)$ &  $\mathcal{O}(n)$ \\ \hline
New algorithm & $\mathcal{O}(\sqrt{n})$ & $\mathcal{O}(\sqrt{n})$ & $\mathcal{O}(\sqrt{n})$ &  $\mathcal{O}(\sqrt{n})$ \\ \hline
\end{tabular}}
\caption{\label{tb:TheoryComplex} The amortized runtime complexity of the quantum operations performed during syndrome extraction of a medial surface-code patch on $n$ qubits.}
\end{table}

While I have presented the runtime of performing syndrome extraction, one might be curious about the runtime required for initializing the state. There are at least two possible methods. One option is to prepare and store a state representation ahead of time. The construction of the state would then be a one-time cost. After the state has been prepared, for each initialization and simulation of the state, one could copy the stored state. The cost to copy the state is on the order of the space complexity of the data-structure of the state, which for the algorithms discussed, is at most the cost of one round of syndrome extraction. Alternatively, one could measure the syndrome extraction circuit to project the state to the stabilizer state. Such a procedure would, again, require at most the time it took to run one round of syndrome extraction.

In my runtime analysis, I considered the simulation of syndrome extraction without errors. When simulating QECC protocols below or near the threshold of a code, error events are fairly infrequent and stabilizer measurements will rapidly project the state back into the stabilizer subspace of the code (up to stabilizer signs); therefore, for such simulations, the runtimes of circuit operations are expected to not deviate significantly from the runtimes without noise. In the next section, I will give numerical evidence in support of this conclusion.

\section{\label{numer-exp}Implementation and Numerical Experiments}

In this section, I will present an implementation of the new algorithm and the results of numerical experiments, which closely follow the predicted runtimes in Table~\ref{tb:TheoryComplex}.

\subsection{Implementation}

I have developed a \pack{Python} package called ``Performance Estimator of Codes On Surfaces'' (\PECOS) to study and evaluate stabilizer codes (see Chapter~\ref{ch.pecos} for a discussion of the software). As part of this package I implemented two versions of my stabilizer simulation algorithm. One was written in pure \pack{Python} \cite{PythonDoc} and the other in \pack{C++} \cite{Cpp2017}. \pack{Cython} \cite{behnel2010cython} was used to wrap the \pack{C++} implementation. Since \pack{CHP} is written in \pack{C} \cite{Cbook} and \pack{GraphSim} is written in \pack{C++}, I will focus our discussion on the \pack{C++} implementation and refer to my implementation of the new algorithm as \pack{SparseSim} (see also Section~\ref{sec.pecos.state_sim}).

An example of using \pack{SparseSim} can be seen in Code Block~\ref{code:surf-init}.

\begin{minipage}{0.95\linewidth}
\begin{lstlisting}[label={code:surf-init},caption={An example of using (\texttt{PECOS} Chapter~\ref{ch.pecos}) to initialize a logical-zero state by measuring the checks of a distance-three surface-code. Here the \texttt{Cython} wrapped \texttt{C++} implementation of \texttt{SparseSim} (\texttt{cysparsesim}) is used. A pure \texttt{Python 3} implementation (\texttt{sparsesim}) can be imported as well.},style=pystyle]
from pecos.simulators import cysparsesim

state = cysparsesim(16)
state.run_gate('H', {10, 4, 12, 6})
state.run_gate('CNOT', 
	{(15, 16), (10, 7), (4, 2), 
	(8, 11), (12, 9), (3, 5)})
state.run_gate('CNOT', 
	{(12, 15), (10, 13), (14, 16), 
	(4, 8), (7, 11), (2, 5)})
state.run_gate('CNOT', 
	{(6, 3), (2, 0), (14, 11), 
	(9, 5), (4, 1), (12, 8)})
state.run_gate('CNOT', 
	{(6, 9), (4, 7), (12, 14), 
	(1, 0), (8, 5), (13, 11)})
state.run_gate('H', {10, 4, 12, 6})
results = state.run_gate('measure Z', 
	{0, 16, 4, 5, 6, 10, 11, 12}, random_outcome=0)
\end{lstlisting}
\end{minipage}

\texttt{PECOS} also includes the planar, medial surface-code as a QECC class; therefore, the Code Block~\ref{code:surf-init2} is also equivalent to Code Block~\ref{code:surf-init}.

\begin{minipage}{0.95\linewidth}
\begin{lstlisting}[label={code:surf-init2},caption={An equivalent example to Code Block~\ref{code:surf-init}. Here a class representing a planar, medial surface-code is used. See Chapter~\ref{ch.pecos} for further discussion on the code seen in this block.},style=pystyle]
import pecos as pc
# Create QECC object
qecc = pc.qeccs.SurfaceMedial4444(distance=3)
# Create logical circuit
init_circ = pc.circuits.LogicalCircuit()
init_circ.append(qecc.gate('ideal init |0>'))
# Initialize the state with SparseSim
sparsesim = pc.simulators.cysparsesim
state = sparsesim.State(qecc.num_qudits)
# Run logical circuit
circ_runner = pc.circuit_runners.Standard()
results = circ_runner.run_logic(state, init_circ)
\end{lstlisting}
\end{minipage}

To help determine if the implementation of the algorithm presented in this work contains errors, I followed a similar procedure used in \cite{GraphSim} to determine if the simulations of \pack{GraphSim} were consistent with \pack{CHP}. To do this I wrote a script to compare the measurement results of \pack{CHP}, \pack{GraphSim}, and \pack{SparseSim}. For each simulator I initialized a state of 500 qubits. I then had each simulator concurrently apply the same ten million random gates and measurements to their states. If a random outcome occurred during a measurement, \pack{GraphSim} and \pack{SparseSim} were forced to ``randomly'' choose the same outcome as \pack{CHP}. Deterministic outcomes of the three simulators were checked against each other and were found to always be the same. Given these results, one may have a high degree of confidence that simulations by \pack{SparseSim} are consistent with both \pack{CHP} and \pack{GraphSim}.

Both \pack{GraphSim} and \pack{SparseSim} use hash-table data-structures to represent sets. The data structure used in the original implementation of \pack{GraphSim} by Anders and Briegel in 2004 was \pack{hash\_set}, which is not part of the ISO \pack{C++} standard library. The data-structure \texttt{unordered\_set} was later introduced to the standard library and replaced \texttt{hash\_set}. Since \texttt{SparseSim} uses \texttt{unordered\_set}, \texttt{hash\_set} was replaced with \texttt{unordered\_set} in \texttt{GraphSim} to provide a more direct comparison between the implementations in the numerical experiments presented in this paper. In practice, I found that \texttt{unordered\_set} improved the performance of \texttt{GraphSim}.

\subsection{Numerical Experiments}

I now introduce the results of numerical experiments, which were run on a 64-bit Linux system and were compiled using the \texttt{GNU} compiler version 4.8.5 \cite{GNU} with the compiler option ``O3.''
	
\begin{figure}[ht]
		\centering
			\includegraphics[width=0.75\textwidth]{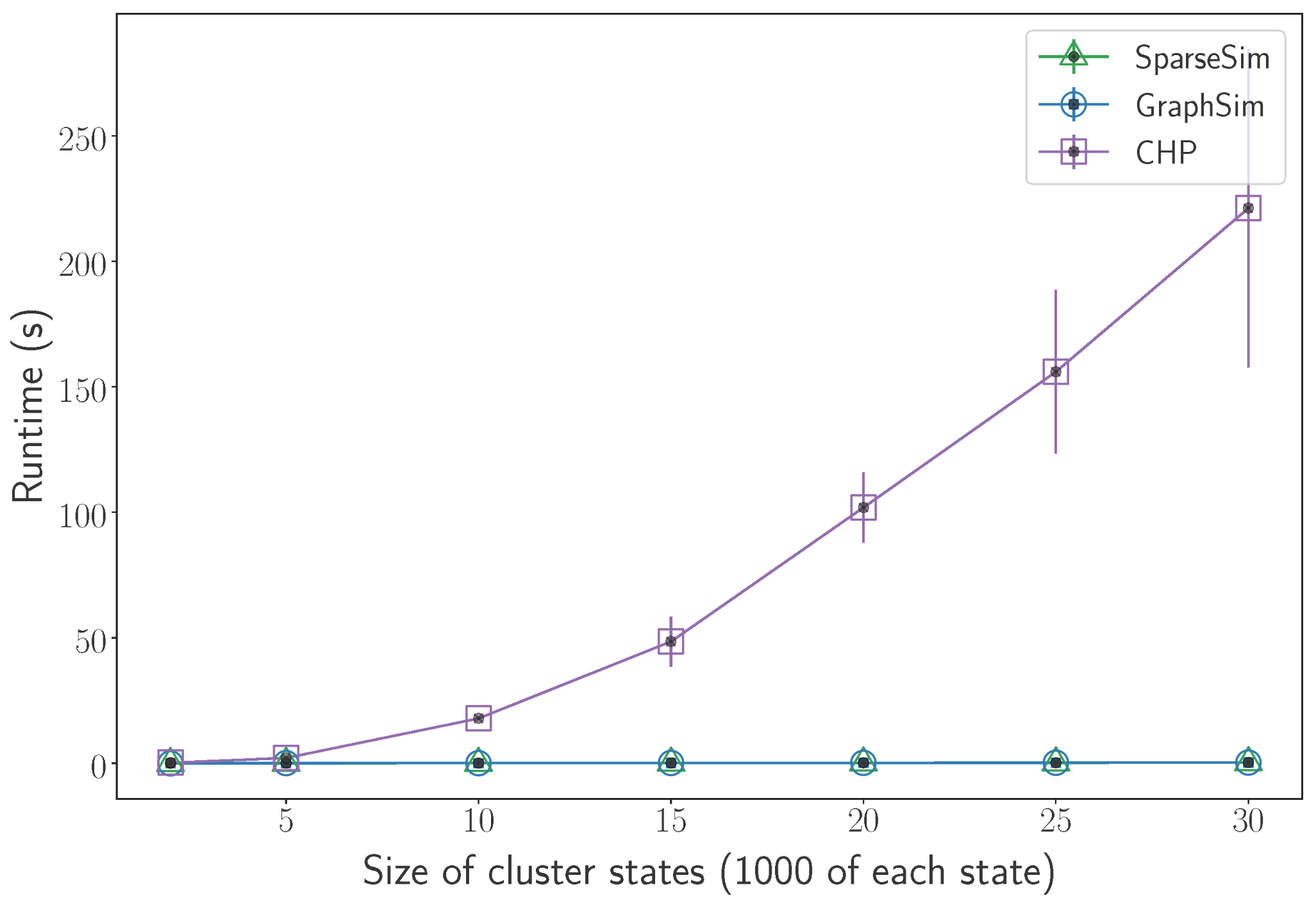}	
		\caption{Total runtime for Entanglement Purification. Note that the \texttt{GraphSim} line lies on top of the \texttt{SparseSim} line. Also, the $x$-axis indicates the size of the linear cluster states. 1000 cluster states of each size is stored in the simulation; therefore, the rightmost data point is a simulation of 30,000 qubits. \label{fig:entang1}}
\end{figure}

\begin{figure}[ht]
		\centering
			\includegraphics[width=0.75\textwidth]{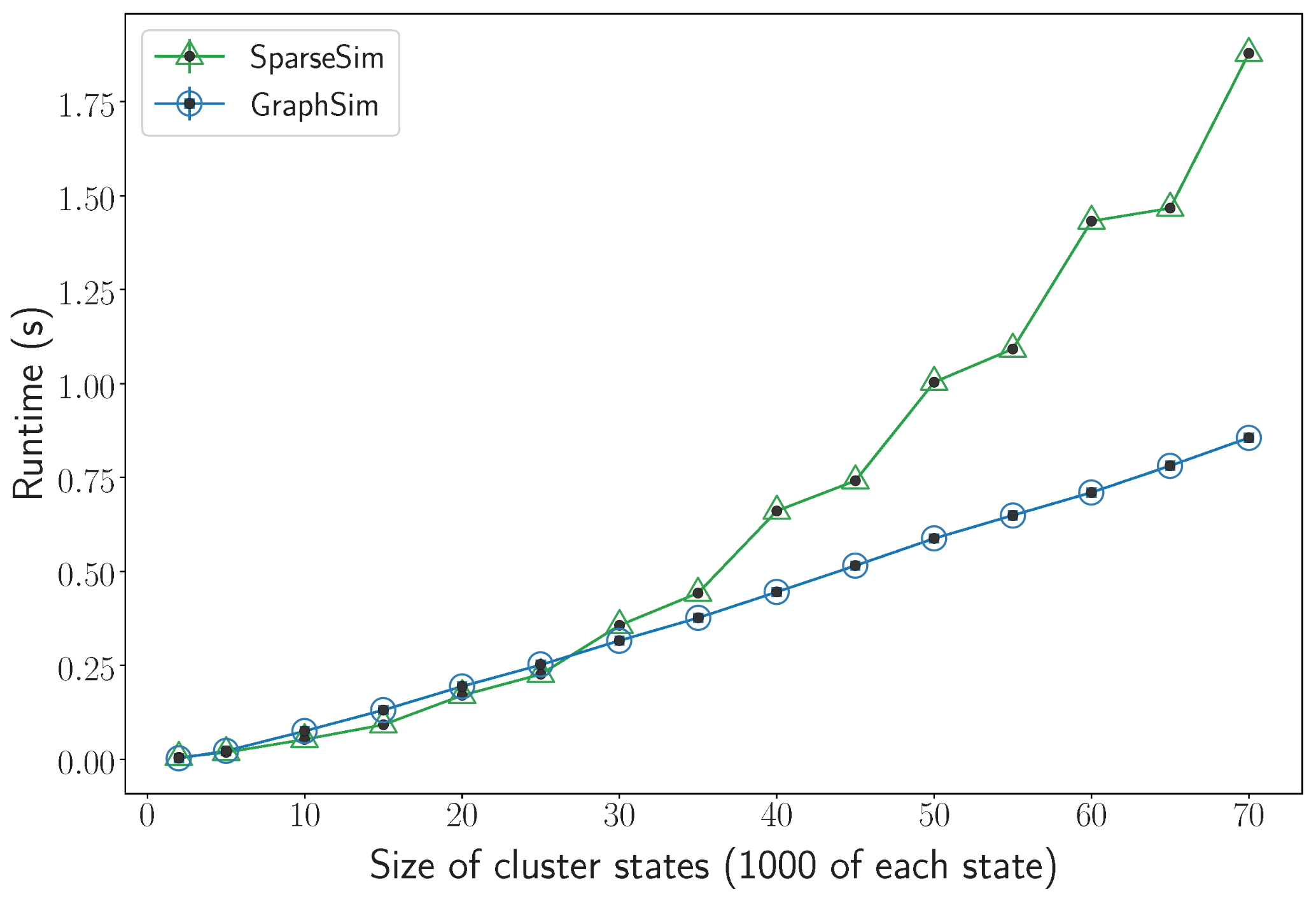}	
		\caption{Total runtime for entanglement purification for just GraphSim and SparseSim.\label{fig:entang2}}
\end{figure}

In the paper that introduced \texttt{GraphSim} \cite{GraphSim}, to compare the performance of \texttt{CHP} and \texttt{GraphSim}, numerical results of the entanglement purification of linear cluster states according to the protocols in \cite{DAB03} were presented. I repeated this experiment and show the results in Fig.~\ref{fig:entang1} and Fig.~\ref{fig:entang2}. As reported previously in \cite{GraphSim}, for this application, \texttt{GraphSim} demonstrates superior performance over \texttt{CHP}. 

In Fig.~\ref{fig:entang2} we see that \texttt{SparseSim} is more competitive with \texttt{GraphSim}; however, for cluster states of large enough sizes \texttt{GraphSim} outperforms \texttt{SparseSim}. This difference in performance is because for this application, the stabilizer generators of both implementations are on average of constant weight; however, the weights of the destabilizer generators chosen by \texttt{SparseSim} scale linearly with the size of the cluster state. There are $\mathcal{O}(n)$ operations for the entanglement purification procedure we studied; therefore, for this application the total runtime of \texttt{GraphSim} is $\mathcal{O}(n)$ and \texttt{SparseSim} is $\mathcal{O}(n^2)$.

\begin{figure}[ht]
\vspace{0.5mm}
		\centering
			\includegraphics[width=0.75\textwidth]{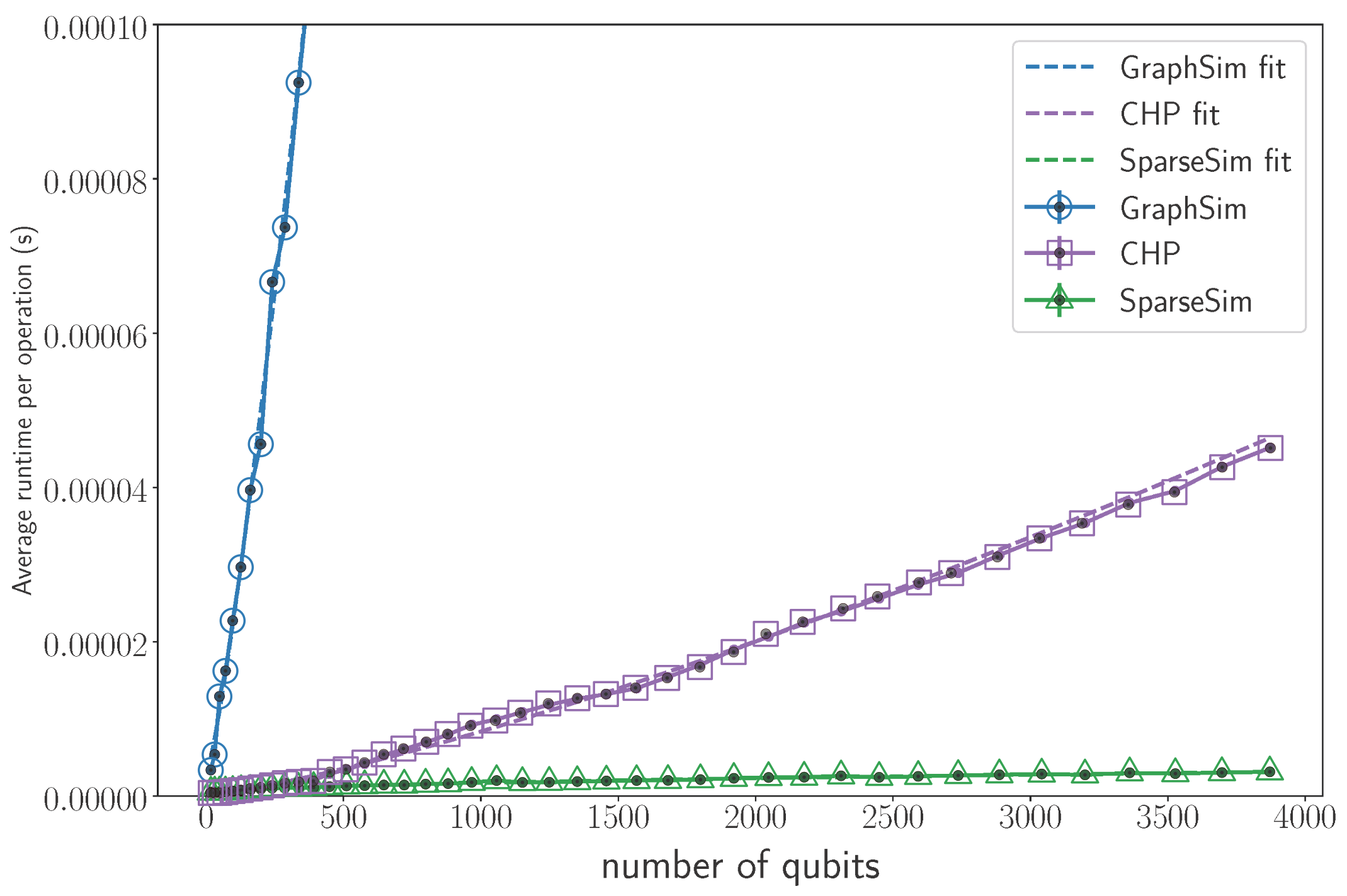}	
		\caption{Average runtime per operation for a round of syndrome extraction of the medial surface-code. The rightmost data points correspond to distance 45. \label{fig:op-time}}
\end{figure}

\pagebreak[4]
However, the application that is the focus of this work is syndrome extraction of the surface code and, by extension, the syndrome extraction of other topological codes and quantum LDPC codes. In Fig.~\ref{fig:op-time}, the average runtime for a circuit operation performed during a single round of syndrome extraction is plotted. Power-law fits to the curves seen in Fig.~\ref{fig:op-time} are given in Table~\ref{tb:Fit}. Both the Aaronson-Gottesman and the Anders-Briegel algorithms have power-law fits that are slightly greater than predicted by the amortized analysis results seen in Table~\ref{tb:TheoryComplex}. These differences may be due to hidden costs in the implementation of the algorithms. We see that experimental run of \texttt{SparseSim}, however, does match the amortized analysis of the new algorithm.

\begin{table}[H]
    \centering
\resizebox{14.5cm}{!}{
\begin{tabular}[b]{|l|c|c|c|}
\hline
Implementation & Fit (seconds) & $R^2$ & Predicted Performance\\ \hline
CHP &  $(1.202 \pm 0.010)10^{-9} n^{1.278\pm 0.001} + (1.766\pm0.030)10^{-7}$ & $99.7\%$ & $\mathcal{O}(n)$ \\ \hline
GraphSim & $(6.92\pm 0.04)10^{-8} n^{1.233\pm 0.001} + (2.690\pm 0.040)10^{-6}$ & $99.3\%$ & $\mathcal{O}(n)$ \\ \hline
SparseSim & $(4.61\pm 0.09)10^{-8} n^{0.499\pm 0.002} + (2.710\pm 0.060)10^{-7}$ & $99.1\%$ & $\mathcal{O}(\sqrt{n})$ \\ \hline
\end{tabular}}
\caption{\label{tb:Fit} Fits of the power-law equation $y = a n^b + c$ to the average runtime of operations performed during syndrome extraction of a medial surface-code patch up to distance 45 as seen in Fig.~\ref{fig:op-time}. Slight deviation from the predicted performance may be due to hidden costs in the implementation of the algorithms.}
\end{table}

\begin{figure}[H]
		\centering
			\includegraphics[width=0.75\textwidth]{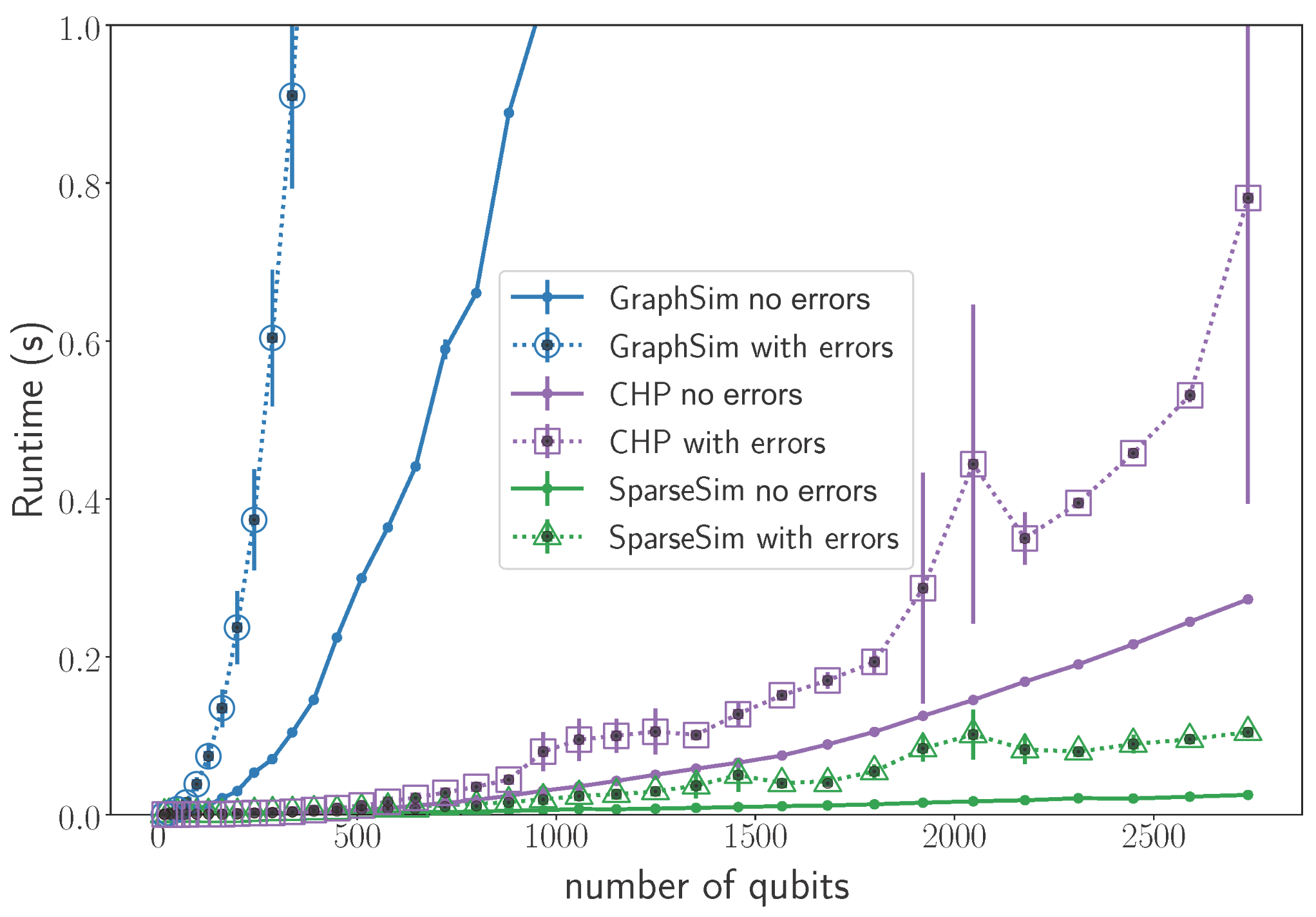}	
		\caption{Average runtime for a single round of syndrome extraction under symmetric single-qubit Clifford noise. Solid (dotted) lines correspond to simulations without (with) error. For each distance $d$, $d$ rounds of syndrome extraction were run, during which single-qubit Clifford errors were applied with probability $p=1.46\%$ for each quantum-circuit element. The runtimes shown here are the average runtime of a single round of syndrome extraction averaged over $d$ rounds. Note, the large error bars are due to varying times encountered during Monte Carlo simulations. \label{fig:c1-error}}
\end{figure}

So far, the theoretical and experimental analysis of stabilizer-simulation runtimes has focused on error-free simulations. Since Pauli noise only flips the signs of stabilizers, depolarizing noise does not affect the runtime of the simulations. Error models that can move the stabilizer subspace outside of the original code subspace (modulo stabilizer signs), may degrade runtimes. To this end, Fig.~\ref{fig:c1-error} shows the runtime of stabilizer simulations that inact a stochastic error-channel that has an equal probability of applying any one of the 23 non-trivial, single-qubit Clifford-gates for single-qubit operations. For two-qubit operations, the error channel has an equal probability of applying the tensor product of any two of the 24 single-qubit Clifford-gates, excluding the trivial gate $I \otimes I$. I apply this symmetric single-qubit Clifford error-channel to each quantum operation during syndrome extraction. If a qubit is idle during a quantum-circuit time-step, then I also apply the error channel to such an idle qubit.

Typically one is interested in determining the logical error-rates at or below the threshold; therefore, one may be interested in how much do stabilizer simulation runtimes slow down when simulating the single-qubit Clifford error-channel with physical error-rates equivalent to those found near the threshold. To determine this value we first consider that it is straightforward to show that for single-qubit gates this error model is equivalent to the error channel

\begin{equation}
\rho \rightarrow (1 - \frac{18}{23}p) \rho + \frac{6}{23} p (X\rho X + Y \rho Y + Z \rho Z), 
\end{equation}
\noindent where $p$ is the probability of a single-qubit Clifford error occurring. Since the surface-code threshold has been found to be between 0.502(1)\% and 1.140(1)\% for the depolarizing channel when applied to at the circuit level \cite{Stephens:2014b}, I chose $p = 1.46\%$ ($\frac{23}{18} 1.140\% \approx 1.46\%$) for the simulation show in Fig.~\ref{fig:c1-error}. Thus, this error-rate ($p = 1.46\%$) is near the highest error rates that one would typically simulate. As $p$ deceases, the runtimes approach the runtimes of simulations with no errors.

We see that Fig.~\ref{fig:c1-error} indicates that while Clifford noise degrades the runtime of the stabilizer simulations, \texttt{SparseSim} continues to obtain faster runtimes.

\section{\label{sec:concl}Conclusions}

In summary, I have presented a new stabilizer-simulation algorithm that makes use of the sparsity present in stabilizer/destabilizer representations of stabilizer states. I presented an amortized runtime analysis of the algorithm, as well as amortized runtime analyses of previous stabilizer-simulation algorithms to provide tighter upper bounds on the practical runtime for simulating syndrome extraction for planar surface-codes. These bounds show that circuit operations run in $\mathcal{O}(\sqrt{n})$ steps for the new algorithm while circuit operations run in $\mathcal{O}(n)$ steps for both the Aaronson-Gottesman and the Anders-Briegel algorithms. Thus, my analysis did not improve the bound on the circuit-operation runtime for the Gottesman algorithm, which remains $\mathcal{O}(n^3)$. I also introduced an implementation of the new algorithm and provided results of numerical experiments that demonstrate that the theoretical time complexities were closely achieved in practice. Because the error analysis of QECC procedures typically involve millions of trials of Monte Carlo simulations, the improvement in average runtime complexity given by the new algorithm will improve the runtime needed to perform stabilizer-simulation-based error-analysis of topological codes and other quantum LDPC codes.

\section{Acknowledgments}

I would like to thank Andrew Landahl and Jaimie Stephens for helpful discussions.
\chapter{Color and Surface Code Lattice-Surgery\label{ch.lattice_surgery}}

\setlength\epigraphwidth{6.5cm}
\epigraph{\textit{``Trust me. I'm the Doctor.''}}{--- \textup{The Doctor \cite{TrustMe}}}

In the previous chapter, I introduced stabilizer codes as well as how to simulate them efficiently. While stabilizer codes provide relatively large reductions in noise, when developing the architecture of a quantum device at the logical level it is important to consider how logical qubits must be laid out for universal computation. Thus, in this chapter I present joint work with Andrew Landahl in which we develop a QEC protocol called \textit{color-code lattice-surgery} as well as reduce the resources required for  \textit{surface-code lattice-surgery}, originally developed by Horsman \textit{et al.} \cite{Horsman:2012a}.

This work appears on the arXiv and is found here: \cite{colorsurgery}. As far as my technical contributions are concerned, I was the primary developer of all the QEC protocols discussed in this chapter. While also active in the design of these procedures, Andrew performed the resource analysis. He also chiefly wrote the paper, while I provided edits.

Since appearing on the arXiv, this work has had an impact on quantum-computing research. For example, color-code lattice-surgery was an important ingredient in development of doubled color-codes \cite{Bravyi:2015a,Jochym-OConnor:2016a,Jones:2015a}, which allows for universal computation in planar architectures with only transversal gates. Several papers have since considered our color-code protocols when developing and/or evaluating fault-tolerant architectures \cite{1503.02065,NFB16,BT16,CT17,LKEO17,LO17,GMB18,LO18} as well as our optimized surface-code protocols \cite{TriangleSurf,Moussa16,BLKW16,CT17,LO17,LWASKBA18}.

\section{Introduction}

Planar topological quantum error-correcting codes have emerged as promising
substrates for fault-tolerant quantum computing because of their high
thresholds~\cite{Stephens:2014b}, compatibility with two-dimensional (2D)
local quantum processing~\cite{Dennis:2002a}, low quantum circuit overheads
\cite{Raussendorf:2007a}, efficient decoding algorithms~\cite{Dennis:2002a,
Duclos-Cianci:2009a, Stephens:2014a}, and the ability to smoothly
interpolate between desired effective error rates, which concatenated codes
cannot do~\cite{Cross:2009a}.

In principle, fault-tolerant quantum computing with surface codes can be
achieved with transversal methods~\cite{Dennis:2002a}, defect-based
methods~\cite{Raussendorf:2006a, Raussendorf:2007a, Raussendorf:2007b}, or
lattice-surgery-based methods~\cite{Horsman:2012a}.  On 2D arrays of qubits
restricted to local quantum processing and local qubit movements,
transversal methods require an amount of information swapping that scales
with the system size.  Defect and lattice-surgery methods avoid this,
improving both their runtime and their accuracy
threshold~\cite{Spedalieri:2009a}.  Of these latter two, lattice surgery
uses substantially fewer qubits to achieve a desired error rate.  For
example, the fewest-qubit fault-tolerant distance-three $\CNOT$ method in a
topological code reported to date uses surface-code lattice surgery and only
requires 53 qubits~\cite{Horsman:2012a}.

Extending transversal surface-code methods to color codes is
straightforward.  Fowler has also extended defect-based surface-code methods to
defect-based color-code methods~\cite{Fowler:2008c}.  Notably absent are
extensions of surface-code lattice-surgery methods to color-code
lattice-surgery methods.  Developing such methods is especially important
because not only are lattice-surgery methods more qubit-efficient than
defect-based methods, but also color codes are significantly more
qubit-efficient than surface codes---for example, 4.8.8 color codes use
about half the qubits as the qubit-optimal medial surface
code~\cite{Bombin:2007d} to achieve the same code
distance~\cite{Landahl:2011a}.

Going beyond the application of a topological quantum
memory~\cite{Dennis:2002a}, color-codes offer additional advantages.  While
transversal two-qubit operations incur penalties for swapping information
around, one-qubit transversal operations do not; these advantages carry over
to lattice-surgery methods.  Two especially noteworthy methods are those for
the encoded, or ``logical,'' Hadamard gate ($H$) and those for the logical
phase gate ($S$) on planar color codes on the 4.8.8 lattice---both can be
implemented in a single parallelized transversal step~\cite{Bombin:2006b}.
For surface codes, neither of these gates have transversal implementations
on any lattice.  Current surface-code solutions for these gates include
elaborate multi-step code deformation procedures to implement the Hadamard
gate~\cite{Fowler:2012c, Horsman:2012a} and lengthy multi-gate teleportation
procedures from (previously distilled) magic states to implement the
phase gate~\cite{Raussendorf:2006a, Aliferis:2007b}.

The only downside to color codes versus surface codes is their lower
accuracy threshold, whose value has been estimated to be $0.143\%$ against
depolarizing circuit-level noise using a perfect-matching
decoder~\cite{Stephens:2014a}.  Surface codes have an accuracy threshold
whose value has been estimated to be in the range $0.502(1)\%$ to
$1.140(1)\%$~\cite{Stephens:2014b} in the same setting.  That said, surface
codes have enjoyed far greater study than color codes and we expect that
there are opportunities to close the gap.  We will show later that, even as
things stand now, at sufficiently low error rates and sufficiently low
desired error rates to be achieved by encoding, color codes still use fewer
qubits, despite their lower accuracy threshold.

Bolstered by the possibility of significant time and qubit reductions for
fault-tolerant operations, in this article we develop methods for universal
fault-tolerant quantum computation using color-code lattice surgery.  We
show that our methods use fewer qubits per logical operation than
surface-code lattice-surgery methods, including the smallest distance-three
$\CNOT$ in a topological code---our color-code lattice-surgery methods only
use 30 qubits when one allocates one syndrome qubit per face (or 22 if one
uses a single mobile syndrome qubit).  Along the way, we also improve the
surface-code lattice-surgery methods so that the distance-three $\CNOT$ now
only uses 39 qubits when one allocates one syndrome qubit per face (or 28 if
one uses a single mobile syndrome qubit).

In Sec.~\ref{sec:background}, we provide a brief background on triangular
4.8.8 color codes to help make our exposition better self-contained.  In
Sec.~\ref{sec:universal}, we describe fault-tolerant color-code
lattice-surgery methods for performing each element in a universal set of
operations.  In Sec.~\ref{sec:resource-analysis}, we calculate the circuit
width and depth overheads required by these methods and compare them to the
corresponding overheads required by surface-code lattice-surgery methods.
Sec.~\ref{sec:conclusion} concludes.

\section{Background}
\label{sec:background}

\begin{figure}[ht!]
\center{
    \begin{subfigure}[t]{0.3\textwidth}
        \centering
        \includegraphics[width=0.75\linewidth]{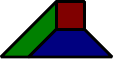}\hspace{0.1em} 
        \caption{ $d = 3$ lattice}
    \end{subfigure}
     \begin{subfigure}[t]{0.3\textwidth}
        \centering
        \includegraphics[width=0.75\linewidth]{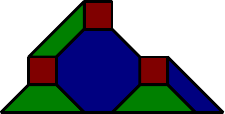}\hspace{0.1em} 
        \caption{ $d = 5$ lattice}
    \end{subfigure}
         \begin{subfigure}[t]{0.3\textwidth}
        \centering
        \includegraphics[width=0.75\linewidth]{{\SurgeryFigs/4_8_8-rgb-d7}}
        \caption{ $d = 7$ lattice}
    \end{subfigure}
}
\caption{Triangular 4.8.8 color codes of distances 3, 5, and 7.  The number
of data qubits for distance $d$ is $(d^2-1)/2+d$.  The number of faces
(which is half the number of checks) is $(d^2+2d-3)/4$.
\small{\label{fig:color-codes} }}
\end{figure}

Our color-code lattice-surgery methods are valid for any color code, but for
concreteness we focus on lattice surgery of triangular color codes on the
4.8.8 lattice, namely the semiregular lattice that has a square and two
octagons surrounding each vertex.  These quantum stabilizer codes
\cite{Gottesman:1997a} exist for any odd code distance $d$ and can be
depicted graphically as in Fig.~\ref{fig:color-codes}.  Each vertex in this
figure corresponds to one (``data'') qubit in the code.  Each face in the
figure corresponds to two code checks, or stabilizer generators; one check
acts as Pauli $X$ on all qubits incident on the face and one check acts as
Pauli $Z$ on all qubits incident on the face.  The collection of qubits and
checks encode a single ``logical'' qubit.  Representatives of the logical
$X$ and $Z$ operators are strings of $X$ and $Z$ operators acting on the
qubits along the bottom side of the triangle.  By multiplying by a suitable
collection of check operators, two other equivalent representatives are
similar strings along either of the other two triangle sides.

Syndrome qubits are associated with the faces in the graph; how many
syndrome qubits are associated with each face is a nuanced function of the
syndrome extraction protocol one uses.  At a minimum, one can use a single
syndrome qubit over and over again, but it would have to be moved either
physically or by $\SWAP$ gates in such a way that it interacted with every
data qubit on the interior six times, every data qubit on the edge four
times, and every data qubit on a corner twice, because that is the number of
checks each of these types of data qubits are involved in.  A faster
syndrome extraction is possible by allocating one syndrome qubit per face so
that each syndrome qubit is used to measure both the $X$ and the $Z$ check
on each face.  By allocating two qubits per face, syndrome extraction can
run faster still, with the $X$ and $Z$ check measurements scheduled in an
interleaved fashion~\cite{Landahl:2011a}.

Adding more syndrome qubits can lead to better performance, such as a higher
accuracy threshold or less error propagation; we examine these tradeoffs in
greater detail in Sec.~\ref{sec:resource-analysis}.  One way to increase the
number of syndrome qubits is to allocate five syndrome qubits per each
octagonal face and two per each square face, extracting the syndrome into
two-qubit and verified four-qubit cat states~\cite{Fowler:2008c,
Stephens:2014a}.  By doubling this number of syndrome qubits, two cat states
per face can be prepared in parallel and used in the interleaved schedule
for $X$ and $Z$ check measurements.  Going even further, one can allocate
one syndrome qubit for every data qubit to enact even more robust
Shor-style~\cite{Shor:1996a} or Steane-style~\cite{Steane:1998a} syndrome
extraction.  This number of qubits can be doubled further to enact
Knill-style syndrome extraction with the same robustness but a faster
extraction circuit~\cite{Knill:2004a}.  We are not aware of any schemes that
use even more syndrome qubits to any advantage, so the number of syndrome
qubits can range anywhere from one to twice the number of data qubits.  In
this article, we will generally restrict attention to schemes which use
either one syndrome qubit per face or one syndrome qubit per check (two per
face), as we believe these offer the closest comparison to the most
widely-studied surface-code syndrome layout scheme, namely the one with one
syndrome qubit per check (one per face)~\cite{Dennis:2002a}.

Color codes are frequently considered in one of three broad classes of error
models~\cite{Landahl:2011a}.  In code-capacity models, data qubits are
subject to error but syndrome qubits are not.  In phenomenological models,
both data and syndrome qubits are subject to error.  In circuit-level
models, data qubits, syndrome qubits, and the individual quantum gates that
act upon them are subject to error.  This latter class is the most realistic
and is the one we focus on in this article.  However, because the available
operations at the circuit level are very hardware-dependent, we abstract
away the specifics of the hardware-level gate basis wherever possible.

Even when the physical circuit gate basis is known, it can be the case that
the error model on that gate basis is not well known.  In the absence of an
experimentally-informed circuit-level error model, a frequently used
surrogate is the independent identically distributed (iid) depolarizing
noise model, as it is kind of a ``worst case'' noise model for iid
stochastic errors.  In the iid depolarizing noise model, noise acts
independently and identically on the outputs of each quantum circuit
element, including the identity gate.  Depolarizing noise causes an error to
occur with probability $p$, and it selects the error equiprobably among the
possible non-identity Pauli operators on the outputs.  For single-qubit
measurement operations, it also flips the classical bit output with
probability $p$ (because a measurement error is a disagreement between the
recorded measurement outcome and the actual state).  While this noise model
is not without its flaws even for iid stochastic errors (see, for example,
Refs.~\cite{Magesan:2013a, Gutierrez:2013a, Stephens:2014b}), it is widely
used.

The syndrome extracted from a color code can be decoded in myriad ways.  For
the best performance, one could use the optimal decoder.  Although optimal
decoding of stabilizer codes is \#P-hard in general~\cite{Iyer:2013a}, it is
possible that an efficient optimal decoder (or one that approximates it
arbitrarily well) for color codes will be found.  For example, the
optimal-decoder-approximating PEPS decoder for surface-codes might be
extended to color codes~\cite{Bravyi:2014a}.  Alternatively, one could use a
slightly weaker integer-program-based decoder that identifies the most
likely error given the syndrome~\cite{Landahl:2011a}.  Weaker still but
faster yet, one could use a matching-based decoder, such as a minimum-weight
perfect matching decoder~\cite{Wang:2009b, Stephens:2014a}, a
renormalization-group matching decoder~\cite{Sarvepalli:2011a,
Duclos-Cianci:2011a, Duclos-Cianci:2014a}, a local greedy matching
decoder~\cite{Dennis:2003a, Bravyi:2013a, Wootton:2013a, Anwar:2014a}, or a
``global attractive-force'' local cellular automaton matching
decoder~\cite{Harrington:2004a, Herold:2014a}.  It is also possible to
exploit the local equivalence between a color code and a finite number of
copies of the surface code to arrive at a decoding solution from mulitple surface-code
decoders~\cite{Bombin:2012a, Delfosse:2014a}.  Developing new color-code
decoders is an active research front, where the trade space between decoding
complexity and decoding performance is being explored.

%
\section{Universal Gate Set}
\label{sec:universal}

In this section, we describe how to fault-tolerantly perform a universal set
of operations by lattice surgery on 4.8.8 triangular color codes.  We use
the notation from Ref.~\cite{QCQI} to denote gates, states,
measurements, and quantum circuits.  The universal set we effect in
encoded form by lattice surgery is as follows:
\begin{align}
\left\{I, |0\>, |+\>, M_Z, M_X, S, H, T|+\>, \CNOT \right\}.
\end{align}

In the absence of hardware-informed circuit-level details, we imagine that
the same set of operations is available on the physical qubits as well, with
the $\CNOT$ gates restricted to nearest-neighbor data-ancilla qubit pairs.

With this gate basis, Pauli operators never need to be applied or even
synthesized from the other gates.  By the Gottesman-Knill
theorem~\cite{GotKnill}, Pauli operators can be propagated through
all stabilizer operations (Clifford gates plus Pauli preparations and
measurements) efficiently classically and used solely to reinterpret
measurement results.  Since this gate basis consists solely of stabilizer
operations and the $T|+\>$ preparation, and because Pauli operators never
need to be propagated through preparations, no Pauli operators are ever
needed.  Importantly, this means that if a decoding algorithm calls for
Pauli operators to be applied as a corrective action, the data need not be
touched by the Pauli operators and the classical ``Pauli frame'' can be
updated instead.  That said, to avoid polynomial-time classical computation,
it might be useful to implement the Pauli-frame updates from time to time.
For example, if errors are not corrected but only tracked, then the observed
syndrome bit rate will climb until it reaches a steady state close to 50\%,
at which point decoding may take longer than if the tracked Pauli
errors had been actually reversed.

\vfill
\pagebreak[4]
In our fault-tolerant constructions, all but the $T|+\>$ preparation allow for the
exponential suppression of errors as distance is increased.  To
increase the fidelity of $T|+\>$ preparations, any of a number of
magic-state distillation protocols can be used~\cite{Bravyi:2005a,
Meier:2012a, Bravyi:2012a, Landahl:2013a}.  These protocols use
high-fidelity operations from the rest of the set to ``distill'' multiple
$T|+\>$ preparations into fewer $T|+\>$ preparations of higher fidelity.

%
\subsection{The Identity Gate \texorpdfstring{$I$}{I}}

To fault-tolerantly implement the encoded identity gate on a triangular
color code, we simply perform fault-tolerant quantum error correction by
measuring the syndrome for $d$ rounds and run a classical decoding algorithm
on the data, such as one of the decoders described in
Refs.~\cite{Landahl:2011a, Duclos-Cianci:2014a, Stephens:2014a, Wang:2009b,
Sarvepalli:2011a}, to infer a corrective action.

%
\subsection{Preparation of \texorpdfstring{$|0\>$}{|0>} and
\texorpdfstring{$|+\>$}{|+>} States}

To fault-tolerantly prepare an encoded $|0\>$ state (the $+1$ eigenstate of
the encoded $Z$ operator), we first prepare each data qubit in a triangular
color code in the state $|0\>$ (the $+1$ eigenstates of the physical $Z$
operators).  We then perform fault-tolerant quantum error correction by
measuring the syndrome $d$ times and running it through a decoder.  The
process of measuring all of the code checks transforms the set of
single-qubit $Z$ checks into a set consisting of~($a$)~the $Z$ checks of the
color code and~($b$)~the encoded $Z$ operator for the color code.

The process for fault-tolerantly preparing an encoded $|+\>$ state (the $+1$
eigenstate of the encoded $X$ operator) is identical, except that the
individual qubits are initially prepared in $+1$ $X$ eigenstates instead of
$+1$ $Z$ eigenstates.

%
\subsection{Measurement \texorpdfstring{$M_Z$}{MZ} and
\texorpdfstring{$M_X$}{MX}}

To fault-tolerantly measure the encoded $Z$ operator, $M_Z$, on a logical
qubit, we measure each of the data qubits in the logical qubit in the $Z$
basis in a single round and perform classical error correction on the
result.  This measurement is ``destructive'' in that it takes the logical
qubit out of the code space.  A non-destructive measurement can be
implemented by augmenting this destructive measurement with an encoded
$\CNOT$ gate using Steane's ancilla-coupled measurement
method~\cite{Steane:1998a}.

Fault-tolerantly measuring the encoded $X$ operator, $M_X$, is similar: we
measure each of the data qubits in the logical qubit in the $X$ basis in a
single round and perform classical error correction on the result.  It is
also a destructive measurement, with a nondestructive version achievable
using Steane's method.

%
\subsection{Phase and Hadamard Gates (\texorpdfstring{$S$}{S} and
\texorpdfstring{$H$}{H})}

Because the 2D color codes are \emph{strong} CSS codes (meaning that not
only do the checks factor into $X$-type and $Z$-type classes but also they
have identical support), the transversal Hadamard gate will swap the two
types of checks.  For triangular color codes (but not, \eg, for color codes
on compact surfaces \cite{Bombin:2007d}), the logical $X$ and $Z$ operators
can be made to be coincident so that the transversal Hadmard gate exchanges
these as well.  The net result is that the transversal Hadamard gate is a
fault-tolerant logical Hadamard gate for triangular color codes.

As shown by Bombin in Ref.~\cite{Bombin:2013a}, the $S$ gate is transversal
for 2D color codes as well, with a suitable choice of which physical qubits
to apply $S$ to and which to apply $S^\dagger$ to.  The 2D triangular color
codes on the 4.8.8 lattice have perhaps the simplest allocation choice: use
the transversal $S$ operator if the code distance is congruent to $1 \bmod
4$ and the transversal $S^\dagger$ operator if the code distance is
congruent to $3 \bmod 4$.

%
\subsection{The CNOT Gate}

\begin{figure}[H]
\centerline{
\Qcircuit @C=1em @R=1em {
                         &                       & (-1)^b                 &            &                &     \\
 \lstick{\text{control}} & \qw                   & \multigate{2}{M_{ZZ}}  & \qw        & \gate{Z^{a+c}} & \qw \\
                         & (-1)^a                & \push{\rule{0em}{1.2em}} & (-1)^c     &                &     \\
 \lstick{\ket{0}}        & \multigate{2}{M_{XX}} & \ghost{M_{ZZ}}         & \gate{M_X} & \gate{Z^c}     & \qw \\
                         &                       & \push{\rule{0em}{1.2em}} &            &                &     \\
 \lstick{\text{target}}  & \ghost{M_{XX}}        & \qw                    & \qw        & \gate{X^b}     & \qw \\
} 
} 
\caption{\small{\label{fig:Bell-CNOT}Measurement-based $\CNOT$ circuit.}}
\end{figure}
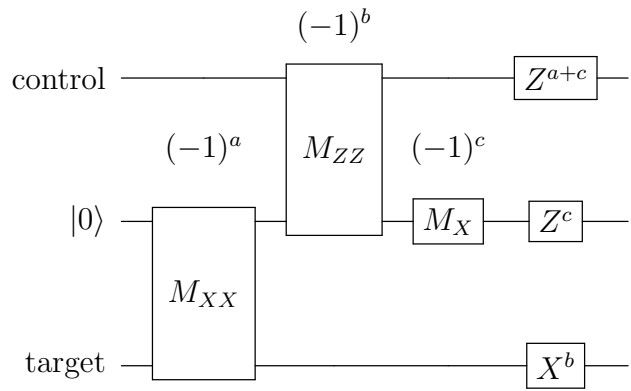

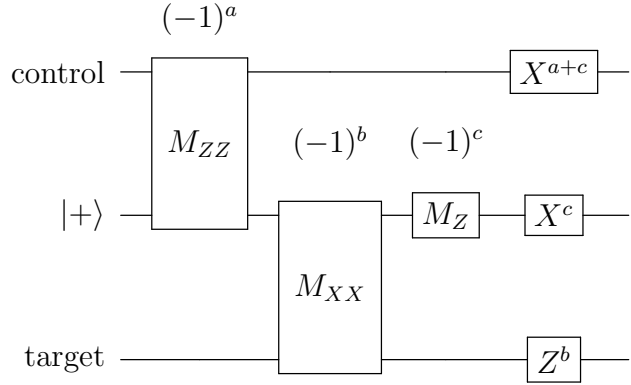
\begin{figure}[H]
\centerline{
\Qcircuit @C=1em @R=1em {
                         & (-1)^a                   &                          &            &                &     \\
 \lstick{\text{control}} & \multigate{2}{M_{ZZ}}    & \qw                      & \qw        & \gate{X^{a+c}} & \qw \\
                         & \push{\rule{0em}{1.2em}} & (-1)^b                   & (-1)^c     &                &     \\
 \lstick{\ket{+}}        & \ghost{M_{ZZ}}           &  \multigate{2}{M_{XX}}   & \gate{M_Z} & \gate{X^c}     & \qw \\
                         & \push{\rule{0em}{1.2em}} &                          &            &                &     \\
 \lstick{\text{target}}  & \qw                      &  \ghost{M_{XX}}          & \qw        & \gate{Z^b}     & \qw \\
} 
} 
\caption{\small{\label{fig:Horsman-CNOT}Alternative measurement-based $\CNOT$ circuit.}}
\end{figure}

To fault-tolerantly implement the encoded $\CNOT$ gate, we use a sequence of
lattice surgery operations.  These operations are intended to mimic either
the circuit in Fig.~\ref{fig:Bell-CNOT} or the circuit in
Fig.~\ref{fig:Horsman-CNOT}, both of which are equivalent to a $\CNOT$ gate;
these circuits were leveraged heavily in Ref.~\cite{Aliferis:2008b} to
combat biased noise.

The Pauli corrections in these circuits can be omitted in our approach 
because of our choice of gate basis; we simply use them to re-interpret
future measurement results as needed.  The only operations depicted in these
circuits that we have not provided methods for yet are the $M_{XX}$ and
$M_{ZZ}$ measurements; with them, we can construct the encoded $\CNOT$
operation.

\begin{figure}[H]
\begin{center}
  \includegraphics[width=0.4\columnwidth]{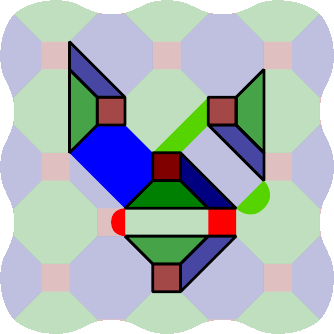}
\caption{\small{\label{fig:MXX-MZZ-d3}To measure $M_{XX}$ ($M_{ZZ}$) between the
central logical qubit and a logical qubit adjacent to one of its sides, measure
only the $X$ ($Z$) checks on the lighter-colored faces on the interface and
the $X$ and $Z$ checks on the full octagons shared across the interface.  (The
figure compresses three separate scenarios into one.)  The outcome is the
product of the lighter-colored check outcomes.}}
\end{center}
\end{figure}

\begin{figure}[H]
\begin{center}
  \includegraphics[width=0.4\columnwidth]{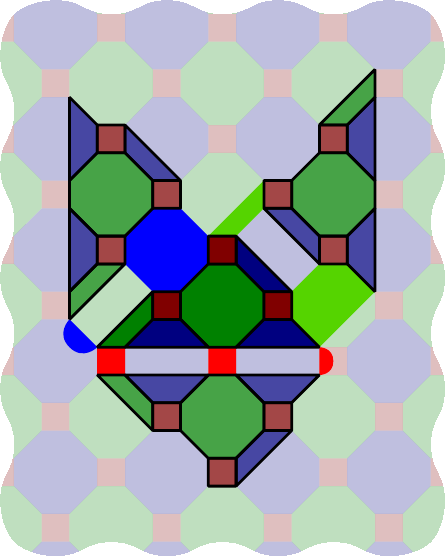}
\caption{\small{\label{fig:MXX-MZZ-d5}The same scenario as
Fig.~\ref{fig:MXX-MZZ-d3}, except with distance-five codes.}}
\end{center}
\end{figure}

To measure $XX$ or $ZZ$ between two triangular color codes, we measure
checks that connect the adjacent logical qubits in an ``osculating'' manner.
Figures~\ref{fig:MXX-MZZ-d3} and \ref{fig:MXX-MZZ-d5} depict how this can be
done for every side of a 4.8.8 triangular color code for code distances 3
and 5; the pattern generalizes in a straightforward way. 

\begin{figure}[ht]
\begin{center}
  \includegraphics[width=0.4\columnwidth]{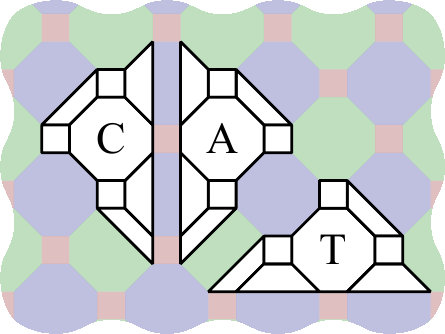}
\caption{\small{\label{fig:CAT-layout-d5}Regions outlined and filled with
white indicate where the control (C), ancilla (A), and target (T) qubits are
located for a distance-five example.}}
\end{center}
\end{figure}

Using these methods for $M_{XX}$ and $M_{ZZ}$ measurements, we describe
step-by-step how to implement a fault-tolerant $\CNOT$ gate by lattice
surgery using a simulation of the circuit in Fig.~\ref{fig:Bell-CNOT}; the
simulation of the circuit in Fig.~\ref{fig:Horsman-CNOT} is similar.  While
our construction works for arbitrary code distances, we depict an example of
each step for $d=5$, with the layout of control, ancilla, and target regions
as depicted in Fig.~\ref{fig:CAT-layout-d5}; other choices of orientation
are possible.

\begin{figure}[H]
\begin{center}
  \includegraphics[width=0.35\columnwidth]{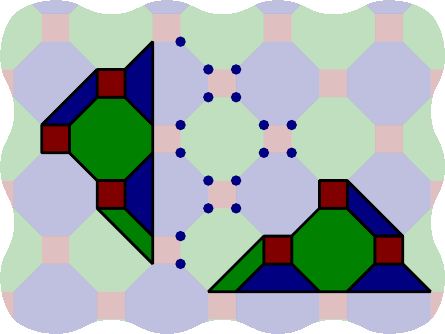}
\caption{\small{\label{fig:CAT-PZhat}(Step 1.) The qubits in the ancilla region (A) are
prepared in $Z = +1$ eigenstates.}}
\end{center}
\end{figure}

\begin{figure}[H]
\begin{center}
  \includegraphics[width=0.35\columnwidth]{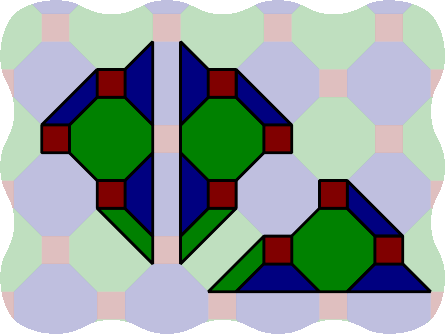}
\caption{\small{\label{fig:CAT-P0}(Step 2.) The checks in the ancilla region (A) are
measured for $d$ rounds and errors are corrected fault-tolerantly.}}
\end{center}
\end{figure}

\begin{figure}[H]
\begin{center}
  \includegraphics[width=0.35\columnwidth]{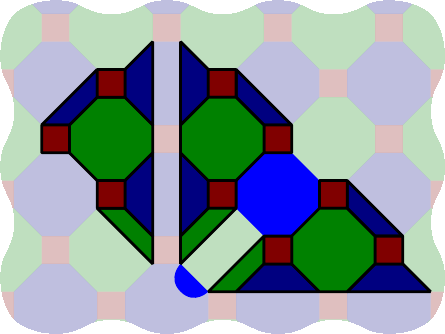}
\caption{\small{\label{fig:CAT-MXX}(Step 3.) The checks that fuse the target and
ancilla logical qubits in an $M_{XX}$ measurement are measured for $d$
rounds and errors are corrected fault-tolerantly.}}
\end{center}
\end{figure}

\begin{figure}[H]
\begin{center}
  \includegraphics[width=0.35\columnwidth]{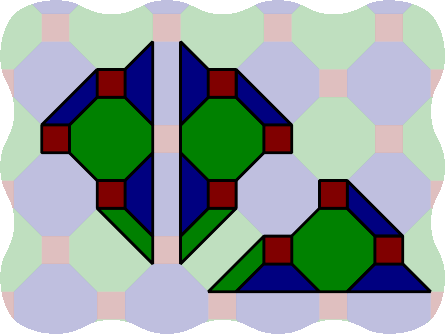}
\caption{\small{\label{fig:CAT-MXX-split}(Step 4.) The $M_{XX}$-fusing checks stop
being measured.  Instead, the target and ancilla logical qubits checks are
measured for $d$ rounds and errors are corrected fault-tolerantly.}}
\end{center}
\end{figure}

\begin{figure}[H]
\begin{center}
  \includegraphics[width=0.35\columnwidth]{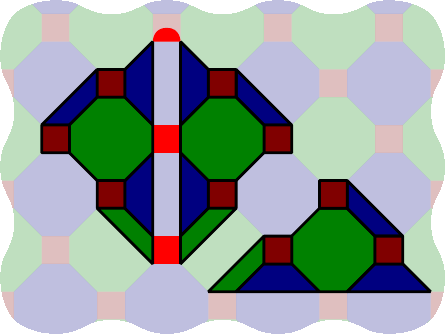}
\caption{\small{\label{fig:CAT-MZZ}(Step 5.) The checks that fuse the control and
ancilla logical qubits in an $M_{ZZ}$ measurement are measured for $d$
rounds and errors are corrected fault-tolerantly.}}
\end{center}
\end{figure}

\begin{figure}[H]
\begin{center}
  \includegraphics[width=0.35\columnwidth]{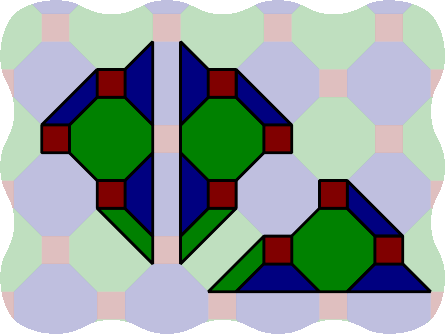}
\caption{\small{\label{fig:CAT-MZZ-split}(Step 6.) The $M_{ZZ}$-fusing checks stop
being measured.  Instead, the control and ancilla logical qubits checks are
measured for $d$ rounds and errors are corrected fault-tolerantly.}}
\end{center}
\end{figure}

\begin{figure}[H]
\begin{center}
  \includegraphics[width=0.35\columnwidth]{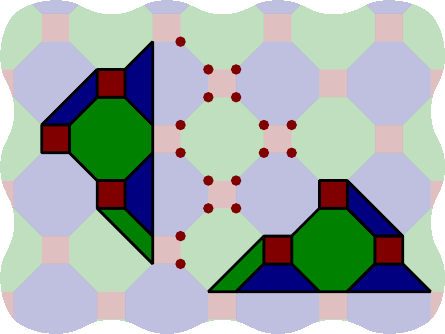}
\caption{\small{\label{fig:CAT-MX}(Step 7.) The qubits in the ancilla region are
measured in the $X$ basis, implementing a destructive $M_X$ measurement.
The result is error-corrected classically.  The control and target logical
qubit checks are measured for $d$ rounds and errors are corrected
fault-tolerantly.}}
\end{center}
\end{figure}

\begin{enumerate}
\item Prepare the data qubits in the ancilla region in $|0\>$ states ($Z =
+1$ eigenstates), as depicted in Fig.~\ref{fig:CAT-PZhat}.
\item Measure the checks in the ancilla region for $d$ rounds and correct
errors fault-tolerantly, as depicted in Fig.~\ref{fig:CAT-P0}.
\item Measure the checks that fuse the target and ancilla logical qubits in
an $M_{XX}$ measurement for $d$ rounds and correct errors fault-tolerantly,
as depicted in Fig.~\ref{fig:CAT-MXX}.
\item Stop measuring the $M_{XX}$-fusing checks and measure the checks for
the target and ancilla logical qubits separately, splitting them apart
again, for $d$ rounds and correct errors fault-tolerantly, as depicted in
Fig.~\ref{fig:CAT-MXX-split}.
\item Measure the checks that fuse the control and ancilla logical qubits in
an $M_{ZZ}$ measurement for $d$ rounds and correct errors fault-tolerantly,
as depicted in Fig.~\ref{fig:CAT-MZZ}.
\item Stop measuring the $M_{ZZ}$-fusing checks and measure the checks for
the control and ancilla logical qubits separately, splitting them apart
again, for $d$ rounds and correct errors fault-tolerantly, as depicted in
Fig.~\ref{fig:CAT-MZZ-split}.
\item Measure the data qubits in the ancilla region in the $X$ basis
destructively and perform classical error correction on the result, as
depicted in Fig.~\ref{fig:CAT-MX}.
\end{enumerate}

%
%

%
%

%
%

%
%

%
%

%
%

%
%

As described, this method takes one round of data-qubit preparation, $5d$
rounds of syndrome extraction, and one round of data-qubit measurement.
However, this time can be sped up considerably.

As a starter, a preparation operation on a logical qubit and a fusing
operation between that logical qubit and another logical qubit can be
combined into a single step---instead of thinking of the operations as
``prepare-then-fuse,'' one can think of them as a single ``grow one of the
logical qubits'' operation.  Step 2 can therefore be eliminated and, without
loss of generality, we can omit step 1 and use the state it prepares as the
initial state of the method.  This reduces the number of rounds of
parallelized measurements to $4d+1$.

Next, a splitting operation between two logical qubits that ``heals'' the
interface between them can happen simultaneously with a fusing operation
acting on a different side of one of the logical qubits and a side of a
third logical qubit.  Running these operations simultaneously does not
hamper the fault-tolerance of the method---the code distances do not drop by
this kind of parallelization.  This observation allows us to eliminate step
4, reducing the number of rounds of parallelized measurements to $3d+1$.  It
also means that the target logical qubit is free to use one of its other
sides after just $d$ rounds of measurements.

Finally, a splitting operation between two logical qubits can happen
simultaneously with a destructive measurement operation that follows on one of them;
again, the operations do not interfere with one another.  Because the
destructive measurement operation only takes one round of parallelized
measurements, the time savings is not very great---the number of rounds is
reduced to $3d$ with this observation.

%
\subsection{Preparation of \texorpdfstring{$T|+\>$}{T|+>} States}

\begin{figure}[H]
\center{
    \begin{subfigure}[t]{0.45\textwidth}
        \centering
        \includegraphics[width=0.75\linewidth]{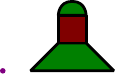}\hspace{1em}
        \caption{Step~1}
    \end{subfigure}
    \begin{subfigure}[t]{0.45\textwidth}
        \centering
        \includegraphics[width=0.75\linewidth]{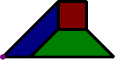}
        \caption{Step~1}
    \end{subfigure}
}
\caption{\small{\label{fig:d3-injection}Injection of $T|+\>$ qubit state
(purple dot) into  $d=3$ triangular 4.8.8 color code (image on right).  In steps 1
and 2, the indicated code checks are measured three times each. }}
\end{figure}

\begin{figure}[H]
\center{
    \begin{subfigure}[t]{0.45\textwidth}
        \centering
        \includegraphics[width=0.75\linewidth]{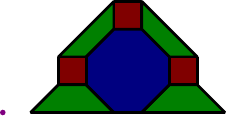}\hspace{1em}
        \caption{Step~1}
    \end{subfigure}
    \begin{subfigure}[t]{0.45\textwidth}
        \centering
        \includegraphics[width=0.75\linewidth]{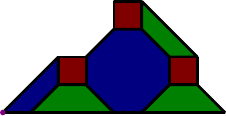}
        \caption{Step~1}
    \end{subfigure}
}
\caption{\small{\label{fig:d5-injection}Same as Fig.~\ref{fig:d3-injection},
but for a $d=5$ triangular 4.8.8 color code. }}
\end{figure}

\begin{figure}[H]
\center{
    \begin{subfigure}[t]{0.45\textwidth}
        \centering
        \includegraphics[width=0.75\linewidth]{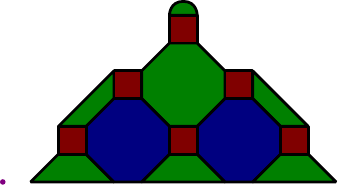}\hspace{1em}
        \caption{Step~1}
    \end{subfigure}
    \begin{subfigure}[t]{0.45\textwidth}
        \centering
        \includegraphics[width=0.75\linewidth]{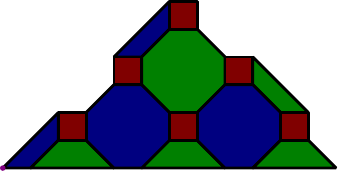}
        \caption{Step~1}
    \end{subfigure}
}
\caption{\small{\label{fig:d7-injection}Same as Fig.~\ref{fig:d3-injection},
but for a $d=7$ triangular 4.8.8 color code. }}
\end{figure}

To fault-tolerantly prepare an encoded $T|+\>$ state, we use the process of
code injection.  Figures~\ref{fig:d3-injection}--\ref{fig:d7-injection}
depict the injection process for distances $d=3$, $5$, and $7$.  The
coloring in these figures is chosen so that the blue side of the final
triangular code is always on the left for ease of discussion.  The top two
rows of qubits in these figures represent two isolated Bell pairs for $d=3$ and $d=7$,
even though they look like they are connected to the rest of the surface via
a square and a digon.

In the first step, we prepare a single qubit in the state $T|+\>$ and we
prepare an adjacent region in an auxillary state that consists of a distance
$d-1$ color-code stabilizer state, along with two additional Bell pairs if
$d \equiv 3 \bmod 4$.  For $d > 3$, we prepare the two Bell pairs to
$\bigO(p^2)$ error by post-selection, with a mean waiting time of
$(1-p)^{-4} \cong 1 + 4p$ rounds of measurement.  In parallel, we measure
the rest of the checks three times and use a classical decoding algorithm to
suppress errors in the distance $d-1$ code state to $\bigO(p^2)$.  We handle
the case of $d=3$ separately; the auxillary state is just three Bell states
in this case, so we prepare it by post-selection to $\bigO(p^2)$ error with
a mean waiting time of $(1-p)^{-6} \cong 1 + 6p$ rounds of measurement.

To inject the state, in the second step we measure the new blue $X$ and $Z$
checks along the left side, accepting whatever syndrome values we obtain as
being ``correct.'' This causes the Pauli $X$ and $Z$ operators on the single
qubit being injected to extend to distance-$d$ logical Pauli $X$ and $Z$
operators along that edge of the triangle.  In parallel, we cease measuring
the green checks along the left side, including the digon operator if one is
present.  However, in parallel we \emph{do} measure all of the other checks
for the code.

For $d > 3$, the checks that persist are capable of detecting up to two
errors on any pair of data qubits, excluding the state to be injected.  Any
single or two-qubit error on the interior data qubits will be detected
because the code distance is sufficiently high.  Any single-qubit error on
data qubits along the left boundary will be incident on a red check or the
bottom-left green check, so it will be detected as well.  If a two-qubit
error afflicts two data qubits on different red checks on the left side or a
red check and the bottom-left green check, they will also be detected.  If a
two-qubit error afflicts two data qubits on a single red check, at least one
other persistent check will detect it, by inspection.  Since the persistent
checks can detect up to two errors, one can use a classical decoding
algorithm on three rounds of extracted syndrome to correct any single error,
suppressing errors to $\bigO(p^2)$.  The case of $d=3$ can be handled as a
special case with, \eg, postselection on the entire injection process.

The total number of rounds of syndrome extraction in the state-injection
process is six: three to prepare the ancillary state and three to decode the
full distance-$d$ code.  The error in the process is $\bigO(p)$, where the
multiplicative constant is solely a function of the circuit elements in the
check measurement circuit that act on the state to be injected.
Importantly, this constant does not grow with the distance of the code.  To
reduce this error further once it is encoded, an encoded magic-state
distillation protocol may be used.

%
\section{Resource Analysis}
\label{sec:resource-analysis}

%
\subsection{Overhead per Code Distance}
\label{sec:overhead-per-code-distance}

Table~\ref{tab:color-code-resources-2qubit} summarizes the space and time
resource overheads used by our color-code lattice-surgery methods for the
scenario in which one syndrome qubit is allocated per check (two per face).
\begin{center}
\begin{table}[ht]
\centering
\begin{tabular}{c||c|c|c|c|c|c|c|c|c}
\multicolumn{10}{c}{Color-code lattice surgery (1 syndrome qubit/check)} \\[1ex] \hline \hline
Gate   & $T|+\>$ & $I$ & $|0\>$ & $|+\>$ & $M_Z$ & $M_X$ & $H$ & $S$ & $\CNOT$ \\[1ex] \hline
Depth  & $6$ & \multicolumn{3}{c|}{$d$} & \multicolumn{2}{c|}{$1$} & \multicolumn{2}{c|}{$0$}
       & $3d$ \\[1ex]  \hline
Qubits & \multicolumn{8}{c|}{$d^2 + 2d - 2$}
       & $3d^2 + 6d - 6$ \\[1ex] \hline
Error  & $\bigO(p)$ & \multicolumn{8}{c}{$\bigO(p^{(d+1)/2})$} \\[1ex] \hline \hline
\end{tabular}
\caption{Resources used by fault-tolerant 4.8.8 triangular color-code
lattice surgery on distance-$d$ codes when two syndrome bits per face are
allocated.  Depth is measured in number of measurement rounds.
Qubit counts include both data and syndrome qubits.  Error is reported in
big-$\bigO$ notation because syndrome-extraction-circuit implementation details
can change the constants.}
\label{tab:color-code-resources-2qubit}
\end{table}
\end{center}

\begin{figure}[ht]
\begin{center}
  \includegraphics[width=0.4\columnwidth]{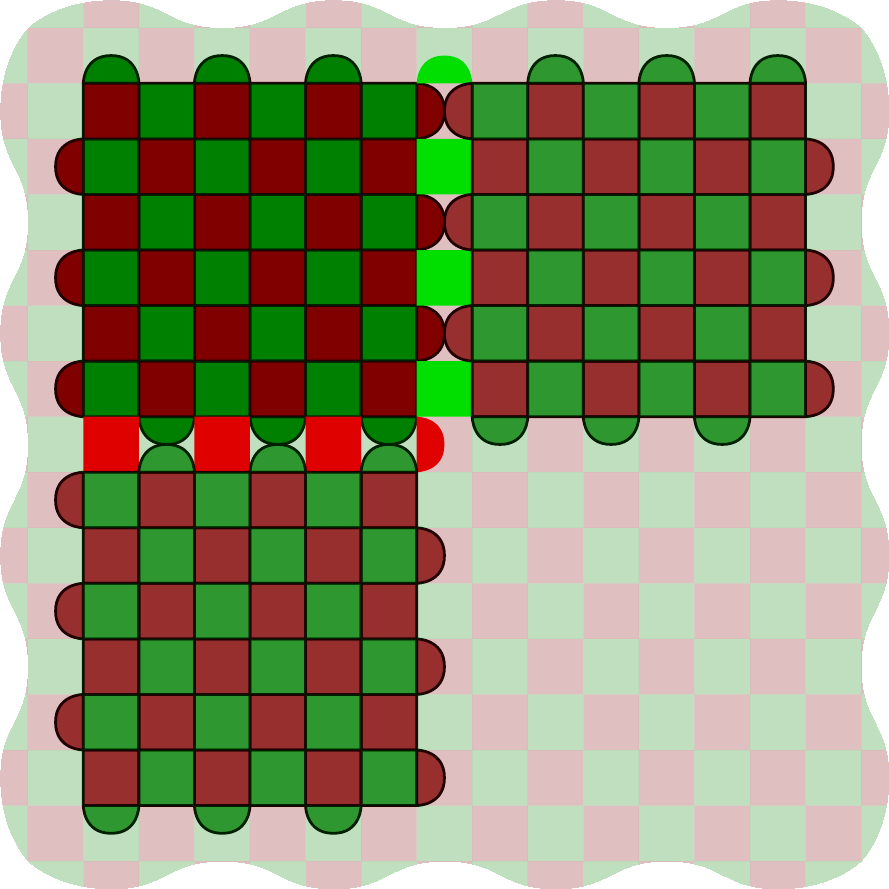}
\caption{\small{\label{fig:surface-CNOT}Layout for the $\CNOT$ gate on
surface codes as in Ref.~\cite{Horsman:2012a}, except with the intermediate
row of data qubits in the osculant regions removed.  The same layout is used
for the Hadamard gate, which grows and shrinks around the corner to
change the orientation of its boundary coloring. }}
\end{center}
\end{figure}

\begin{figure}[ht]
\begin{center}
  \includegraphics[width=0.4\columnwidth]{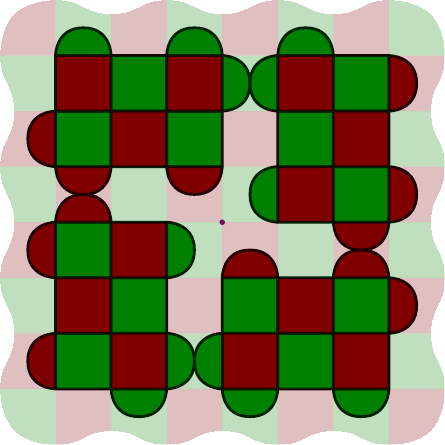}
\caption{\small{\label{fig:surf-code-chiral-injection}Injection procedure
for surface codes similar to the one in Fig.~\ref{fig:d7-injection}.  In
steps 1 and 2, the indicated code checks are measured three times each. }}
\end{center}
\end{figure}

While surface-code lattice-surgery was first explored in by Dennis \etal\ in
the context of state injection~\cite{Dennis:2002a}, the first exploration of
a universal set of logical gates on surface codes using lattice-surgery
methods was performed by Horsman {\textit{et al.}}~\cite{Horsman:2012a}.
Inspired by our color-code lattice surgery methods, we improved the methods
presented in Ref.~\cite{Horsman:2012a} so that they now use fewer qubits for
the $\CNOT$, $H$, and $S$ gates, using the layout depicted in
Fig.~\ref{fig:surface-CNOT}.  We also developed a new six-step surface-code
state-injection method similar to our color-code state-injection method; the
surface-code layout is depicted in
Fig.~\ref{fig:surf-code-chiral-injection}.
Table~\ref{tab:surface-code-resources} lists the resources used by these
improved surface-surgery methods on the ``rotated'' or ``medial'' surface
code, with an allocation of one syndrome per check (one per face).

\begin{center}
\begin{table}[ht]
\centering
\begin{tabular}{c||c|c|c|c|c|c|c|c|c}
\multicolumn{10}{c}{Surface-code lattice surgery (1 syndrome qubit/check)} \\[1ex] \hline \hline
Gate   & $T|+\>$ & $I$ & $|0\>$ & $|+\>$ & $M_Z$ & $M_X$ & $H$ & $S$ & $\CNOT$ \\[1ex] \hline
Depth  & $6$ & \multicolumn{3}{c|}{$d$} & \multicolumn{2}{c|}{$1$}
       & {$6d$}
       & {$12d$} & {$3d$} \\[1ex]  \hline
Qubits & \multicolumn{6}{c|}{$2d^2 -2d + 1$}
       & \multicolumn{3}{c}{$6d^2 - 6d+3$} \\[1ex] \hline
Error  & $\bigO(p)$ & \multicolumn{8}{c}{$\bigO(p^{(d+1)/2})$} \\[1ex] \hline \hline
\end{tabular}
\caption{Resources used by fault-tolerant medial surface-code lattice
surgery on distance-$d$ codes when one syndrome bit per face is allocated.
Depth is measured in number of measurement rounds.  Qubit counts include both
data and syndrome qubits.  Error is reported in big-$\bigO$ notation because
syndrome-extraction-circuit implementation details can change the constants.
The logical $S$ gate is implemented by catalytic teleportation from the $HS|+\>$
state, which requires two logical $\CNOT$ gates and a logical Hadamard gate
\cite{Aliferis:2007b}.  The logical $H$ gate is performed by lattice surgery as in
Ref.~\cite{Horsman:2012a}, but qubits are shifted $d$ sites horizontally and
$d$ sites vertically in the method to ensure that the size of the logical
operators do not drop below $d$, making the operation fault-tolerant.}
\label{tab:surface-code-resources}
\end{table}
\end{center}

From these tables, we see that color codes use approximately half as many
qubits as surface codes to achieve the same order of error suppression.
Color-code lattice surgery also performs encoded gates in essentially the
same time or faster than they are performed via surface-code lattice
surgery.  Even when both models are optimized for qubits by exploiting a
single roving syndrome qubit, the color-code $\CNOT$ uses $(3d^2 + 6d -
1)/2$ qubits whereas the surface-code $\CNOT$ uses $3d^2 + 1$
qubits---again about half as many.

%
\subsection{Overhead per Desired Level of Error Suppression}
\label{sec:overhead-per-error-suppression}

Because the accuracy threshold against circuit-level depolarizing noise is
smaller for color codes than for surface codes, a color code will need a
larger code distance than a surface would need to achieve the same
level of error suppression (\ie, to achieve the same logical failure
probability $p_{\text{fail}}$).  This erodes the factor-of-two qubit
savings that color codes provide at the same code distance, and could
possibly eliminate the savings entirely.

To compute the qubit overhead $\Omega$ to achieve a given $p_{\text{fail}}$
for a logical operation, one inverts the relationship $p_{\text{fail}}(d)$
and plugs the solution $d(p_{\text{fail}})$ into the appropriate expression
for the number of qubits per operation, \eg, from the ``Qubits'' entry in
Table~\ref{tab:color-code-resources-2qubit} or
Table~\ref{tab:surface-code-resources}.  The analytic expression best-suited
for $p_{\text{fail}}(d)$ depends on the relative magnitudes of $d$ and the
depolarizing probability $p$~\cite{Dennis:2002a, Wang:2003a,
Raussendorf:2007b, Fowler:2013a, Watson:2013a, Bravyi:2013a}; for example,
Watson and Barrett have shown that the scaling of $p_{\text{fail}}$ with $d$
is qualitatively different in the regime $d < 1/4p$ and $d > 1/4p$ for
code-capacity and phenomenological error models~\cite{Watson:2013a}.  Since
overhead comparisons are most relevant for non-asymptotic $d$ and for $p$
below the relevant pseudothreshold (\ie, the $p$ at a \emph{fixed} code
distance below which $p_{\text{fail}} < p$), and because we are most
interested in the scaling for circuit-level error models, we use the
expression for fixed $d$ and low $p$ for these models that Fowler found fit
well to surface-codes in Ref.~\cite{Fowler:2013a}, namely
\begin{align}
\label{eq:combinatoric-bound}
p_{\text{fail}} = A(d)\left(\frac{p}{p_{\text{th}}}\right)^{d/2}.
\end{align}

It is an interesting question as to whether color codes can exhibit the same
scaling at this in the low-$p$ regime.  Stephens has noted that his
color-code matching decoder in Ref.~\cite{Stephens:2014a} does not attain
the full algebraic code distance, suggesting that the exponent in
Eq.~(\ref{eq:combinatoric-bound}) using his decoder will be $\alpha d$,
where $\alpha < 1/2$.  In contrast, the integer-program (IP) decoder in
Ref.~\cite{Landahl:2011a} should attain the full code distance at the cost
of running more slowly.  If only one syndrome qubit per face or one per
check is used with the IP decoder, though, errors may spread badly, cutting
in to the effective code distance.  Using Shor-, Steane-, or Knill-style
syndrome extraction should eliminate this problem at the cost of many extra
syndrome qubits.  It may suffice to use the verificed four-cat and two-cat
states per octagonal and square faces respectively as used in
Refs.~\cite{Fowler:2008c, Stephens:2014a} with the IP decoder to achieve
this scaling, but currently that is an open question.  Although the IP
decoder appears to be inefficient at high error rates, at low error rates it
can be expected to run quickly.  Moreover, the recent linear-time PEPS
decoder for surface codes \cite{Bravyi:2014a} gives hope that a truly
efficient color-code decoder that achieves the scaling of
Eq.~(\ref{eq:combinatoric-bound}) will be found.  For the purposes of
comparision, and with this optimism in mind, we will assume that the scaling
law in Eq.~(\ref{eq:combinatoric-bound}) holds for both surface and color
codes.  However, we urge caution in reading too much into the results
derived from this assumption.

Using Eq.~(\ref{eq:combinatoric-bound}), the color-code distance $d_c$ that
gives the same error-suppression power as a surface code with distance $d_s$
is 
\begin{align}
\label{eq:dc-to-ds}
d_c &=
  d_s\left(\frac{\log p/p_{\text{th}}^{(s)}}
     {\log p/p_{\text{th}}^{(c)}}\right)
  + 2\left(\frac{\log A_s(d)/A_c(d)}
     {\log p/p_{\text{th}}^{(c)}}\right).
\end{align}
Fowler's numerical simulations suggest that $A_s(d)$ is approximately a
constant function of $d$ for $d$ up to 10~\cite{Fowler:2013a}; there is no
reason to expect that $A_c(d)$ is not also a comparably-sized constant
function of $d$ in the same range, or indeed that $A_s(d)$ and $A_c(d)$
should scale substantially differently for any $d$.  The numerator in the
second term of Eq.~(\ref{eq:combinatoric-bound}) should therefore be quite
small because of the logarithm.  Moreover, the denominator gets larger as
$p$ is reduced below the color-code (pseudo)threshold, making the overall
term even smaller.  For these reasons, we will neglect the second term in
Eq.~(\ref{eq:combinatoric-bound}) in our subsequent analysis.

\begin{figure}[ht]
\begin{center}
  \includegraphics[width=0.7\columnwidth]{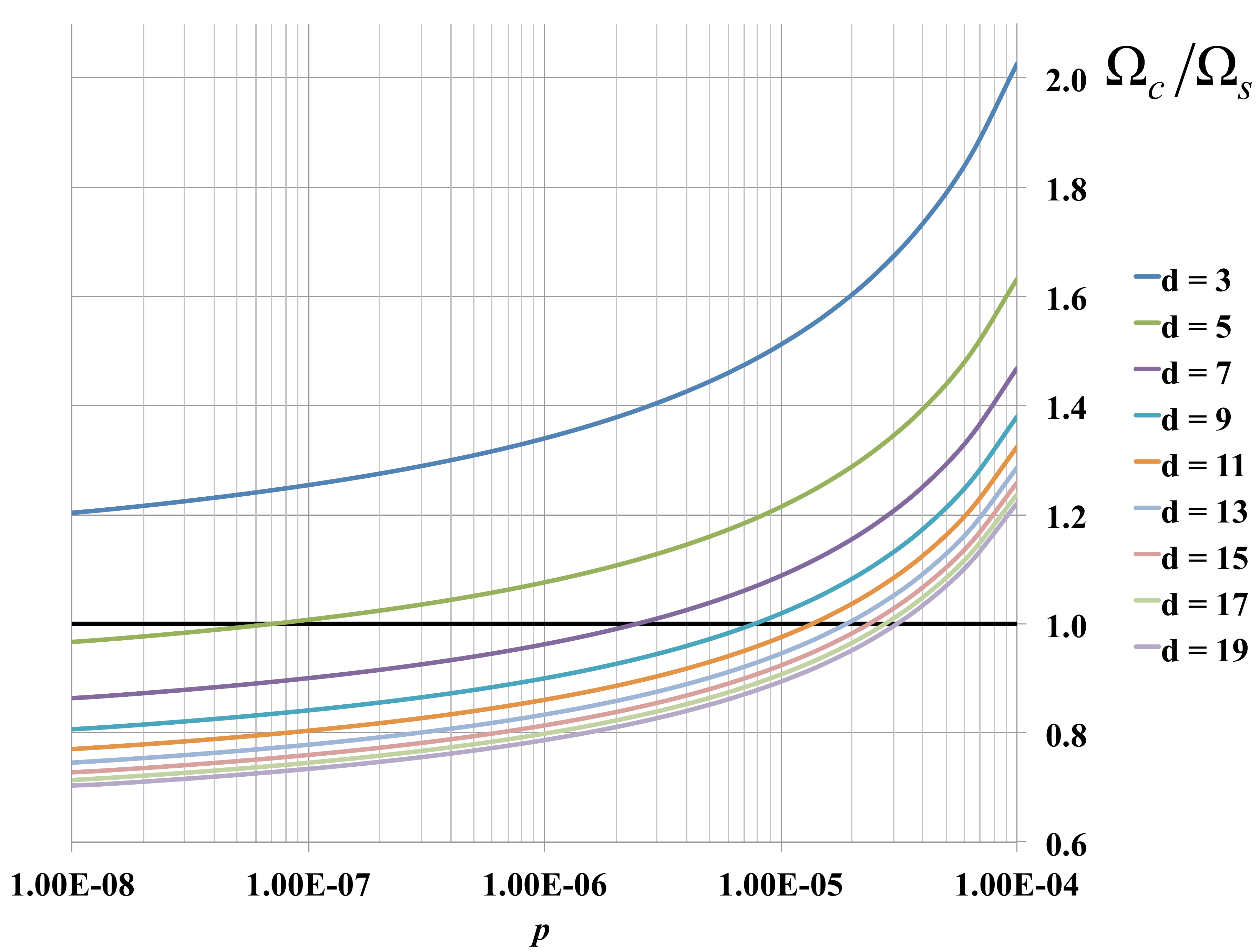}
\caption{\small{\label{fig:overhead-ratio-best-for-color}Ratio of color-code
to surface-code qubit overhead $\Omega_c/\Omega_s$ versus circuit-level
depolarizing probability $p$ when both codes are tuned via
Eq.~(\ref{eq:dc-to-ds}) to achieve the same logical qubit failure
probability.  Plots assume a color-code accuracy threshold of $0.143\%$ and
a surface-code accuracy threshold of $0.502\%$. }}
\end{center}
\end{figure}

\begin{figure}[ht]
\begin{center}
  \includegraphics[width=0.7\columnwidth]{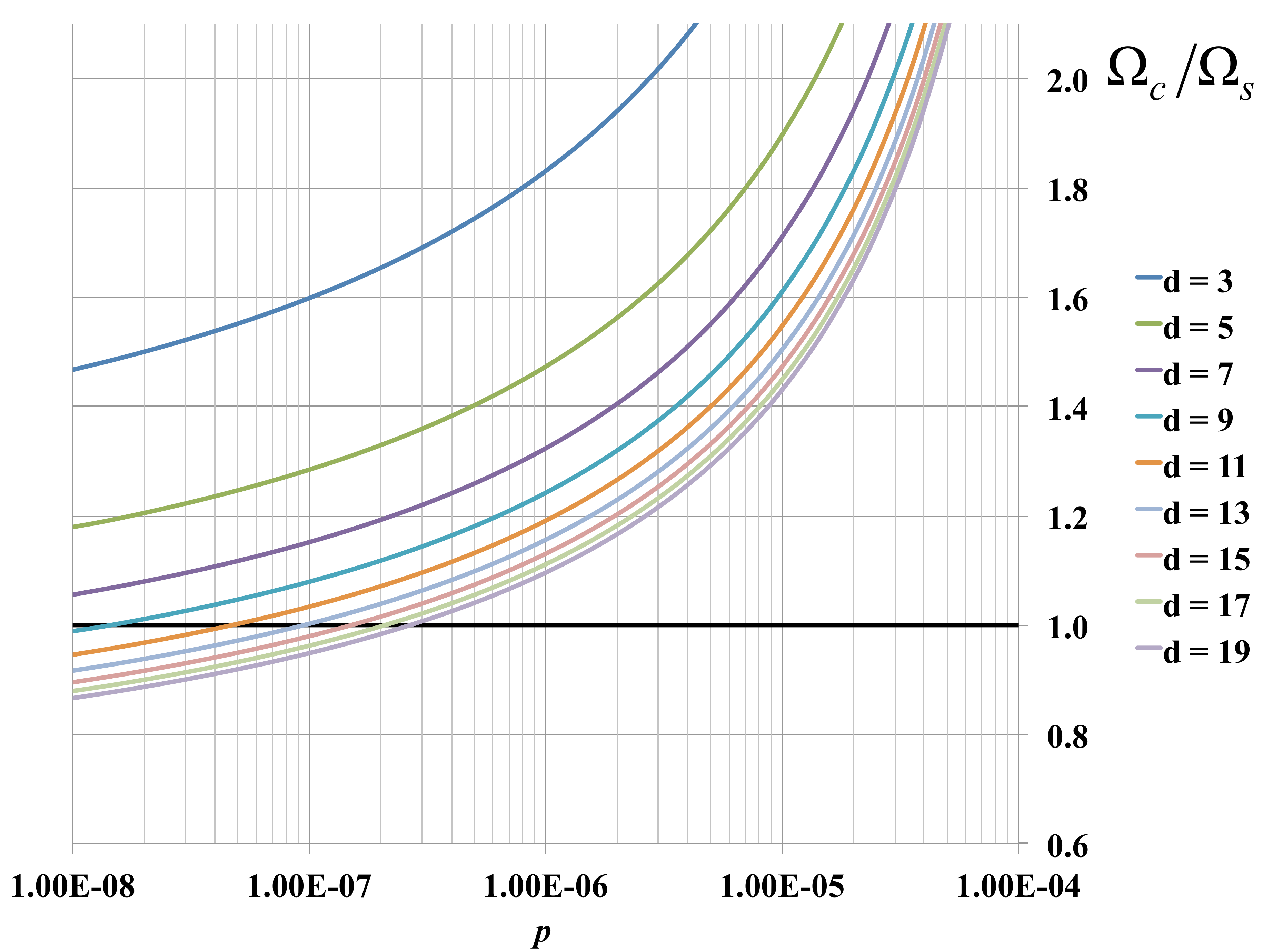}
\caption{\small{\label{fig:overhead-ratio-best-for-surface}Ratio of color-code
to surface-code qubit overhead $\Omega_c/\Omega_s$ versus circuit-level
depolarizing probability $p$ when both codes are tuned via
Eq.~(\ref{eq:dc-to-ds}) to achieve the same logical qubit failure
probability.  Plots assume a color-code accuracy threshold of $0.082\%$ and
a surface-code accuracy threshold of $1.140\%$.}}
\end{center}
\end{figure}

Using the expressions in Tables~\ref{tab:color-code-resources-2qubit} and
\ref{tab:surface-code-resources} for the color-code and surface-code qubit
overheads, which we denote by $\Omega_c(d)$ and $\Omega_s(d)$, and the
relationship in Eq.~(\ref{eq:dc-to-ds}), we plot the ratio
$\Omega_c(d_c(d_s))/\Omega_s(d_s)$ versus $p$ for several values of $d_s$.
This ratio is sensitive to the estimates for $p_{\text{th}}^{(c)}$ and
$p_{\text{th}}^{(s)}$, so we present two plots at the extremes of the
estimates. Figure~\ref{fig:overhead-ratio-best-for-color} is the plot using
the highest estimate for the color-code accuracy threshold ($0.143\%$) and
the lowest estimate for the color-code accuracy threshold ($0.502\%$).
Figure~\ref{fig:overhead-ratio-best-for-surface} is the plot using the
lowest estimate for the color-code accuracy threshold ($0.082\%$) and the
highest estimate for the surface-code accuracy threshold ($1.140\%$).

From these plots, we see that for distances greater than 11, as long as $p$
is below a value bracketed approximately somewhere between $10^{-5}$ to
$10^{-7}$, color codes use fewer qubits to achieve the same level of error
suppression.

This conclusion could be sharpened by direct numerical simulations, which we
believe would be an interesting future research project.  Rather than
assuming a phenomenological scaling law as in
Eq.~(\ref{eq:combinatoric-bound}) for the failure probability and using it
to infer the overhead, one could perform direct numerical estimation of the
overhead as a function of $d$ and $p$ and compare the results for color
codes and surface codes.  In addition to removing the need to fit an assumed
scaling law, this approach would also remove the need to estimate accuracy
thresholds, or even pseudothresholds, because it gets directly at the
question at hand.

%
\subsection{Overhead for Small Logical CNOT Gates}
\label{sec:small-CNOTs}

Because of the interest expressed in Ref.~\cite{Horsman:2012a} in designing
the fewest-qubit implementation of a $\CNOT$ gate with a topological
stabilizer code, we thought it would be valuable to list the qubit overheads
required by the methods described here for small distances.  As mentioned in
Sec.~\ref{sec:background}, the number of syndrome qubits used by an
implementation of a topological stabilizer code is design dependent: a
single roving syndrome qubit would suffice, but one could use a number of
syndrome qubits up to twice the number of data qubits to some advantage.

In Table~\ref{tab:CNOT-qubits-small-d}, we list the qubit overhead required
for the low end of the syndrome-allocation spectrum for the following
methods: (a) color-code transversal methods, (b) our color-code
lattice-surgery methods, (c) surface-code transversal methods, (d) our
surface-code lattice-surgery methods, and (e) the surface-code
lattice-surgery methods described in Ref.~\cite{Horsman:2012a}.

As noted in our introduction, transversal methods are not well-suited to
local quantum processing on two-dimensional arrays of qubits restricted to
local movements; we list the overheads here despite this because at small
distances, one might be able to exploit nonlocal processing and/or nonlocal
qubit movement.  For example, a recent demonstration in a trapped-ion
quantum computer of a single-round of error correction on a distance-three
color code exploited the fact that all seven $^{40}\text{Ca}^+$ ions
involved were trapped in a single Paul trap~\cite{IonColor}.  (The minimal
extra ``roving'' syndrome qubit was not used in the experiment because the
protocol was not fault-tolerant---instead of repeating syndrome measurements
into one or more auxillary qubits, the data-qubit ions were measured
destructively once via resonance-fluorescence.)

For all methods in Table~\ref{tab:CNOT-qubits-small-d}, we consider the
allocations of (i) a single roving syndrome qubit, (ii) one syndrome qubit
per face, and (iii) one syndrome qubit per check, which is the same as one
per face for surface codes but is two per face for color codes.  For
transversal methods, we also consider an in-between variant with (iv) one
syndrome qubit per two faces, because one might want to share the syndrome
qubits transversally between the two logical qubits.  (The case of sharing
two syndrome qubits per face between the two logical qubits has the same
overhead count as having both logical qubits use one syndrome qubit per
face.)

\begin{table}[ht!]
\centering

\noindent\begin{tabular*}{0.9\columnwidth}{@{\extracolsep{\stretch{1}}}*{6}{l|rrrrr}@{}}
\hline\hline
\hspace{5em} $d$\hspace{1em} & 3 & 5 & 7 & 9 & 11  \\
\hline
Color transversal: 1 total & 15 & 35 & 63 & 99 & 143 \\
Color transversal: $\text{faces}/2$ & 17 & 42 & 77 & 122 & 177 \\
Color transversal: faces & 20 & 50 & 92 & 146 & 212 \\
Color transversal: $2\times$faces & 26 & 66 & 122 & 194 & 282 \\
\hline
Color surgery: 1 total & 22 & 52 & 94 & 148 & 214 \\
Color surgery: faces & 30 & 75 & 138 & 219 & 318 \\
Color surgery: $2\times$faces & 39 & 99 & 183 & 291 & 423 \\
\hline
Surface transversal: 1 total & 19 & 51 & 99 & 163 & 243 \\
Surface transversal: faces/2 & 22 & 66 & 134 & 226 & 342 \\
Surface transversal: faces & 26 & 82 & 170 & 290 & 442 \\
\hline
Surface surgery: 1 total & 28 & 76 & 148 & 244 & 364 \\
Surface surgery: faces & 39 & 123 & 255 & 435 & 663 \\
\hline
Surface surgery~\cite{Horsman:2012a}: 1 total & 34 & 86 & 162 & 262 & 386 \\
Surface surgery~\cite{Horsman:2012a}: faces & 53 & 149 & 293 & 485 & 725 \\
\hline \hline
\end{tabular*}
\caption{Number of qubits needed to implement a logical $\CNOT$ gate for
several color-code and surface-code methods for small values
of the code distance $d$, assuming that the number of syndrome qubits used
is as indicated.}
\label{tab:CNOT-qubits-small-d}
\end{table}
%

%
\section{Conclusions}
\label{sec:conclusion}

Our color-code lattice-surgery methods open new possibilities for achieving
fault-tolerant quantum computation using fewer resources.  Per code
distance, they are manifestly superior to surface-code lattice-surgery
methods, using approximately half the qubits and the same time or less to
perform logical quantum operations.  Although we did not discuss it, they
also use fewer qubits and the same time or less than defect-based
``spacetime braiding'' methods for both
surface-codes~\cite{Raussendorf:2006a} and color-codes~\cite{Fowler:2008c}.
Transversal methods do use fewer qubits per code distance than color-code
lattice surgery to perform logical operations~\cite{Dennis:2002a}, but
transversal methods cannot be implemented in systems utilizing local quantum
processing on two-dimensional arrays of qubits restricted to local
movements.

Because color codes are estimated to have a lower accuracy threshold than
surface codes against uncorrelated circuit-level depolarizing noise
\cite{Landahl:2011a, Stephens:2014a, Stephens:2014b}, the superiority of
color codes only becomes manifest at sufficiently low depolarizing error
probabilities and sufficiently large code distances.  Subject to an assumed
scaling law given by Eq.~(\ref{eq:combinatoric-bound}) for both surface
codes and color codes, the depolarizing probability cutoff is approximately
somewhere in the range $p = 10^{-5}$ to $p = 10^{-7}$ with a corresponding
distance cutoff of $d = 11$.  Color-code decoder research is only in its
infancy, and we believe that the regime of superiority can be expanded with
further study.  For example, the recent linear-time PEPS decoder by Bravyi
{\textit{et al.}}~\cite{Bravyi:2014a} might be extended to color codes,
allowing one to approximate the optimal decoder quite well with only
linear-time processing.  The close relationship between color codes and
surface codes at the topological-phase level~\cite{Bombin:2012a} means that
the decoding complexity, if not the performance, can always be made
comparable for the two classes of codes~\cite{Bombin:2012a,
Duclos-Cianci:2014a, Delfosse:2014a}.

It would seem then, color codes are equal to or superior to surface codes,
at least insofar as space and time overhead considerations are concerned,
for systems that are sufficiently mature, meaning that they have
sufficiently low error rates and sufficiently many qubits available.  When
technology brings us to this point, we believe the transition from (two-colorable)
surface-codes to (three-colorable) color codes will resemble the transition
of television broadcasts from black-and-white to color: perhaps a little
bumpy at first, but inevitable.  Until then, the mandate for color-code
research is to bring that horizon closer to the present.

\section{Acknowledgments}

The authors would like to thank Eric Bahr, Chris Cesare, Anand Ganti, Setso Metodi, and Uzoma Onunkwo for helpful
discussions.
\chapter{{P}erformance {E}stimator of {C}odes {O}n {S}urfaces\label{ch.pecos}}

\setlength\epigraphwidth{13cm}
\epigraph{\textit{``I mean, making simulations of what you're going to build is tremendously useful if you can get feedback from them that will tell you where you've gone wrong and what you can do about it.''}}{--- \textup{Christopher Alexander \cite{Baker06}}}

\setlength\epigraphwidth{13cm}
\epigraph{\textit{``Software is a great combination between artistry and engineering.''}}{--- \textup{Bill Gates \cite{BillGates2008}}}

Much of the field of quantum error-correction (QEC) involves developing and evaluating the performance of QEC protocols. Often a scientist, after developing some new idea, will quickly cobble together code to verify and validate their discoveries. Rarely does one create a robust, well-written package that can be called on, with minimum extension, to enable the quick exploration of ideas. 

\vfill
\pagebreak[4]
In recent years there has been a shift, though. As noisy intermediate-scale quantum (NISQ) computers are in development \cite{PreskillNISQ}, more and more software packages for quantum computing have been released; however, much of the focus has been on circuit-level modeling and quantum algorithms (see \cite{LaRose18,Fingerhuth18,CQC18}).

In this chapter, I describe a \pack{Python} package that I created called ``Performance Estimator of Codes On Surfaces'' (\PECOS). This package provides a framework for studying, developing, and evaluating QECC protocols.

As of the writing of this dissertation, I am not aware of any similar frameworks that are publicly available. However, Microsoft's \pack{Q\#} allows one to represent quantum error-correcting codes (QECCs) \cite{Qsharp:Error}. Also, Chris Granade and Ben Criger have a \pack{Python} package called \pack{QuaEC} that serves as a library for describing and manipulating QECCs \cite{QuaEC}.  Although not currently publicly available, David Tuckett is developing a similar \pack{Python} library for simulating stabilizer code \cite{Tuckett17}.

\PECOS is an attempt at balancing simplicity, usability, functionality, and extendibility, and future-proofing. In the spirit of extendibility, \PECOS is agnostic to quantum simulators, quantum operations, and QECCs. Of course, it is difficult to eloquently represent all QEC techniques. While agnostic to QECCs, the primary focus of \PECOS has been the simulation and evaluation of lattice-surgery for topological stabilizer codes.

\PECOS is a culmination of four years of on-and-off development. The first incarnation was some Python code I cobbled together to verify the lattice-surgery protocols presented in Chapter~\ref{ch.lattice_surgery}. Further development of my code eventually resulted in the new stabilizer-simulation algorithm discussed in Chapter~\ref{ch.stab_sim}, which has improved runtimes for the simulation of topological stabilizer codes.  

\vfill
\pagebreak[4]
To facilitate the use of \PECOS, this chapter will present \PECOS in the style of a user's manual. Therefore, this chapter will be terse with discussions but heavy with examples. I will assume that the reader is familiar with QEC and \pack{Python}.

Because I developed \PECOS independently, all the technical and writing contributions in this chapter are my own.

The chapter is organized as follows: In Section \ref{sec.pecos.gettingstarted}, I present package requirements, how to install and uninstall \PECOS, the organization of the \PECOS package, and an example script. In Section \ref{sec.pecos.quantum_circuits}, I present the data structure used for representing quantum circuits. In Section \ref{sec.pecos.qeccs}, I present the class used by \PECOS to represent QECCs, their logical gates, and logical instructions. In Section \ref{sec.pecos.state_sim}, I briefly discuss the quantum-state simulators included with \PECOS. In Section \ref{sec.pecos.circ_run}, I present the classes used to run quantum simulations. In Section \ref{sec.pecos.logical_circuit}, I discuss a data structure for representing quantum circuits at the logical level. In Section \ref{sec.pecos.error_gens}, I present the classes used to generate noise for quantum circuits. Finally, in Section \ref{sec.pecos.decode} I discuss decoders, which interpret measurements of syndrome extraction rounds to determine recovery operations.


\section{Getting Started\label{sec.pecos.gettingstarted}}

This chapter addresses version 0.1.0 of \PECOS. In the following section, we will discuss how to get started with \PECOS. If you would like to jump straight to examples, see the list in Section~\ref{sec:pecos:examplelist}.

\vspace{2in}

\subsection{Requirements}

\PECOS requires \pack{Python} 3.5 or greater to run. If you currently do not have Python, I recommended to installing a scientific Python distribution, a list of which is given in \cite{SciPy:Distributions,Python:Distributions}.

The following packages are required:

\begin{itemize}
\item \pack{NumPy} 1.15+
\item \pack{SciPy} 1.1+
\item \pack{Matplotlib} 2.2+
\item \pack{NetworkX} 2.1+
\end{itemize}

These are optional packages:

\begin{itemize}
\item \pack{Cython} (to compile \pack{C} and \pack{C++} extensions)
\item \pack{PyTest} (to run tests)
\item \pack{Sphinx} 2.7.6+ (to compile documentation)
\end{itemize}

\subsection{Installing and Uninstalling of PECOS}

\PECOS has been developed to run equally well on Windows and Linux-based systems. 

To install via pip run:

\begin{minipage}{0.95\linewidth}
\begin{lstlisting}[style=pystyle,caption={Installing the \PECOS package through pip.}]
>>> pip install quantum-pecos
\end{lstlisting}
\end{minipage}

Alternatively, to install \PECOS from its source, download or clone the package from: 

https://github.com/PECOS-packages/PECOS

One can then navigate to the package's root directory where the file \textcode{setup.py} is located and run the following line:

\begin{minipage}{0.95\linewidth}
\begin{lstlisting}[style=pystyle,caption={Installing the \PECOS package from source for non-developers.}]
>>> pip install .
\end{lstlisting}
\end{minipage}

However, if one wishes to develop \PECOS, instead run:

\begin{minipage}{0.95\linewidth}
\begin{lstlisting}[style=pystyle,caption={Installing the \PECOS package from source for developers.}]
>>> pip install -e .
\end{lstlisting}
\end{minipage}

\noindent This will allow you to edit \PECOS and have those changes take effect. 

To uninstall:

\begin{minipage}{0.95\linewidth}
\begin{lstlisting}[style=pystyle,caption={Uninstalling the \PECOS package.}]
>>> pip uninstall quantum-pecos
\end{lstlisting}
\end{minipage}

\subsection{Tests}

After installing \PECOS, you may want to run tests, which come with the package. Tests help verify the expected behavior of the current version of \PECOS. To run the tests, navigate to you \PECOS installation directory and run:

\begin{minipage}{0.95\linewidth}
\begin{lstlisting}[label={code.pecos.test}, style=pystyle,caption={Running tests.}]
>>> py.test
\end{lstlisting}
\end{minipage}

\pack{PyTest} will then automatically detect and run any tests included in \PECOS. 

If developing \PECOS, these tests can also be used to determine if any functionality has been broken by alterations. 

\subsection{Importing PECOS}

To begin using \PECOS, import the package by entering:

\begin{minipage}{0.95\linewidth}
\begin{lstlisting}[style=pystyle, caption={Importing \PECOS.}]
import pecos as pc
\end{lstlisting}
\end{minipage}

\noindent In the rest of this chapter, it will be assumed that \PECOS has been imported in this manner.

\vfill
\pagebreak[4]
\subsection{Package Organization}

Concepts in PECOS are organized around the following namespaces (Table~\ref{tb.pecos.namespace}):

\begin{table}[H]
\centering
\caption{Namespaces in \PECOS. \label{tb.pecos.namespace}}
\begin{tabular}{@{}ll@{}}
\hline
Name & Description \\
\hline
\code{circuits} & Circuits of different abstraction levels. \\
\code{qeccs} & Represent QEC protocols. \\
\code{error\_gens} & Used to specify error models and generate errors. \\
\code{simulators} &Simulate states and operations. \\
\code{circuit\_runners} & Coordinate \code{circuits} and \code{error\_gens} with \code{simulators}. \\
\code{decoders} & Produce recovery operations given syndromes. \\
\code{tools} & Tools for studying and evaluating QEC protocols. \\
\code{misc} & 
A catchall namespace. \\
\hline
\end{tabular}
\end{table}

In this chapter, I will use the convention that classes or functions designed to implement the purpose of the namespace are referred to by the singular form of the namespace. For example, a QECC class belonging to the namespace \code{qeccs} is referred to as a \code{qecc}.

\vfill
\pagebreak[4]
\subsection{Basic PECOS Script}

The following is simple example of using the {\PECOS} application programming interface (API): 

\begin{minipage}{0.95\linewidth}
\begin{lstlisting}[label=code.pecos.example, language=Python, style=pystyle, caption={Example of finding the threshold of a surface code.}]
"""Example: A simple script using PECOS"""
# Getting PECOS:
import pecos as pc
# User defined tool:
from . import some_tool
# User chosen parameters:
from .params import some_parameters
# Quantum error-correcting class:
surface_code = pc.qeccs.Surface4444
# Error generating class:
depolar = pc.error_gens.DepolarGen(model_level='code_capacity')
# Decoder class (creates recovery operations from measurements):
mwpm = pc.decoders.MWPM2D
# Quantum-state simulator class:
stab_sim = pc.simulators.sparsesim.State
# Combines classes together to run the simulation:
circuit_sim = pc.circuit_runners.Standard(simulator=stab_sim, seed=0)
# Running a user defined tool called ``some_tool``:
results = some_tool(surface_code, depolar, mwpm, circuit_sim, some_parameters)
\end{lstlisting}
\end{minipage}

\vfill
\pagebreak[4]
\subsection{List of Examples\label{sec:pecos:examplelist}}

For some examples of using \PECOS, see the appendices:

\begin{itemize}
\item Appendix~\ref{app.pecos.stabeval}: Verifying/developing a stabilizer code.
\item Appendix~\ref{app.pecos.qecc}: Creating a \pack{Python} class to represent a QECC and extend \PECOS.
\item Appendix~\ref{app.pecos.errors}: Defining simple error models.
\item Appendix~\ref{app.pecos.monte}: Writing a basic Monte Carlo script to determine logical error rates.
\end{itemize}

\section{Quantum Circuits\label{sec.pecos.quantum_circuits}}

Gate-based protocols, such as QEC procedures, are described in terms of quantum circuits. In \PECOS, the data structure used to represent quantum circuits is simply called \QuantumCircuit. This class was designed with similar methods as the commonly used data structures in Python such as \plist, \pdict, and \code{set}. This choice was made so that users accustomed to Python data structures would find \QuantumCircuit familiar and, hopefully, easy to use.

The \QuantumCircuit data-structure was particularly designed to efficiently represent the quantum circuits of QEC protocols. During each time step (tick) in QEC circuits, many gates of just a few gate-types are applied to most of the qubits in the QECC. \QuantumCircuit is a data structure that represents a sequence of ticks, where for each tick there is an associated data structure that keeps track of what few gate-types are being applied and, for each of these types, what qubits are being acted on. We will see examples of this in the following.

Note, in the following I will refer to ``qudits'' rather than ``qubits'' (see definition in Section~\ref{intro.sec.qstate}) since a \QuantumCircuit could represent a sequence of qudit operations.

\subsection{Attributes and Methods}

For convenience, a list of attributes and methods of \QuantumCircuit are described in tables \ref{tb.pecos.attr} and \ref{tb.pecos.methods}.

\begin{table}[H]
\centering
\caption{QuantumCircuit attributes. \label{tb.pecos.attr}}
\begin{tabular}{@{}ll@{}}
\hline
Name & Description\\
\hline
\code{active\_qudits} & A \code{list} of \code{set}s of active qudits per tick. \\
\code{params} & A dictionary used to store extra information about the circuit. \\
\hline
\end{tabular}
\end{table}

\begin{table}[H]
\centering
\caption{QuantumCircuit methods. \label{tb.pecos.methods}}
\begin{tabular}{@{}ll@{}}
\hline
Name & Description \\
\hline
\code{append()} & Adds a collection of gates representing a single tick \\
& to the end of the \QuantumCircuit. \\
\code{update()} & Adds more gates to a tick. \\
\code{discard()} & Removes gates based on gate locations. \\
\code{items()} & A generator used to loop over the gates in the circuit. \\
\hline
\end{tabular}
\end{table} 

\subsection{An Instance}

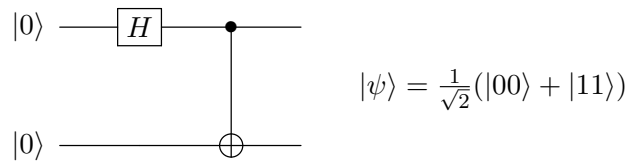
\begin{figure}[ht]
{\small \begin{center}
    \begin{tabular}{l} 
			\Qcircuit @C=2em @R=0.8em {
\lstick{\ket{0}} & \gate{H} & \ctrl{2} & \qw & \\
 & & & & \push{\ket{\psi}=\frac{1}{\sqrt{2}}(\ket{00}+\ket{11})} \\
\lstick{\ket{0}} &  \qw  & \targ & \qw & \\
}
    \end{tabular}
\end{center}}
\caption{\label{fig.bell}A quantum circuit that prepares the Bell state $\frac{1}{\sqrt{2}}(\ket{00}+\ket{11})$.}
\end{figure}

To represent a quantum circuit, such as as the preparation of the Bell state $\frac{1}{\sqrt{2}}(\ket{00}+\ket{11})$ as seen in Fig. \ref{fig.bell}, we first begin by creating an instance of \QuantumCircuit. This is done in Code Block~\ref{code:qcinst}.

\begin{minipage}{0.95\linewidth}
\begin{lstlisting}[style=pystyle,label={code:qcinst}, caption={Creating an instance of a QuantumCircuit}]
>>> qc = pc.circuits.QuantumCircuit()
\end{lstlisting}
\end{minipage}

A string representation of the \QuantumCircuit can then be obtained:

\begin{minipage}{0.95\linewidth}
\begin{lstlisting}[label={code:rep},style=pystyle,caption={An empty QuantumCircuit.}]
>>> # Following from the previous example.
>>> qc
QuantumCircuit([])
\end{lstlisting}
\end{minipage}

\noindent Here, that the object is a instance of the \QuantumCircuit class is indicated by \code{QuantumCircuit()}. The brackets \code{[]} indicate an empty sequence. 

If needed, empty ticks can be reserved when the \QuantumCircuit is first created: 

\begin{minipage}{0.95\linewidth}
\begin{lstlisting}[style=pystyle,caption={An empty QuantumCircuit with three ticks.}]
>>> qc = pc.circuits.QuantumCircuit(3)
>>> qc
QuantumCircuit([{}, {}, {}])
\end{lstlisting}
\end{minipage}

\noindent Here, each tick is represented by a pair of braces \code{\{\}} and is separated by a comma. We will see later that the method \pupdate can used to add gates to empty ticks.

\subsection{Modifying a QuantumCircuit\label{sec.pecos.modqc}} 

Next, we will see how alter a \QuantumCircuit by using the methods \pappend, \pupdate, and \code{discard}.

\subsubsection{Append}

We can add a tick containing some gates to the end of a \QuantumCircuit by using the method \pappend. Doing so allows us to represent the quantum circuit in Fig.~\ref{fig.bell}:

\begin{minipage}{0.95\linewidth}
\begin{lstlisting}[label=code.bell, style=pystyle,caption={Example of using the append method to construct a representation of Fig.~\ref{fig.bell} using QuantumCircuit.}]
>>> qc = pc.circuits.QuantumCircuit()
>>> qc.append('init |0>', {0, 1})
>>> qc.append('H', {0})
>>> qc.append('CNOT', {(0,1)})
>>> qc
QuantumCircuit([{'init |0>': {0, 1}}, {'H': {0}}, {'CNOT': {(0, 1)}}])
\end{lstlisting}
\end{minipage}

\noindent Here in the final line we see a string representation of the quantum circuit in Fig~\ref{fig.bell}. As indicated by the string, gates of the same type are grouped together. Each gate-type is indicated by a symbol (string). The standard symbols use for qubit gates in \PECOS are given in appendix \ref{append.pecos.symbols}. Other symbols can be used by \PECOS so long as the symbols are hashable and recognized by the state-simulator used in applying the quantum circuit (see Section~\ref{sec.pecos.state_sim} for state simulators).

Paired with each gate symbol is set of gate locations, which are integers or tuples of integers. Integers are used to index qudits. Tuples are used to indicate qudits that are acted on by multi-qudit gates. The order of the qudit indices in a tuple may matter. For example, for a CNOT the first qubit is the control qubit while the second, is the target.

Code Block~\ref{code.bell} shows how to append a tick that consists of only one gate type. We can also append multiple gate-types per tick:

\begin{minipage}{0.95\linewidth}
\begin{lstlisting}[style=pystyle,caption={Using the append method to add multiple gate-types.}]
>>> qc = pc.circuits.QuantumCircuit()
>>> qc.append({'init |0>': {0, 1, 2, 3}})
>>> qc.append({'H': {0, 2}, 'X': {1, 3}})
>>> qc.append('CNOT', {(0,1), (2, 3)})
>>> qc
QuantumCircuit([{'init |0>': {0, 1, 2, 3}}, {'H': {0, 2}, 'X': {1, 3}}, {'CNOT': {(0, 1), (2, 3)}}])
\end{lstlisting}
\end{minipage} 

Both {\QuantumCircuit}s and gates may have extra information that we wish to include. Such information can be added to the \QuantumCircuit by including extra keywords as seen here:

\begin{minipage}{0.95\linewidth}
\begin{lstlisting}[label=code.params, style=pystyle,caption={Example of using params.}]
>>> qc = pc.circuits.QuantumCircuit(a_var=3.0)
>>> qc.append('init |0>', {0, 1}, duration=5)
>>> qc.append({'H': {0}, 'X': {1}}, duration=1)
>>> qc
QuantumCircuit(params={'a_var': 3.0}, ticks=[{'init |0>': loc: {0, 1} - params={'duration': 5}}, {'H': loc: {0} - params={'duration': 1}, 'X': loc: {1} - params={'duration': 1}}])
\end{lstlisting}
\end{minipage}

\noindent As we can see in this example, extra keyword arguments are gathered into the {\pdict}s referred to as \textit{params}. We will see later how the information in the params can be retrieved.

Note, the \pappend method associates the extra keywords with all the gates in the tick. This limitation can be overcome by the \code{update}, which is discussed next.

\subsubsection{Update\label{sec.pecos.update}}

The \pupdate method of \QuantumCircuit adds additional gates to a pre-existing tick. An example of using \pupdate is seen in the following:

\begin{minipage}{0.95\linewidth}
\begin{lstlisting}[label=code.update, style=pystyle,caption={Example of using the update method.}]
>>> qc = pc.circuits.QuantumCircuit()
>>> qc.append({'X': {0, 1}, 'Z': {2, 3}})
>>> qc.append({'H': {0, 2}})
>>> qc.append({'Y': {1, 3}})
>>> qc.update({'CNOT': {(6, 7), (8, 9)}, 'H': {10, 11}}, tick=0)
>>> qc.update('X', {4, 5})  # updates the currently last tick
>>> qc.append({'Q': {1, 2, 3}})
>>> qc.update({'H': {7}, 'S': {8}}, tick=1)
>>> qc.update({'R': {5}})
>>> qc
QuantumCircuit([{'X': {0, 1}, 'Z': {2, 3}, 'CNOT': {(8, 9), (6, 7)}, 'H': {10, 11}}, {'H': {0, 2, 7}, 'S': {8}}, {'Y': {1, 3}, 'X': {4, 5}}, {'Q': {1, 2, 3}, 'R': {5}}])
\end{lstlisting}
\end{minipage}

By default, \pupdate adds gates to the current last tick of the \QuantumCircuit. The \code{tick} keyword can be used to specify a tick. Each tick is index by an integer starting with 0.

Note, \pupdate will not override gate symbol-location pairs in the tick; instead, it will only add additional gate locations.

Like \pappend, \pupdate accepts other keyword arguments and stores such information in the params \pdict: 

\begin{minipage}{0.95\linewidth}
\begin{lstlisting}[style=pystyle,caption={An example of using update with params.}]
>>> qc = pc.circuits.QuantumCircuit(1)
>>> qc.update('X', {0, 1}, duration=3)
>>> qc.update('H', {2, 3}, duration=2)
>>> qc
QuantumCircuit([{'X': loc: {0, 1} - params={'duration': 3}, 'H': loc: {2, 3} - params={'duration': 2}}])
\end{lstlisting}
\end{minipage}

Note, since gates in a tick should be parallel operations, if more than one gate acts on a single qudit during a gate, an \code{Exception} is raised.

\subsubsection{Discard}

If needed, gate locations can be removed using the method \code{discard}. This can be seen in the following:

\begin{minipage}{0.95\linewidth}
\begin{lstlisting}[style=pystyle,caption={Example of using discard to remove gates.}]
>>> qc = pc.circuits.QuantumCircuit()
>>> qc.append('X', {0, 1, 2})
>>> qc.discard({1})
>>> qc
QuantumCircuit([{'X': {0, 2}}])
\end{lstlisting}
\end{minipage}

A \code{tick} keyword can be used to specify which tick the gate is discarded from. If no tick is specified, then \code{discard} removes gates from the last tick.

\subsection{Retrieving Information\label{sec.pecos.retrieve}}

\subsubsection{Number of Ticks}

The number of ticks in a \QuantumCircuit can be obtained using Python's \code{len} function:

\begin{minipage}{0.95\linewidth}
\begin{lstlisting}[label={pecos.code.len},style=pystyle,caption={Example of using len to determine the number of ticks in a QuantumCircuit.}]
>>> qc = pc.circuits.QuantumCircuit(5)
>>> len(qc)
5
>>> qc
QuantumCircuit([{}, {}, {}, {}, {}])
\end{lstlisting}
\end{minipage}

\subsubsection{Active Qudits\label{sec.pecos.active}}

The \QuantumCircuit data structure keeps track of which qudits have been acted on during a tick. These qudits are known as \textit{active qudits}. The \code{active\_qudits} attribute can be used to retrieve a list of these qudits:

\begin{minipage}{0.95\linewidth}
\begin{lstlisting}[style=pystyle,caption={Example of retrieving a list of active qudits.}]
>>> qc = pc.circuits.QuantumCircuit()
>>> qc.append({'X': {0}, 'Z': {2, 3}})
>>> qc.append({'CNOT': {(0, 2), (1, 3)}})
>>> qc.append('H', {2})
>>> qc.active_qudits
[{0, 2, 3}, {0, 1, 2, 3}, {2}]
\end{lstlisting}
\end{minipage}

This information can be useful if one wants to apply errors to inactive qudits.

\pagebreak[4]
\subsubsection{For Loops\label{sec.pecos.loop}}

The \QuantumCircuit class has the generator \code{items}, which can be used to iterate over the circuit and obtain a sequence of gate symbols, locations, and params:

\begin{minipage}{0.95\linewidth}
\begin{lstlisting}[label=code.loop, style=pystyle,caption={Example of using a for loop to retrieve gates from a QuantumCircuit.}]
>>> qc = pc.circuits.QuantumCircuit()
>>> qc.append({'X': {3, 5}, 'Z': {0, 1, 2}}, duration=1)
>>> qc.append({'H': {0, 1, 2, 3}})
>>> qc.append({'measure Z': {0, 3, 5}})
>>> for gate, gate_locations, params in qc.items():
...     print('%s -> %s, params: %s' % (gate, gate_locations, params))
X -> {3, 5}, params: {'duration': 1}
Z -> {0, 1, 2}, params: {'duration': 1}
H -> {0, 1, 2, 3}, params: {}
measure Z -> {0, 3, 5}, params: {}
\end{lstlisting}
\end{minipage}

One can loop over a single tick by using the keyword \code{tick}:

\begin{minipage}{0.95\linewidth}
\begin{lstlisting}[label={code.loop_tick}, style=pystyle,caption={Example of looping over a single tick in a QuantumCircuit.}]
>>> # Following the previous example
>>> for gate, gate_locations, params in qc.items(tick=0):
...     print('%s -> %s, params: %s' % (gate, gate_locations, params))
X -> {3, 5}, params: {'duration': 1}
Z -> {0, 1, 2}, params: {'duration': 1}
\end{lstlisting}
\end{minipage}

\section{QECCs\label{sec.pecos.qeccs}}

Each QECC or family of QECCs can be represented by a class. The classes available in \PECOS are in the namespace \code{qeccs}. We will refer to such a QECC class as a \code{qecc}. In this section we will discuss the methods, attributes, and structure of a \code{qecc}. In Appendix~\ref{app.pecos.qecc}, an example is given of how to construct a new QECC class, which can be used by \PECOS.

The primary role of a \code{qecc} is to provide the quantum circuits of QEC protocols associated with the \code{qecc} such as the logical-state initialization, logical gates, and logical measurements. In the following, we will look at some examples of \code{qecc} classes and how they encapsulate QEC procedures.

\subsection{Attributes and Methods}

The minimally expected methods and attributes for a \code{qecc} are listed in the tables \ref{tb.pecos.qecc_methods} and \ref{tb.pecos.qecc_attrib}.

\begin{table}[H]
\centering
\caption{Expected methods of a QECC class. \label{tb.pecos.qecc_methods}}
\begin{tabular}{@{}ll@{}}
\hline
Name & Description \\
\hline
\code{gate()} & Returns an instance of a requested logical gate. \\
\code{instruction()} & Returns an instance of a requested logical instruction. \\
\code{plot()} & Plots a qudit layout. \\
\hline
\end{tabular}
\end{table}

\begin{table}[H]
\centering
\caption{Expected attributes of a QECC class. \label{tb.pecos.qecc_attrib}}
\begin{tabular}{@{}ll@{}}
\hline
Name & Description \\
\hline
\code{name} & Name of the QECC. \\
\code{qecc\_params} & Dictionary of parameters. \\
\code{distance} & Minimum number of single qudit operations that results \\
 & in a logical error. \\
\code{num\_logical\_qudits} & Number of logical qudits. \\
\code{num\_data\_qudtis} & Number of data qudits. \\
\code{num\_ancilla\_qudits} & Number of ancillas. \\
\code{num\_qudits} & Number of qudits. \\
\code{qudit\_set} & Set of qudit labels used internally in the \code{qecc}.\\
\code{data\_qudit\_set} & Set of data qudit labels used internally. \\
\code{ancilla\_qudit\_set} & Set of ancilla qudit labels used internally. \\
\code{layout} & A dictionary of qudit label to position (\ptuple).\\
\code{sides} & A dictionary describing the geometry of the \code{qecc}. \\
\hline
\end{tabular}
\end{table} 

\subsection{An Instance}

Currently, the namespace \code{qeccs} contains classes representing the surface code on the 4.4.4.4 lattice (\code{Surface4444}) \cite{Kitaev:1996a}, the medial surface-code on the 4.4.4.4 lattice (\code{SurfaceMedial4444}) \cite{Bombin:2007d}, and the color-code on the 4.8.8 lattice (\code{Color488}) \cite{Bombin:2006b}.

\vfill
\pagebreak[4]
An example instance of a \code{Surface4444} is given here:

\begin{minipage}{0.95\linewidth}
\begin{lstlisting}[label=code.surf, style=pystyle,caption={Creating an instance of a surface code.}]
>>> # A distance-three surface-code:
>>> surface = pc.qeccs.Surface4444(distance=3)
\end{lstlisting}
\end{minipage}

\noindent As seen in Code Block~\ref{code.surf}, parameters are used to identify a member of the code family. For \code{Surface4444}, either the keyword \code{distance} or the keywords \code{height} and \code{width} are used to specify a member. If \code{distance} is used, then a representation of a square surface-code patch will be created. The \code{SurfaceMedial4444} class will take the same keywords as code family parameters as the \code{Surface4444} class. The \code{Color488} class only accepts \code{distance} as a keyword.

\subsection{Logical Gate\label{sec.pecos.logical_gate}}

The class \code{LogicalGate} represent a collection of quantum circuits that act on logical qubits. Each \code{LogicalGate}s is identified by a symbol (string). Using this symbol, the \code{gate} method of a \code{qecc} can be used to obtain an instance of a corresponding \code{LogicalGate} instance:

\begin{minipage}{0.95\linewidth}
\begin{lstlisting}[style=pystyle,caption={Example of requesting a logical gate.}]
>>> surface = pc.qeccs.Surface4444(distance=3)
>>> identity = surface.gate('I')
\end{lstlisting}
\end{minipage}

\noindent In the above code, the symbol \code{'I'} is used to retrieve a logical gate corresponding to identity (syndrome extraction).

\vfill
\pagebreak[4]
Keyword arguments may be used to modify \code{LogicalGate}s:

\begin{minipage}{0.95\linewidth}
\begin{lstlisting}[style=pystyle,caption={Example of specifying a gate through keywords.}]
>>> surface = pc.qeccs.Surface4444(distance=3)
>>> # Get an identity gate with only one round of syndrome extraction.
>>> identity = surface.gate('I', num_syn_extract=1)
\end{lstlisting} 
\end{minipage}

\noindent Here the keyword argument \code{num\_syn\_extract} is used to explicitly request an identity with only one round of syndrome extraction. Typically, the number of rounds of syndrome extraction for an identity gate is equal to the QECC's distance.

The main use for \code{LogicalGate} instances is as logical operations in the logical analogs of quantum circuits, which are described in Section~\ref{sec.pecos.logical_circuit}.

\subsection{Logical Instruction\label{sec.pecos.logical_instr}}

A \code{LogicalGate} is composed of a sequence of \code{LogicalInstruction}s. A \code{LogicalInstruction} represents a collection of quantum circuits. Often these collections are repeated or used in multiple \code{LogicalGate}s. An example of a \code{LogicalInstruction} is one round of error correction.

Like \code{LogicalGate}s, \code{LogicalInstuction}s are represented by symbols (strings). The \code{instr\_symbols} attribute of a \code{LogicalGate} can be use to retrieve a list of symbols corresponding to the \code{LogicalInstuction}s that form the \code{LogicalGate}:

\begin{minipage}{0.95\linewidth}
\begin{lstlisting}[style=pystyle,caption={Example of obtaining a logical gate's list of logical instruction symbols.}]
>>> surface = pc.qeccs.Surface4444(distance=3)
>>> identity = surface.gate('I')
>>> identity.instr_symbols
['instr_syn_extract', 'instr_syn_extract', 'instr_syn_extract']
>>> # Request an identity with a single round of syndrome extraction.
>>> identity = surface.gate('I', num_syn_extract=1)
>>> identity.instr_symbols
['instr_syn_extract']
\end{lstlisting} 
\end{minipage}

In the following, we see how to retrieve an instance of the \code{'instr\_syn\_extract'} instruction and then see what \QuantumCircuit it represents:

\begin{minipage}{0.95\linewidth}
\begin{lstlisting}[style=pystyle,caption={Example of requesting obtaining a quantum circuit of a logical instruction.}]
>>> surface = pc.qeccs.SurfaceMedial4444(distance=3)
>>> # Get the LogicalInstruction instance representing the syndrome-extraction instruction.
>>> instr = surface.instruction('instr_syn_extract')
>>> instr.circuit
QuantumCircuit([{'init |0>': {0, 16, 4, 5, 6, 10, 11, 12}}, {'H': {0, 16, 11, 5}}, {'CNOT': {(15, 12), (11, 14), (8, 6), (5, 7), (13, 10), (0, 2)}}, {'CNOT': {(9, 12), (2, 6), (7, 10), (11, 15), (0, 3), (5, 8)}}, {'CNOT': {(7, 4), (16, 13), (14, 10), (11, 8), (5, 1), (9, 6)}}, {'CNOT': {(3, 6), (16, 14), (11, 9), (5, 2), (8, 10), (1, 4)}}, {'H': {0, 16, 11, 5}}, {'measure Z': {0, 16, 4, 5, 6, 10, 11, 12}}])
\end{lstlisting} 
\end{minipage}

\subsection{Plotting}

Both \code{qecc}s and \code{LogicalInstuction}s have a method called \code{plot} that will generate a plot that represents the object. These plots can be useful in understanding the structure of a QECC and its logical instructions.

The following is an example of using the \code{plot} method for a \code{qecc}:

\begin{minipage}{0.95\linewidth}
\begin{lstlisting}[style=pystyle,caption={Plotting the layout of a non-medial surface-code.}]
>>> surface = pc.qeccs.SurfaceMedial4444(distance=3)
>>> surface.plot()
\end{lstlisting}
\end{minipage}

\begin{figure}[H]
	\centering
		\includegraphics[width=0.75\textwidth]{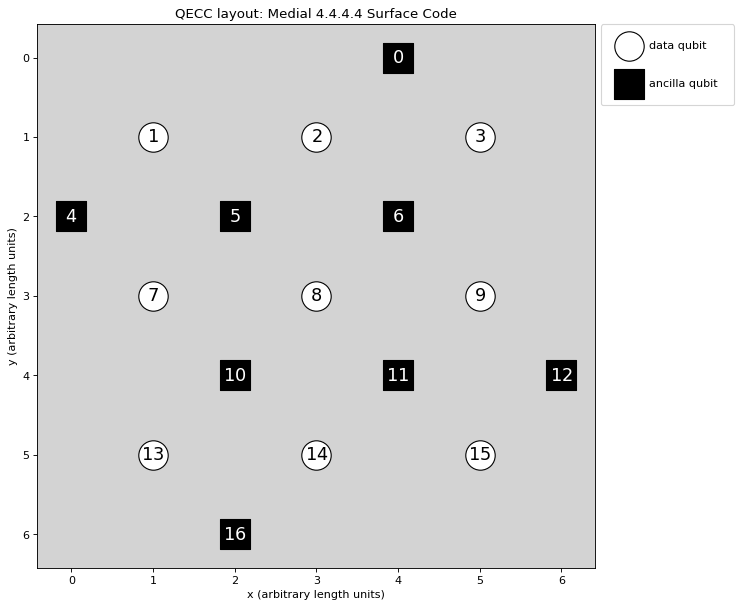}
	\caption{Qubit layout of the distance-three, medial surface-code.}
	\label{fig:pecos.qecc_medial}
\end{figure}

\noindent This results in the plot seen in Fig. \ref{fig:pecos.qecc_medial}. 

The plot of \code{LogicalInstruction}s often indicates the sequence of gate operations. An example of a plot of the syndrome extraction instruction of \code{surface} be obtained by the following lines:

\begin{minipage}{0.95\linewidth}
\begin{lstlisting}[style=pystyle,caption={Example of plotting the medial surface-code's syndrome extraction.}]
>>> surface = pc.qeccs.SurfaceMedial4444(distance=3)
>>> syn_extract = surface.instruction('instr_syn_extract')
>>> syn_extract.plot()
\end{lstlisting}
\end{minipage}

\begin{figure}[H]
	\centering
		\includegraphics[width=0.75\textwidth]{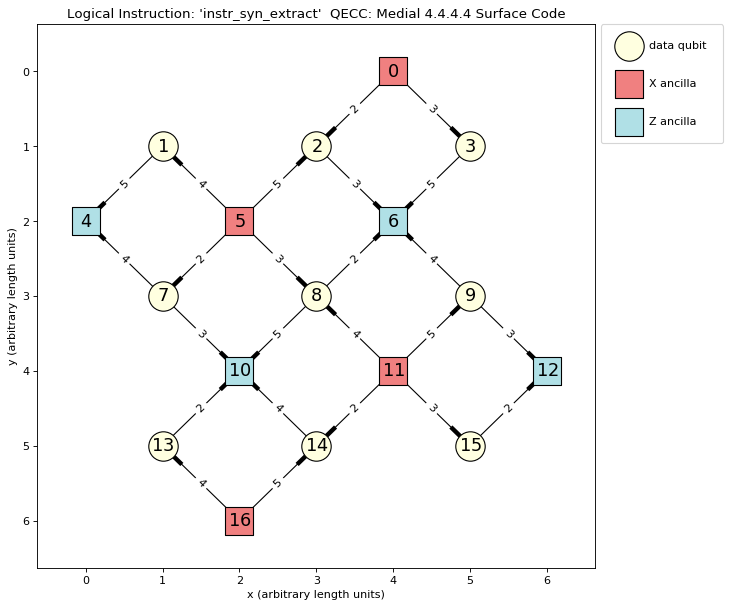}
	\caption{Syndrome-extraction logical-instruction of a distance-three, medial surface-code. Red squares represent the ancillas involved in X checks, the blue squares represent the ancillas involved in Z checks, and the cream circles represent the data qubits. The numbers inside the squares and circles are labels of the qubits as used in QuantumCircuits. The edges indicate the CNOTs used in the checks. The thicker end of each edge denotes the qubit that is the target of the CNOT. The numbers labeling the edges are ticks when the corresponding CNOT is applied.}
	\label{fig:pecos.qecc_medial_syn_extract}
\end{figure}

\noindent The resulting plot is seen in Fig. \ref{fig:pecos.qecc_medial_syn_extract}.

\section{Logical Circuits\label{sec.pecos.logical_circuit}}

The class \code{LogicalCircuit}, which is found in the \code{circuits} namespace, is a logical analog of the class \QuantumCircuit. The \code{LogicalCircuit} class has the same methods and attributes as \QuantumCircuit; however, there are a few changes in the behavior of some of the methods. As the two classes are very similar, I will give a few examples of using the \code{LogicalCircuit} class to illustrate their differences.

\subsection{Example Usage}

An instance of a \code{LogicalCircuit} can be created using the following lines:

\begin{minipage}{0.95\linewidth}
\begin{lstlisting}[style=pystyle,caption={Creating an instance of a \code{LogicalCircuit}.}]
>>> logic = pc.circuits.LogicalCircuit()
\end{lstlisting}
\end{minipage}

Instead of gate symbols, the \code{append} method of the \code{LogicalCircuit} class accepts \code{LogicalGate}s directly. Also, if a \code{LogicalCircuit} contains a single \code{qecc} then a gate location is not needed:

\begin{minipage}{0.95\linewidth}
\begin{lstlisting}[style=pystyle,caption={Example of creating a \code{LogicalCircuit}.}]
>>> surface = pc.qeccs.Surface4444(distance=3)
>>> logic = pc.circuits.LogicalCircuit()
>>> logic.append(surface.gate('ideal init |0>'))
>>> logic.append(surface.gate('I'))
\end{lstlisting}
\end{minipage}


\section{State Simulators\label{sec.pecos.state_sim}}

Quantum states and their dynamics are simulated by classes belonging to the namespace \code{simulators}. \PECOS includes two stabilizer simulators: \code{pySparseSim} and \code{cySparseSim}. \code{pySparseSim} is a pure Python implementation of the algorithm described in Chapter~\ref{ch.stab_sim}, while \code{cySparseSim} is a \pack{C++} implementation that has been wrapped using \pack{Cython}. 

To compile the \pack{C++} implementation navigate, from the \PECOS installation directory, to the namespace \code{simulators}. Then run:

\begin{minipage}{0.95\linewidth}
\begin{lstlisting}[style=pystyle,label={pecos.sim.cycompile},caption={Running the script to compile Cython wrapped simulators.}]
>>> python compile_cython.py
\end{lstlisting}
\end{minipage}  

The script will let you know if compilation was successful. If \code{cySparseSim} is compiled then \code{SparseSim} will be an alias of \code{cySparseSim} is if \code{cySparseSim} is compiled, otherwise \code{SparseSim} is an alias of \code{pySparseSim}.

\subsection{Expected Methods}

The set of gates allowed by a simulator may differ (the standard set for \PECOS is given in Appendix~\ref{append.pecos.symbols}); however, each simulator is expected to have a set of standard methods. I will describe them in this section.

When initializing a simulator, the first argument is expected to be the number of qudits to be simulated. This reserves the size of the quantum registry:

\begin{minipage}{0.95\linewidth}
\begin{lstlisting}[style=pystyle,label={pecos.sim.code.init},caption={Importing and state initialization with pySparseSim.}]
>>> from pecos.simulators import pySparseSim
>>> state = pySparseSim(4)
\end{lstlisting}
\end{minipage}    

Note, for all simulators, the initial state of each qudit is the state $\ket{0}$.

\pagebreak[4]
The only other method expected is the \code{run\_gate} method. This method can be used to apply gates to a \code{simulator} instance by using the \code{run\_gate} method:

\begin{minipage}{0.95\linewidth}
\begin{lstlisting}[style=pystyle,caption={Applying Pauli $X$ with pySparseSim.}]
>>> # Continuing from the previous Code Block.
>>> state.run_gate('X', {0, 1})
\end{lstlisting}
\end{minipage}

\noindent Here the first argument is a gate symbol that is recognized by the \code{simulator} and the second argument is a set of gate locations. Other keywords and arguments may be supplied if it is allowed by the \code{simulator}. Such arguments could be used to change the behavior of the gate. For example, arguments could be used to define gate rotation-angles.

If measurements are made then a dictionary indicating the measurement results is returned by \code{run\_gate}:

\begin{minipage}{0.95\linewidth}
\begin{lstlisting}[style=pystyle,label={pecos.sim.code.rungates},caption={Measuring Pauli-$Z$ with pySparseSim.}]
>>> # Continuing from the previous Code Block.
>>> state.run_gate('measure Z', {0, 1, 3})
{0: 1, 1: 1}
\end{lstlisting}
\end{minipage}

\noindent Here we see that the keys of the results dictionary are the qudit locations of the measurements, and the values are the corresponding measurement results except that zero results are not returned.

Classes in the \code{circuit\_runners} namespace combine {\QuantumCircuit}s and simulators to apply gates to simulated quantum states. For a discussion about these classes, see Section~\ref{sec.pecos.circ_run}.

\subsection{pySparseSim/cySparseSim}

Both \code{pySparseSim} and \code{cySparseSim} have additional useful methods. I will now describe them.    

The \code{print\_stabs} method prints a stabilizer table corresponding to the state currently store in the simulator:

\begin{minipage}{0.95\linewidth}
\begin{lstlisting}[style=pystyle,caption={Using the \code{print\_stabs} method to print a list of stabilizer and destabilizer generators.}]
>>> state = pySparseSim(3)
>>> state.run_gate('CNOT', {(0, 1)})
>>> state.run_gate('X', {0})
>>> state.print_stabs()
 -ZII
 -ZZI
  IIZ
-------------------------------
  XXI
  IXI
  IIX
([' -ZII', ' -ZZI', '  IIZ'], ['  XXI', '  IXI', '  IIX'])
\end{lstlisting}
\end{minipage}

\noindent Here in the print output, stabilizer generators are indicated by the strings above the dashed lines, while destabilizer generators are indicated by the strings below.  

The \code{logical\_sign} method can be used to determine the sign of stabilizer generators. As the stabilizer simulators represent stabilizer states, logical basis-states are stabilized by logical operators. Therefore, this method is useful in Monte Carlo simulations to determine if logical errors have flipped the sign of logical operators.

\pagebreak[4]
An example of using the \code{logical\_sign} method is seen in the following:

\begin{minipage}{0.95\linewidth}
\begin{lstlisting}[style=pystyle,caption={Using \code{logical\_sign} to determine the sign of a logical operator.}]
>>> # Continuing with the following example:
>>> from pecos.circuits import QuantumCircuit
>>> stab = QuantumCircuit([{'Z': {0, 1}}])
>>> state.logical_sign(stab)
1
>>> stab = QuantumCircuit([{'Z': {2}}])
>>> state.logical_sign(stab)
0
\end{lstlisting}
\end{minipage}   

A $1$ is returned if the phase of the stabilizer is $-1$, and a $0$ is returned if the phase is $+1$. If the stabilizer supplied to \code{logical\_sign} is not a stabilizer of the state, then an exception will be raised.


\section{Circuit Runners\label{sec.pecos.circ_run}}

Classes belonging to the \code{circuit\_runners} namespace apply the gates of \code{LogicalCircuit}s and \code{QuantumCircuit}s to states represented by simulators. \code{circuit\_runner}s are also responsible for applying error models to quantum circuit; however, we will discus this in Section~\ref{sec.pecos.error_gens}.

The main \code{circuit\_runner} is simply called \code{Standard}. There is another call \code{TimingRunner}, which is essentially the same as \code{Standard} except that it is used to time how long it takes simulators to apply gates and can be used to compare the runtime of simulators. I will now discuss these two \code{circuit\_runner}s.
 
\pagebreak[4]
\subsection{Standard}

For convenience, tables \ref{tb.pecos.stand.methods} and \ref{tb.pecos.stand.attr} list the attributes and methods of \code{Standard}:

\begin{table}[H]
\centering
\caption{The methods of Standard\label{tb.pecos.stand.methods}}
\begin{tabular}{@{}ll@{}}
\hline
Name & Description \\
\hline
\code{init} & Adds a collection of gates to the end of \code{ticks}. \\
\code{run\_circuit} & Applies a \QuantumCircuit. \\
\code{run\_logic} & Applies a \code{LogicalCircuit}. \\
\hline
\end{tabular}
\end{table} 

\begin{table}[H]
\centering
\caption{The attributes of Standard\label{tb.pecos.stand.attr}}
\begin{tabular}{@{}ll@{}}
\hline
Name & Description \\
\hline
\code{seed} & The integer used as a seed for random number generators.\\
\hline
\end{tabular}
\end{table}

To create an instance of \code{Standard} one can simply write:

\begin{minipage}{0.95\linewidth}
\begin{lstlisting}[style=pystyle,caption={Creating an instance of the \code{Standard} \code{simulator}.}]
>>> circ_runner = pc.circuit_runners.Standard()
\end{lstlisting}
\end{minipage}

By default, a \code{Standard} uses the \code{SparseSim} as a simulator. This can be changed as follows:

\begin{minipage}{0.95\linewidth}
\begin{lstlisting}[style=pystyle,caption={Changing the default simulator.}]
>>> from pecos.simulators import cySparseSim
>>> circ_runner = pc.circuit_runners.Standard(simulator=cySparseSim)
\end{lstlisting}
\end{minipage}    

\pagebreak[4]
The \code{init} method is used to (re)initialize a \code{SparseSim}/\code{cySparseSim} instance. An example of using this method to create a four-qubit registry is seen here:

\begin{minipage}{0.95\linewidth}
\begin{lstlisting}[style=pystyle,caption={Initializing a qubit registry with a \code{circuit\_runner}.}]
>>> # Following from previous Code Block.
>>> circ_runner = pc.circuit_runners.Standard()
>>> state = circ_runner.init(4)
\end{lstlisting}
\end{minipage} 

The \code{run\_circuit} method is used to apply a \QuantumCircuit to a state in the following:

\begin{minipage}{0.95\linewidth}
\begin{lstlisting}[style=pystyle,caption={Running a circuit with Standard.},label={pecos.code.run.qc}]
>>> # Continuing with the previous Code Block.
>>> qc = pc.circuits.QuantumCircuit()
>>> qc.append('X', {0, 1})
>>> qc.append('measure Z', {0, 1, 3})
>>> circ_runner.run_circuit(state, qc)
{1: {0: 1, 1: 1}}
\end{lstlisting}
\end{minipage}

\noindent In the last line of this example, we see the measurement record produced by the \code{circuit\_runner}. The keys of the outer dictionary are tick indices, while for the inner dictionary the keys are the indices of qubits with non-zero measurements and the values are the measurement results.

\pagebreak[4]
The \code{run\_logic} method is used to apply \code{LogicalCircuit}s:

\begin{minipage}{0.95\linewidth}
\begin{lstlisting}[style=pystyle,caption={Using \code{run\_logic} to apply gates from a \code{LogicalCircuit}.}]
>>> surface = pc.qeccs.Surface4444(distance=3)
>>> logic = pc.circuits.LogicalCircuit()
>>> logic.append(surface.gate('ideal init |0>'))
>>> logic.append(surface.gate('I'))
>>> state = circ_runner.init(surface.num_qudits)
>>> circ_runner.run_logic(state, logic)
({}, {})
\end{lstlisting}
\end{minipage}

\noindent The final line is the output of \code{run\_logic}. The first dictionary is a record of measurement outcomes and the second is a record of the errors generated. In this example, all the measurement results are zero and we have not applied any error models. In Section~\ref{sec.pecos.error_gens}, there are examples of where this is not the case; therefore, refer to that section if you are curious about the output of \code{run\_logic}.
   
\subsection{TimingRunner}

As mention, \code{TimingRunner} is essentially the same as \code{Standard} except the runtime for applying gates is recorded. The attribute \code{total\_time} stores this value and is used in the following:

\begin{minipage}{0.95\linewidth}
\begin{lstlisting}[style=pystyle,caption={Example using TimingRunner.}]
>>> circ_runner = pc.circuit_runners.TimingRunner()
>>> state = circ_runner.init(4)
>>> qc = pc.circuits.QuantumCircuit()
>>> qc.append('X', {0, 1, 2, 3})
>>> circ_runner.run_circuit(state, qc)
>>> circ_runner.total_time
7.22257152574457e-06
\end{lstlisting}
\end{minipage}

\code{TimingRunner} times the execution of gates by using \pack{Python}'s \code{perf\_counter} method. The time recorded by \code{total\_time} continues to accumulate until it is reset by the \code{reset\_time} method:

\begin{minipage}{0.95\linewidth}
\begin{lstlisting}[style=pystyle,caption={Resetting the recorded time of a TimingRunner instance.}]
>>> # Continuing from previous Code Block.
>>> circ_runner.reset_time()
>>> circ_runner.total_time
0.0
\end{lstlisting}
\end{minipage}


\section{Error Generators\label{sec.pecos.error_gens}}

Error models are represented by classes called \textit{error generators} that are in the \code{error\_gens} namespace. They are called upon by \code{circuit\_runner}s to apply noise to ideal quantum circuits.

In this section I will discuss \code{DepolarGen} and \code{GatewiseGen} classes. Both represent \textit{stochastic error models}. That is, error models that apply gates as noise according to classical probability distributions. 

\subsection{GatewiseGen}

The \code{GatewiseGen} class allow one to define custom stochastic error-models where for each ideal gate-type the errors applied to the ideal gate and the classical probability distribution for applying errors can be specified. Since many examples of using the class are given, I have moved the discussion of the \code{GatewiseGen} class to Appendix~\ref{app.pecos.errors}.

For a \code{error\_gen} that is commonly used in the study of QEC, see the following section (Sec.~\ref{pecos.errgen.depolar}). This section also provides examples of how \code{error\_gen}s are used in practice.

\subsection{DepolarGen\label{pecos.errgen.depolar}}

The \code{DepolarGen} class is used to represent the symmetric depolarizing channel, which is commonly studied in QEC. For single-qubit gates, this class is used to apply errors at probability $p$ from set $\{X, Y, Z\}$. For two-qubit gates, errors also occur with probability $p$ but errors are chosen uniformally from the set $\{I, X, Y, Z\}^{\otimes 2} \; \setminus \; I\otimes I$. Errors are always applied after ideal gates except for measurements. In which case, the errors are applied before.

An example of creating an instance of \code{DepolarGen} is seen here:

\begin{minipage}{0.95\linewidth}
\begin{lstlisting}[style=pystyle,caption={Creating an instance of the \code{DepolarGen} class.}]
>>> depolar = pc.error_gens.DepolarGen(model_level='code_capacity', has_idle_errors=False, perp_errors=True)
\end{lstlisting}
\end{minipage} 

The \code{model\_level} keyword is used to specify to what set of gates the \code{DepolarGen} is applied to. If \code{model\_level} is set to the value of \code{'code\_capacity'}, then the error model is applied before each \code{LogicalInstruction} to each data qubits as if these qubits are acted on by \code{'I'}. The error model is not applied to any other circuit element. If \code{model\_level} is set to the value \code{'phenomenological'}, then the error model applied to data qubits before each \code{LogicalInstruction} as well as to any measurement. If \code{model\_level} is set to the value \code{'circuit'}, then the error model is applied to all the gates in the \QuantumCircuit. The default value of \code{model\_level} is \code{'circuit'}. Since simulators (see Section \ref{sec.pecos.state_sim}) may simulate QEC codes and procedures by applying circuits, when using \code{model\_level=='code\_capacity'} or \code{model\_level=='phenomenological'} one needs to take special care that the errors applied spread through the code as expected. See, \textit{e.g.}, \cite{Landahl:2011a} for a more detailed definition of code capacity, phenomenological, and circuit-level noise.

The \code{has\_idle\_errors} is a keyword that is only relevant when \code{model\_level == 'circuit'}. If \code{has\_idle\_errors} is set to \code{True}, then the error model is applied to inactive qubits as if the qubit is acted on by \code{'I'}. If \code{has\_idle\_errors} is set to \code{False}, then this does not occur. The default value of \code{has\_idle\_errors} is \code{True}. 

If the \code{perp\_errors} keyword is set to \code{True}, then errors that are applied to Pauli-basis initializations and measurements are errors that do not include the Pauli-basis of the initializations or measurements. That is, \code{perp\_errors==True} ensures that errors are always chosen so that they are \textit{perpendicular} to the initialization and measurement bases. So, for example, $Z$ is not applied as an error to the \code{'init |0>'} operation. If the \code{perp\_errors} keyword is set to \code{False}, then there is no restriction to the errors. The default value of \code{perp\_errors} is \code{True}.

An example of applying an error model using \code{DepolarGen} to a \code{LogicalCircuit} is seen in the following:

\begin{minipage}{0.95\linewidth}
\begin{lstlisting}[style=pystyle,caption={Applying the error generator \code{DepolarGen} to a logical circuit.},label={list:pecos:depolarapp}]
>>> depolar = pc.error_gens.DepolarGen(model_level='code_capacity')
>>> surface = pc.qeccs.Surface4444(distance=3)
>>> logic = pc.circuits.LogicalCircuit()
>>> logic.append(surface.gate('ideal init |0>'))
>>> logic.append(surface.gate('I'))
>>> circ_runner = pc.circuit_runners.Standard(seed=1)
>>> state = circ_runner.init(surface.num_qudits)
>>> meas, err = circ_runner.run_logic(state, logic, error_gen=depolar, error_params={'p': 0.1})
\end{lstlisting}
\end{minipage}

\noindent Note that the keyword argument \code{error\_params} is used to pass a dictionary that indicates the probability $p$ of the depolarizing error model.

The values returned by the \code{run\_logic} method is recorded in the variables \code{meas} and \code{err}. These variables are dictionaries that record the measurement output and applied errors. 

An example of measurement outcomes is given here:

\begin{minipage}{0.95\linewidth}
\begin{lstlisting}[style=pystyle,caption={Measurement outcomes from an erroneous \code{LogicalCircuit}.}]
# Following the previous example.
>>> meas
{(1, 0): {7: {9: 1, 11: 1}}}
\end{lstlisting}
\end{minipage}

\vfill
\pagebreak[4]
\noindent Here, in the last line, we see the measurement outcome. The key of the outer dictionary is a tuple where the first element is the tick index of the \code{LogicalGate} and the second element is an index corresponding to a \code{LogicalInstance}. That is, the tuple records at what point in the \code{LogicalCircuit} was the measurement made. The value of the outer dictionary is just the measurement-outcome dictionary of a \QuantumCircuit as discussed in Code Blocks~\ref{pecos.code.run.qc}. 

We can see the errors that were generated by the \code{DepolarGen} in Code Block~\ref{list:pecos:depolarapp} in these lines:

\begin{minipage}{0.95\linewidth}
\begin{lstlisting}[style=pystyle, caption={Example error dictionary.}, label={pecos.errgen.errdict}]
# Following the previous example.
>>> err
{(1, 0): {0: {'after': QuantumCircuit([{'X': {4}, 'Z': {10}}])}}}
\end{lstlisting}
\end{minipage}

\noindent In the above Code Block, we see a dictionary that stores what errors were applied to the \code{LogicalCircuit}. The key of the outer dictionary, once again, is a tuple indicating the tick of a \code{LogicalGate} and the index of a \code{LogicalInstance}. The key of the next inner dictionary is \QuantumCircuit tick when the error occurred. The key \code{'after'} of the next inner dictionary indicates that the errors are applied after ideal gates. The key \code{'before'} is used when indicating that errors are applied before gates. The values of both the \code{'after'} and \code{'before'} keys are {\QuantumCircuit}s. These circuits are the errors that are applied.      

The data structure used to describe the errors that are applied to a \code{LogicalCircuit}  as shown in Code Block~\ref{pecos.errgen.errdict} can be directly supplied to a \code{run\_logic} method of a \code{circuit\_runner}. Doing so will cause the \code{run\_logic} method to apply the given error to a \code{LogicalCircuit}. This can be seen in the following:

\begin{minipage}{0.95\linewidth}
\begin{lstlisting}[style=pystyle,caption={Running a \code{LogicalCircuit} with predefined errors.}]
# Continuing the previous examples.
>>> logic2 = pc.circuits.LogicalCircuit()
>>> logic2.append(surface.gate('ideal init |+>'))
>>> logic2.append(surface.gate('I'))
>>> state2 = circ_runner.init(surface.num_qudits)
>>> meas2, err2 = circ_runner.run_logic(state2, logic2, error_circuits=err)
\end{lstlisting}
\end{minipage}

One use for this is to apply the same error to a different logical basis-state. Doing so allows one to determine if a logical error occurs for the logical operations that stabilizer the basis state. 

Note that the \code{circuit\_runners} currently only apply errors to \code{LogicalCircuit}s and not to {\QuantumCircuit}s.


\section{Decoders\label{sec.pecos.decode}}

A decoder in \PECOS is simply a function or other callable that takes the measurement outcomes from error extractions (syndromes) as input and returns a \QuantumCircuit, which is used as a recovery operation to mitigate errors. Decoder classes and functions are in the \code{decoders} namespace.

The \code{MWPM2D} class is an available decoder class, which I will discuss next.

\vspace{2cm}
\pagebreak[4]
\subsection{MWPM2D}

One of the standard decoders used for surface codes is the minimum-weight-perfect-matching (MWPM) decoder \cite{Dennis:2002a}. The \code{MWPM2D} class implements the 2D version of this decoder for \code{Surface4444} and \code{SurfaceMedial4444}, that is, it decodes syndromes for a single round of error extraction:

\begin{minipage}{0.95\linewidth}
\begin{lstlisting}[style=pystyle,caption={An example of generating errors and measurements to decode.}]
>>> depolar = pc.error_gens.DepolarGen(model_level='code_capacity')
>>> surface = pc.qeccs.Surface4444(distance=3)
>>> logic = pc.circuits.LogicalCircuit()
>>> logic.append(surface.gate('ideal init |0>'))
>>> logic.append(surface.gate('I', num_syn_extract=1))
>>> circ_runner = pc.circuit_runners.Standard(seed=1)
>>> state = circ_runner.init(surface.num_qudits)
>>> decode = pc.decoders.MWPM2D(surface).decode
>>> meas, err = circ_runner.run_logic(state, logic, error_gen=depolar, error_params={'p': 0.1})
\end{lstlisting}
\end{minipage}

The errors generated and the measurement outcomes are given in:

\begin{minipage}{0.95\linewidth}
\begin{lstlisting}[style=pystyle,caption={Errors and measurements generated.}]
>>> err
{(1, 0): {0: {'after': QuantumCircuit([{'Y': {4}, 'X': {10}}])}}}
>>> meas
{(1, 0): {7: {3: 1, 5: 1, 9: 1, 15: 1}}}
\end{lstlisting}
\end{minipage}

\vspace{2cm}
\pagebreak[4]
Given only the measurement outcomes, the \code{decoder} returns a recovery operation (\QuantumCircuit):

\begin{minipage}{0.95\linewidth}
\begin{lstlisting}[style=pystyle,caption={Decoding syndromes/measurements with the MWPM2D decoder.}]
>>> decode(meas)
QuantumCircuit([{'X': {10}, 'Y': {4}}])
\end{lstlisting}
\end{minipage}

\section{Acknowledgments}

I would like to thank Andrew Landahl, Jaimie Stephens, Jay Van Der Wall, Setso Metodi, Anand Ganti, Uzoma Onunkwo, and Jonathan Moussa for useful discussions. I would also like to thank Andrew Landahl for employing his exceptional skills in creating acronyms to come up with the name: ``Performance Estimator of Codes On Surfaces'' (\PECOS).

\chapter{Conclusion\label{chp:conclusion}}

\setlength\epigraphwidth{12cm}
\epigraph{\textit{``Where a calculator like ENIAC today is equipped with 18,000 vacuum tubes and weighs 30 tons, computers in the future may have only 1000 vacuum tubes and perhaps weigh only $1 \frac{1}{2}$ tons.''}}{--- \textup{Andrew Hamilton (Popular Mechanics, 1949) \cite{popularvac,}}}

\section{Summary}

While theoretical and technological progress in quantum computation has been accelerating, there is still much work to be done to realize the potential of both NISQ-era and large-scale, universal quantum computation. The work in this dissertation was an attempt contribute to this realization.

I began this dissertation by introducing the language of quantum computation in Chapter~\ref{ch.qi_intro}. Next, in Chapter~\ref{chp:qaoa}, in joint work with Ojas Parekh, I analyzed the performance of QAOA, a quantum-algorithmic framework with a scalable circuit-depth used to find approximate solutions to classical CSPs. I derived a closed-form expression for the expectation value of $\text{QAOA}_1$ for Max Cut on any simple graph. From this result, I argued that $\text{QAOA}_1$ outperforms the know best classical approximation algorithm for Max Cut on $k$-regular triangle-free graphs. Then, in Chapter~\ref{ch.stab_sim}, I introduced a new stabilizer simulation algorithm. I gave an analysis of the runtime complexity and argued that the typical runtime for the new algorithm should outperform previous stabilizer simulator algorithms when simulating topological stabilizer codes. I then presented numerical data that supported this claim. Afterwards, in Chapter~\ref{ch.lattice_surgery}, in joint work with Andrew Landahl, I introduced a set of new QEC operations called color-code lattice surgery. I further improved surface-code lattice surgery in terms of the number of qubits needed for logical operations. I argued that for sufficiently low physical error-rates of depolarizing noise, color-code lattice surgery offers the same degree of error suppression while requiring fewer qubits compared to surface-code lattice surgery. Finally, in Chapter~\ref{ch.pecos}, I described a \pack{Python} package called \PECOS, which is a framework for the systematic study and evaluation of QEC protocols. 

\section{Outlook}
The work in this dissertation naturally leads to other problems to consider: 

An obvious first step to extend our work in $\text{QAOA}_1$ is to generalize our proofs to other classical CSPs and characterize QAOA's performance for those problems. Also, since QAOA can trade solution quality for circuit depth, it is reasonable to consider applying the algorithm to NISQ devices. It would, therefore, be of interest to study how realistic, NISQ-era noise would affect the performance of QAOA as circuit depth is varied. 

In comparing the new stabilizer-simulation algorithm to previous ones, it was made clear that the data structure used to simulate stabilizer states and circuits significantly affect the simulation runtime. Therefore, future work could entail designing an algorithm that chooses the appropriate data structure to minimize the runtime for simulating the action of a given stabilizer circuit on a given stabilizer state. 

Since the development of color-code lattice surgery, other related ideas have arisen such as the doubled color-codes \cite{Bravyi:2015a,Jochym-OConnor:2016a,Jones:2015a} and the triangular surface-codes \cite{TriangleSurf}. Both of which make use of lattice surgery. Different topological stabilizer codes have  various strengths and weaknesses; thus, by applying ideas such as code-switching \cite{NFB16}, one could study whether a planar, logical architecture could be designed to take advantage of multiple QECCs to minimize qubit resources, operation runtimes, and logical error-rates.

Currently, \PECOS only contains a few QECCs and a single decoder. \PECOS could be expanded to include an extensive library of QECCs, error models, and decoders. Such a library would encourage the standardization of QEC evaluation techniques and facilitate direct performance comparisons between QEC protocols. It would also permit \PECOS to serve as a tool to teach different aspects of QEC and gain new insight.

I hope that the work described in this dissertation will assist the progress towards large-scale quantum computation.

\chapter*{Appendices}

\addcontentsline{toc}{chapter}{Appendices}

\appendix

\chapter{Basics of Group Theory}

Group theory is the study of the symmetries that arise when elements are combined under certain simple rules. As the understanding of symmetries can often lead to insights into problems and simplification of notation, group theory has been useful in the study of many fields outside of pure mathematics. This includes quantum information. Much of the discussion of stabilizer codes can be efficiently communicated through the language of basic group theory. For this reason, in this appendix, we will review some introductory concepts of group theory. An elementary understanding of set theory is assumed.
\vfil
\pagebreak[4]
\section{Definitions}
A group is an algebraic structure indicated by a tuple $(G, *)$ that consists of a set $G$ and a binary operator $*$ that satisfy the group axioms. $*$ is a rule that uniquely identifies the two elements to the operator's left and right to a third element. 

\goodbreak
The four group axioms are:
\begin{axiom}[Closure]
For all $a, b \in G,\; a*b = c \in G$.
\end{axiom}

This is simply the statement that the application of $*$ on $G$ will not result in elements outside of $G$. 

\begin{axiom}[Associativity]
For all $a, b, c \in G,\; a*(b*c) = (a*b)*c$.
\end{axiom}

\begin{axiom}[Identity]
There exist $e \in G$ such that for all $a \in G,\; e*a = a*e = a$. 
\end{axiom}

The element $e$ is know as the \textit{identity element} of $G$ and is unique.

\begin{axiom}[Inverse]
For every $a \in G$ there exist a corresponding element $a^{-1} \in G$ such that $a^{-1}*a = a*a^{-1} = e$.
\end{axiom}

The element $a^{-1}$ is known as the \textit{inverse element} of $a$.

There are a few shorthands that are often used. A group is commonly referred to simply by the set $G$ when the group operation is implied or stated. In those cases, whether $G$ is used to mean the group or the set is understood through context. It is also common to use $ab$ to mean $a*b$. The group operator may be referred to as addition, subtraction, multiplication, or division depending on how an author wishes to view the group. In the following, when we refer to ``multiply'' we mean to use the group operator.  

\pagebreak[4]
$(S, *)$ is said to be a \textit{subgroup} of $(G, *)$ if and only if $(S, *)$ is a group in its own right, the two groups share the same group operation, and the set $S$ is a subset of $G$. A \textit{proper subgroup} is a subgroup that is not equal to the group. That is, for a proper subgroup $S$ of $G$, the set $S$ is contained in set $G$ but not equal to $G$. A proper subgroup $S$ of $G$ is indicated as $S \subsetneq G$. If the subgroup $S$ can equal the group $G$, then this is notated as $S \subseteq G$. 

An \textit{Abelian group} is a group where all elements commute with each other. That is, $\forall a,b \in G, a * b = b * a$.

A \textit{finite group} is a group $G$ that has a finite number of elements in the set $G$.

The \textit{order} or \textit{cardinality} of a group or set $G$, is denoted by $\abs{G}$, is the number of elements in $G$. The order of an element in a group is the smallest number of times an element must be multiply by itself to give the identity element. That is, the order $m$ of an element $a$ in the smallest integer such that $a^m=e$.

A \textit{generating set} of a group is a subset of the group for which elements of the subset and their inverses can be multiplied together to form any element in the group. Elements of a generating set are known as \textit{generators}. That a generating set $S$ generates the group $G$ is denoted by $\expval{S} = G$.

The \textit{rank} of a group $G$, denoted by $\rank(G)$, is the minimum number of elements needed to generate $G$. 
\chapter{PECOS Standard: \\ Qubit-Gate Symbols\label{append.pecos.symbols}}

In this appendix, the standard symbols used in \PECOS to represent gates are listed in tables. In particular, Pauli-basis initializations are found in Table~\ref{tb.pecos.symbols1}, single-qubit Clifford gates are found in \cref{tb.pecos.symbols2,tb.pecos.symbols3,tb.pecos.symbols4,tb.pecos.symbols5}, two-qubit Clifford gates are given in Table~\ref{tb.pecos.symbols6}, and Pauli-basis measurements are listed in Table~\ref{tb.pecos.symbols7}. These symbols are compatible with the \code{SparseSim} stabilizer simulator. Other simulators may provide additional gate-symbols. 

\section{Initializations}

{
\renewcommand{\arraystretch}{1.3}
\begin{table}[H]
\centering
\begin{tabular}{@{}ll@{}}
\hline
Symbol & Description \\
\hline
\texttt{'init |+>'} & (Re)initialize the state $\ket{+}$. \\
\texttt{'init |->'} & (Re)initialize the state $\ket{-}$. \\
\texttt{'init |+i>'} & (Re)initialize the state $\ket{+i}$. \\
\texttt{'init |-i>'} & (Re)initialize the state $\ket{-i}$. \\
\texttt{'init |0>'} & (Re)initialize the state $\ket{0}$.\\
\texttt{'init |1>'} & (Re)initialize the state $\ket{1}$. \\
\hline
\end{tabular}
\caption{Pauli-basis Initializations. 
\label{tb.pecos.symbols1}}
\end{table}
}

\section{One-qubit Cliffords}

\begin{figure}[H]
    \centering
    \includegraphics[width=0.3\linewidth]{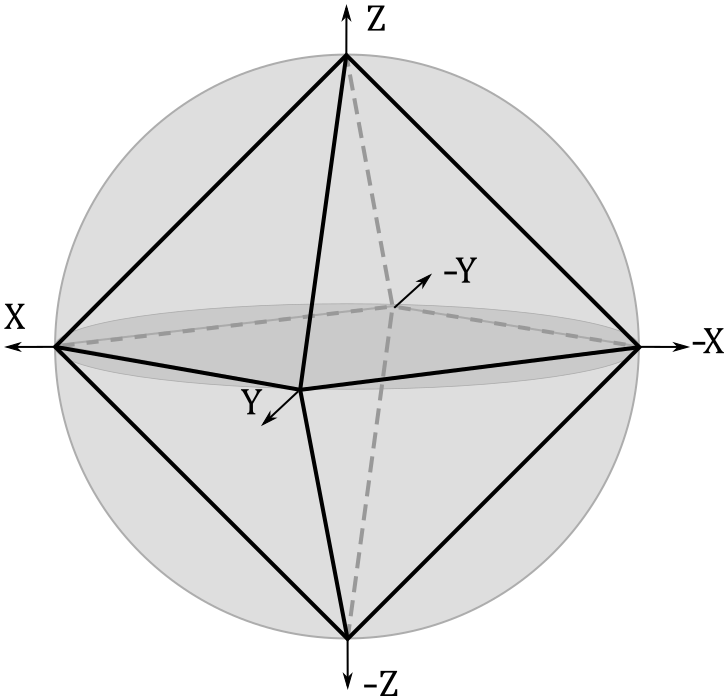} 
        \caption{The octahedron formed from connecting the Pauli stabilizers associated with pure states ($\ket{0}$, $\ket{1}$, $\ket{\pm}$, and $\ket{\pm i}$) on the Bloch sphere with edges. Since the Cliffords take Pauli operators to Pauli operators, the one-qubit Cliffords can be identified with the symmetries of the octahedron.} \label{fig:oct}
\end{figure}

Note that tables \cref{tb.pecos.symbols2,tb.pecos.symbols3,tb.pecos.symbols4,tb.pecos.symbols5} together list the compete set (up to a phase) of all  twenty-four one-qubit Clifford gates. For convenience, a figure similar to Fig.~\ref{fig:oct} is included with each table. These figures indicate how the Clifford gates of the tables rotate the Pauli operators to other Pauli operators. 

{
\renewcommand{\arraystretch}{1.3}
{\centering
\begin{minipage}{0.45\textwidth}
\vspace{4mm}
   \centering
\begin{tabular}{@{}ll@{}}
\hline
Symbol & Transformation \\
\hline
\texttt{'I'} &  $X\rightarrow X$, $Z\rightarrow Z$.\\
\texttt{'X'} & $X\rightarrow X$, $Z\rightarrow -Z$.\\
\texttt{'Y'} & $X\rightarrow -X$, $Z\rightarrow -Z$.\\
\texttt{'Z'} & $X\rightarrow -X$, $Z\rightarrow Z$.\\
\hline
\end{tabular}
\vspace{3mm}
   \captionof{table}{Pauli gates.\label{tb.pecos.symbols2}}
\end{minipage}
\begin{minipage}{0.45\textwidth}
   \centering
    \includegraphics[width=0.90\linewidth]{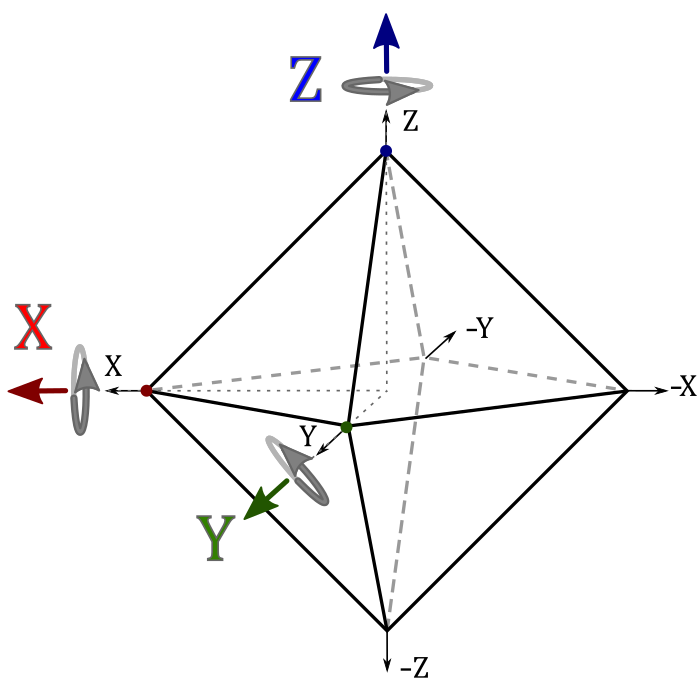} 
   \captionof{figure}{Pauli gates (excluding identity) are rotations by $\pi$ around the Pauli axes.}
\end{minipage}
}\hfill
\vspace{1cm}
{\centering
\begin{minipage}{0.45\textwidth}
\vspace{5mm}
   \centering
\begin{tabular}{@{}ll@{}}
\hline
Symbol & Transformation \\
\hline
\texttt{'Q'} & $X \rightarrow X$, $Z \rightarrow -Y$\\
\texttt{'R'} & $X \rightarrow -Z$, $Z \rightarrow X$\\
\texttt{'S'} & $X \rightarrow Y$, $Z \rightarrow Z$\\
\texttt{'Qd'} & $X \rightarrow X$, $Z \rightarrow Y$\\
\texttt{'Rd'} & $X \rightarrow Z$, $Z \rightarrow -X$\\
\texttt{'Sd'} & $X \rightarrow -Y$, $Z \rightarrow Z$\\
\hline
\end{tabular}
   \captionof{table}{Square-root of Pauli gates. \label{tb.pecos.symbols3}}
\end{minipage}
\begin{minipage}{0.45\textwidth}
   \centering
    \includegraphics[width=0.90\linewidth]{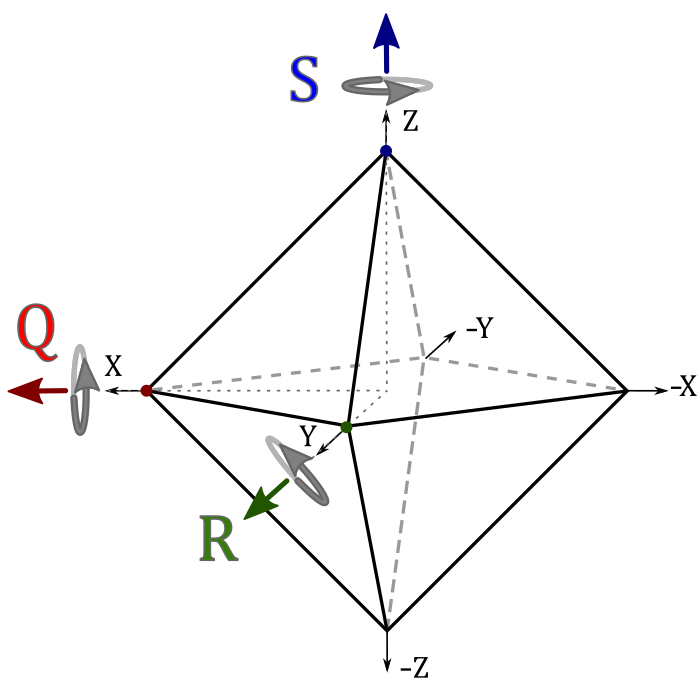}
		\vspace{0mm}
   \captionof{figure}{Square-root of Pauli gates are clockwise or anticlockwise rotations by $\pi/2$ around the indicated axes.}
\end{minipage}
}\hfill
\vspace{1cm}
{\centering
\renewcommand{\arraystretch}{1.3}
\begin{minipage}{0.45\textwidth}
\vspace{4mm}
   \centering
\resizebox{0.95\textwidth}{!}{%
\begin{tabular}{@{}ll@{}}
\hline
Symbol & Transformation \\
\hline
\texttt{'H'}, \texttt{'H+z+x'}, or \texttt{'H1'} & $X\leftrightarrow Z$.\\
\texttt{'H-z-x'} or \texttt{'H2'} & $X\leftrightarrow -Z$.\\
\texttt{'H+y-z'} or \texttt{'H3'} & $X\rightarrow Y$, $Z\rightarrow -Z$.\\
\texttt{'H-y-z'} or \texttt{'H4'} & $X\rightarrow -Y$, $Z\rightarrow -Z$.\\
\texttt{'H-x+y'} or \texttt{'H5'} & $X\rightarrow -X$, $Z\rightarrow Y$.\\
\texttt{'H-x-y'} or \texttt{'H6'} & $X\rightarrow -X$, $Z\rightarrow -Y$.\\
\hline
\end{tabular}%
}
\vspace{4mm}
   \captionof{table}{Hadamard-like gates. \label{tb.pecos.symbols4}}
\end{minipage}
\begin{minipage}{0.45\textwidth}
   \centering
    \includegraphics[width=0.90\linewidth]{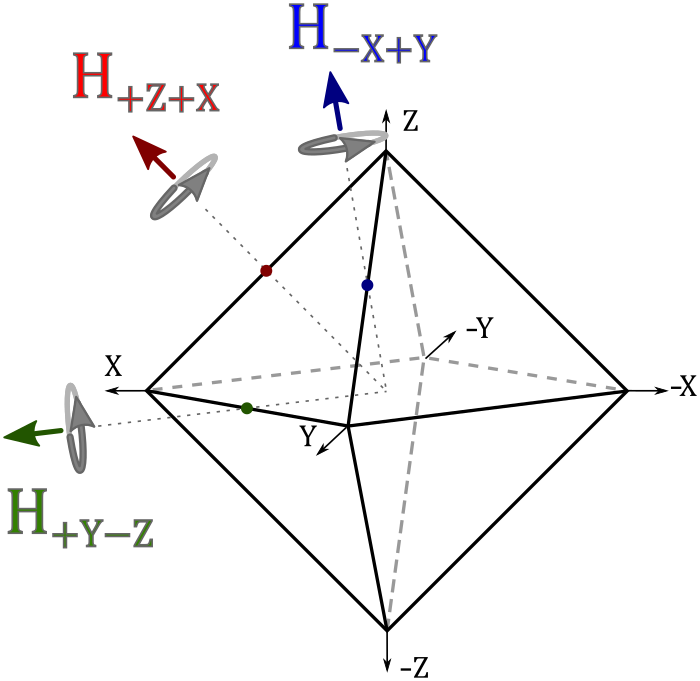}
   \captionof{figure}{Hadamard-like gates are rotations by $\pi$ around the indicated and similar axes.}
\end{minipage}
}\hfill
\vspace{1cm}
{\centering
\renewcommand{\arraystretch}{1.0}
\begin{minipage}{0.45\textwidth}
\vspace{6mm}
   \centering
\begin{tabular}{@{}ll@{}}
\hline
Symbol & Transformation \\
\hline
\texttt{'F1'} & $X\rightarrow Y\rightarrow Z \rightarrow X$\\
\texttt{'F2'} & $X \rightarrow -Z$, $Z \rightarrow Y$\\
\texttt{'F3'} & $X \rightarrow Y$, $Z \rightarrow -X$ \\
\texttt{'F4'} & $X \rightarrow Z$, $Z \rightarrow -Y$\\

\texttt{'F1d'} & $X\rightarrow Z\rightarrow Y \rightarrow X$\\
\texttt{'F2d'} & $X \rightarrow -Y$, $Z \rightarrow -X$\\
\texttt{'F3d'} & $X \rightarrow -Z$, $Z \rightarrow -Y$\\
\texttt{'F4d'} & $X \rightarrow -Y$, $Z \rightarrow X$\\
\hline
\end{tabular}
   \captionof{table}{Rotations about the faces of the octahedron. \label{tb.pecos.symbols5}}
\end{minipage}
\begin{minipage}{0.45\textwidth}
   \centering
    \includegraphics[width=0.90\linewidth]{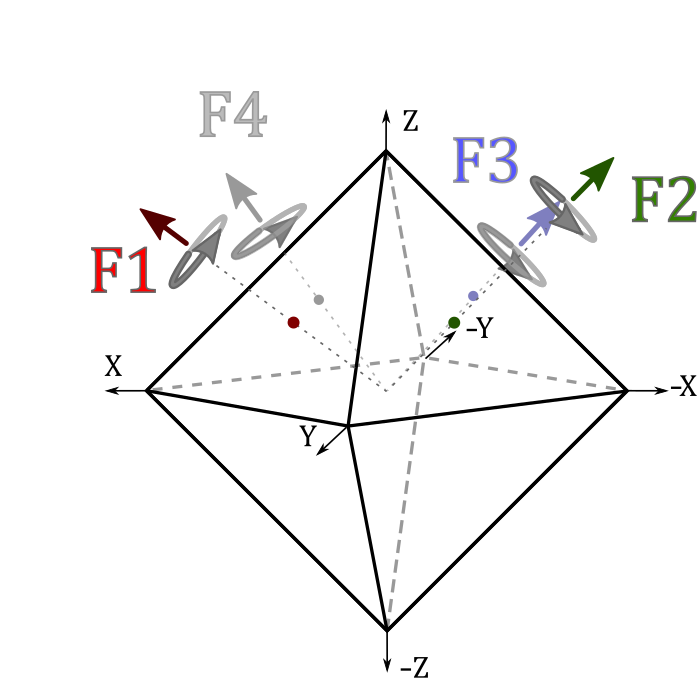}
		\vspace{3mm}
   \captionof{figure}{Face rotations are clockwise or anticlockwise rotations by $2\pi/3$ around the indicated axes.}
\end{minipage}
}
}

\section{Two-qubit Cliffords}

{
\renewcommand{\arraystretch}{1.5}
\begin{table}[H]
\centering
\begin{tabular}{@{}ll@{}}
\hline
Symbol & Description \\
\hline
\texttt{'CNOT'} & The controlled-X gate.\\
\texttt{'CZ'} & The controlled-Z gate.\\
\texttt{'SWAP'} & Swap two qubits.\\
\hline
\end{tabular}
\caption{Two-qubit Clifford gates. \label{tb.pecos.symbols6}}
\end{table}
}

\section{Measurements}

{
\renewcommand{\arraystretch}{1.5}
\begin{table}[H]
\centering
\begin{tabular}{@{}ll@{}}
\hline
Symbol & Description \\
\hline
\texttt{'measure X'} & Measure in the $X$-basis.\\
\texttt{'measure Y'} & Measure in the $Y$-basis.\\
\texttt{'measure Z'} & Measure in the $Z$-basis.\\
\hline
\end{tabular}
\caption{Pauli-basis Measurements. 
\label{tb.pecos.symbols7}}
\end{table}
} 
\chapter{PECOS Example: \\Stabilizer Code Verification\label{app.pecos.stabeval}}

Many times, I have had the experience of developing a new QECC only to find that the code has a much smaller distance than I thought, or finding that not all the checks commute. Thus, I developed the class \code{VerifyStabilizers} in the namespace \code{tools} to verify that a stabilizer code is of the correct form (all stabilizer generators commute) and to determine the distance of the code. 

\begin{figure}[H]
    \centering
    \includegraphics[width=0.45\linewidth]{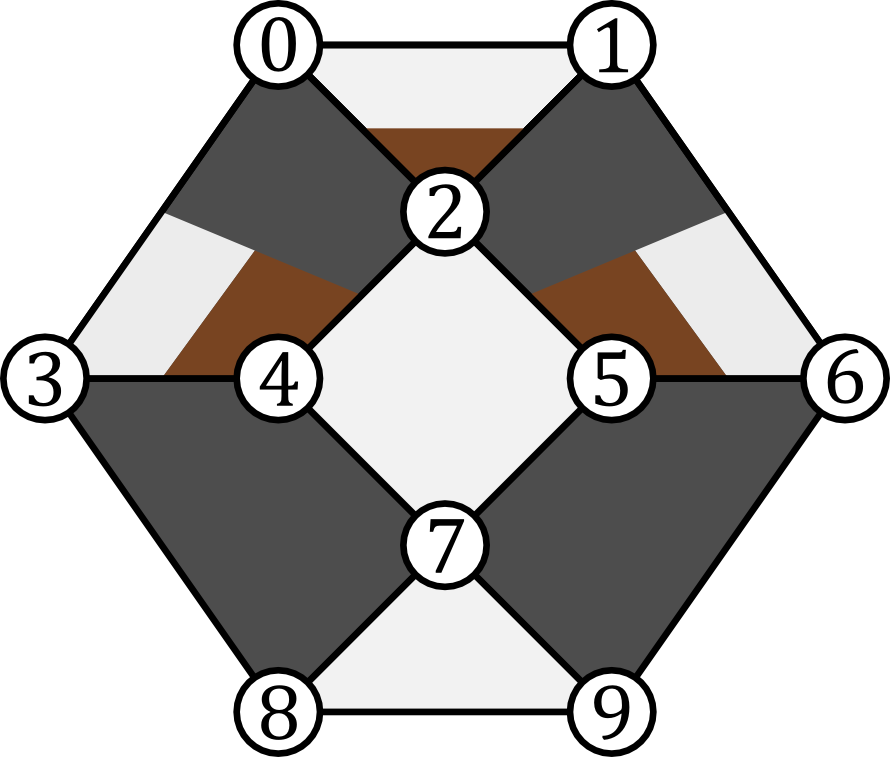} 
        \caption{An example stabilizer code. Data qubits are indicated as white circles and are labelled by numbers. The black outlined polygons represent checks. The interior coloring of the polygons indicate how the associated check acts on qubits it touches. When the check is white on a qubit, the check acts as $Z$. When black, $X$. When brown, $Y$. Note: Some polygons in the figure have multiple colors associated with them, such as the check $X_1X_2Y_5Z_6$.} \label{fig:ex.stabcode1}
\end{figure}

We will see how \code{VerifyStabilizers} can be used to develop a simple, distance-three code. We begin by considering the generators in Fig.~\ref{fig:ex.stabcode1}.

We now use \code{VerifyStabilizers} to represent the checks in Fig~\ref{fig:ex.stabcode1}:

\begin{minipage}{0.95\linewidth}
\begin{lstlisting}[style=pystyle,caption={Specify checks with VerifyStabilizers.}]
>>> qecc = pc.tools.VerifyStabilizers()
>>> qecc.check('X', (3, 4, 7, 8))
>>> qecc.check('X', (5, 6, 7, 9))
>>> qecc.check('Z', (2, 4, 5, 7))
>>> qecc.check('Z', (7, 8, 9))
>>> qecc.check(('Z', 'Z', 'Y'), (0, 1, 2))
>>> qecc.check(('X', 'X', 'Z', 'Y'), (0, 2, 3, 4))
>>> qecc.check(('X', 'X', 'Z', 'Y'), (1, 2, 6, 5))
\end{lstlisting}
\end{minipage}

\noindent Here we see that the \code{check} method can be used to specify a generator. If the first argument is a string, then this indicates the Pauli-type of the check. The second argument then indicates which qubits the check acts on. If the first argument is a tuple, then the tuple is a sequence of strings which indicate how the generator acts on the corresponding qubits indicated in the tuple of the second argument.  

Once one has finished specifying the generators of the code, the \code{compile} method should be used:

\begin{minipage}{0.95\linewidth}
\begin{lstlisting}[style=pystyle,caption={Checking if checks anticommute.}]
>>> # Continuing the last Code Block.
>>> qecc.compile()
Check:
check(('Z', 'Z', 'Y'), (0, 1, 2))
anticommutes with:
check(Z, (2, 4, 5, 7))
\end{lstlisting}
\end{minipage}

\noindent Once \code{compile} is called, \code{VerifyStabilizers} checks to see if all the generators anticommute. If any do \code{VerifyStabilizers} prints out a message indicating which checks anticommute. 

\begin{figure}[H]
    \centering
    \includegraphics[width=0.45\linewidth]{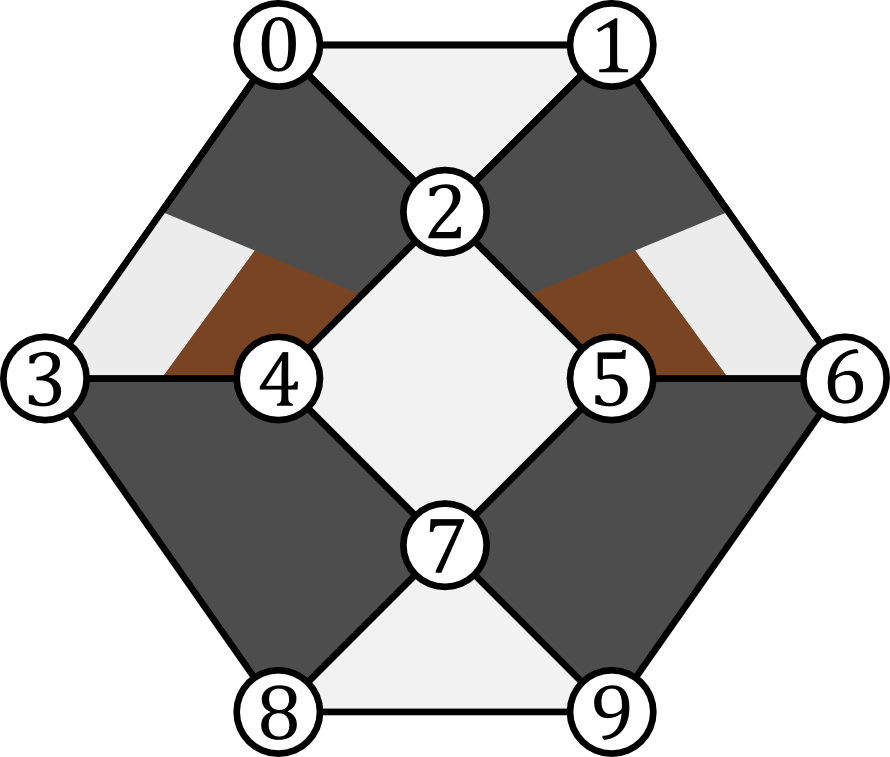} 
        \caption{A slightly modified version of the code seen Fig.~\ref{fig:ex.stabcode1} where all the checks commute.} \label{fig:ex.stabcode2}
\end{figure}

Finding that our original stabilizer code design does have anticommuting generators, we can modify the QECC slightly to address the issue as seen in Fig.~\ref{fig:ex.stabcode2}.

\pagebreak[4]
Re-specifying the generators according to Fig.~\ref{fig:ex.stabcode2}, we run \code{compile} again and see that we have solved the commutation problem:

\begin{minipage}{0.95\linewidth}
\begin{lstlisting}[label={lst.pecos.eval2}, style=pystyle,caption={Compiling a collection of commuting checks.}]
>>> qecc = pc.tools.VerifyStabilizers()
>>> qecc.check('X', (3, 4, 7, 8))
>>> qecc.check('X', (5, 6, 7, 9))
>>> qecc.check('Z', (2, 4, 5, 7))
>>> qecc.check('Z', (7, 8, 9))
>>> qecc.check(('Z', 'Z', 'Z'), (0, 1, 2))
>>> qecc.check(('X', 'X', 'Z', 'Y'), (0, 2, 3, 4))
>>> qecc.check(('X', 'X', 'Z', 'Y'), (1, 2, 6, 5))
>>> qecc.compile()
\end{lstlisting}
\end{minipage}

It is possible that we have specified a set of generators with redundant elements. That is, some of the generators can be written as products of the others. To check if this is the case, we use the method \code{generators}:

\begin{minipage}{0.95\linewidth}
\begin{lstlisting}[label={Eval3}, style=pystyle]
>>> # Following the last Code Block.
>>> qecc.generators()
Number of data qubits: 10
Number of checks: 7
Number of logical qubits: 3

Stabilizer generators:
  XIXZYIIIII
  IXXIIYZIII
	IIZIZZIZII  
  ZZZIIIIIII
  IIIIIIIZZZ
\end{lstlisting}
\end{minipage}

\vfill
\begin{minipage}{0.95\linewidth}
\begin{lstlisting}[nolol, style=pystyle, firstnumber=13,caption={Running the generators method.}]
  IIIXXIIXXI
  IIIIIXXXIX
  
Destabilizer generators:
  ZIIIIIIIII
  IZIIIIIIII
  ZIIIXIIIII
  ZIXIXIIIII
  ZIIIXIIXII
  IIIIIIIIZI
  IIIIIIIIIZ
  
Logical operators:
. Logical Z #1:
  IIIZIIIIZI
. Logical X #1:
  ZIIXIIIIII
. Logical Z #2:
  IZIIIZIIIZ
. Logical X #2:
  ZZIIXXIIII
. Logical Z #3:
  IIIIIIZIIZ
. Logical X #3:
  IZIIIIXIII
\end{lstlisting}
\end{minipage}

\pagebreak[4]
If we had redundant generators then \code{generators} would alert us. Luckily, we do not and \code{generators} has printed out some useful information including number of logical qubits, destabilizers, and a possible set of logical operators. All of which is seen in the Code Block above. 

We can then use the \code{distance} method to determine the distance of the code. Note, to find the distance of a code, this method will try all combinations of possible Pauli errors. It starts with the smallest weight and evaluating larger and larger weights until a logical error is detected. Since this is a combinatorial search, the algorithm is not efficient and the runtime quickly grows with the size of the code. In practice, for smaller code of less than 20 or so qubits, the runtime is manageable.

We now run the \code{distance} method:

\begin{minipage}{0.95\linewidth}
\begin{lstlisting}[label={lst.pecos.eval4}, style=pystyle,caption={Using the \code{distance} method to find the distance of the QECC.}]
>>> # Following the last Code Block.
qecc.distance()
----
Checking errors of length 1...
Checking errors of length 2...
Logical error found: Xs - {0, 1} Zs - set()
This is a [[10, 3, 2]] code.
\end{lstlisting}
\end{minipage}

\begin{figure}[H]
    \centering
    \includegraphics[width=0.45\linewidth]{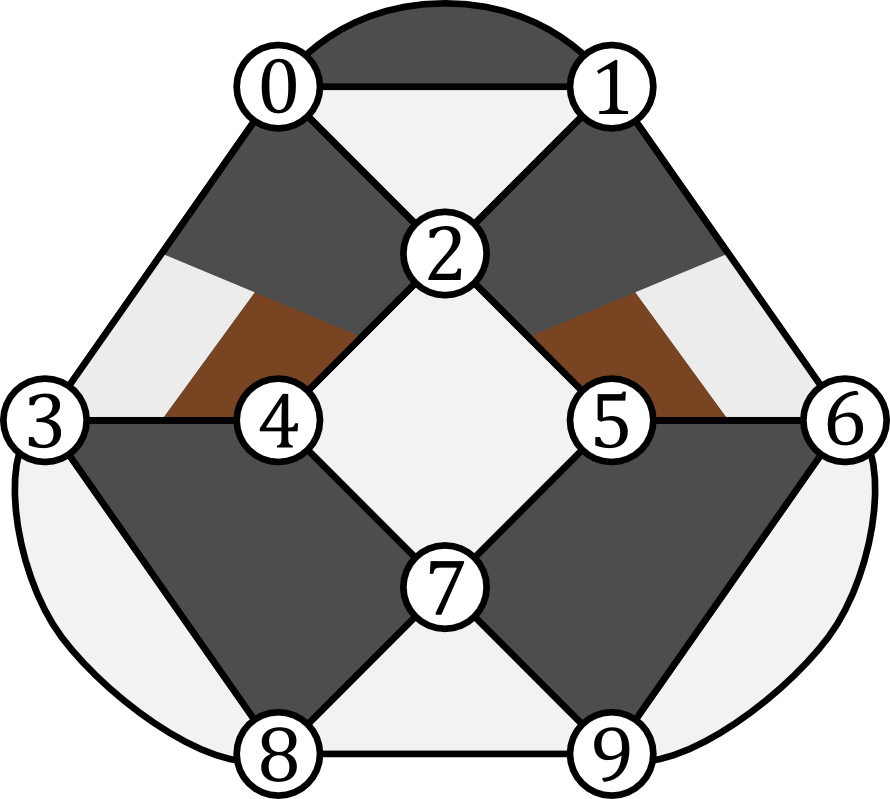} 
        \caption{A slightly modified version of the code seen Fig.~\ref{fig:ex.stabcode2} with additional weight-two checks.} \label{fig:ex.stabcode3}
\end{figure}

\noindent The last line of the code block indicates what type of QECC we have. The notation $[[n, k, d]]$ indicates that the code encodes $k$ qubits into $n$ physical qubits and has a distance of $d$. Since the number of errors a QECC can correct is $t=\left \lfloor{(d-1)/2}\right\rfloor$ and the distance of our code is two, this means the QECC can only detect but not correct errors. Because the \code{distance} method indicates the smallest logical error it found, we can use this information to mitigate the error by either introducing another check to detect the error or by including the logical error as a check. We do the later. Doing this, we find that we have not increased the distance of the code. If we repeat the process two more times we will end up with a code (as seen in Fig.~\ref{fig:ex.stabcode3}) that has no logical qubits and, therefore, encodes a stabilizer state.

\begin{figure}[H]
    \centering
    \includegraphics[width=0.45\linewidth]{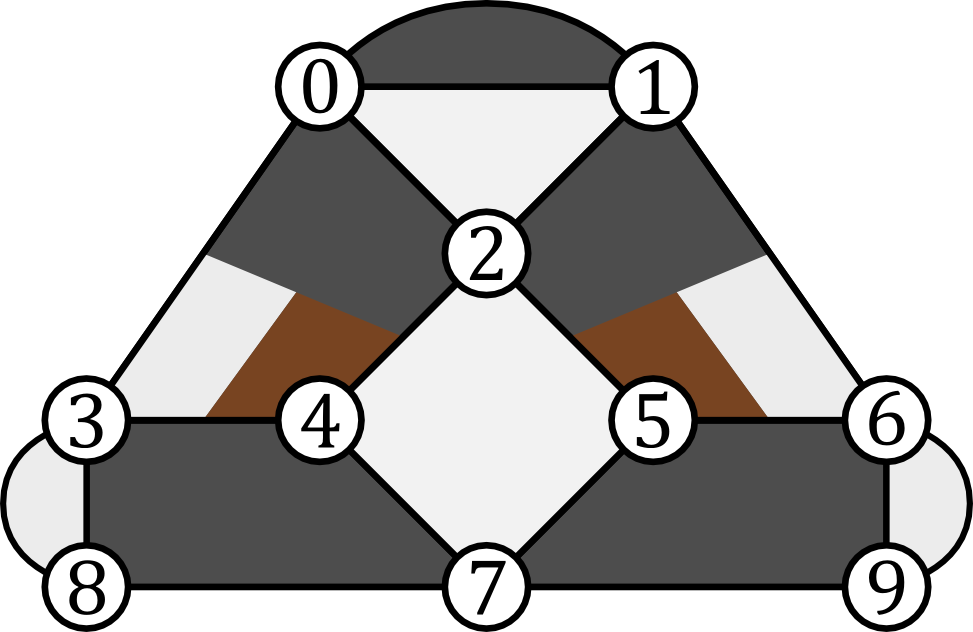} 
        \caption{The QECC in Fig.~\ref{fig:ex.stabcode3} with a weight three $Z$ check on qubits 7, 8, and 9 removed.} \label{fig:ex.stabcode4}
\end{figure}

We seemingly failed to create a higher distance code; however, we can persevere by removing a higher-weight stabilizer generator. If we remove the check that acts like Pauli $Z$ on qubits 7, 8, and 9, we will get the stabilizer code in Fig.~\ref{fig:ex.stabcode4}.   

\pagebreak[4]
Evaluating the distance of this new version of the code:

\begin{minipage}{0.95\linewidth}
\begin{lstlisting}[label={lst.pecos.eval5}, style=pystyle]
>>> qecc = pc.tools.VerifyStabilizers()
>>> qecc.check('Z', (2, 4, 5, 7))
>>> # qecc.check('Z', (7, 8, 9))
>>> qecc.check('X', (3, 4, 7, 8))
>>> qecc.check('X', (5, 6, 7, 9))
>>> qecc.check(('X', 'X', 'Z', 'Y'), (0, 2, 3, 4))
>>> qecc.check(('X', 'X', 'Z', 'Y'), (1, 2, 6, 5))
>>> qecc.check('Z', (0, 1, 2))
>>> qecc.check('X', (0, 1))
>>> qecc.check('Z', (3, 8))
>>> qecc.check('Z', (6, 9))
>>> qecc.compile()
>>> qecc.generators()
Number of data qubits: 10
Number of checks: 9
Number of logical qubits: 1
----

Stabilizer generators:
  XIXZYIIIII
  IXXIIYZIII
  IIZIZZIZII
  IIIZIIIIZI
  XXIIIIIIII
  ZZZIIIIIII
  IIIIIIZIIZ
  IIIXXIIXXI
  IIIIIXXXIX
\end{lstlisting}
\end{minipage}

\vfill
\begin{minipage}{0.95\linewidth}
\begin{lstlisting}[nolol, style=pystyle, firstnumber=39,caption={Evaluating a new QECC.}]
  
Destabilizer generators:
  IIIIZIIIZI
  ZZIIZIIIZI
  ZZIIZXIIZI
  IIIXZIIIZI
  ZIIIZIIIZI
  ZZXIZXIIZI
  ZZIIZIXIZI
  IIIIIIIIZI
  IIIIIIIIIZ
  
Logical operators:
. Logical Z #1:
  IIIIIIIZZZ
. Logical X #1:
  ZZIIZXIXZI
>>> qecc.distance()

----
Checking errors of length 1...
Checking errors of length 2...
Checking errors of length 3...
Logical error found: Xs - {0, 2, 7} Zs - set()
This is a [[10, 1, 3]] code.
\end{lstlisting}
\end{minipage}

Thus, we have used the \code{VerifyStabilizers} tool to arrive at a simple, distance-three QECC.

\chapter{PECOS Example: \\ Creating a QECC Class\label{app.pecos.qecc}}

\section{Introduction}

To facilitate the evaluation of QECC protocols not included in \PECOS, this appendix shows how to represent a QECC with \pack{Python} so can be used with \PECOS. In particular, we look at representing the repetition code.

To begin, we create an empty Python file called \code{zrepetition.py} and import some useful classes:

\begin{minipage}{0.95\linewidth}
\begin{lstlisting}[style=pystyle,caption={The initial script for the QECC class.}]
"""
A representation of the Z-check repetition code. 
"""
from pecos.circuits import QuantumCircuit
from pecos.qeccs import QECC, LogicalGate, LogicalInstruction
\end{lstlisting}
\end{minipage}

\pagebreak[4]
Subclasses of \code{QECC}, \code{LogicalGate}, and \code{LogicalInstruction} inherit numerous methods and attributes that simplify the creation of new \code{qecc}s. If some of the inherited methods and attributes are not appropriate for a QECC, one can typically override them.

\section{The QECC class}

We now create a class \code{ZReptition} to represent our \code{qecc}:  

\begin{minipage}{0.95\linewidth}
\begin{lstlisting}[style=pystyle,caption={The QECC initialization method.}]
class ZRepetition(QECC):
    def __init__(self, **qecc_params):
    	# Pass qecc_params to the parent class:         
        super().__init__(**qecc_params)
						
        # Set variables that describe the QECC:
        self._set_qecc_description()
						
        # Create a lattice for placing qubits:
        self.layout = self._generate_layout()
				
        # Identify the sides of the QECC:
        self._determine_sides()
				
        # Identify symbols with gate/instruction classes:
        self._set_symbols() 
\end{lstlisting}
\end{minipage}

\noindent Here, the \pdict called \code{qecc\_params} will be used to specify parameters that identify a member of the QECC's family. We will discuss later the method calls see in the \code{\_\_init\_\_} method. 

Next, we write the \code{\_set\_qecc\_description}, which sets class attributes that describe the QECC:

\begin{minipage}{0.95\linewidth}
\begin{lstlisting}[style=pystyle,caption={The method describing the QECC.}]
def _set_qecc_description(self):
    self.name = 'Z Repetition Code'
    # Size of the repetition code:
    self.length = self.qecc_params['length']
    self.distance = 1
    self.num_data_qudits = self.length
    self.num_logical_qudits = 1
    self.num_ancillas = self.num_data_qudits - 1
\end{lstlisting}
\end{minipage}

\noindent  The \code{name} attribute identifies the code. The \code{length} attribute we will use to define how long the QECC is. We use \code{distance} to determine the size of the QECC. We will be describing a repetition code that only has $Z$ checks; therefore, the code will not detect any $Z$ errors. For this reason, the distance is one no matter the length of the QECC. \code{num\_data\_qudits} is the number of data qubits. The attribute \code{num\_logical\_qudits} is the number of logical qubits we will encode with this QECC. The total number of ancillas used in all the \code{qecc}'s procedures is equal to \code{num\_ancillas}. The total number of qubits is equal to the \code{num\_qudits} attribute. This attribute is determined by the parent class \code{QECC}.

Next, we construct \code{\_set\_symbols}, which contain dictionaries that associate symbols to \code{LogicalInstruction}s and \code{LogicalGate}s. We will describe these classes later. 

\begin{minipage}{0.95\linewidth}
\begin{lstlisting}[style=pystyle,caption={Identifying symbols with \code{LogicalGate}s \code{LogicalInstruction}s.}]
def _set_symbols(self):
    self.sym2instruction_class = {
        'instr_syn_extract': InstrSynExtraction,
        'instr_init_zero': InstrInitZero, }
    self.sym2gate_class = {
        'I': GateIdentity,
        'init |0>': GateInitZero, }
\end{lstlisting}
\end{minipage}

Now we write the method \code{\_generate\_layout}, which generates the physical layout of qubits. As we will see later, a physical layout is useful for defining the quantum circuits of the QECC protocol. 

\begin{minipage}{0.95\linewidth}
\begin{lstlisting}[style=pystyle,caption={Creating the physical layout of the QECC.}]
def _generate_layout(self):
    self.lattice_width = self.num_qudits
    data_ids = self._data_id_iter()
    ancilla_ids = self._ancilla_id_iter()
    y = 1
    for x in range(self.lattice_width):    
        if x%2 == 0: # Even (ancilla qubit)
            self._add_node(x, y, data_ids)        
        else: # Odd (data qubit)
            self._add_node(x, y, ancilla_ids)        
    return self.layout
\end{lstlisting}
\end{minipage}

Finally for the \code{qecc}, we will add the method \code{\_determine\_sides} to create a dictionary that defines the physical boundary of the QECC. This information can be used by decoders to understand the geometry of the code.

\begin{minipage}{0.95\linewidth}
\begin{lstlisting}[style=pystyle,caption={Defining the sides of the QECC.}]
def _determine_sides(self):
		# Useful geometric description for decoders
    self.sides = { 
		'length': set(self.data_qudit_set) 
		}
\end{lstlisting}
\end{minipage}

\section{Logical Instruction Classes}

Now that we have created a class to represent the QECC, we will now create classes to represent logical instructions. First create an logical instruction class, called \code{InstrSynExtraction}, that represents one round of syndrome extraction. Similar to the \code{ZRepitition} class, we will subclass our class off of the \code{LogicalInstruction}, which is provided by \PECOS. After we do this, we will write an initialization method that receives as arguments the \code{qecc} instance the instruction belongs to, the associated symbol, and a dictionary of logical gate parameters called \code{gate\_params}. This dictionary will come from the \code{LogicalGate} that contains the \code{LogicalInstruction} and may alter the \code{LogicalGate} and the \QuantumCircuit contained in the \code{LogicalInstruction}.

\vspace{2cm}
\begin{minipage}{0.95\linewidth}
\begin{lstlisting}[style=pystyle,caption={The syndrome extraction instruction.}]
class InstrSynExtraction(LogicalInstruction):
    def __init__(self, qecc, symbol, **gate_params):
        super().__init__(qecc, symbol, **gate_params)
        self.ancilla_x_check = set()
        self.ancilla_z_check = qecc.ancilla_qudit_set        
        self._create_checks()
        self.set_logical_ops()
        self._compile_circuit(self.abstract_circuit) # Call at end
\end{lstlisting}
\end{minipage}

We now include the \code{\_create\_checks} method, which we will use to define the checks of the QECC:

\begin{minipage}{0.95\linewidth}
\begin{lstlisting}[style=pystyle,caption={The method for creating checks.}]
def _create_checks(self):
    self.abstract_circuit = QuantumCircuit(**self.gate_params)
    for qid in self.qecc.ancilla_qudit_set:    
        x, y = qecc.layout[qid]
        
				# Get the data qubits to each side.
        d1 = qecc.position2qudit[(x-1, y)]
        d2 = qecc.position2qudit[(x+1, y)]    
        self.abstract_circuit.append('Z check', {qid, d1, d2}, 
				datas=[d1, d2], ancillas=[qid])
\end{lstlisting}
\end{minipage}

\noindent Here we use the physical layout of the QECC to construct checks. A \QuantumCircuit called \code{abstract\_circuit} is used to register each $Z$-type check, the qubits it acts on, and whether the qubits are used as data or ancilla qubits. Note, check circuits such as the ones seen in Fig~\ref{fig:surf-checks} are used to implement the checks. The order of the data qubits in the \code{datas} keyword indicates the order which the data qubits are acted on by the check circuits. The checks registered by \code{abstract\_circuit} are later compiled into quantum circuits.  

Now we will write the method \code{set\_logical\_ops}, which define the logical operators of the QECCs.        

\begin{minipage}{0.95\linewidth}
\begin{lstlisting}[style=pystyle,caption={Defining the logical operators of the QECC.}]
def set_logical_ops(self):
    data_qubits = set(self.qecc.data_qudit_set)
    logical_ops = [
        {
				    'X': QuantumCircuit([{'X': {0}}]),
            'Z': QuantumCircuit([{'Z': data_qubits}])
				}
		]
    self.initial_logical_ops = logical_ops
    self.final_logical_ops = logical_ops
    
		# The final logical sign and stabilizer
    self.logical_stabilizers = None
    self.logical_signs = None
\end{lstlisting}
\end{minipage}

\noindent Here, the variables \code{initial\_logical\_ops} and \code{final\_logical\_ops} that represent the initial and final logical operators, respectively, are set. Each of these variables are a list where each element represents a collection of logical operators of an encoded qudit. In particular, each element is a dictionary where the keys are symbols identified with the logical operator and the values are {\QuantumCircuit}s representing the unitaries of logical operators. 

If a logical operator encodes a stabilizer state then \code{logical\_stabilizers} is a list of the strings representing the logical operators that stabilizer the state. If the logical operator does not specifically encode a stabilizer state, then \code{logical\_stabilizers} is set to \code{None}. The variable \code{logical\_signs} is a list of signs the corresponding logical operators in \code{logical\_stabilizers}. If the phase of the operators is $+1$, then the element of \code{logical\_signs} is 0. If the phase of the operators is $-1$, then the element of \code{logical\_signs} is 1.  If \code{logical\_stabilizers} is \code{None}, then \code{logical\_signs} is \code{None}. 

We now define the initialization of the logical zero-state:

\begin{minipage}{0.95\linewidth}
\begin{lstlisting}[style=pystyle,caption={The logical zero-state initialization instruction.}]
class InstrInitZero(LogicalInstruction):
    def __init__(self, qecc, symbol, **gate_params):
        super().__init__(qecc, symbol, **gate_params)
        
				# The following are convienent for plotting:
        self.ancilla_x_check = set()
        self.ancilla_z_check = qecc.ancilla_qudit_set
        self._create_checks()
        self.set_logical_ops()
        
				# Must be called at the end of initiation.
        self._compile_circuit(self.abstract_circuit)
\end{lstlisting}
\end{minipage}

\noindent Here, the method \code{\_create\_checks} is used to create check by first making a shallow copy of the \code{abstract\_circuit} of the \code{InstrSynExtraction} class. After doing this we add $\ket{0}$ initialization of the data qubits on the 0th tick.

\pagebreak[4]
The \code{\_create\_checks} method is as follows: 

\begin{minipage}{0.95\linewidth}
\begin{lstlisting}[style=pystyle,caption={A method for creating the checks for the logical initialization instruction.}]
def _create_checks(self):
   
    # Get an instance of the syndrome extraction instruction
    syn_ext = qecc.instruction('instr_syn_extract', **self.gate_params)
		
    # Make a shallow copy of the abstract circuits.
    self.abstract_circuit = syn_ext.abstract_circuit.copy()
		
    # Add initialization of the data qubits
    data_qudits = set(qecc.data_qudit_set)
    self.abstract_circuit.append('init |0>', locations=data_qudits, tick=0)
\end{lstlisting}
\end{minipage}

The \code{set\_logical\_ops} method is similar to the of method of the same name in \code{InstrSynExtraction}. The difference for this class is that a logical zero-state is encoded by the logical operator. Because of this, \code{logical\_stabilizers} is set to \code{['Z']} and \code{logical\_signs} is set to \code{[0]}. 

\vspace{4cm}
\begin{minipage}{0.95\linewidth}
\begin{lstlisting}[style=pystyle,caption={Declaring the logical instructions.}]
def set_logical_ops(self):
    data_qubits = set(self.qecc.data_qudit_set)
    self.initial_logical_ops = [
        {
				   'X': QuantumCircuit([{'X': {0}}]), 
           'Z': QuantumCircuit([{'Z': {0}}])
				}  
		]
    self.final_logical_ops = [
        {
				   'X': QuantumCircuit([{'X': {0}}]), 
           'Z': QuantumCircuit([{'Z': data_qubits}])
				} 
		]
    self.logical_stabilizers = ['Z']
    self.logical_signs = [0]
\end{lstlisting}
\end{minipage}

\section{Logical Gate Classes}

We now construct the \code{LogicalClass} classes. The construction of these classes is relatively simple compared to the create of \code{LogicalInstruction} classes.

\pagebreak[4]
To begin, we write the class representing the logical identity called \code{GateIdentity}:

\begin{minipage}{0.95\linewidth}
\begin{lstlisting}[style=pystyle,caption={The identity gate class.}]
class GateIdentity(LogicalGate):
    def __init__(self, qecc, symbol, **gate_params):
        super().__init__(qecc, symbol, **gate_params)
        self.expected_params(gate_params, {'num_syn_extract', 'error_free', 'random_outcome'})
        self.num_syn_extract = gate_params.get('num_syn_extract', qecc.length)
        self.instr_symbols = ['instr_syn_extract'] * self.num_syn_extract
\end{lstlisting}
\end{minipage}

\noindent Here, the initialization method includes the argument \code{qecc} and the argument \code{symbol}. These are the \code{qecc} instance of the \code{LogicalGate} class and the string used to represent the \code{LogicalGate}, respectively. The initialization method also accepts a keyword arguments, which are stored in the dictionary \code{gate\_params} and may be used to alter the \code{LogicalGate} and associated \code{LogicalInstruction}s. 

The method \code{expected\_params} determines the keyword arguments that are accepted from \code{gate\_params}. The number of syndrome extraction rounds equal to \code{'num\_syn\_extract'}. in the \code{gate\_params} dictionary. Finally, a list of \code{LogicalInstruction} symbols is stored in the variable \code{instr\_symbols}. The \code{instr\_symbols} indicates the order of \code{LogicalInstruction}s that the gate represents. The correspondence between the \code{LogicalInstruction} classes and symbols was established by the \code{sym2instruction\_class} method of the \code{ZRepetition} class. 

\pagebreak[4]
We will also create a \code{LogicalGate} class the represents the initialization of logical zero:

\begin{minipage}{0.95\linewidth}
\begin{lstlisting}[style=pystyle,caption={The logical-zero initialization gate class.}]
class GateInitZero(LogicalGate):
    def __init__(self, qecc, symbol, **gate_params):
        super().__init__(qecc, symbol, **gate_params)
        self.expected_params(gate_params, {'num_syn_extract', 'error_free', 'random_outcome'})
        self.num_syn_extract = gate_params.get('num_syn_extract', 0)
        self.instr_symbols = ['instr_init_zero']
        syn_extract = ['instr_syn_extract'] * self.num_syn_extract
        self.instr_symbols.extend(syn_extract)
\end{lstlisting}
\end{minipage}

\noindent Here, all the methods function the same way as those in the \code{GateIdentity} class.

\section{Example Usage}

Now we will look at a small example of using the \code{ZRepetition} class that we created. We begin by importing the class from the \code{zrepetition.py} script and creating an instance of length three:

\begin{minipage}{0.95\linewidth}
\begin{lstlisting}[style=pystyle,caption={Importing the repetition code example.}]
from zrepetition import ZRepetition
qecc = ZRepetition(length=3)
\end{lstlisting}
\end{minipage}

Now that we have created an instance, we will use the \code{plot} method that is inherited by the syndrome-extraction instruction:

\begin{minipage}{0.95\linewidth}
\begin{lstlisting}[style=pystyle,caption={Running the plotting method for the newly defined QECC.}]
qecc.instruction('instr_syn_extract').plot()
\end{lstlisting}
\end{minipage}

This code results in the plot of the length three repetition code:

\begin{figure}[H]
	\centering
		\includegraphics[width=0.7\textwidth]{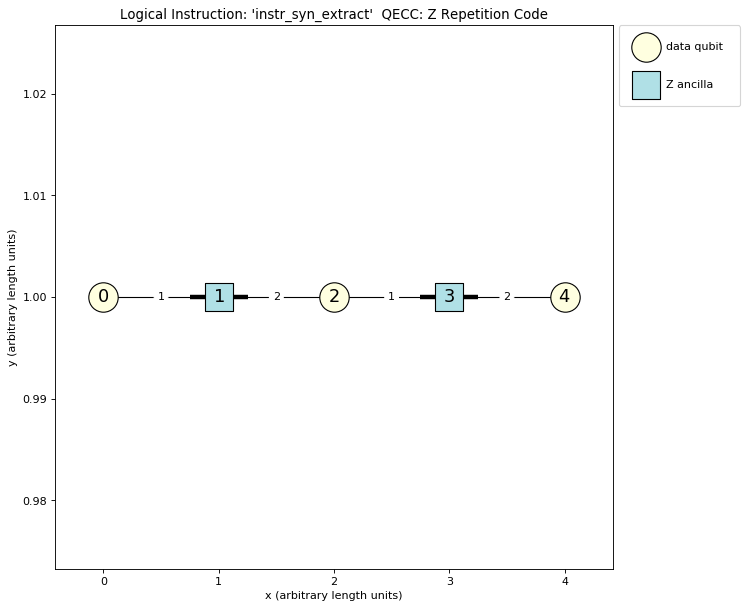}
	\caption{Syndrome-extraction logical-instruction of a length-three repetition-code.}
	\label{fig:app.pecos.qecc_zrep_syn_extract}
\end{figure}

\pagebreak[4]
The \code{ZRepetition} class can be used just like any other \code{qecc} that comes with \PECOS. For example, we can run the following simulation:

\begin{minipage}{0.95\linewidth}
\begin{lstlisting}[style=pystyle,caption={Running a simulation with the repetition code.}]
>>> import pecos as pc
>>> depolar = pc.error_gens.DepolarGen(model_level='code_capacity')
>>> logic = pc.circuits.LogicalCircuit()
>>> logic.append(qecc.gate('ideal init |0>'))
>>> logic.append(qecc.gate('I'))
>>> circ_runner = pc.circuit_runners.Standard(seed=3)
>>> state = circ_runner.init(qecc.num_qudits)
>>> meas, err = circ_runner.run_logic(state, logic, error_gen=depolar, error_params={'p': 0.1})
>>> meas
{(1, 2): {3: {3: 1}}}
>>> err
{(1, 2): {0: {'after': QuantumCircuit([{'X': {4}}])}}}
\end{lstlisting}
\end{minipage}
  
\chapter{PECOS Example: \\Defining Error-models with the GatewiseGen Class\label{app.pecos.errors}}

The \code{GatewiseGen} is an \code{error\_gen} that allows users to design error models where gates can be applied according to classical probability distributions that are specified for individual ideal gates or groups of ideal gates. To being we write the following:

\begin{minipage}{0.95\linewidth}
\begin{lstlisting}[style=pystyle,caption={Creating an instance of the \code{GatewiseGen} class.}]
>>> myerrors = pc.error_gens.GatewiseGen()
\end{lstlisting}
\end{minipage}

To randomly add an $X$ error after every Hadamard we write:

\begin{minipage}{0.95\linewidth}
\begin{lstlisting}[style=pystyle,caption={Adding a random $X$ error for the $H$ gate.}]
>>> # Continuing from last example.
>>> myerrors.set_gate_error('H', 'X')
\end{lstlisting}
\end{minipage}

\pagebreak[4]
\noindent Here, the probability of a $X$ error occurring will, by default, equal to the value of the key \code{'p'} in an \code{error\_params} dictionary that is passed to the \code{run\_logic} method of a \code{circuit\_runner}.

To test the error model we are creating, we can use the \code{get\_gate\_error} method to generate errors. The first argument of the method is the ideal gate-symbol. The second, is a set of qudit locations the errors may occur on. The third, is a \code{error\_params} dictionary used to specify the probability of errors. An example of using this method is seen here:

\begin{minipage}{0.95\linewidth}
\begin{lstlisting}[style=pystyle,caption={Adding a random $X$ error for the $H$ gate.}]
>>> # Continuing from last example.
>>> myerrors.get_gate_error('H', {0, 1, 2, 3, 4}, error_params={'p':0.5})
(QuantumCircuit([{'X': {0, 1, 3}}]), QuantumCircuit([]), set())
\end{lstlisting}
\end{minipage}

\noindent Here, the method returns a tuple. The first element is the error circuit that is applied after the ideal gates. The second, before the ideal gates. The third element is the set of qudit locations corresponding to gate locations of ideal gates to be removed from the ideal quantum-circuit.

Note, by default errors specified by the \code{set\_gate\_error} method will be generated after the ideal quantum-gates. To generate errors before the gates, one can set the keyword \code{after} to \code{False} when using the \code{set\_gate\_error} method. 

\pagebreak[4]
The probability-parameter used (default being \code{'p'}) can be changed by using the keyword \code{error\_param}:

\begin{minipage}{0.95\linewidth}
\begin{lstlisting}[style=pystyle,caption={Specifying the variable in \code{error\_param} that defines the probability of error for a gate.}]
>>> # Continuing from last example.
>>> myerrors.set_gate_error('H', 'X', error_param='q', after=False)
>>> myerrors.get_gate_error('H', {0, 1, 2, 3, 4}, error_params={'q':0.5})
(QuantumCircuit([]), QuantumCircuit([{'X': {0, 3}}]), set())
\end{lstlisting}
\end{minipage}

\noindent Here we used the keyword \code{error\_param} to declare that \code{'q'} will be used to set the probability of an $X$ error occurring. We also see an example of the keyword \code{after} being used to indicate that errors should be applied before the ideal gates rather than after.

Besides specifying errors of a single gate-type, we can declare a set of errors to be uniformly drawn from:

\begin{minipage}{0.95\linewidth}
\begin{lstlisting}[style=pystyle,caption={Defining a gate error that is that draws from a uniform distribution of errors.}]
>>> # Continuing from last example.
>>> myerrors.set_gate_error('X', {'X', 'Y', 'S'}, error_param='r')
>>> myerrors.get_gate_error('X', {0, 1, 2, 3, 4}, error_params={'r':0.5})
(QuantumCircuit([{'S': {3, 4}}]), QuantumCircuit([]), set())
\end{lstlisting}
\end{minipage}

\pagebreak[4]
Such uniform error-distributions can be made for two-qubit gates as well:

\begin{minipage}{0.95\linewidth}
\begin{lstlisting}[style=pystyle,caption={An example of uniform distributions for two-qubit gates.}]
>>> # Continuing from last example.
>>> myerrors.set_gate_error('CNOT', {('I', 'X'), ('X', 'X'), 'CNOT'}, error_param='r')
>>> myerrors.get_gate_error('CNOT', {(0,1), (2,3), (4,5), (6,7), (8,9)}, error_params={'r':0.8})
(QuantumCircuit([{'CNOT': {(0, 1), (4, 5), (8, 9)}, 'X': {3, 6, 7}, 'I': {2}}]), QuantumCircuit([]), set())
\end{lstlisting}
\end{minipage}

\noindent Here we see that two-qubit gates or tuples of single-qubit gates can be supplied as errors.

Other distributions besides the uniform distribution can be specified by passing a callable, such as a function or a method. An example is seen in the following:

\begin{minipage}{0.95\linewidth}
\begin{lstlisting}[style=pystyle]
>>> # Continuing from last example.
>>> import random
>>> def error_func(after, before, replace, location, error_params):
...     s = error_params['s']
...     rand = random.triangular(0, 1, 0.6)
...     if rand < 0.6:
...         err = 'Q'
...		elif rand < 0.7:
...		    err = 'S'
...     else:
...         err ='R'
...     before.update(err, {location}, emptyappend=True)
\end{lstlisting}
\end{minipage}

\noindent Here, callables that are used to create unique error distributions must take the arguments \code{after}, \code{before}, \code{replace}, \code{location}, and \code{error\_params}. The variables \code{after} and \code{before} are {\QuantumCircuit}s representing the errors that are applied after and before the ideals gates of a single tick, respectively. The variable \code{replace} is the set of qubit gate-locations of the ideals gates to be removed from the ideal quantum-circuit. These callables are called only if error occurs according to the probability of an associated error parameter, which we will see later how to set. The \code{location} variable is the qudit index or tuple of qudit indices where the error has occurred. The variable \code{error\_params} is the dictionary of error parameters that are being used to determine the probability distribution of errors. In the above callable, we see a triangular distribution being used to apply quantum errors. Note that the callable is responsible for updating {\QuantumCircuit}s  \code{after}, \code{before}, \code{replace} as appropriate. 

To use callables to generate errors, we can call the \code{set\_gate\_error} method in the following manner:  

\begin{minipage}{0.95\linewidth}
\begin{lstlisting}[style=pystyle]
>>> # Continuing from last example.
>>> myerrors.set_gate_error('Y', error_func, error_param='s')
>>> myerrors.get_gate_error('Y', {0, 1, 2, 3, 4}, error_params={'s':0.5})
(QuantumCircuit([]), QuantumCircuit([{'R': {0, 4}, 'Q': {1, 2}}]), set())
\end{lstlisting}
\end{minipage}

\noindent Here we set the probability of \code{error\_func} being called to generate errors using the \code{error\_param} keyword argument.

There are two special gate-symbols for which error distributions can be assigned to. These special symbols are \code{'data'} and \code{'idle'}. The error distribution associated with \code{'data'} is used to generate errors at the beginning of each \code{LogicalInstruction} for each data qudit. An error distribution associated with the \code{'idle'} symbol is used to generate errors whenever a qubit is not acted on by a quantum operation during a \code{LogicalCircuit}. 

An example of setting the errors of a \code{'data'} and \code{'idle'} can see here:

\begin{minipage}{0.95\linewidth}
\begin{lstlisting}[style=pystyle, caption=={Setting idle and data errors}]
>>> # Continuing from last example.
>>> myerrors.set_gate_error('data', 'X', error_param='q')
>>> myerrors.set_gate_error('idle', 'Y', error_param='s')
\end{lstlisting}
\end{minipage}

Besides specifying errors for individual gate-types, one can specify errors for a group of gates. To do this one may define a gate group and set the error distribution for this group:

\begin{minipage}{0.95\linewidth}
\begin{lstlisting}[style=pystyle, caption={Setting a gate group and a gate group's error.}]
>>> # Continuing from last example.
>>> myerrors.set_gate_group('measurements', {'measure X', 'measure Y', 'measure Z'})
>>> myerrors.set_group_error('measurements', {'X', 'Y', 'Z'}, error_param='m')
\end{lstlisting}
\end{minipage}

Note, \code{set\_group\_error} will override the error distribution of any gate belonging to the gate group.

\vspace{6cm}
\pagebreak[4]
The gate groups that are defined by default can be found by running:

\begin{minipage}{0.95\linewidth}
\begin{lstlisting}[style=pystyle,caption={Default gate groups.}]
>>> newerrors = pc.error_gens.GatewiseGen()
>>> newerrors.gate_groups
{'measurements': {'measure X', 'measure Y', 'measure Z'},
 'inits': {'init |+>', 'init |+i>', 'init |->', 'init |-i>', 'init |0>', 'init |1>'},
 'two_qubits': {'CNOT', 'CZ', 'G', 'SWAP'},
 'one_qubits': {'F1', 'F1d', 'F2', 'F2d', 'F3', 'F3d', 'F4', 'F4d', 'H', 'H+y-z', 'H+z+x', 'H-x+y', 'H-x-y', 'H-y-z', 'H-z-x', 'H1', 'H2', 'H3', 'H4', 'H5', 'H6', 'I', 'Q', 'Qd', 'R', 'Rd', 'S', 'Sd', 'X', 'Y', 'Z'}}
\end{lstlisting}
\end{minipage}

\noindent Here the keys are symbols representing the gate groups and the values are the set of gate symbols belong to the corresponding gate group. These gate groups (\code{'measurements'}, \code{'inits'}, \code{'two\_qubits'}, and \code{'one\_qubits'}) can be redefined by the user.

\vspace{6cm}
\pagebreak[4]
\section{Example: The Symmetric Depolarizing-channel}

As an example, the circuit-level symmetric depolarizing-channel, which was discussed in Section~\ref{sec:intro:errors} and is modeled by \code{DepolarGen} as discussed in Section~\ref{pecos.errgen.depolar}, can be represented by the \code{GatewiseGen} class as follows: 

\begin{minipage}{0.95\linewidth}
\begin{lstlisting}[style=pystyle,caption={Example of implementing a circuit-level depolarizing error-channel with perpendicular initialization and measurement errors.}]
depolar_circuit = pc.error_gens.GatewiseGen()
set_gate_group('Xinit', {'init |+>', 'init |->'})
set_gate_group('Yinit', {'init |+i>', 'init |-i>'})
set_gate_group('Zinit', {'init |0>', 'init |1>'})
depolar_circuit.set_group_error('Xinit', 'Z')
depolar_circuit.set_group_error('Yinit', 'Z')
depolar_circuit.set_group_error('Zinit', 'X')
depolar_circuit.set_gate_error('measure X', 'Z', after=False)
depolar_circuit.set_gate_error('measure Y', 'Z', after=False)
depolar_circuit.set_gate_error('measure Z', 'X', after=False)
depolar_circuit.set_group_error('one_qubits', {'X', 'Y', 'Z'})
depolar_circuit.set_group_error('two_qubits', 
{('I', 'X'), ('I', 'Y'), ('I', 'Z'), 
('X', 'I'), ('X', 'X'), ('X', 'Y'), ('X', 'Z'), 
('Y', 'I'), ('Y', 'X'), ('Y', 'Y'), ('Y', 'Z'), 
('Z', 'I'), ('Z', 'X'), ('I', 'Y'), ('Z', 'Z')})                     
\end{lstlisting}
\end{minipage}

\pagebreak[4]
\section{Example: The Amplitude-damping Channel}

The stochastic circuit-level amplitude-damping channel, which was discussed in Section~\ref{sec:intro:errors}, can be described as: 

\begin{minipage}{0.95\linewidth}
\begin{lstlisting}[style=pystyle,caption={Example of describing a circuit-level amplitude-damping channel.}]
amp_damp = pc.error_gens.GatewiseGen()
amp_damp.set_group_error('inits', 'init |0>')
amp_damp.set_gate_error('measurements', 'init |0>', after=False)
amp_damp.set_group_error('one_qubits', 'init |0>')
amp_damp.set_group_error('two_qubits', {('I', 'init |0>'), ('init |0>', 'I'), ('init |0>', 'init |0>')})                     
\end{lstlisting}
\end{minipage}
 
\chapter{PECOS Example: \\Monte Carlo Script for Finding Logical Error-rates\label{app.pecos.monte}}

In this appendix, I present how \PECOS can be used to create a script to run a Monte Carlo simulation that determines logical error-rates as a function of physical error-rates for a fixed distance of a medial surface-code patch. The break-even point where the physical error-rate equals the logical error-rate is known as the \textit{pseudo-threshold}. The \textit{threshold} is the value the pseudo-threshold converges to as the distance of the code approaches infinity.

We begin by creating a Python script \code{error\_rates.py} and importing \pack{NumPy} and \PECOS:

\begin{minipage}{0.95\linewidth}
\begin{lstlisting}[style=pystyle,caption={The beginnings of a Monte Carlo script.}]
import numpy as np
import pecos as pc
\end{lstlisting}
\end{minipage}

\pagebreak[4]
For this example, we will evaluate the identity gate of \code{SurfaceMedial4444} and start in the ideal logical zero-state:

\begin{minipage}{0.95\linewidth}
\begin{lstlisting}[style=pystyle,caption={Starting initializations for the Monte Carlo simulation.}]
surface = pc.qeccs.SurfaceMedial4444(distance=3)
logic = pc.circuits.LogicalCircuit(layout=surface.layout)
logic.append(surface.gate('ideal init |0>'))
logic.append(surface.gate('I', num_syn_extract=1))
circ_runner = pc.circuit_runners.Standard(seed=0)
logical_ops = surface.instruction('instr_syn_extract').final_logical_ops[0]
\end{lstlisting}
\end{minipage}

\noindent Here we also initialize the \code{circuit\_runner} we will use and create the variable \code{logical\_ops}, which stores the logical operations of the QECC. This can be used to determine the logical error-rate since we can track whether errors flip the signs of the logical operators. 

Now we choose the depolarizing channel as our noise model (see Section \ref{sec.pecos.error_gens}) and the MWPM decoder (see Section~\ref{sec.pecos.decode}) to interpret syndromes and determine recovery operations:

\begin{minipage}{0.95\linewidth}
\begin{lstlisting}[style=pystyle,caption={Choosing the error model and decoder.}]
depolar = pc.error_gens.DepolarGen(model_level='code_capacity')
decode = pc.decoders.MWPM2D(surface).decode
\end{lstlisting}
\end{minipage}

We next create the function  \code{determine\_fails} to decide if logical error occurs by examining whether, after applying a recovery operation to the state, errors have flipped logical $Z$. Note, since we are just protecting a logical zero-state we are only concerned with errors that flip the sign of the logical $Z$ operator. $Z$ errors do not affect the state.

\pagebreak[4]
The \code{determine\_fails} function is:
 
\begin{minipage}{0.95\linewidth}
\begin{lstlisting}[style=pystyle,caption={Function to determine if a failure has occured.}]
def determine_fails(meas, decoder, circ_runner, state, logical_ops, fails):
    if meas:
        recovery = decoder(meas)
        circ_runner.run_circuit(state, recovery)
				
    sign = state.logical_sign(logical_ops['Z'], logical_ops['X'])
    fails += sign    
    return fails
\end{lstlisting}
\end{minipage} 

We are now almost ready to define the Monte Carlo loop. First, however, we set \code{runs} to represent the number of evaluations we will make per physical error-rate. Next, we add the variable \code{ps}, which is set to an array of  10 linearly space points between 0.1 and 0.3 to serve as the physical error-rates that we will evaluate. This array is created by NumPy's \code{linspace} function. Finally, we include the variable \code{plog}, which stores the logical error-rates we find corresponding to the physical error-rates in \code{ps}. All of this is done in the following lines:

\begin{minipage}{0.95\linewidth}
\begin{lstlisting}[style=pystyle,caption={Some variables used in the Monte Carlo script.}]
runs = 10000
ps = np.linspace(0.1, 0.4, 10)
plog = []
\end{lstlisting}
\end{minipage}

We now create the Monte Carlo loop, which prepares a fresh initial state, applies depolarizing noise with a probability chosen by looping over \code{ps}, and counts the number of failures (logical flips) to determine the logical error-rate, which is stored in \code{plog}:

\begin{minipage}{0.95\linewidth}
\begin{lstlisting}[style=pystyle,caption={The Monte Carlo loop.}]
for p in ps:
    fails = 0
    for i in range(runs):
        state = circ_runner.init(surface.num_qudits)
        meas, _ = circ_runner.run_logic(state, logic, error_gen=depolar, error_params={'p': p})
        fails = determine_fails(meas, decoder, circ_runner, state, logical_ops, fails)
    plog.append(fails / runs)
print('ps=', list(ps))
print('plog=', plog)
\end{lstlisting}
\end{minipage}

When this script is run, an example output is:

\begin{minipage}{0.95\linewidth}
\begin{lstlisting}[style=pystyle,label={append:pecos:monte:data},caption={Output from a Monte Carlo simulation.}]
ps= [0.1, 0.13333333333333336, 0.16666666666666669, 0.2, 0.23333333333333336, 0.2666666666666667, 0.30000000000000004, 0.33333333333333337, 0.3666666666666667, 0.4]
plog= [0.0588, 0.102, 0.1497, 0.1835, 0.2241, 0.2702, 0.3052, 0.3485, 0.3783, 0.4017]
\end{lstlisting}
\end{minipage}

One can then use plotting packages such as \pack{Matplotlib} to produce plots as appropriate for the data. \PECOS provides a tool for quickly plotting and evaluating logical vs physical error-rates:

\begin{minipage}{0.95\linewidth}
\begin{lstlisting}[style=pystyle,caption={Quick plotting tool.}]
from pecos.tools import plot_pseudo
plot_pseudo(deg=2,plist=ps,plog=plog)
\end{lstlisting}
\end{minipage}

\pagebreak[4]
Running this tool for the data given in Code Block~\ref{append:pecos:monte:data} results in the plot:

\begin{figure}[H]
	\centering
		\includegraphics[width=0.75\textwidth]{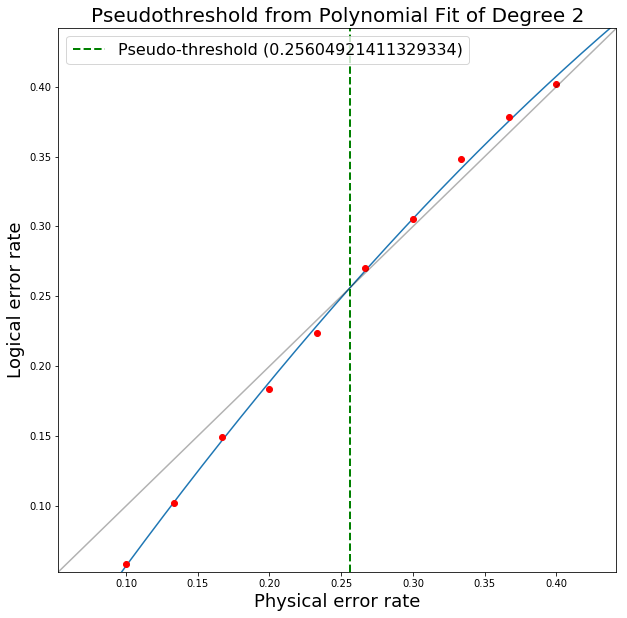}
	\caption{Code-capacity pseudo-threshold of a distance-three, medial surface-code.}
	\label{fig:pecos.tool_nonmedial_pseudothres}
\end{figure}

The script described in this appendix can be used as a basis for developing other Monte Carlo simulation scripts for evaluating pseudo-thresholds or thresholds. 

\printbibliography
\addcontentsline{toc}{chapter}{Bibliography}

\end{document}